%% file: 02_SP_Revision_03.tex
\documentclass[namedreferences]{SolarPhysics}
 \pdfoutput=1					% For ArXiv 
\usepackage{totcount}			% ADDITIONAL PACKAGE Counters (Needs KEYVAL)
\newcounter{magicrownumbers}
\regtotcounter{magicrownumbers} % In order to Use the ToTCount
\newcommand\rownumber{\stepcounter{magicrownumbers}\arabic{magicrownumbers}}
\usepackage[optionalrh]{spr-sola-addons} % For Solar Physics
\usepackage{rotating}
\usepackage{caption}
\usepackage{longtable}
\usepackage{lscape}		       	% ADDITIONAL PACKAGE FOR LANDSCAPE ENVIRONMENT

\usepackage{graphicx}   		% For eps figures, newer & more powerfull
\usepackage{array}				% For Rugged Alignment
\usepackage{courier}   			% Change the \texttt command to courier style
\usepackage{natbib}        	%  for SP
\usepackage{amssymb}			% useful mathematical symbols
\usepackage{color}       		% For color text: \color command
\usepackage{url}               	% For breaking URLs easily through lines
\newcommand{\etal}{et al.}
\newcommand{\be}{\begin{equation}}
\newcommand{\ee}{\end{equation}}
\newcommand{\beq}{\begin{eqnarray}}
\newcommand{\eeq}{\end{eqnarray}}
\newcommand{\RSUN}{$R_{\odot}$}

\newcommand{\bl}{\color{blue}} %   
\renewcommand{\bl}{} %   
\newcommand{\event}		{--}
\newcommand{\FIER}		{{\it first impulsive energy release}}
\newcommand{\adv}		{{\it Adv. Space Res.}}

\newcommand{\aap}		{{\it Astron. Astrophys.}}
\newcommand{\aaps}		{{\it Astron. Astrophys. Suppl.}}
\newcommand{\aapr}		{{\it Astron. Astrophys. Rev.}}

\newcommand{\apj}		{{\it Astrophys. J.}}

\newcommand{\cjaa}		{{\it Chinese J. Astron. Astrophys.}}

\newcommand{\jgr}		{{\it J. Geophys. Res.}}

\newcommand{\memsai}	{{\it Mem.S.A.It.}}

\newcommand{\solphys}	{{\it Solar Phys.}}

\newcommand{\ssr}		{{\it Space Sci. Rev.}}
\graphicspath{ {./images/} {./images_B/}}
\begin{document}
\begin{article}
\begin{opening}
\title{Fine Structure of Metric Type-IV Radio Bursts Observed with the ARTEMIS-IV Radio--Spectrograph: Association with Flares and Coronal Mass Ejections}
%-------------------------------------------------
%% Authors Names
%
\author{C.~\surname{Bouratzis}$^{1}$, A.~\surname{Hillaris}$^{1}$, C.~E.~\surname{Alissandrakis}$^{2}$,
        P.~\surname{Preka-Papadema}$^{1}$, X.~\surname{Moussas}$^{1}$, C.~\surname{Caroubalos}$^{2}$,
        P.~\surname{Tsitsipis}$^{3}$, A.~\surname{Kontogeorgos}$^{3}$}
%-------------------------------------------------
% Runningheads
\runningauthor{Bouratzis \etal}
\runningtitle{Association of Metric Type-IV Fine Structures with Flares and CMEs}
%-------------------------------------------------
%% Affilations
\institute{$^{1}$ University of Athens,GR-15784 Athens, Greece\\
           $^{2}$ University of Ioannina, GR-45110 Ioannina, Greece\\
           $^{3}$ Technological Educational Institute of Lamia, 35100 Lamia\\}
%-------------------------------------------------
\begin{abstract}
{Fine structures embedded in type-IV burst continua may be used as diagnostics of the magnetic field restructuring and the corresponding energy release associated with the low corona development of flare/CME events.} % context
 {A catalog of 36 type-IV bursts observed with the SAO receiver of the ARTEMIS-IV solar radio-spectrograph in the 450--270 MHz range at high cadence (0.01 sec)  was compiled; the fine structures were classified into five basic classes with two or more sub-classes each. The time of fine structure emission was compared with the injection of energetic electrons as evidenced by HXR and microwave emission, the SXR light-curves and the CME onset time.~}% methods
{Our results indicate a very good temporal association between energy release episodes and pulsations, spikes, narrow-band bursts of the type-III family and zebra bursts. Of the remaining categories, the featureless broadband continuum starts near the time of the first energy release, between the CME onset and the SXR peak, but extends for several tens of minutes after that, covering almost the full extent of the flare--CME event. The intermediate drift bursts, fibers in their majority,  mostly follow the first energy release but have a wider distribution, compared to other fine structures. }
%-------------------------------------------------
\end{abstract}
%-------------------------------------------------
%% Keywords
\end{opening}
%-------------------------------------------------
\section{Introduction}\label{Intro}

Solar metric radio bursts provide a unique diagnostic of the development of flare/CME events in the low corona; their onset and evolution coincides with an extended opening of the magnetic field, accompanied by energetic-particle acceleration and injection into interplanetary space as well as shocks \citep[e.g. review by~][]{Pick08}. Their signatures at metric--decimetric and longer waves trace disturbances propagating from the low corona to interplanetary space.

The complexity of the above mentioned processes is reflected in a diversity of forms in dynamic spectra, which exhibit a variety of fine structures in time and frequency; these are characterized by a wide range in period, bandwidth, amplitude, temporal and spatial signatures. The fine structures may be used for the detailed study of the magnetic field restructuring and the corresponding energy release associated with solar flare/CME events \citep[e.g. reviews by][]{Benz03,Nindos07}. A number of morphological taxonomy schemes, mostly in the microwaves and the decimetric frequency range, have been presented {\bl \citep{Bernold80, Slottje1981, Guedel88, Allaart90, Isliker94, Jiricka01, Fu04}} of which the most recent are also the most comprehensive. 

In this work we examine fine structures observed during type-IV solar radio events observed with the  ARTEMIS-IV solar radio-spectrograph from the beginning of 1999 until the end of 2005; to these we added two well observed events with rich fine structure recorded in 2010. Some of these events were first catalogued in \citet{Caroubalos04}. Our study is concentrated on a statistical analysis of the fine structures and  their association with the various phases of the flare/CME phenomenon, which could be useful in understanding details of the evolution of the solar energetic phenomena through their radio signatures. In developing our classification scheme we {\bl have built upon} the Ondrejov catalogue  \citep{Jiricka01,Jiricka02,Meszarosova05} which was based on data in the 0.8--2.0 GHz range.

In Section \ref{Obs} we discuss the instrumentation and the data selection. The results of our morphological analysis and classification are given in Section \ref{analysis}, the relative timing of fine structures with respect to the flare evolution is discussed in Section \ref{timing} and the conclusions are presented in Section \ref{discussion}.

\section{Observations and Data Selection} \label{Obs}

The basic data used in this study are the high and medium resolution dynamic spectra recorded by the ARTEMIS\footnote{Appareil de Routine pour le Traitement et l'Enregistrement Magnetique de l' Information Spectral}--IV solar radio-spectrograph at Thermopylae \citep{Caroubalos01, Caroubalos06, Kontogeorgos}. It consists of a 7\,m parabolic antenna covering the metric range; to this a dipole aerial adapted to the decametric range was added in October 2002. Two receivers operate in parallel, a sweep frequency analyzer (ASG) covering the 650-20 MHz range in 630 channels with a cadence of 10~samples/sec and a high sensitivity multi-channel acousto-optical analyzer (SAO), which covers the 270-450 MHz range in 128 channels with a high time resolution of 100~samples/sec. The narrow band, high time resolution SAO recordings are used in the analysis of the fine temporal and spectral structures and they constitute the major data set of this work. The broad band, medium time resolution data of the ASG, on the other hand, are used for the detection and analysis of radio emission from the base of the corona to $\sim2$ \RSUN.

For the study of the association of the fine structures with the flare evolution,  we used:
\begin{itemize}
\item{CME data from the LASCO coronograph \citep{Brueckner95} on line\footnote{http://cdaw.gsfc.nasa.gov/CMElist}  event list \citep{Yashiro04,Gopalswamy09}; the  CME onset times used in this study were estimated from the LASCO movies using the linear regression by \citet{Yashiro01} and are included in the on--line LASCO event list. We have supplemented this data set with information from the CACTUS\footnote{http://sidc.oma.be/cactus/} CME Catalogue \citep{Robbrecht2004,Robbrecht2009}}

\item{The NOAA \textit{Solar Geophysical Data} catalogues  and  Soft X--Ray (SXR) on line\footnote{http//www.sel.noaa.gov/ftpmenu/indices} light curves from GOES. The SXR observations provide a fairly accurate estimate of the start of solar flares and a less accurate one of their end time; the corresponding source locations are also included in the NOAA catalogue and are used in this work, in addition to the EIT images, for the establishment of the spatial association of the SXR flare--CME--radio emission (see below)}

\item{Hard X--ray (HXR) light curves were obtained from the RHESSI \citep{Lin01B,Lin02} archive for the events after the beginning of 2003. For the events prior to 2003 we have used data from the MTI/HXRS \citep{Farnik2001} and BATSE/GRP \citep{Fishman1982,Fishman1984} experiments.}

\item Microwave data from  the \textit{Radio Solar Telescope Network} \citep[RSTN,~][] {Guidice81} at 4.995 GHz; in a few events the 2.695 GHz channel of the \textit{Trieste Solar Radio System} \citep[TSRS~][]{Messerotti2001} was used instead. 

\item Two-dimensional images of the Sun at five frequencies (164, 236.6, 327, 410.5, and 432 MHz) from the \textit{Nan\c cay Radio Heliograph} (NRH) \citep{Kerdraon97}. All five frequencies are within the spectral range of the ASG, while the last three are also within the range of the SAO; they hence, supplement the dynamic spectra with positional information on the radio emission. A detailed investigation of the positions of fine structures with respect to the bulk of the type-IV emission has not been attempted in this work, but will be the subject of subsequent publications.

\item Images from the \textit{Extreme Ultraviolet Imaging Telescope} (EIT) on-board SOHO \citep{Delaboudiniere95}; they were used in order to provide information on the position of the associated flare.
\end{itemize}

%-------------------------------------------------
\begin{figure}
\begin{center}% trim=0cm 1cm  0cm 1cm,clip,
\includegraphics[width=\textwidth]{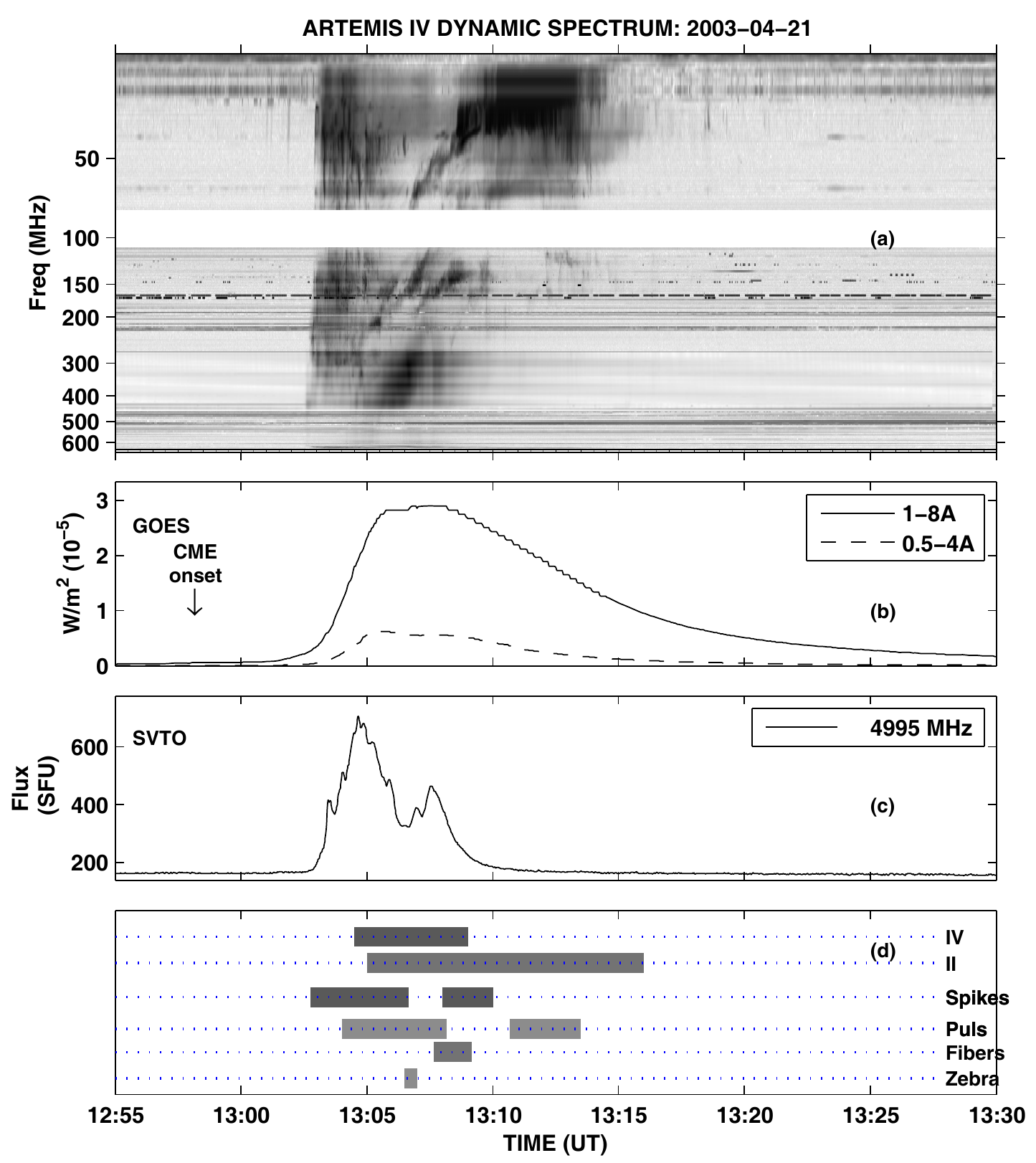}
\caption{Example of Type-IV and fine structure--Flare--CME  temporal relationship from combined data for the 21 April 2003 Event. (a): ARTEMIS-IV ASG dynamic spectrum. (b): The GOES SXR Flux; the CME onset time, from the LASCO lists is marked with arrow. (c): Microwave (4995 MHz) flux from the RSTN (SVTO). (d): Time range of the type-IV, II, spikes, fibers, pulsating and zebra structures.}
\label{Timeline}
\end{center}
\end{figure}

%-------------------------------------------------
\begin{figure}
\begin{center}% trim=0cm 1cm  0cm 1cm,clip,
\includegraphics[trim=0.8cm 0.0cm  0cm 0cm,clip,width=0.95\textwidth]{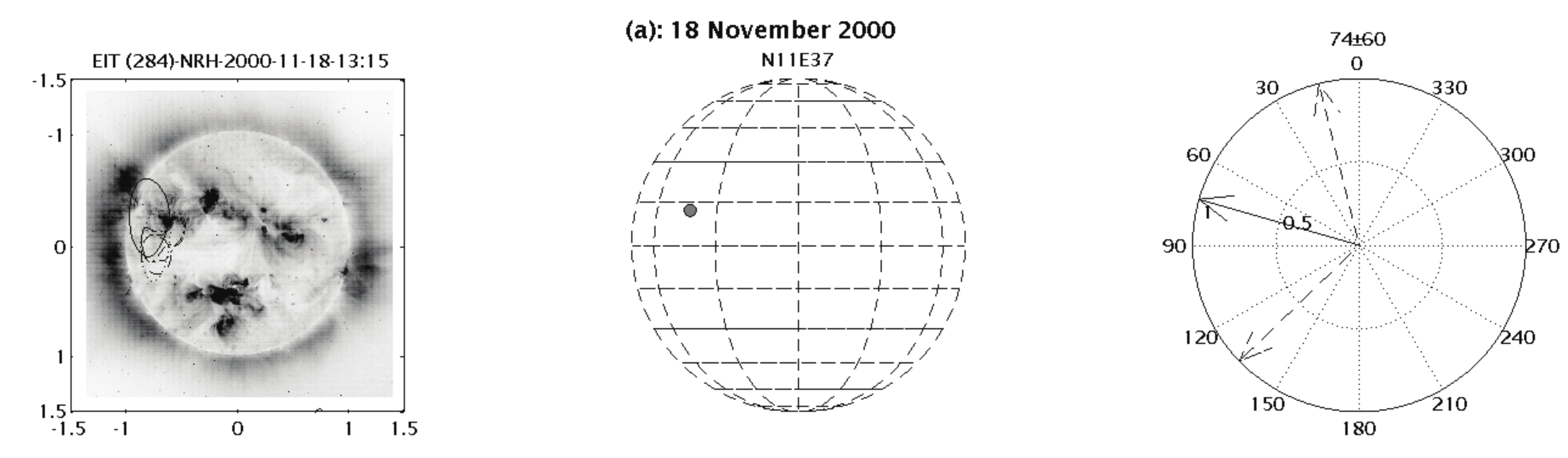}
\includegraphics[trim=0.8cm 0.0cm  0cm 0cm,clip,width=0.95\textwidth]{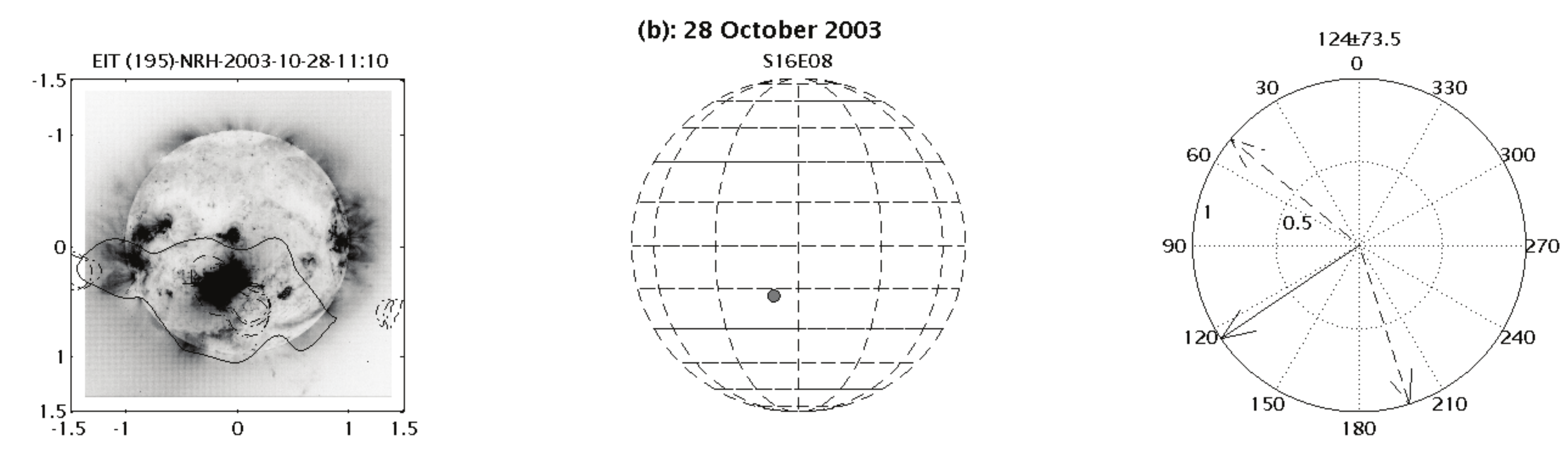}
\includegraphics[trim=0.8cm 0.0cm  0cm 0cm,clip,width=0.95\textwidth]{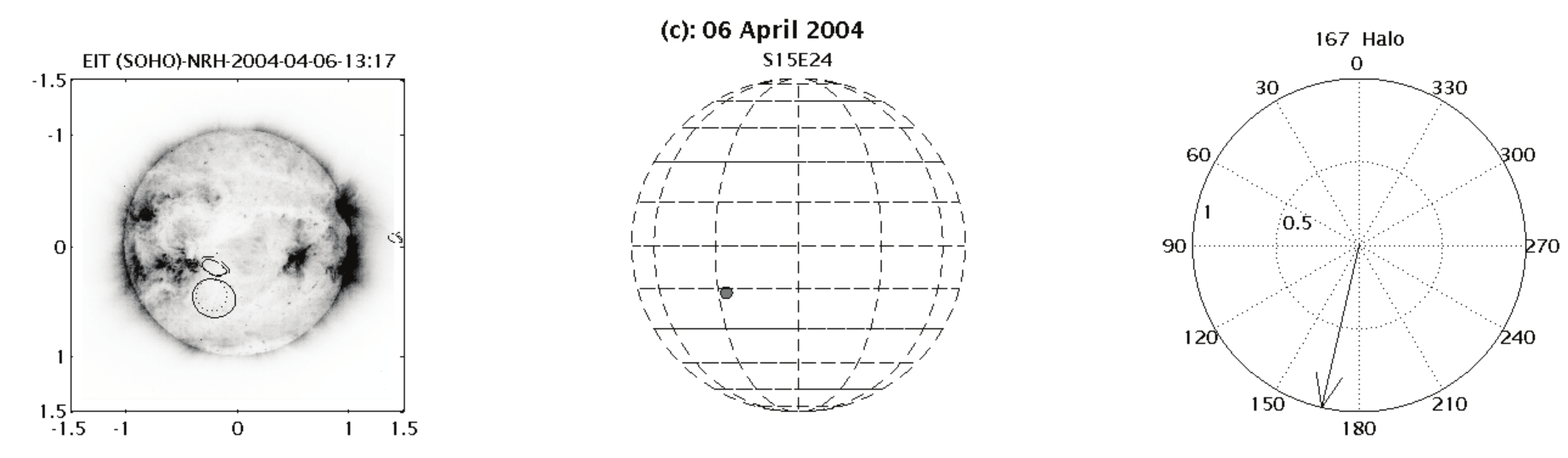}
\includegraphics[trim=0.8cm 0.0cm  0cm 0cm,clip,width=0.95\textwidth]{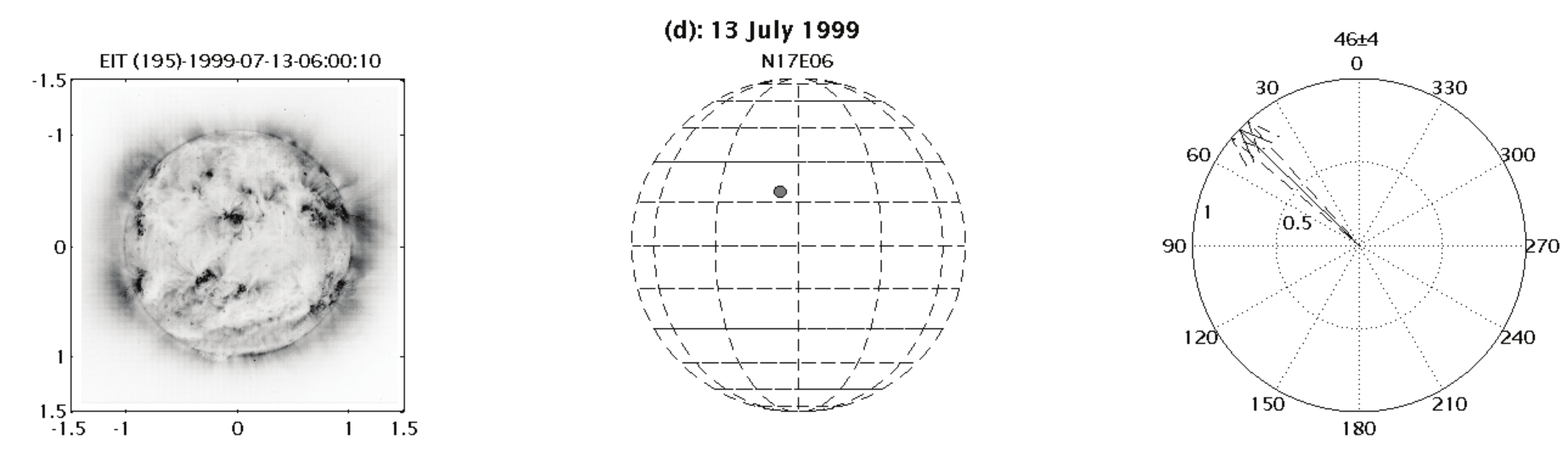}
\caption{Examples of type-IV--flare--CME  spatial association from combined data. In the left column the NRH half power contours are overlaid on EIT images. The middle and right columns show the flare position from the NOAA/SGD catalogues and the direction of the CME launch (except for halo); this is marked graphically by the CME position angle and the angular width from the LASCO lists. The events of 18 November 2000 (a) and 28 October 2003 (b) for which excellent spatial association was established among all data sets. (c):  The event of 06 April  2004 accompanied by a halo CME exhibits a good association with the active region and the type-IV position (see text for details). (d): The event of 13 July 1999 at 06:00 UT, outside the NRH observation window.}
\label{FlareCME}
\end{center}
\end{figure}
%-------------------------------------------------

The soft X-ray light curves provided the overall time evolution of the flares; the microwave and hard X-ray light curves, which are very good signatures of electron acceleration, were used to estimate the time of the first major episode of energy release. This time was used as a reference for the relative timing of fine structures. For the association of the type-IV radio events with flare emissions and CMEs we used spatial and temporal criteria as follows \citep[see also~][~where similar criteria were employed]{Caroubalos04}.:

\begin{itemize}
\item For the type-IV--flare association: for the temporal association we required the overlap, at least in part, of the total duration of the flare with the total duration of the radio emission (see Figure \ref{Timeline} for an example).  For the spatial association we used positional data from the Nan\c cay Radio Heliograph images and movies; we examined coincidence with flares using their position recorded in the NOAA \textit{Solar Geophysical Data} catalogues. If both criteria were satisfied we classified the association as excellent; if we could not establish a spatial association due to lack of NRH positional data, the association was characterized good. 

\item {Once relationship between type IV and SXR flare was established, we determined the time of the first peak of the HXR and/or the microwave bursts, which we treated as the signature of the {\it first impulsive energy release}. This was used as a reference for comparing the relative timing of the fine structure.} 

\item{For the CME--flare  association: from the time--height diagrams in the CME lists we defined a time window of 60 minutes between the CME onset time and the peak of the accompanying SXR flare; we use the flare peak rather than the onset, since the former is more easily identifiable. To establish spatial association, we required that the flare and the CME originated in the same quadrant as schematically depicted in Figure \ref{FlareCME}. For this we compared the flare location with the CME position angle, which refers to the fastest moving segment of the CME leading edge, and the angular width; the latter was only used for non-halo CMEs. }

\end{itemize}

For twelve events the spatial criterion failed to lead to the acceptance or rejection of association, due to either the lack of positional data for the radio burst or the flare or to the appearance of more than one accompanying flares as candidates for association. These events were not eliminated from the data set, but were only used in deriving the basic statistics of the fine structure parameters.

The basic characteristics of the type-IV continuum, the fine structure, the accompanying type-II and type-III bursts and the associated flare and CME of the 36 selected events are presented in Table~\ref{Table02}. Their dynamic spectra, light curves and the relative timing of fine structures are given in Appendix B, while four examples of the CME--flare--type-IV event spatial association, of varying quality, are presented in figure \ref{FlareCME}. 

%-------------------------------------------------
\begin{figure}
\begin{center}% trim=0cm 1cm  0cm 1cm,clip,
\includegraphics[trim=0cm 0.0cm  0cm 0cm,clip,width=\textwidth]{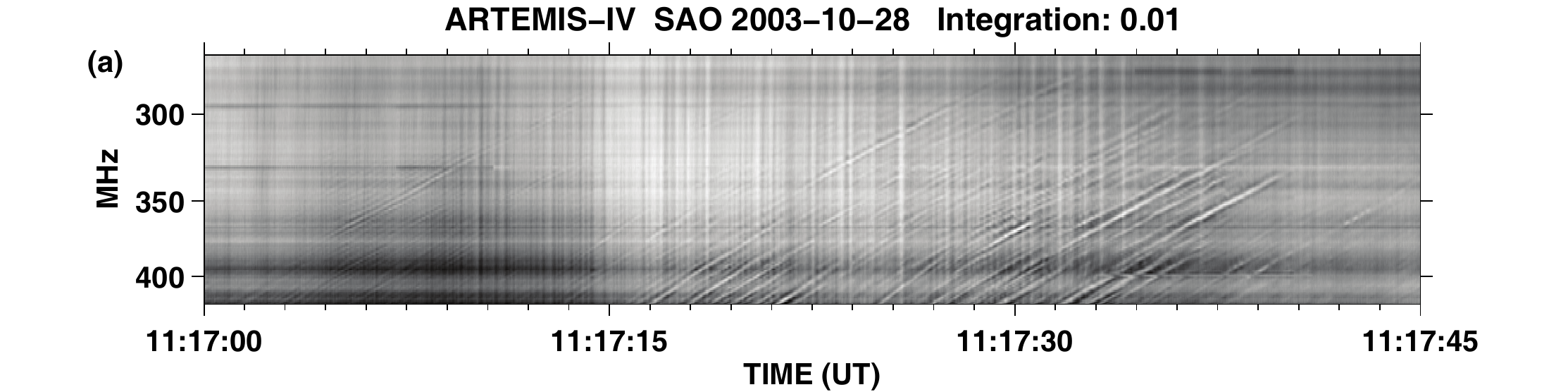}
\includegraphics[trim=0cm 0.0cm  0cm 0cm,clip,width=\textwidth]{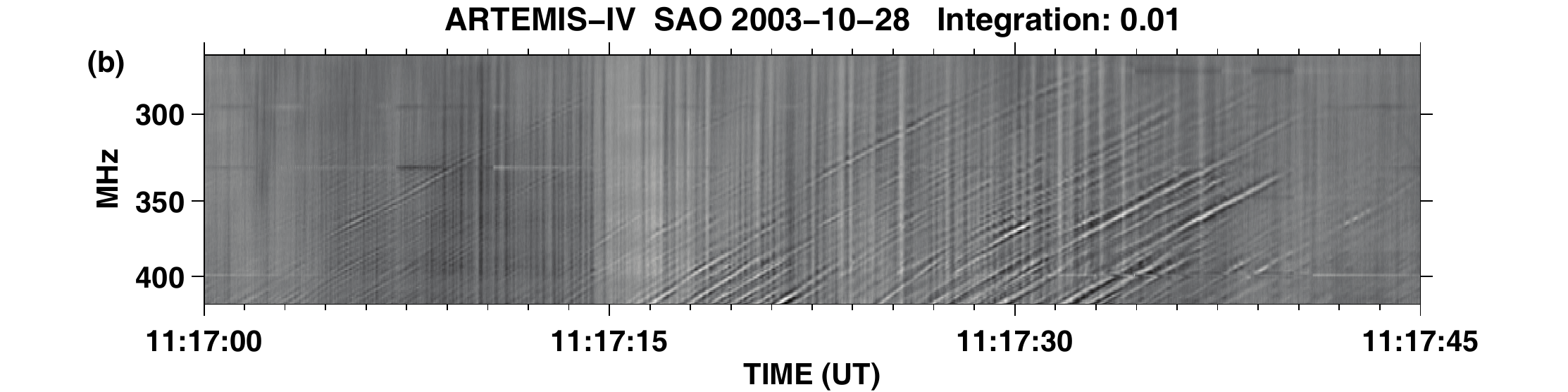}
\includegraphics[trim=0cm 0.0cm  0cm 0cm,clip,width=\textwidth]{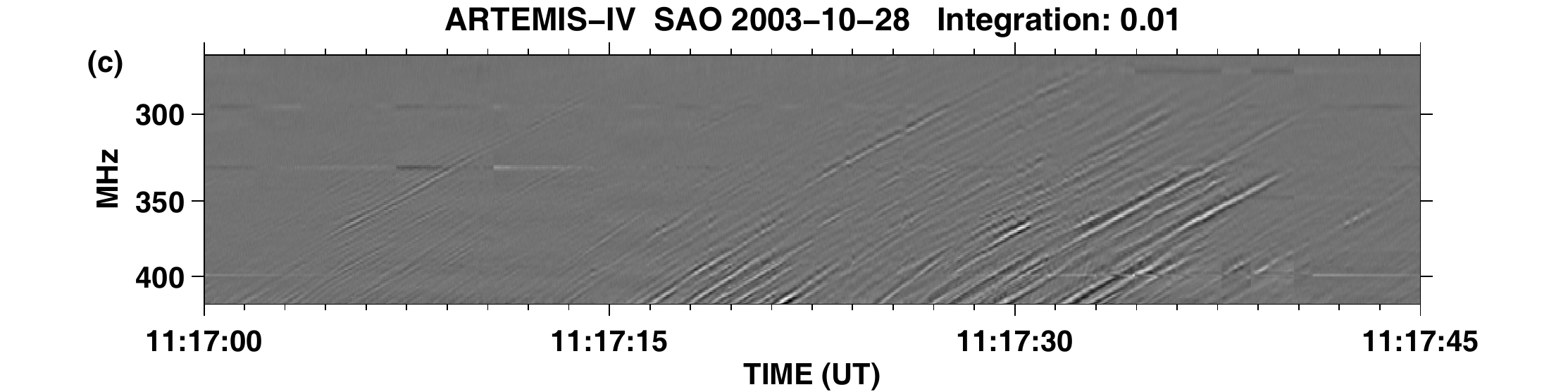}
\includegraphics[trim=0cm 0.0cm  0cm 0cm,clip,width=\textwidth]{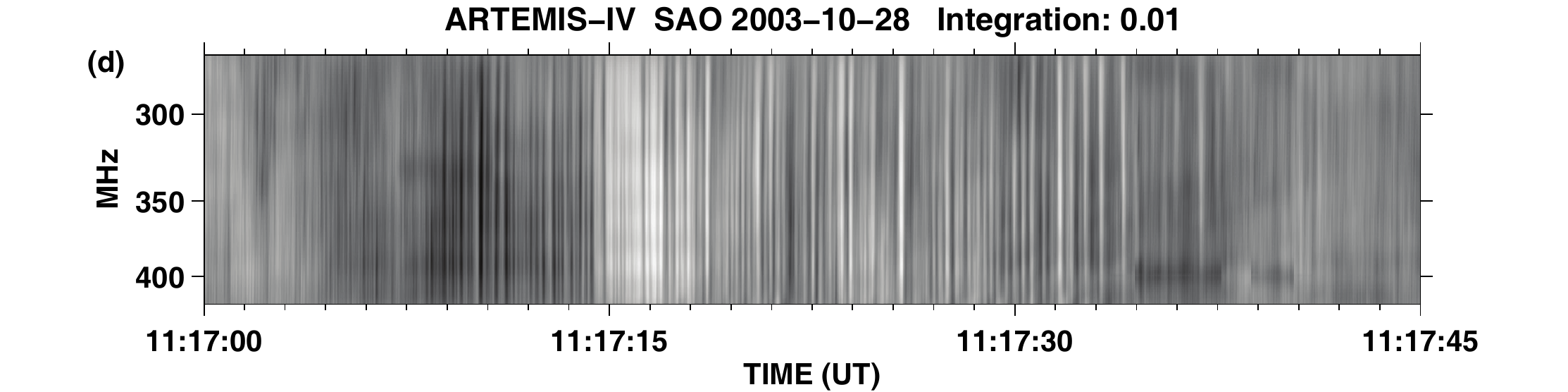}
\caption{Example of high resolution type IV recording of the ARTEMIS-IV/SAO with fine structure enhancement by means of filtering. (a): The original dynamic spectrum of the type-IV continuum. (b): High pass filtering along the time axis reveals underlying pulsations and fibers. These are disentangled by means of high pass filtering along the frequency axis which removes pulsations (c) and low pass filtering along the frequency axis which suppresses fibers (d).}
\label{Filter}
\end{center}
\end{figure}
%-------------------------------------------------

\section{Morphology and Classification}\label{analysis}
The 36~type-IV bursts selected as described in Section \ref{Obs} were further examined for the existence of fine structures, which were classified on the basis of their morphology in the SAO dynamic spectra. 

As the various types of fine structure overlapped, the use of a number of high and low pass filters on dynamic spectra became necessary. The suppression of the continuum background by high pass filtering in time resulted in enhancement of fast time-varying spectral structures such as pulsations, fibers, etc. The disentanglement of pulsations from other types of structures with medium to low frequency variation, such as fibers, was obtained by low--pass filtering of the dynamic spectra along the frequency axis; the opposite was obtained with the complementary high-pass filtering providing pulsation suppression and facilitating the detection of fibers and similar structures (see figure \ref{Filter}). Occasionally the fine structure onset appears to precede the start of the type-IV continuum on the synoptic chronological evolution diagrams ({\it e.g.} Figure \ref{Timeline}); this effect results from the above-mentioned processing of the dynamic spectra which permits detection of fine-structure even if the continuum is not detectable, being too close to the background. 

In all events the radio continuum was found to exhibit a combination of three or more types of fine structures embedded within longer or shorter periods of smooth continuum. The classification of fine structures was based on phenomenological characteristics, grouping these bursts by the similarity of their form in the dynamic spectra. The criteria jointly employed were: bandwidth, duration, drift rate, substructures, impulsive behavior, etc. \citep[see the review by~][]{Benz03}. This work focuses mainly on basic classes of fine structures that have been identified and documented from observations by a number of radio-spectrographs during a rather extended period of time; further division into sub-classes has been known to depend on receiver sensitivity as well as on the time and frequency resolution \citep[see~][~for example]{Elgaroy1986}. Examples of the effect of improved resolution are zebras and fibers which, at high cadence observations,  turn out  to be patterns of dot-bursts \citep{Meszarosova2008} or spikes \citep{Chernov2003} in zebra-like or fiber-like chains. Most of the basic classes examined, include more than one sub-classes corresponding to the  Ondrejov classification \citep{Jiricka01} and the earlier catalog by \citet{Isliker94}. This approach introduces a two level hierarchy of  basic classes  and sub-classes. 

We identified the following basic classes of fine structures embedded in the type-IV continua:

\subsection{Featureless Broadband Continuum}
This class of structures, known also under the \emph{diffuse continuum} label, comprises both featureless segments of type-IV bursts as well as structures of smaller frequency bandwidth and duration such as {\emph{slowly drifting bursts}} and {\emph{patches}}. Our data indicate long periods of  broadband continuum overlapping in time with the SXR emission, with embedded intermittent periods of pulsation, fiber bursts, spike groups and zebra bands. Intensity variations within the smooth periods of the type-IV bursts, which could qualify as patches  were also recorded. On the average, the smooth periods of absence of fine structure varied between 0\% to 60\% of the type-IV continuum duration; the relatively longest smooth periods were found to increase with the duration of the type-IV burst. We note that, although featureless periods have been recorded in the past, our high cadense data indicate that fine structure may still be revealed at adequately high resolution and sensitivity.

\subsection{Pulsating Structures}\label{Puls}
{This class includes drifting and stationary pulsating structures \citep[see review by \,][]{Nindos07}; the shortest groups of pulsating structures with duration of the order of 10 seconds appear as Isolated Broadband Pulses in  \citet{Jiricka01, Jiricka02, Meszarosova05}. The pulsations are considered as the radio-signature of kinetic plasma instabilities, induced by energetic electron populations from quasi-periodic acceleration episodes in reconnecting current sheets \citep{Aurass2007}.

%-------------------------------------------------
%\input{tableCategories_Revision.tex} % Label {FSCategories}
%-------------------------------------------------

In our sample 59 periods of pulsations were detected in 33 events; an auto-correlation analysis of their intensity--time profiles indicated periodicities in the 0.6--3 sec range and individual pulse widths in the 0.35--1.3 sec range.} Their bandwidth exceeds that of the SAO receiver (180 MHz).

%-------------------------------------------------
\begin{figure}
\begin{center}% trim=0cm 1cm  0cm 1cm,clip,
\includegraphics[width=\textwidth]{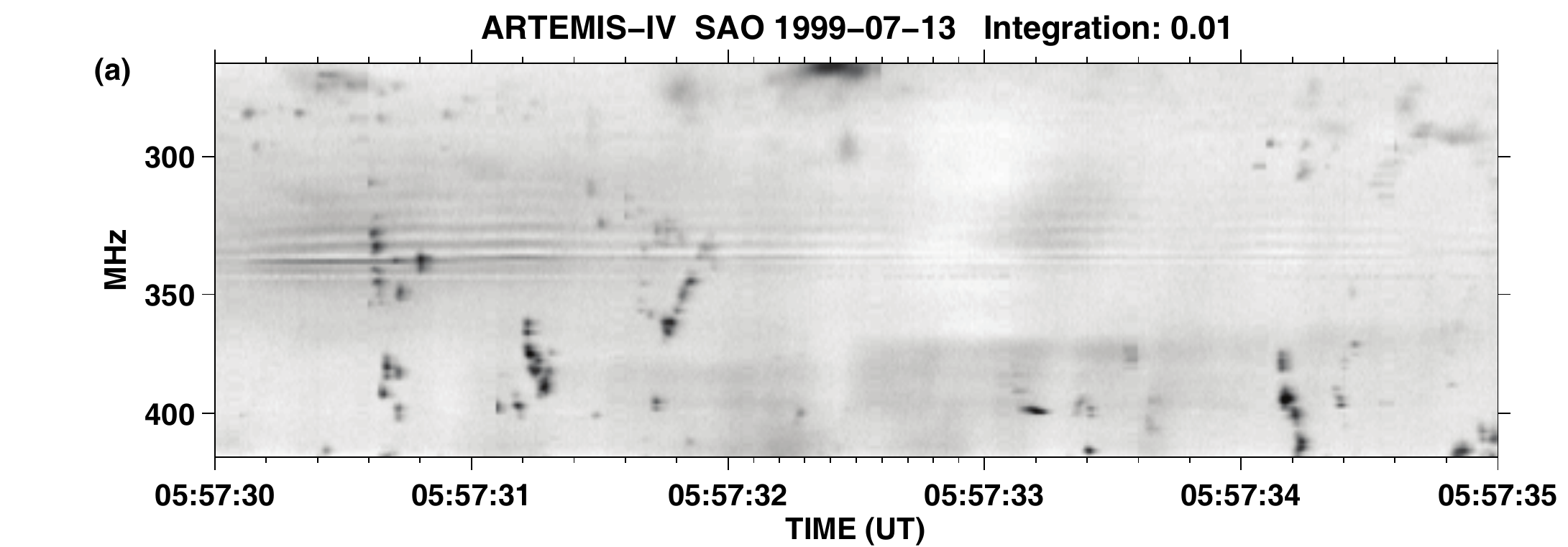}
\includegraphics[width=\textwidth]{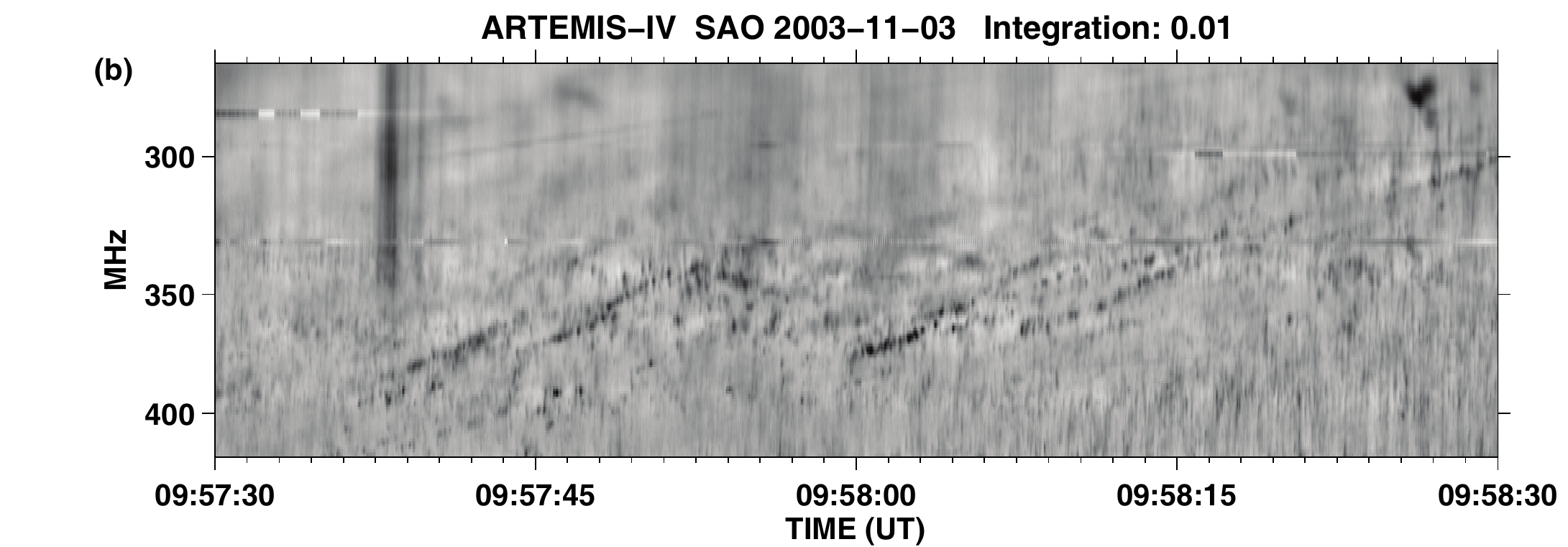}
\includegraphics[width=\textwidth]{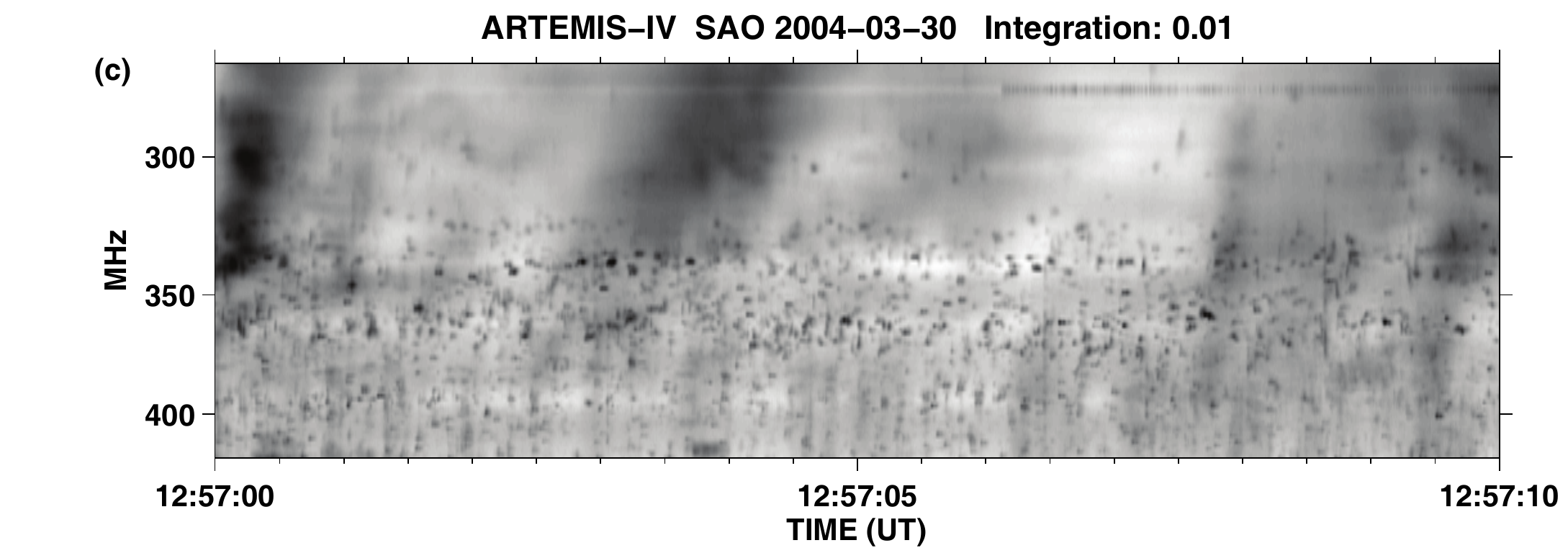}
\includegraphics[width=\textwidth]{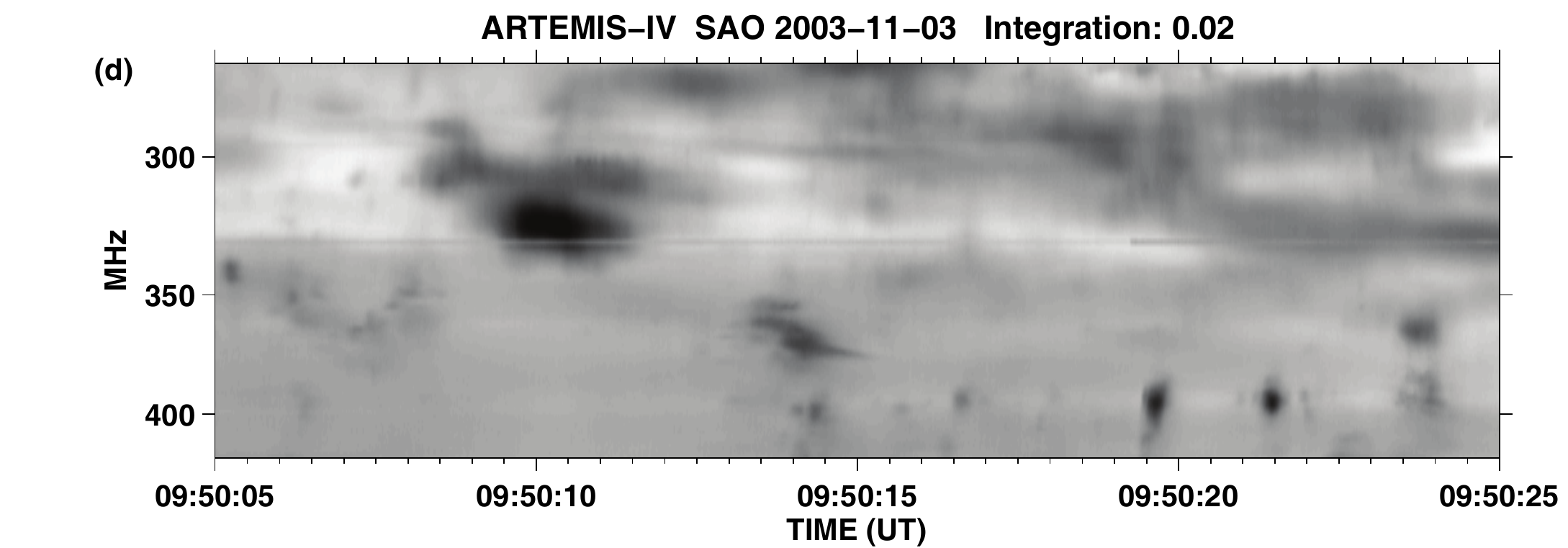}
\caption{Examples of Narrow Band Bursts observed by ARTEMIS-IV. (a): Spike clusters on 13 July 1999 at 0.01 sec resolution. (b): Spike drifting Chains on November, 03 2003. (c): Subsecond Narrow Band Bursts. (d): {\bl Example of patch on 03 November 2003.}}
\label{Spikes}
\end{center}
\end{figure}
%-------------------------------------------------

\subsection{Narrow Band Bursts}\label{Narrow}
These are reported as narrow-band bursts of the type-III family, spikes, dots and sub--second patches, depending on their shape on the dynamic spectra, \citep[part of the same family are the III(U) and III(J) narrow-band bursts reported by~][]{Fu04, Bouratzis10}. They have been interpreted as signatures of small scale acceleration episodes \citep{Nindos07}. In this basic class we might also include the \emph{sawtooth oscillations} by \citep{Klassen01} although associated with type-II shocks. Figure \ref{Spikes} shows {\bl some} examples from our data set.

In the ARTEMIS-IV/SAO data set the average duration of individual spikes was found to be $\approx 70$ ms; the relative bandwidth on the dynamic spectra was $df/f\approx 2\%$. In a subset of the recorded spikes a positive or negative frequency drift rate was measurable; typical values were found to be $df/fdt\approx\pm(0.3-0.6)\,sec^{-1}$ comparable to the type-III frequency drift rate~\citep[see Table 1 of~][~where $df/fdt=0.31$\,sec$^{-1}$ at 328\,MHz]{Benz1996}. More often than not, the bursts of this class were grouped in clusters of individual spikes close in time and/or frequency. {\bl A particular class of cluster are the spike-chains which exhibit overall frequency drift; the majority of these exhibits negative drift $df/fdt\approx$\,--0.021 while few drift towards higher frequencies at a rate  $df/fdt\approx$\,-0.033\,sec$^{-1}$. The average chain duration in our data set was within the 2--20\,sec range}. 

\subsection{Intermediate Drift Bursts}\label{IDB}
These include the typical fibers and the narrow-band \emph{Rope-like fibers} \citep{Mann1989, Chernov1990,Chernov2006, Chernov2008}; there are also variants of these sub-classes such as the \emph{broadband fibers} \citep{Chernov2007} which were observed in the wake of type-II bursts. The fiber bursts are thought to be the result of whistler--Langmuir or Alfven--Langmuir wave interaction in, mainly, postflare loops; a more recent interpretation \citep{Karlicky2013} resorts to fast magnetoacoustic wave trains. Due to their origin they qualify as, model dependent, magnetic field diagnostics \citep[see~][]{Kuijpers1975,Aurass2005, Rausche07}.

%-------------------------------------------------
\begin{figure}
\begin{center}% trim=0cm 1cm  0cm 1cm,clip,
\includegraphics[trim=0cm 0.5cm  0cm 0cm,clip,width=\textwidth]{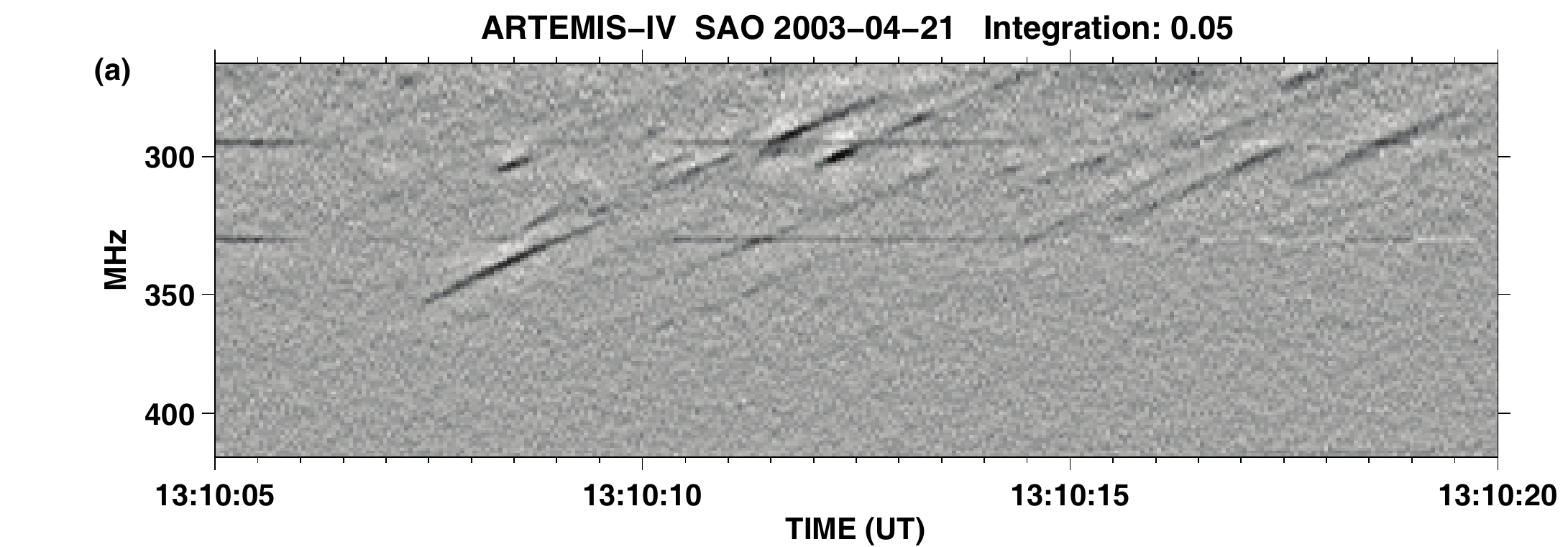}
\includegraphics[trim=0cm 0.5cm  0cm 0cm,clip,width=\textwidth]{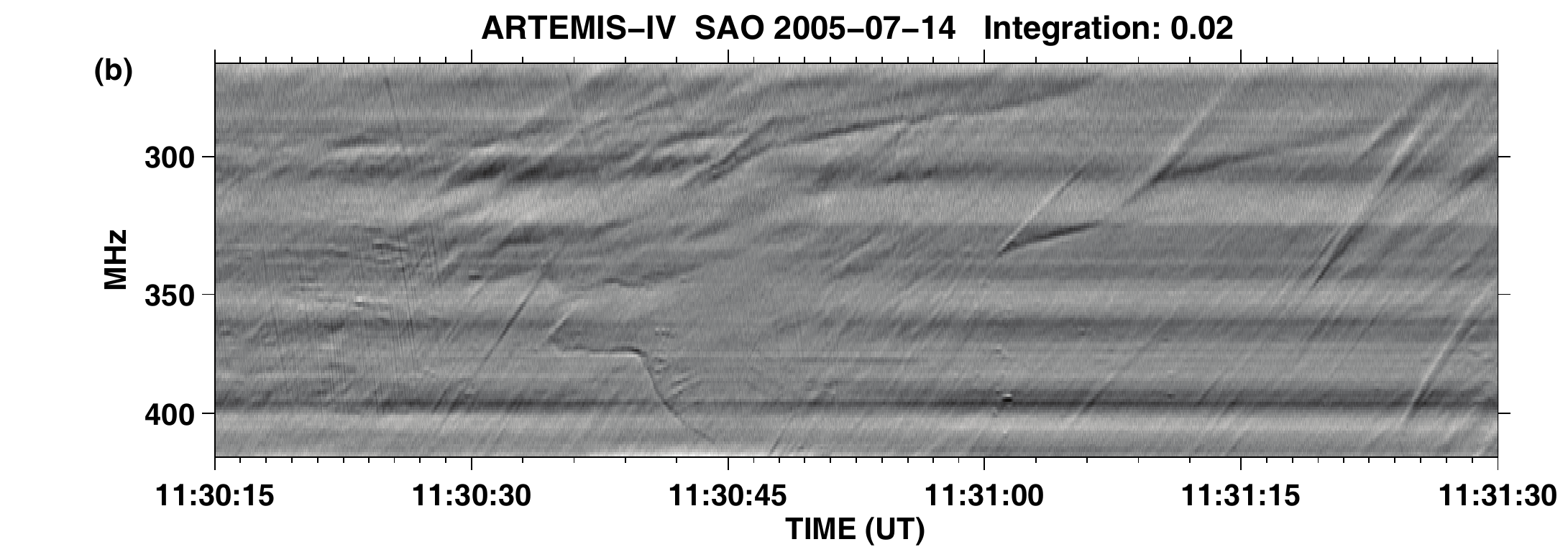}
\includegraphics[trim=0cm 0.5cm  0cm 0cm,clip,width=\textwidth]{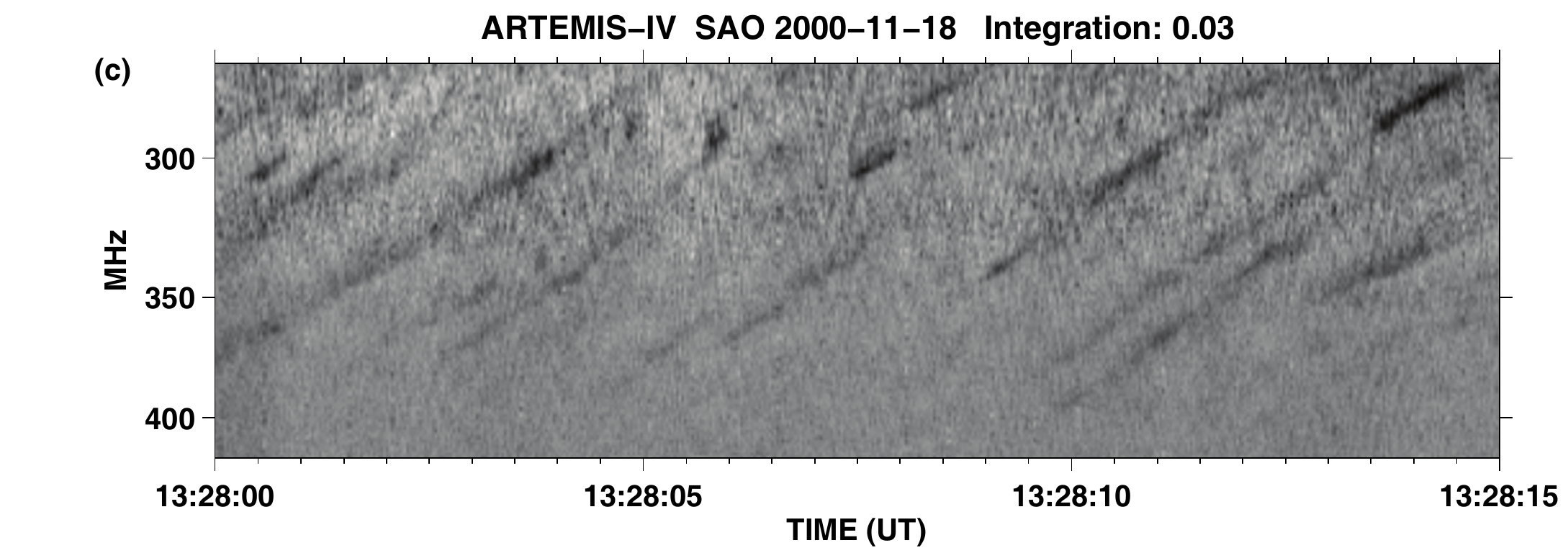}
\includegraphics[trim=0cm 0.0cm  0cm 0cm,clip,width=\textwidth]{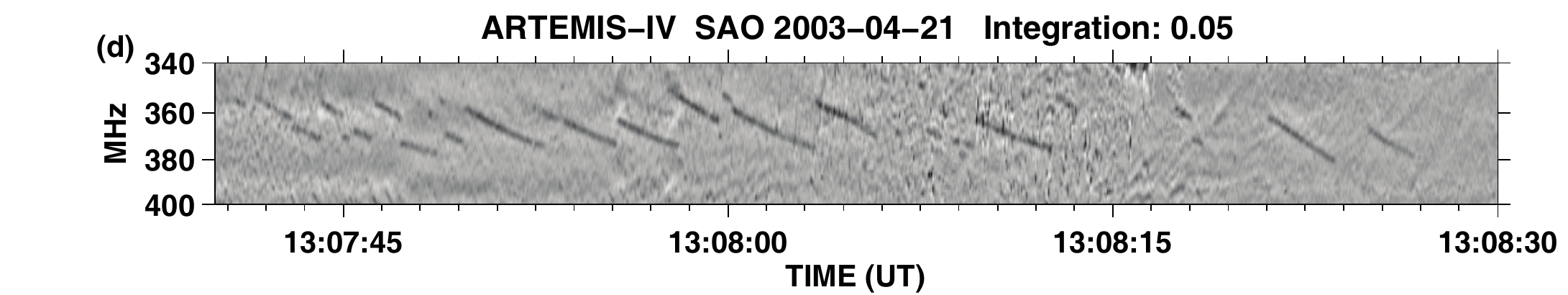}
\caption{Example of Intermediate Drift Bursts. (a): Fiber bursts. (b): A complex group of fibers observed by ARTEMIS-IV on 14 July 2005. (c): Fast drift structures with drift rate approximately double that of the typical fiber. (d): Ropes.}
\label{FiberComplex}
\end{center}
\end{figure}
%-------------------------------------------------

On the average, the fiber bursts recorded by the ARTEMIS-IV/SAO (see examples in Figure \ref{FiberComplex}) exhibit normalized drift rates $df/fdt\approx$0.03\,sec$^{-1}$. Some outliers, however, of the drift rate distribution reached $\approx$0.4\,sec$^{-1}$ which implies exciter speeds comparable to the type-III bursts; these were dubbed  Fast Drift Bursts by  \citet{Jiricka01,Jiricka02,Meszarosova05}. In this work, these outliers were provisionally retained in the Intermediate Drift Bursts category.

%-------------------------------------------------
\begin{figure}
\begin{center}% trim=0cm 1cm  0cm 1cm,clip,
\includegraphics[trim=0cm 0.5cm  0cm 0cm,clip,width=\textwidth]{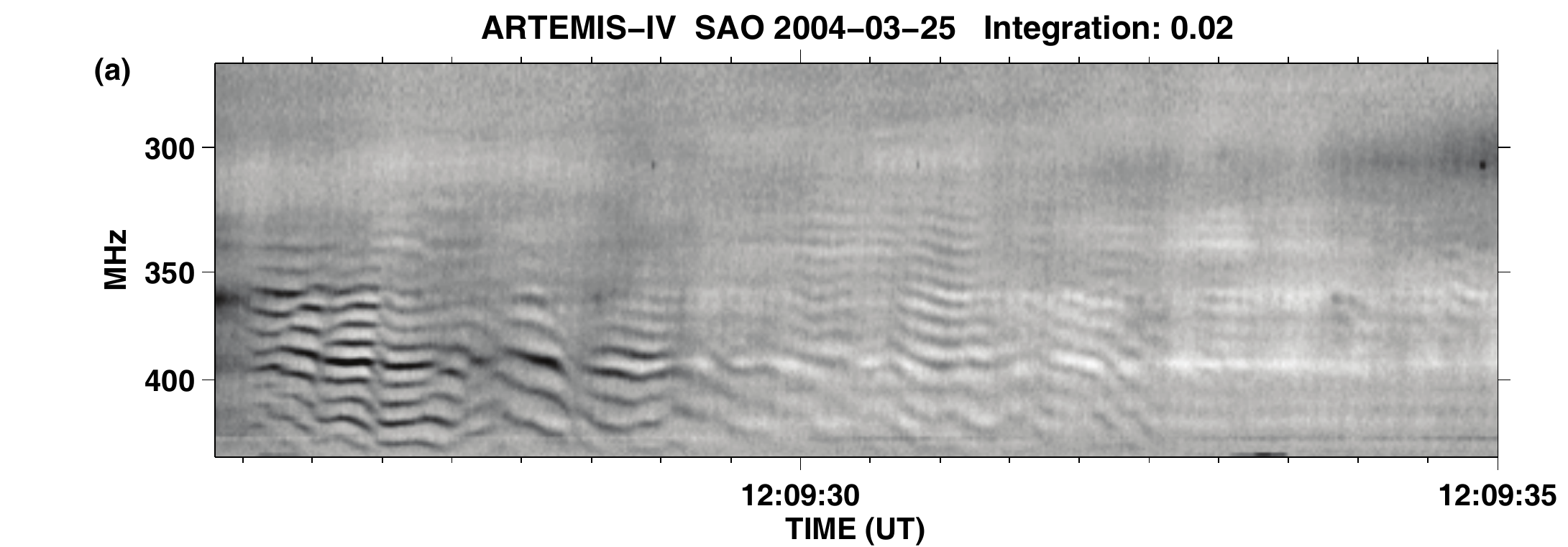}
\includegraphics[trim=0cm 0.5cm  0cm 0cm,clip,width=\textwidth]{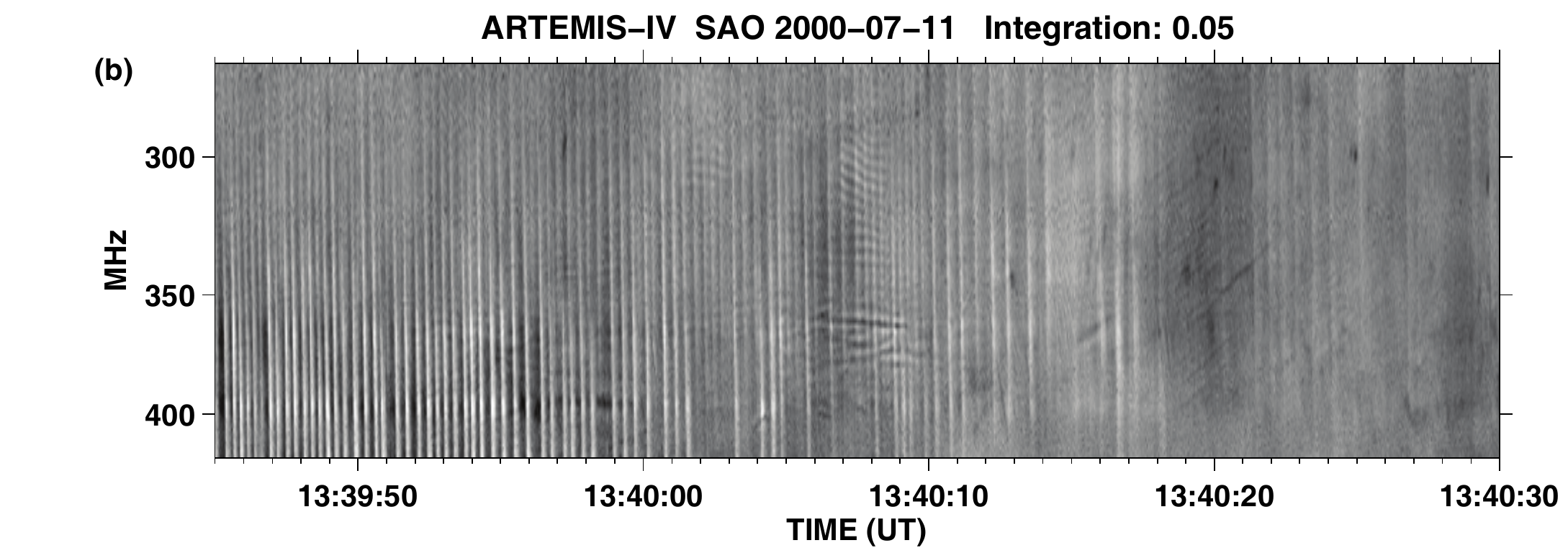}
\includegraphics[trim=0.7cm 0.5cm  1cm 0cm,clip,width=\textwidth]{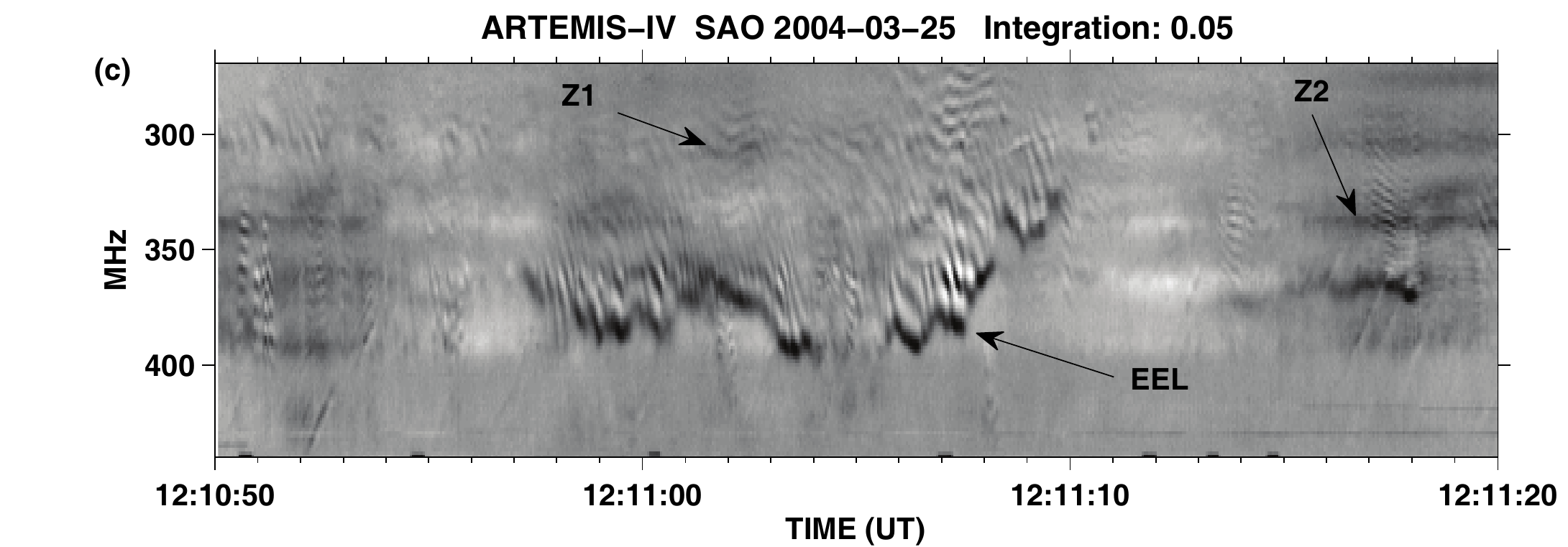}
\includegraphics[trim=0.7cm 0.0cm  1cm 0cm,clip,width=\textwidth]{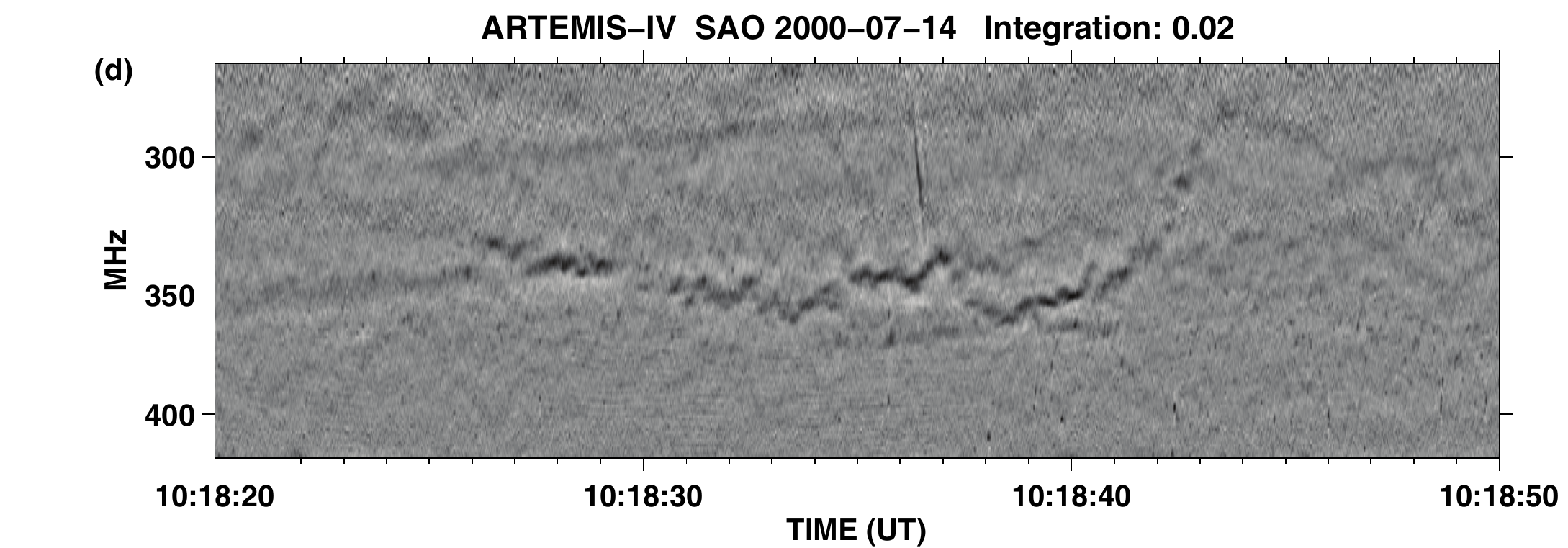}
\caption{ARTEMIS-IV high--resolution dynamic spectra  of Emission bands: (a): Zebra structure. (b): Zebra superposed on pulsations. (c): Recording of Zebra structures with possible Evolving Emission Line (EEL on the figure) at the high frequency limit of the Event; Z1 and Z2 mark typical and fiber-associated zebra patterns. (d): Lace Burst  on 14 July 2000.}
\label{ZebraLace}
\end{center}
\end{figure}
%-------------------------------------------------

In our data set nineteen events had multiple groups of fibers and in two events rope-like fibers were recorded; the fiber rate period was in the range 0.46--2.3\,sec. The individual fibers had a duration of $\approx$\,0.4\,sec  and instantaneous bandwidth of $\approx$\,3.5\,MHz \citep[consistent with observations of\,][\,in the 1--3 GHz range]{Benz98}; the  group extent in frequency was, on the average, $\approx$\,40\,MHz which corresponds to $df/f\approx0.11$. The majority of the fibers had negative drift rates $df/fdt\approx-0.023$\,sec$^{-1}$, while those with positive drift reached rates $df/fdt\approx 0.034$\,sec$^{-1}$; the latter always coexisted with fibers of negative drift rate. A comparison of these results with the spike-chain characteristics in subSection~\ref{Narrow} indicates that the fiber frequency drift rates and instantaneous band-widths are, on the average, equal to the spike-chain frequency drift rate and the individual spike bandwidth. 

Finally, groups of fibers with different drift rates were found overlapping on dynamic spectra (see Figure \ref{FiberComplex} for an example); it is not clear whether these might constitute a separate sub-class or if they originate from different regions; further analysis employing NRH images may clarify this question.

\subsection{Emission Bands}\label{Bands}
{These comprise the various sub-types of the zebra family (classic or pulsation--superposed zebra patterns, fiber-associated zebras  \citep[][]{Chernov2005, Chernov2006}, zebras with drifting emission envelope \citep{Zlotnik2009}),  the rare lace-bursts \citep[first reported by~][~see  bottom panel of Figure \ref{ZebraLace}~for an example recorded by ARTEMIS-IV]{Karlicky01}, the equally rare single emission band, dubbed \textit{Evolving Emission Line~(EEL)},  first reported by \citet{Chernov98}~in the decimetric frequency range and \citet{Fu04,Ning2008} in the GHz range. In the same basic class we may include some unusual bursts consisting of short ($\approx$ 2--4 ms), parallel  stripes with a relative delay as frequency decreases; they are characterized by an overall frequency drift \citep{Oberoi2009}.  The zebra-burst emission mechanism has been attributed to a number of different interpretations, some based on double plasma resonance \citep{Zlotnik03,Aurass03}, Bernstein modes or plasma wave trapped in resonators, to mention but a few \citep[see~][]{Chernov2005,Nindos08}; the double resonance interpretation has been also proposed for the lace-bursts \citep{Fernandes2003,Barta2005}.  \citet{Chen2011} have provided observational evidence in support of this interpretation.}

Twenty two events of our data-set exhibited one or more patches of zebra structures; they appeared in almost equal numbers before and after the flare maximum. The examination of the energy release episodes indicated a good correspondence between these two; all {\textit{patches of zebra bands}} were within 5 minutes of the time of the release episode, provided that the {\textit{frequency}} of the latter, estimated from the type-III feet or the type-IV burst high frequency boundary, was within the SAO range. The majority were pulsation and fiber associated zebra (12) within overlapping pulsation and fiber activity and their association could not be resolved. Five zebra patches appear clearly pulsation-associated. Seven periods of lace-bursts were also recorded; they were found to coincide in time and frequency, mostly,  with pulsations and spikes. Only one EEL was recorded. Examples of zebras, laces and the above-mentioned EEL are shown in Figure \ref{ZebraLace}.

A summary of the properties of all fine structures is presented in Table \ref{FSCategories}.  

%-------------------------------------------------
\begin{landscape}
\begin{table}
%\raggedright
\caption{Summary of the Fine Structure Properties. The last column gives the median and the width of the distribution.}
\label{FSCategories}
%\begin{tabular}{p{0.25\textwidth}p{0.35\textwidth}p{0.3\textwidth}}
%
\begin{tabular}{>{\raggedright}p{.30\textheight}
				>{\raggedright}p{.50\textheight}
				>{\raggedright}p{.48\textheight}
%				>{\raggedright\arraybackslash}p{.15\textheight}}
				>{}p{.15\textheight}}
\hline
Category or Subcategory			& Characteristics																									&	Remarks																	& 	T$_{md}$, FWHM \\
								& 																												&																				&     min \\
\hline\hline                                                                                                                                               
Featureless Broadband Continuum		& Lack of fine structure. 																						& Part of type-IV Burst.													& 	-- \\
\hline
Pulsating Structures  				& Periodicities 0.6--3 sec.																							& Embedded within the type-IV Burst Continuum (Moving or Stationary)	& 	1.5 (8.0)\\
\textit{drifting}						& Drift rate $|df/fdt|\approx0.003 sec^{-1}$                                       								& Part of moving type-IV Burst.											& 			 \\
\textit{stationary}					& Drift Rate negligible																								& 																		& 			 \\
\textit{Isolated Broadband Pulses}	& Duration $\approx$10 seconds 																						& Shortest pulsating structures. 										& 			 \\
\hline                                                                                                                                                                                                                                 		
Narrow Band Bursts				& Sub--second Narrow Band Bursts, near the time resolution limits of SAO. 												&																		&	6.5 (18.0)\\
\textit{Narrow--Band Type III}		& $df/f\approx 10\%$,	$|df/fdt|\approx0.4 sec^{-1}$																	& Including III(U)--III(J) narrow-band bursts.									&	\\
\textit{Spikes}						& Individual Spike: Bandwidth $df/f\approx 2\%$, duration $\approx$ 70 ms, $|df/fdt|\approx 0.45 sec^{-1}$			& Often Clustered, Occasionally organized in fiber-like or type-III-like sequences.		&	\\
\textit{patches}					& Bandwidth $f/f\approx 3-4\%$, Drift Rate $|df/fdt|\lesssim0.03 sec^{-1}$, duration $\approx$ 1--5 sec.		& 																		&\\
\hline                                                         				                                                                            
Intermediate Drift Bursts			& Drift Rate between type II--type III bursts.																	& 																			& 	-- \\
\textit{Fibers}						& $|df/fdt|\approx0.03 sec^{-1}$, Bandwidth $df/f\gtrsim10\%$														& 																			&		\\
\textit{Ropes}						& Similar to fibers, bandwidth $df/f\lesssim10\%$																& 																			&		\\
\textit{Fast Drift Bursts}			& Frequency drift rate up to $|df/fdt|\approx0.4 sec^{-1}$															& Drift rate comparable to type IIIs. 												&			\\
\hline                                                         				                                                                            
Emission Bands					& 																													& 																			&	6.8 (10.0)\\ 							 							 							
\textit{Zebra}						& Single band: $df/f\approx 2.5\%$, Duration 1--10 sec; Total Bandwidth $df/f$ up to $35\%$.						& Most of the emission bands.											&	\\
\textit{lace--bursts}				&																													& 																			&	\\ 							 							 							
\textit{Evolving Emis. Line} (EEL)	&																													& Only one was recorded.												&	\\ 							 							 							
\hline\hline
\end{tabular}
\end{table}
\end{landscape}
%-------------------------------------------------

%-------------------------------------------------
\begin{figure}
\begin{center}
\includegraphics[width=.333\textwidth]{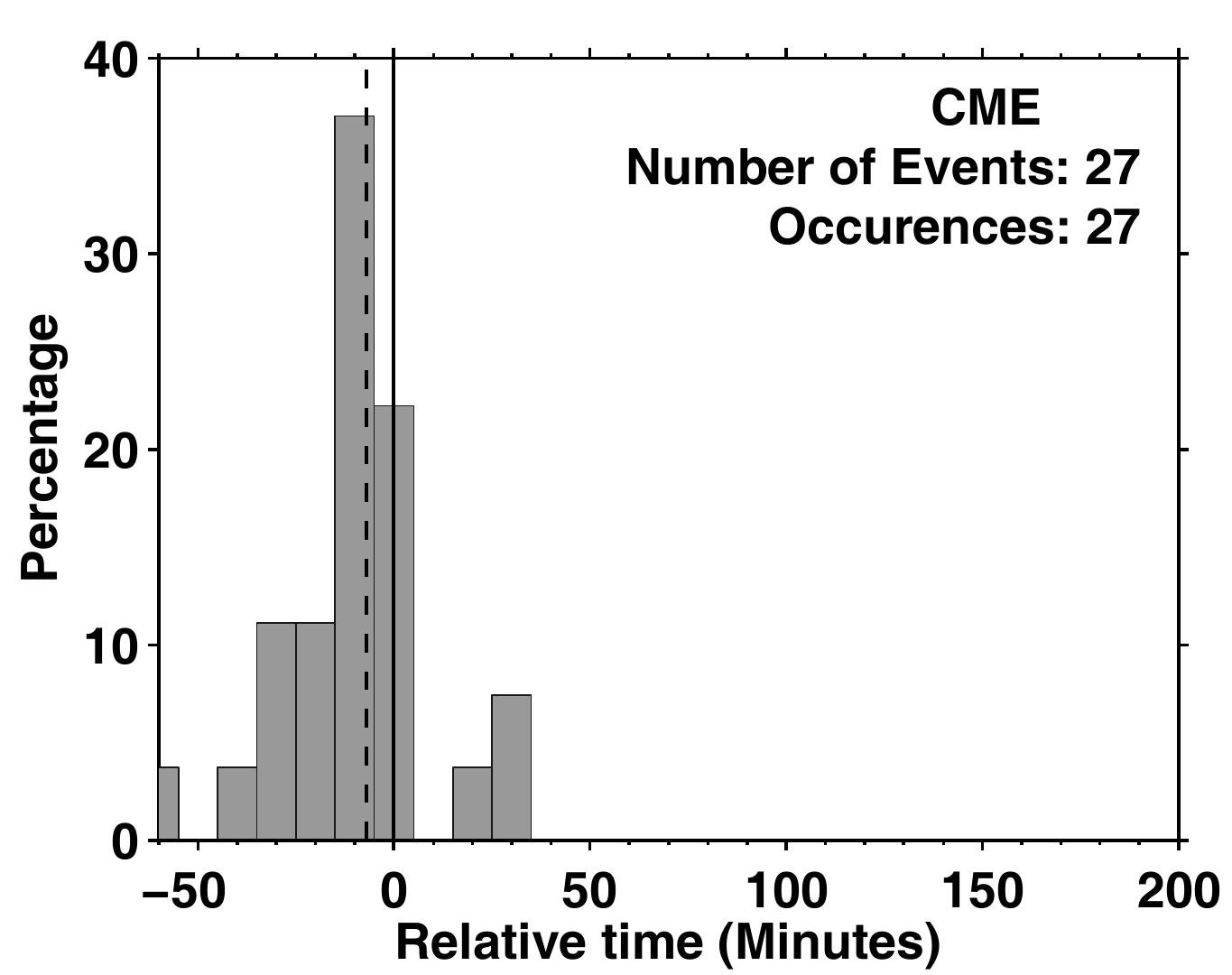}%
\includegraphics[width=.333\textwidth]{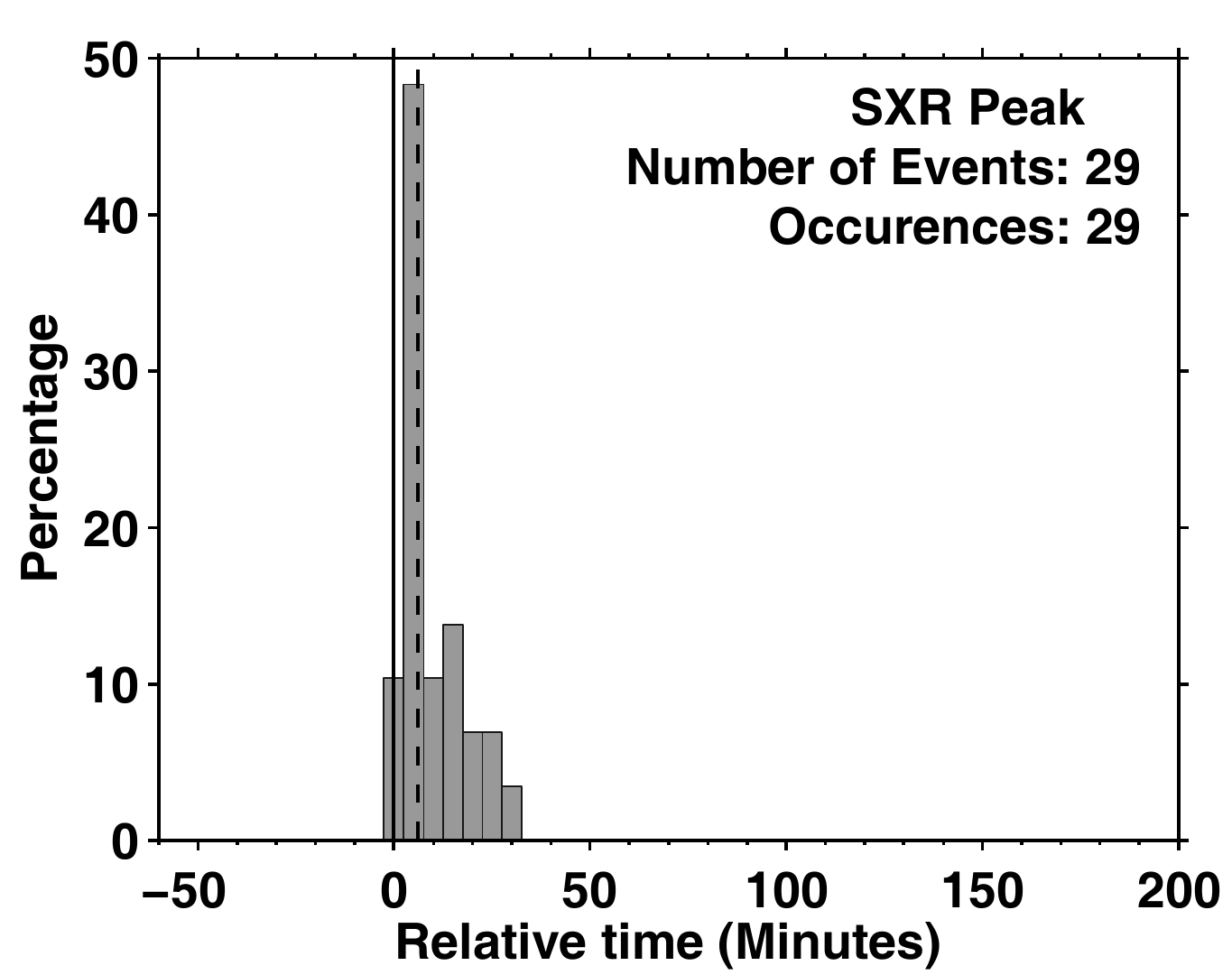}%
\includegraphics[width=.333\textwidth]{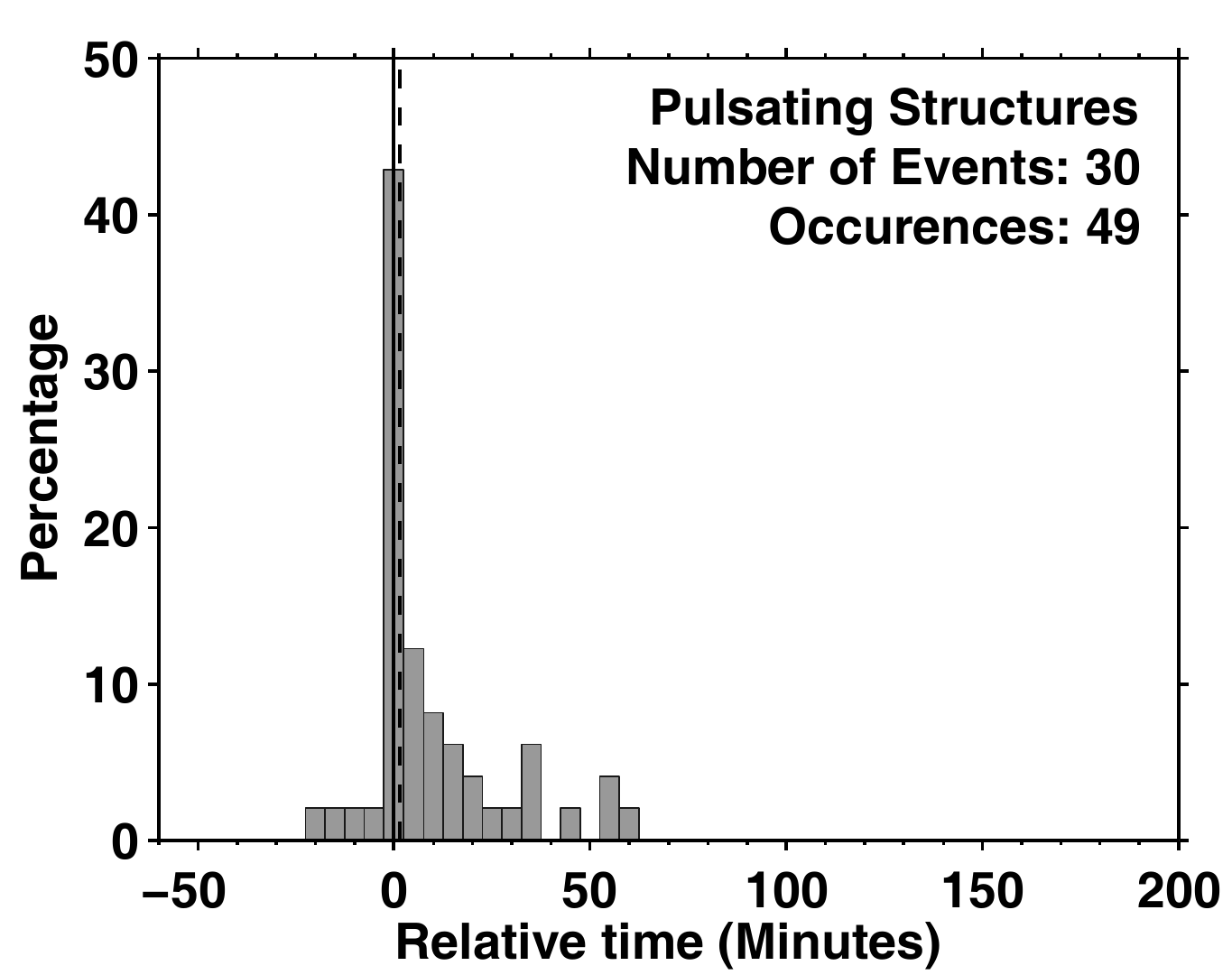}
\includegraphics[width=.333\textwidth]{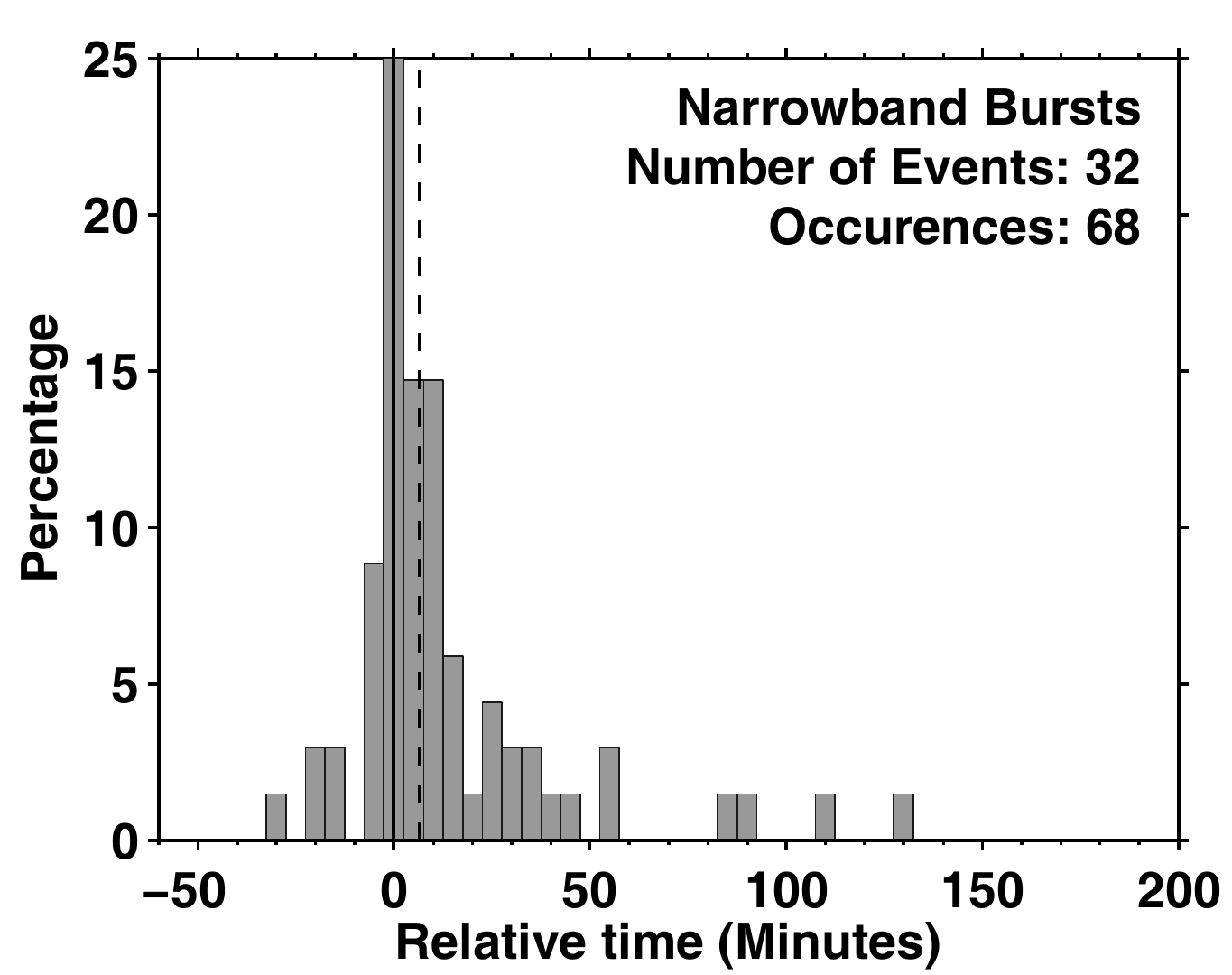}%
\includegraphics[width=.333\textwidth]{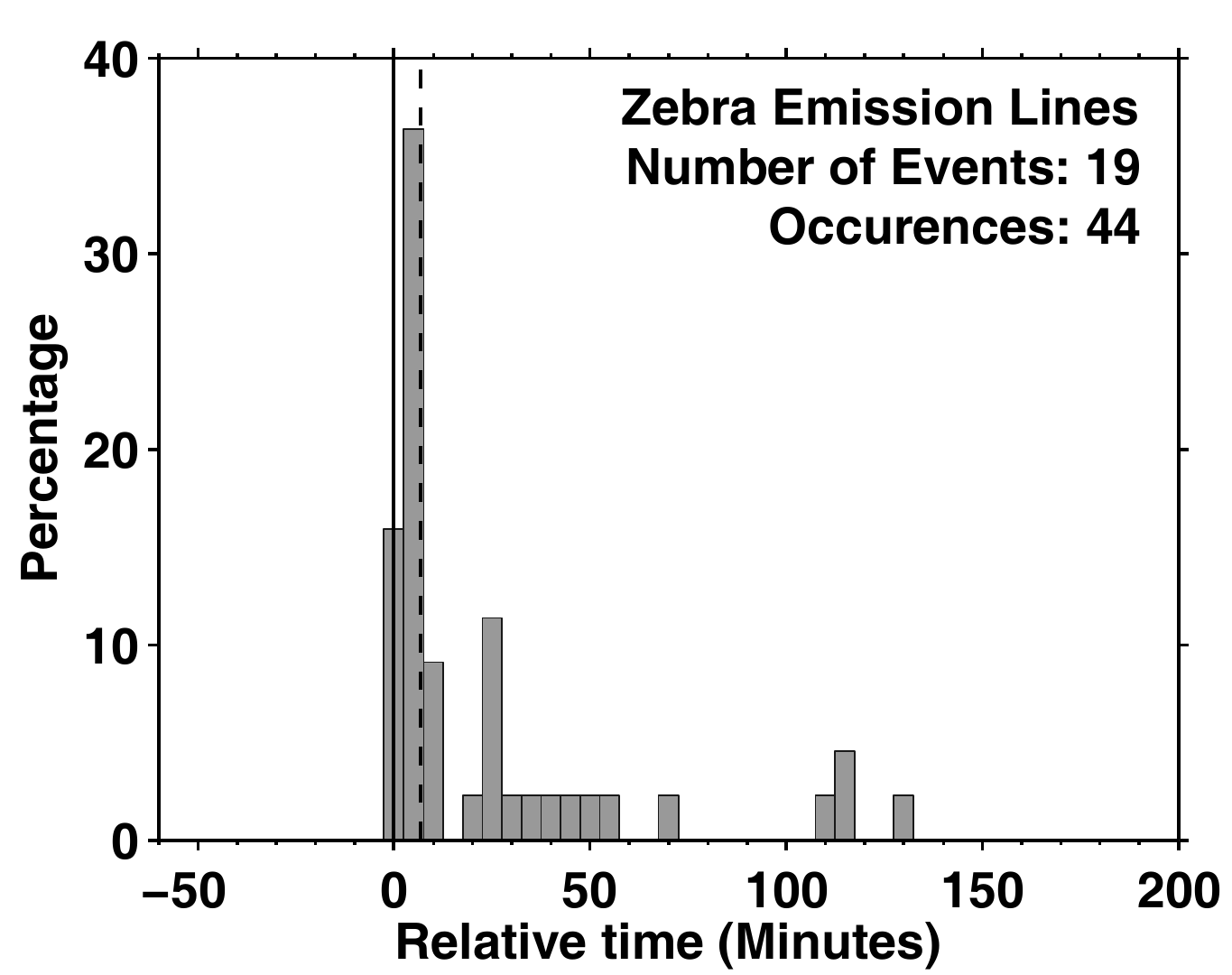}%
\includegraphics[width=.333\textwidth]{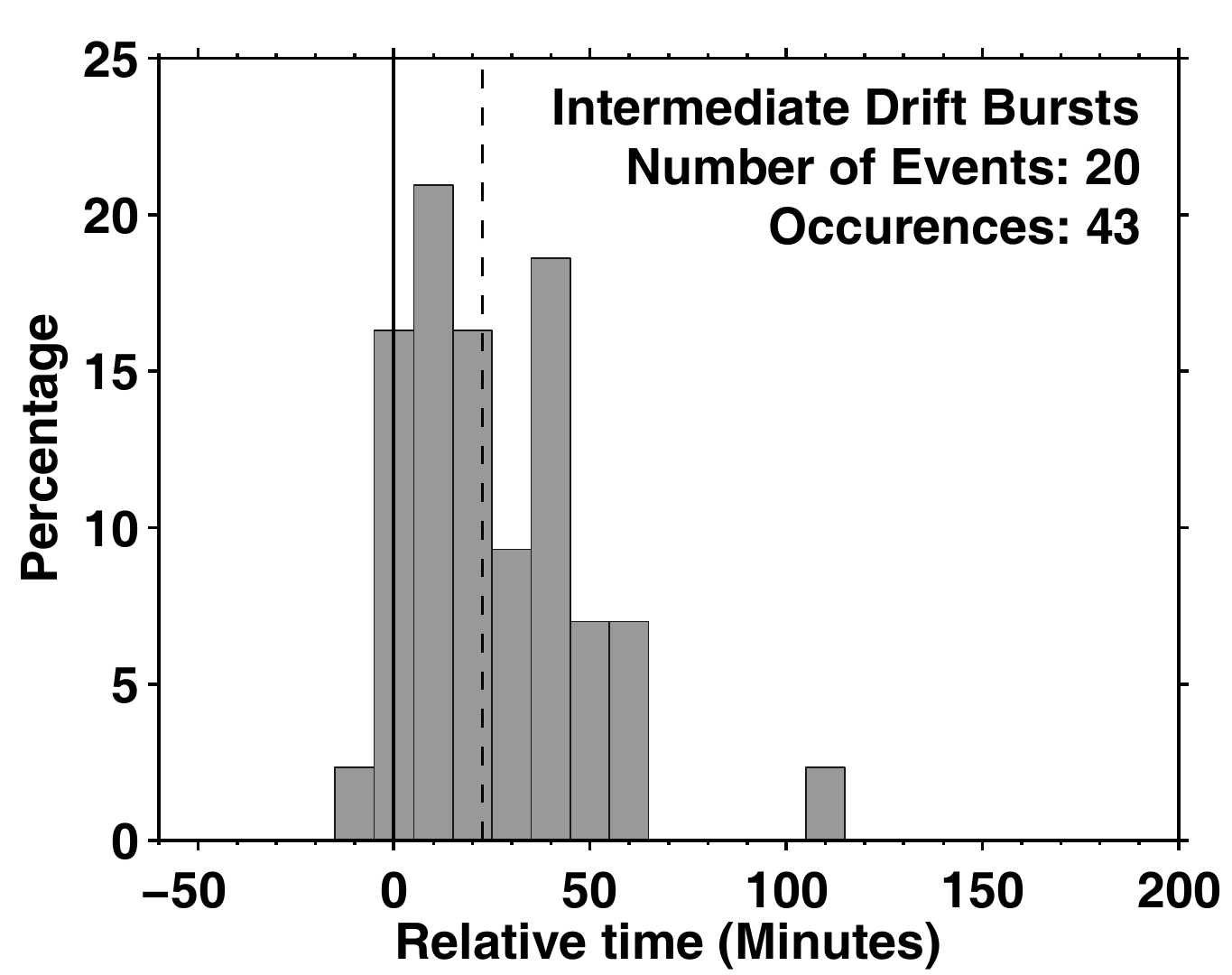}
\end{center}
\caption{Histograms of the time of occurrence of the CME onset, SXR peak, pulsations, spikes (narrowband bursts), zebra patterns and fiber bursts (intermediate drift bursts) with respect to the time of the first impulsive energy release. Note that the bin size is 10 min for the CMEs and the Intermeniate Drift Bursts and 5 min for the others. Dashed vertical lines mark the median of the distributions.
% for type-IIIs, zebras and pulsations and 10 min for the others.
}
\label{hist}
\end{figure}
%-------------------------------------------------
\begin{figure}
\begin{center}
\includegraphics[trim=2.5cm 0.0cm  0.0cm 0.0cm,clip,width=\textwidth]{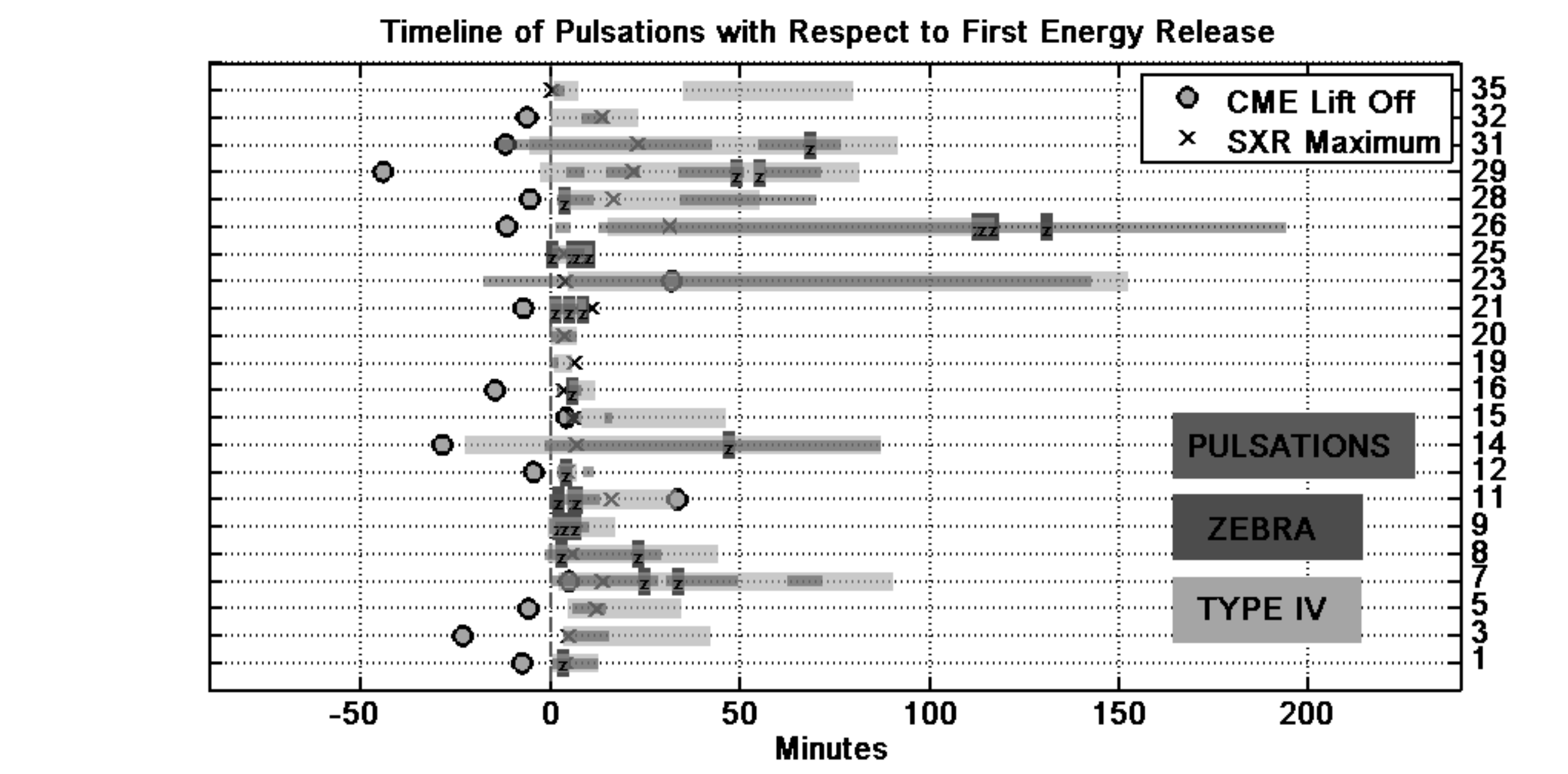}
\end{center}
\caption{Relative timing of pulsations and zebra stripes associated with the~Type-IV~continua (gray); {\bl the {\textbf{z} labels} mark the short periods of zebra stripes}. The  circles mark the CME onset {\bl and the X symbols the SXR flux maximum}. Time is in minutes from the first impulsive energy release. The event number is marked at the right ({\it cf.} Table \ref{Table02}).}
\label{AllFiberTimeLinePuls}
\end{figure}
%-------------------------------------------------
\begin{figure}
\begin{center}
\includegraphics[trim=2.5cm 0.0cm  0.0cm 0.0cm,clip,width=\textwidth]{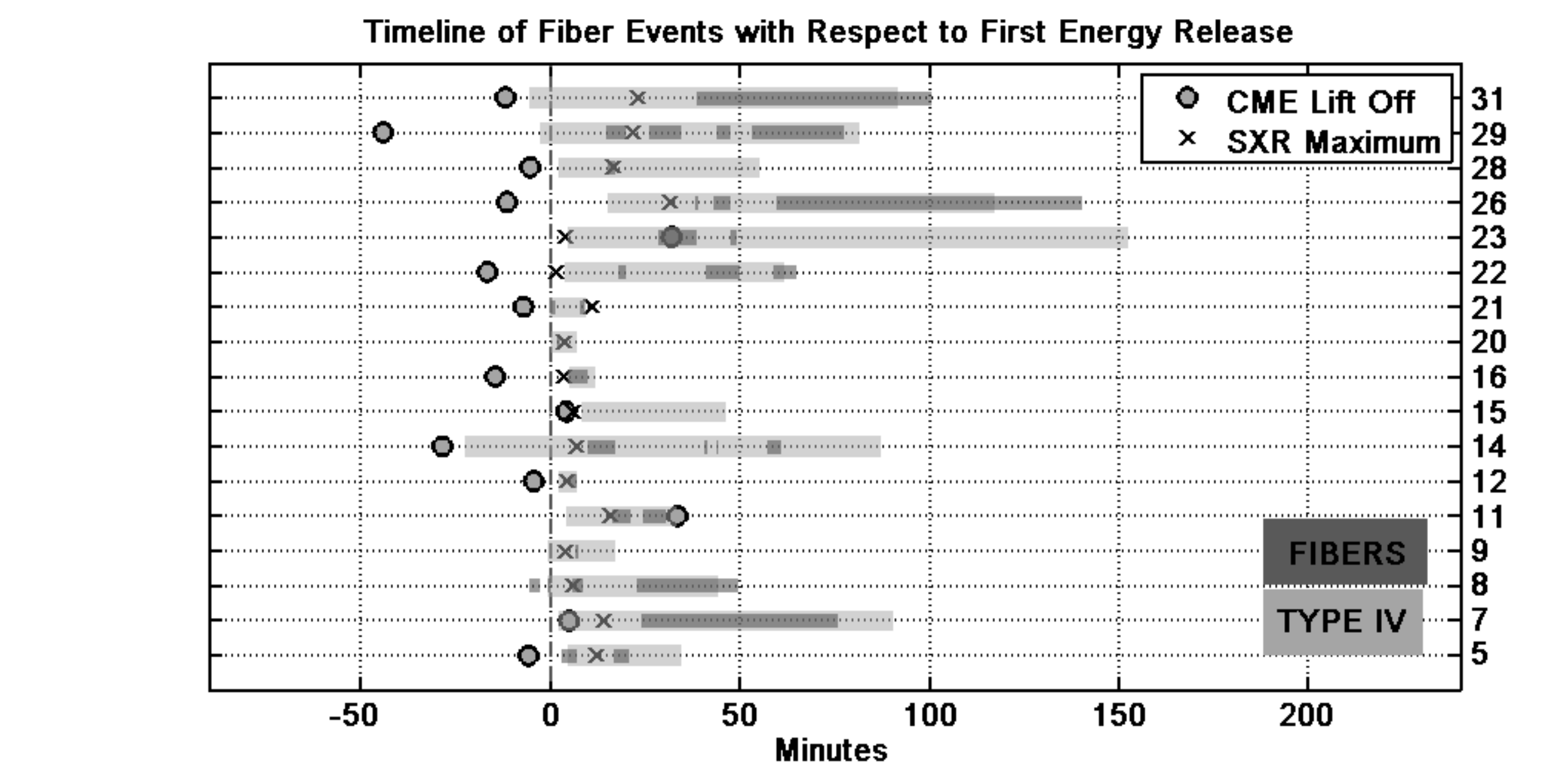}
\end{center}
\caption{Same as Figure \ref{AllFiberTimeLineFb} for fiber bursts.}
\label{AllFiberTimeLineFb}
\end{figure}
%-------------------------------------------------
\begin{figure}
\begin{center}
\includegraphics[trim=2.5cm 0.0cm  0.0cm 0.0cm,clip,width=\textwidth]{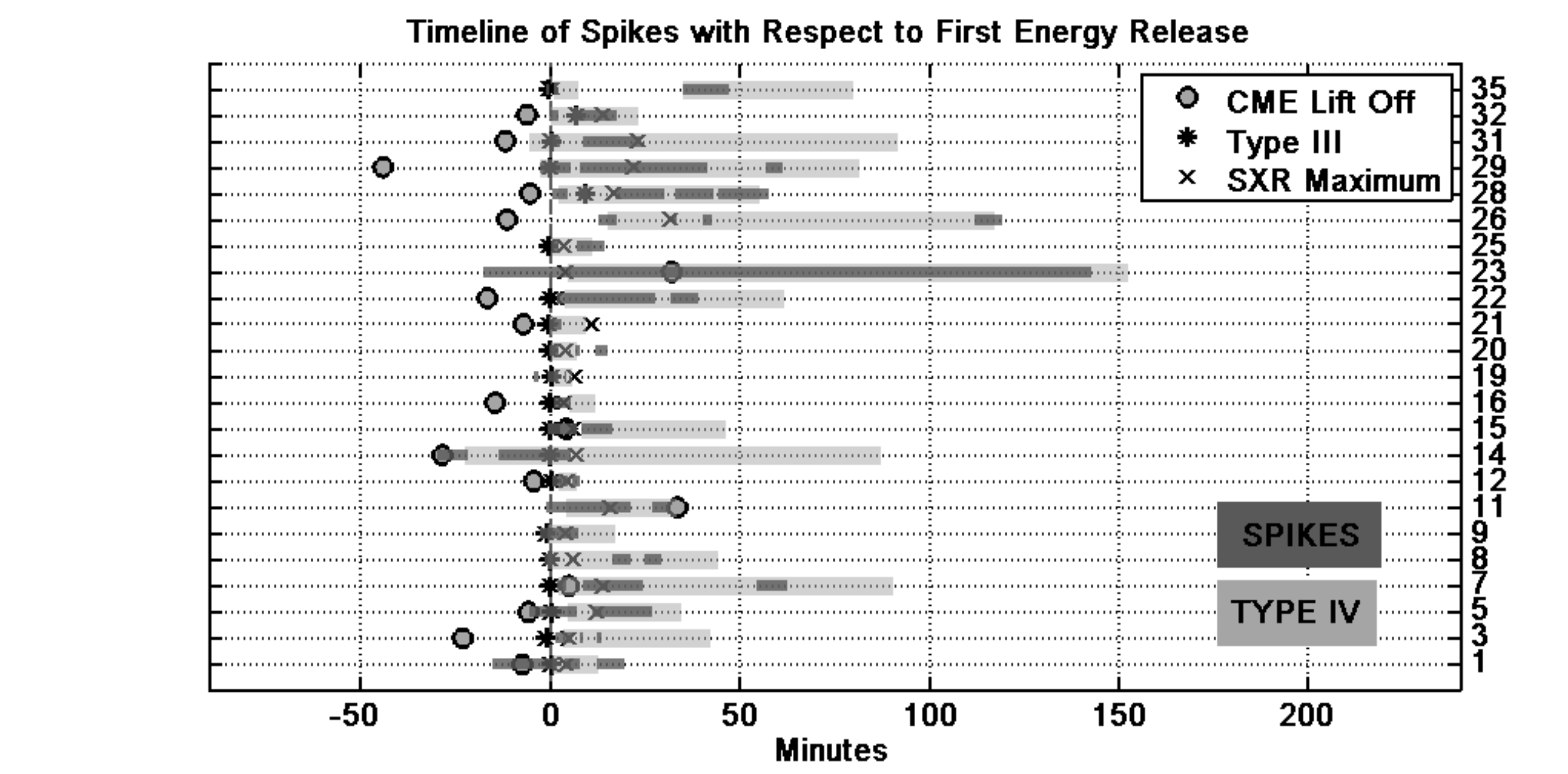}
\end{center}
\caption{Same as Figure \ref{AllFiberTimeLineFb} for Spike groups and type-IIIs {(\bl Standard type-IIIs)}. }
\label{AllFiberTimeLineSp}
\end{figure}
%-------------------------------------------------

\section{Relative Timing of Fine Structures} \label{timing}
As mentioned in Section~\ref{Obs}, the time of \FIER, evidenced from the first HXR/microwave peak, was used as  reference for timing the appearance of fine structures with respect to the evolution of the flare process. Figure~\ref{hist} shows the distributions of the relative time of occurrence of various fine structures, together with those of the CME onset and SXR peaks. As each event shows multiple instances of fine structure, we used all of them for the computation of the histograms. A number of events exhibit composite structure, with clearly distinct HXR/microwave peaks or groups, within the same SXR peak and the same type-IV burst (see for example event 27, Figure~\ref{B27}). Some of the associated fine structures could not be identified with a particular HXR/microwave peak; these were not used for the computation of the histograms in Figure \ref{hist}.% 18 March 2014

The CME onset preceeds the \FIER by several minutes (Figure \ref{hist}) in agreement with  \citet{Zhang2001,Webb2012}. The histogram of SXR peaks shows a sharp maximum 5 min after the first impulsive energy release, as expected. A more detailed chronological evolution is schematically depicted in figures \ref{AllFiberTimeLinePuls}, \ref{AllFiberTimeLineFb}  and \ref{AllFiberTimeLineSp}; {\bl in the latter we have included the times of the standard type-III bursts}. We note that, in the case of the continuum, the determination of the onset is threshold dependent so the times reported in the figures as well as in Table \ref{Table02} are approximate only.

Among our 36 events, small periods of broadband diffuse continuum were recorded in 8 and patches were found in 7. With very few exceptions, they started between the CME onset and the SXR peak, extending for several tens of minutes after that. This time interval of continuum emission, well past the impulsive phase, was named by \citet{Benz2006} \textit{extended decimetric emission}, based on dynamic spectra obtained during the active period October--November 2003. 

As shown in Figure \ref{hist}, pulsations show a narrow distribution around the time of the \FIER, with the median of their histogram at 1.5\,min and a full width at half maximum (FWHM) of 8.0\,min. Their total duration is smaller than that of the continuum, though in event 14 they extended for more than 150 min after the flare maximum (Figure~\ref{AllFiberTimeLinePuls}). 

Narrow band bursts also cluster around the \FIER; their histogram peaks at 0\,min, with a median value of 6.5\,min and a FWHM of 18.0\,min. Although they concentrate mostly around the flare maximum, they occasionally cover longer periods within the flare decay phase, {probably associated with subsequent energy releases. Furthermore, narrow band fine structures may appear before the impulsive phase of the flare \citep[see for example\,][]{Aurass2007}.}

The peak of the histogram for Zebra patterns is 5\,min after the \FIER, with a median value of at 6.8\,min and a FWHM of 10.0\,min. {The distribution exhibits a secondary peak around 120 min due to the contribution of the longest events which are characterized by secondary energy releases accompanied by Zebra patterns. It appears that the energy release episodes, initial and secondary, provide the energetic electrons required in triggering this type of fine structure within loops \citep[see\,][]{Zlotnik05}. Characteristic examples of multiple energy release episodes are events 13 (26 October 2003) and 19 (30 March 2004 (B)) in Appendix B.}

Intermediate drift bursts (fibers), {which are known to appear often in postflare loops, \citep[see\,][]{Chernov2006}}, are the most dispersed of all, with their distribution showing two peaks, 10 and 40\,min after the \FIER. 
%\clearpage
\section{Discussion and Conclusions} \label{discussion}

Using the high time resolution SAO receiver, operating in the frequency range of 450--270 MHz on the ARTEMIS-IV radio-spectrograph, we observed a number of fine structure busts embedded in metric type-IV radio continua; these were compared with the associated HXR bursts, the GOES/SXR flares and SOHO/LASCO CMEs in order to establish a relationship between this type of radio-bursts and the evolution of solar energetic phenomena. Our study started with the examination of the characteristics of each type of fine structure, such as bandwidth, duration, frequency drift rate, shape on the dynamic spectrum etc. This necessitated an appropriate taxonomy, due to the diversity of form; we introduced a two level hierarchy of basic classes and sub-classes.

At the top level, this two-level hierarchy adopted the basic classes of fine structures which have been recognized and documented from multiple observations  over an long period of time. The second level division into sub-classes permits the inclusion of new and, probably, rare types of bursts alongside the well known; the ensuing subdivisions match, more or less, the Ondrejov taxonomy  \citep{Jiricka01}, but include other types of fine structure as well. We note at this point, that the subdivision into types and sub-types, based solely on morphology and not the underlying radiation process, remains an artificial construct, often dependent on the resolution of observational data. Furthermore the classes thus defined, are rather broad and may, at times, include bursts morphologically similar yet originating from different radiation processes. It is, however, necessary as a background for theoretical work, as already pointed out by \citet{Elgaroy1986} and \citet{Benz03}, despite the fact that the classification, in particular the second level, remains more or less tentative.

The time of appearance of the type-IV fine structures with respect to the evolution of the associated CMEs and flares was subsequently examined. Our data indicate that the type-IV continua, in which the fine structures were embedded, were associated with flare eruptions exhibiting spatial scales from the active region size up to almost one solar radius, thus verifying that bursts accompanied by type-IVs are usually complex events which probably involve multiple components of smaller scale. We therefore used the time of the \FIER, evidenced {\bl from} HXR and/or microwave time profiles, as a reference for timing the fine structures. This reference time is very near the time of the first appearance of type-III emissions, as expected.

In the vast majority of the cases studied, the onset of all classes of fine structure shows a close temporal association with the \FIER, which places their onset between the CME onset and the SXR peak. The closest association was found for the pulsating structures, which show a narrow distribution with a median at 1.5\,min after the \FIER. Narrow band bursts (spikes and narrow band bursts of the type-III family) come next, with a histogram median at 6.5\,min, closely followed by  Zebra patterns at 6.8\,min. Intermediate drift bursts (fibers) are more dispersed, with their distribution showing two peaks, 10 and 40\,min after the \FIER.

Pulsations and zebras show the narrowest distribution with a full width at half maximum (FWHM) of 8.0 and 10.0\,min respectively, followed by narrow-band bursts with FWHM=18.0\,min. More dispersed are the intermediate drift bursts, with two peaks in their distribution at 10 and 40\,min.

More detailed studies of the various fine structures will appear in subsequent publications.

%-------------------------------------------------
\begin{acks}
{\bl We would like to thank Gennady P. Chernov, Lidia van Driel-Gesztelyi, the anonymous first reviewer and the guest editors whose comments and suggestions have improved the quality of this work}. This work was supported in part by the Special Account for Research Grants of the National and Kapodistrian University of Athens. The LASCO CME catalog is generated and maintained at the CDAW Data Center by NASA and The Catholic University of America in cooperation with the Naval Research Laboratory. SOHO is a project of international cooperation between ESA and NASA. The NRH (Nan\c cay Radioheliograph) is operated by the Observatoire de Paris and funded by the French research agency CNRS/INSU. The Radio Solar Telescope Network (RSTN) is a network of solar observatories maintained and operated by the U.S. Air Force Weather Agency.
\end{acks}

%-------------------------------------------------
%\bibliographystyle{spr-mp-sola-cnd}
%\bibliographystyle{spr-mp-sola}
%\bibliography{Ref/P01,Ref/P02,Ref/P03,Ref/P04,Ref/P05,Ref/P06,Ref/P07,Ref/BurstsII,Ref/BurstsIII,Ref/General,Ref/LNP725,Ref/SEE2007,Ref/Books,Ref/Density}

\input{02_SP_Revision_03.bbl}
%-------------------------------------------------
\clearpage
\appendix
%-------------------------------------------------
\input{tableglobal_Revision_03.tex} % Label {Table02}
\clearpage
%-------------------------------------------------
\input{TimelineRevision_03.tex} 
%-------------------------------------------------
\end{article}
\end{document}

%% file: tableglobal_Revision_03.tex
%-------------------------------------------------
\setcounter{table}{0}
\renewcommand{\thetable}{A\arabic{table}}
\setcounter{figure}{0}   
\renewcommand{\thefigure}{A\arabic{figure}}
\section{Comprehensive Catalogue of the ARTEMIS-IV Recordings}
%-------------------------------------------------
{In Table \ref{Table02} we provide a summary of the metric type-IV bursts with fine structure recorded by the ARTEMIS-IV/SAO receiver; accompanying metric busts and explosive events are included for comparison. Column 3 gives the type of activity (for SXR flares we give the flare class, IV corresponds to type-IV continuum). Column 4 gives the extrapolated launch time of CMEs, as specified in section \ref{analysis}. The location of the flare on the disk and the NOAA active region number are given in column 8. In the same column we give the Measurement Position Angle (MPA) of the CMEs  with their angular width in parenthesis as explained in section  \ref{Obs}. Comments and remarks, when necessary, have been added below the appropriate entries}. A collection of observational data, including dynamic spectra, is given in Appendix \ref{Details}.

%-------------------------------------------------
%\begin{landscape}
\newcommand{\NRHGAP}{NRH Data GAP}
\newcommand{\NRHOUT}{Outside NRH Daily Observations}
\nopagebreak
\begin{longtable}[c]{l l p{3.5cm} lll l}
\caption{ARTEMIS-IV Observations: Metric Radio Bursts and Associated LASCO CMEs and GOES SXR Flares.}
\setlength\tabcolsep{0.05pt} % default value: 5pt
\label{Table02}\\ \nopagebreak
%This is the header for the first page of the table...
\hline
\#  					&	Date		&	Activity					&	Start	&	Max	&	End		 		&	Position of 	\\ \nopagebreak
 						&				&						& \multicolumn{3}{c}{~~Universal~Time} 	&	 AR--CME 		\vspace{0.25mm}\\ \nopagebreak
\hline   \nopagebreak              	                                                                                	
\endfirsthead                                                                                                               	
%This is the header for the remaining page(s) of the table...                                                               	
\multicolumn{7}{c}{{\tablename} \thetable{}.:~ARTEMIS-IV Observations~--~Continued} \\ \nopagebreak                         	
\hline                                                                                                                      	
\#  					&	Date		&	Activity					&	Start	&	Max		&	End			&	Position of 	\\ \nopagebreak
 						&				&							& \multicolumn{3}{c}{~~Universal~Time} 	&	 AR--CME 		\vspace{0.25mm}\\ \nopagebreak
\hline    \nopagebreak                                      	
\endhead
%\hline \\ \nopagebreak
\rownumber 				& 30.06.1999	&	M1.9						&	11:24	&	11:30	&	11:45		&	S15E00--8603	\\ \nopagebreak
						&				&	CME 						&	11:18	&			&				&	03 (halo)		\\\nopagebreak
						&				&	SVTO 4995\,MHz				&	11:26	&	11:28	&	11:29		&					\\ \nopagebreak
						&				&	IV							&	11:26	&			&	11:38		&					\\ \nopagebreak
						&				&	II							&	11:26	&			&	11:28		&					\\ \nopagebreak
						&				&	III							&	11:26	&			&				&					\\ \nopagebreak						
						&				&	Spikes						&	11:10	&			&	11:27		&					\\ \nopagebreak
						&				&	Puls.						&	11:26	&			&	11:38		&					\\ \nopagebreak
						&				&	Zebra						&	11:29	&			&	11:29		&					\\ \nopagebreak
						&				&	Spikes						&	11:29	&			&	11:33		&					\\ \nopagebreak
						&				&	Spikes						&	11:37	&			&	11:45		&					 \\ \nopagebreak
\multicolumn{7}{p{12cm}}{Note: In the 11:10--11:22 UT Interval there are intermittent spike bursts; these become chains and clusters after 11:22 UT by the time of the energy release episode marked by the SVTO peak.}		\vspace{0.25mm}\\ \nopagebreak	
\hline\pagebreak[3]
\rownumber 				& 13.07.1999	&	C2.9						&	05:22	&	05:46	&	07:00		&	8628			\\ \nopagebreak
						&				&	CME 						&	05:43	&			&				&	46 (8) 			\\ \nopagebreak
						&				&	IV							&	05:57	&			&	05:59		&					\\ \nopagebreak
						&				&	II							&	06:02	&			&	06:05		&					\\ \nopagebreak
						&				&	SVTO 4995\,MHz				&	05:57	&	05:58	&	06:00		&					\\ \nopagebreak
						&				&	III							&	05:57	&			&				&					\\ \nopagebreak
						&				&	Spikes						&	05:55	&			&	6:05		&					\\ \nopagebreak
						&				&	Puls.						&	05:57	&			&	05:59		&					\\ \nopagebreak
						&				&	Zebra						&	05:57	&			&	05:58		&					\\ \nopagebreak
\multicolumn{7}{p{12cm}}{Note: The event started before the ARTEMIS-IV Observations. \NRHOUT.} \vspace{0.25mm}\\ \nopagebreak		
\hline\pagebreak[3]
\rownumber 				& 15.04.2000	&	M4.3						&	10:09	&	10:18	&	10:35		&	S22E29--8955	\\ \nopagebreak
						&				&	CME 						&	10:04	&			&				&	93 (176)		\\ \nopagebreak
						&				&	IV							&	10:16	&			&	10:55		&					\\ \nopagebreak
						&				&	III							&	10:12	&			&				&					\\ \nopagebreak
						&				&	III							&	10:17	&			&				&					\\ \nopagebreak
						&				&	Spikes						&	10:14	&			&	10:18		&					\\ \nopagebreak
						&				&	Puls.						&	10:16	&			&	10:27		&					\\ \nopagebreak
						&				&	Ropes						&	10:17	&			&	10:17		&					\\ \nopagebreak
						&				&	Spikes						&	10:20	&			&	10:22		&					\\ \nopagebreak
\multicolumn{7}{p{12cm}}{Note: No HXR or Microwave Data.}		\vspace{0.25mm}\\ \nopagebreak
\hline\pagebreak[3]
\rownumber 				& 15.04.2000	& 	C1.0						&	12:13	&	12:17	&	12:23		&	--				\\ \nopagebreak
						&				& 	C3.0						&	13:38	&	13:43	&	13:50		&	S22E29--8955	\\ \nopagebreak
						&				& 	M2.2						&	14:37	&	14:48	&	15:00		&	S23E28--8955	\\ \nopagebreak
						&				&	CME 						&	14:18	&			&				&	165	(43)		\\ \nopagebreak
						&				&	HXRS 45\,KeV				&	14:31	&	14:34	&	14:37		&	 				\\ \nopagebreak			
						&				&	IV							&	12:13	&			&	14:44		&					\\ \nopagebreak
						&				&	III							&	13:41	&			&				&					\\ \nopagebreak
						&				&	III							&	14:40	&			&				&					\\ \nopagebreak
						&				&	III							&	15:10	&			&				&					\\ \nopagebreak
						&				&	HXRS 45\,KeV				&	14:40	&	14:42	&	14:44		&					\\ \nopagebreak	
						&				&	Spikes						&	11:50	&			&	15:10		&					\\ \nopagebreak
						&				&	Puls.						&	14:39	&			&	14:43		&					\\ \nopagebreak
\multicolumn{7}{p{12cm}}{Note: Spikes started during the previous flare at 10:14 UT and lasted till the end of the day. There are also several type-III bursts during this time interval. Propably a type-IV burst was still present from the previous flare (event 03), but it is too faint on dynamic spectrum. \NRHGAP.}	\vspace{0.25mm}\\ \nopagebreak	
\hline\pagebreak[3]
\rownumber 				& 30.04.2000	&	C7.7						&	07:53	&	08:08	&	09:30		&	S11W18--8976	\\ \nopagebreak
						&				&	CME 						&	08:10	&			&				&	186	(104)		\\ \nopagebreak
						&				&	IV							&	08:00	&			&	08:30		&					\\ \nopagebreak
						&				&	SVTO 4995\,MHz				&	07:54	&	08:01	&	08:09		&					\\ \nopagebreak	
						&				&	II							&	07:55	&			&	08:02		&					\\ \nopagebreak
						&				&	III							&	07:56	&			&				&					\\ \nopagebreak
						&				&	Spikes						&	07:50	&			&	08:01		&					\\ \nopagebreak
						&				&	Fiber						&	07:58	&			&	08:02		&					\\ \nopagebreak
						&				&	Puls.						&	08:01	&			&	08:10		&					\\ \nopagebreak
						&				&	Spikes						&	08:08	&			&	08:22		&					\\ \nopagebreak
						&				&	Fiber						&	08:12	&			&	08:16		&					\\ \nopagebreak
						&				&	Ropes						&	08:20	&			&	08:20		&					\\ \nopagebreak
\multicolumn{7}{p{12cm}}{Note: Weak flux enhancement at 4995 MHz, peak uncertain. The type-IV burst in two parts; a drifting type-IV 07:59-08:07 UT from 500--200 MHz and a stationary continuum in the 600--250 MHz range; \NRHGAP.}\vspace{0.25mm}\\ \nopagebreak					
%\multicolumn{7}{p{12cm}}{Note: Weak flux enhancement at 4995 MHz, peak uncertain. The type-IV burst exhibits two parts; a drifting type-IV 07:59-08:07 UT from 500--200 MHz and a stationary continuum within the same interval in the 600--250 MHz range. \NRHGAP.}\vspace{0.25mm}\\ \nopagebreak					
\hline\pagebreak[3]
\rownumber 				& 11.07.2000	&	X1.0						&	12:12	&	13:10	&	14:30		&	N18E42--9077	\\ \nopagebreak
						&				&	CME 						&	12:33	&			&				&	62 (halo)		\\ \nopagebreak
						&				&	IV							&	12:36	&			&	15:20 		&					\\ \nopagebreak
						&				&	HXRS 45\,KeV				& 			&	12:58	&	 			&					\\ \nopagebreak	
						&				&	Spikes						&	12:38	&			&	13:16		&					\\ \nopagebreak
						&				&	Puls.						&	12:47	&			&	12:48		&					\\ \nopagebreak
						&				&	Fiber						&	12:55	&			&	12:56		&					\\ \nopagebreak
						&				&	Puls.						&	12:56	&			&	13:29		&					\\ \nopagebreak
						&				&	Fiber						&	13:03	&			&	13:04		&					\\ \nopagebreak
						&				&	Fiber						&	13:19	&			&	13:26		&					\\ \nopagebreak
						&				&	Spikes						&	13:22	&			&	13:29		&					\\ \nopagebreak
						&				&	Zebra						&	13:29	&			&	13:29		&					\\ \nopagebreak
						&				&	Fiber						&	13:31	&			&	13:40		&					\\ \nopagebreak
						&				&	Puls.						&	13:32	&			&	13:40		&					\\ \nopagebreak
						&				&	Zebra						&	13:35	&			&	13:35		&					\\ \nopagebreak
						&				&	Puls.						&	13:43	&			&	13:53		&					\\ \nopagebreak
						&				&	Fiber						&	13:48	&			&	13:54		&					\\ \nopagebreak						
\multicolumn{7}{p{12cm}}{Note: Double peak in HXR} \vspace{0.25mm}\\ \nopagebreak				
\hline\pagebreak[3]
\rownumber 				& 14.07.2000	&	X5.7						&	10:03	&	10:24	&	11:30		&	N22W07--9077	\\ \nopagebreak
						&				&	CME 						&	10:15	&			&				&	273 (halo)		\\ \nopagebreak
						&				&	IV							&	10:12	&			&	11:41 		&					\\ \nopagebreak
						&				&	II							&	10:11	&			&	10:37		&					\\ \nopagebreak
						&				&	III							&	10:10	&			&				&					\\ \nopagebreak
						&				&	HXRS 45\,KeV				& 	10:08	&			&	10:15		&					\\ \nopagebreak							
						&				&	TSRS 2695\,MHz				&	10:10	&	10:29	&	10:45		& 					\\ \nopagebreak	
						&				&	III							&	10:28	&			&				&					\\ \nopagebreak
						&				&	Puls.						&	10:10	&			&	10:38		&					\\ \nopagebreak
						&				&	Spikes						&	10:11	&			&	10:14		&					\\ \nopagebreak
						&				&	Spikes						&	10:18 	&			&	10:34		&					\\ \nopagebreak
						&				&	Lace						&	10:18	&			&	10:19		&					\\ \nopagebreak
						&				&	Fiber						&	10:33	&			&	11:25		&					\\ \nopagebreak
						&				&	Zebra						&	10:35	&			&	10:35		&					\\ \nopagebreak
						&				&	Puls.						&	10:40	&			&	10:59		&					\\ \nopagebreak
						&				&	Zebra						&	10:43	&			&	10:44		&					\\ \nopagebreak
						&				&	Spikes						&	11:04	&			&	11:12		&					\\ \nopagebreak
						&				&	Puls.						&	11:12	&			&	11:21		&					\\ \nopagebreak
\multicolumn{7}{p{12cm}}{Note: The type-IV burst appears in two parts: a structureless drifting continuum starting at 10:12 and a stationary type-IV starting at 10:10, becoming quite pronounced at 10:27  continuing up to 10:41. The TSRS 2695 MHz flux exhibits a number of peaks between 10:10--10:45 UT} \vspace{0.25mm}\\ \nopagebreak
\hline\pagebreak[3]
\rownumber 				& 14.07.2000	&	M1.7						&	12:50	&	12:57	&	13:10		&	S09W01--9002	\\ \nopagebreak
						&				&	TSRS 2695\,MHz				&	12:50	&	12:52	&	12:56		&	No CME			\\ \nopagebreak	
						&				&	IV							&	12:50	&			&	13:35		&					\\ \nopagebreak
						&				&	Fiber						&	12:45	&			&	12:48		&					\\ \nopagebreak
						&				&	Puls.						&	12:49	&			&	13:20		&					\\ \nopagebreak
						&				&	Fiber						&	12:50	&			&	12:51		&					\\ \nopagebreak
						&				&	Zebra						&	12:54	&			&	12:54		&					\\ \nopagebreak
						&				&	Fiber						&	12:56	&			&	12:59		&					\\ \nopagebreak
						&				&	Fiber						&	13:13	&			&	13:40		&					\\ \nopagebreak
						&				&	Spikes						&	12:49	&			&	12:51 		&					\\ \nopagebreak
						&				&	Spikes						&	13:07 	&			&	13:12		&					\\ \nopagebreak
						&				&	Zebra						&	13:14	&			&	13:14		&					\\ \nopagebreak
						&				&	Spikes						&	13:15	&			&	13:20		&					\\ \nopagebreak
\hline\pagebreak[3]
\rownumber 				& 14.07.2000	&	M3.7						&	13:44	&	14:00	&	14:30		&	N22W07--9077	\\ \nopagebreak
						&				&	III							&	13:47	&			&				&	No CME			\\ \nopagebreak
						&				&	TSRS 2695\,MHz				&	13:46	&	13:50	&	13:51		&		 			\\ \nopagebreak		
						&				&	Spikes						&	13:45	&			&	13:55		&					\\ \nopagebreak
						&				&	Puls.						&	13:49	&			&	13:58		&					\\ \nopagebreak
						&				&	III							&	13:55	&			&				&					\\ \nopagebreak
						&				&	TSRS 2695\,MHz				&	13:53	&	13:54	&	13:56		&	 				\\ \nopagebreak	
						&				&	IV							&	13:47	&			&	14:05		&					\\ \nopagebreak
						&				&	Zebra						&	13:50	&			&	13:50		&					\\ \nopagebreak
						&				&	Zebra						&	13:51	&			&	13:52		&					\\ \nopagebreak
						&				&	Zebra						&	13:54	&			&	13:54		&					\\ \nopagebreak
						&				&	Fiber						&	13:54	&			&	13:55		&					\\ \nopagebreak
\multicolumn{7}{p{12cm}}{Note: Double peak of TSRS 2695\,MHz flux.}		\vspace{0.25mm}\\ \nopagebreak					
\hline\pagebreak[3]
\rownumber 				& 19.09.2000	&	M5.1						&	08:06	&	08:26	&	08:42	&	N14 W46--9165	\\ \nopagebreak
 						&				&	CME 						&	08:08	&			&			&	283	(76)		\\ \nopagebreak
						&				&	IV							&	08:11	&			&	08:27	&					\\ \nopagebreak
						&				&	II							&	08:13	&			&	08:20	&					\\ \nopagebreak
 						&				&	III							&	08:13	&			&			&					\\ \nopagebreak
						&				&	Spikes						&	08:10	&			&	08:12	&					\\ \nopagebreak
						&				&	Puls.						&	08:11	&			&	08:12	&					\\ \nopagebreak
\multicolumn{7}{p{12cm}}{Note: No HXR or Microwave Data.}		\vspace{0.25mm}\\ \nopagebreak
 \hline\pagebreak[3]
\rownumber 				& 18.11.2000	&	M1.5						&	13:02	&	13:25	&	15:00	&	N11E37--9235	\\ \nopagebreak
						&				&	CME 						&	13:42	&			&			&	74 (120)		\\ \nopagebreak
						&				&	HXRS 45\,KeV				&	13:09	&	13:11	&	13:12	& 					\\ \nopagebreak	
						&				&	IV							&	13:13	&			&	13:42	&					\\ \nopagebreak
						&				&	II							&	13:12	&			&	13:17	&					\\ \nopagebreak
						&				&	Spikes						&	13:07 	&			&	13:30	&					\\ \nopagebreak
						&				&	Puls.						&	13:09	&			&	13:21	&					\\ \nopagebreak
						&				&	Zebra						&	13:11	&			&	13:11	&					\\ \nopagebreak
						&				&	Zebra						&	13:15	&			&	13:15	&					\\ \nopagebreak
						&				&	Zebra						&	13:16	&			&	13:16	&					\\ \nopagebreak
						&				&	Lace						&	13:22	&			&	13:22	&					\\ \nopagebreak
						&				&	Spikes						&	13:35 	&			&	13:40	&					\\ \nopagebreak
\hline\pagebreak[3]
\rownumber 				& 21.04.2003	&	M2.8						&	12:54	&	13:07	&	13:30	&	N18E02--10338	\\ \nopagebreak
						&				&	CME 						&	12:58	&			&			&	355 (163)		\\ \nopagebreak
						&				&	IV							&	13:04	&			&	13:09  	&					\\ \nopagebreak
						&				&	II							&	13:05	&			&	13:16	&					\\ \nopagebreak
						&				&	SVTO 4995\,MHz				&	13:03	&	13:04	&	13:10	&					\\ \nopagebreak								
						&				&	III							&	13:03	&			&			&					\\ \nopagebreak
						&				&	Spikes						&	13:03	&			&	13:07	&					\\ \nopagebreak
						&				&	Puls.						&	13:04	&			&	13:08	&					\\ \nopagebreak
						&				&	Lace						&	13:05	&			&	13:05	&					\\ \nopagebreak
						&				&	Zebra						&	13:06	&			&	13:07	&					\\ \nopagebreak
						&				&	Fiber						&	13:07	&			&	13:09	&					\\ \nopagebreak
						&				&	Spikes						&	13:08	&			&	13:10 	&					\\ \nopagebreak
						&				&	Puls.						&	13:11			&				&	13:13	&					\\ \nopagebreak
\multicolumn{7}{p{12cm}}{Note: Position Angle from CACTUS Catalogue. SVTO 4995\,MHz double peak. AR Localization from MDI.} \vspace{0.25mm}\\ \nopagebreak
\hline\pagebreak[3]
\rownumber 				& 26.10.2003	&	X1.2						&	05:57	&	06:54	&	09:00	&	S15E44--10486	\\ \nopagebreak
						&				&	CME 						&	06:13	&			&			&	108 (207)		\\ \nopagebreak
						&				&	IV							&	07:06	&			&	09:10	&					\\ \nopagebreak
						&				&	II							&	06:16	&			&	06:30	&					\\ \nopagebreak
						&				&	SVTO 4995\,MHz				&	06:12	&	 		&	08:00	&					\\ \nopagebreak	
						&				& RHESSI \mbox{50--100~KeV}		&	06:09	&	06:12	&	06:18	&	 				\\ \nopagebreak			
						&				&	Puls.						&	07:07	&			&	07:45	&					\\ \nopagebreak
						&				&	Spikes						&	07:11	&			&	07:12 	&					\\ \nopagebreak
						&				&	Lace						&	07:19	&			&	07:19	&					\\ \nopagebreak
						&				&	Zebra						&	07:22	&			&	07:23	&					\\ \nopagebreak
						&				& RHESSI \mbox{50--100~KeV}		&	08:30	&	08:32	&	08:34	&	 				\\ \nopagebreak
						&				&	Puls.						&	08:35	&			&	08:40	&					\\ \nopagebreak									
						&				& RHESSI \mbox{50--100~KeV}		&	08:40	&	08:41	&	08:42	&	 				\\ \nopagebreak									
\multicolumn{7}{p{12cm}}{Note: Probably Multiple Event which  began before start time of ARTEMIS-IV; SVTO 4995\,MHz flux exhibits multiple peaks. \NRHGAP.}	\vspace{0.25mm}\\ \nopagebreak	
\hline\pagebreak[3]
\rownumber 				& 28.10.2003	&	X17.2						&	09:51	&	11:10	&	12:40	&	S16E08--10486	\\ \nopagebreak
						&				&	CME 						&	10:34	&			&			&	124 (147)		\\ \nopagebreak
						&				&	IV							&	10:40	&			&	15:00 	&					\\ \nopagebreak
						&				&	II							&	11:03	&			&	11:11 	&					\\ \nopagebreak
						&				&	SVTO 4995\,MHz				&	11:03	&	 		&	11:25	&					\\ \nopagebreak	
						&				&	III							&	11:03	&			&			&					\\ \nopagebreak
						&				&	Spikes						&	10:33	&			&	11:41	&					\\ \nopagebreak
						&				&	Spikes						&	10:49	&			&	11:08	&					\\ \nopagebreak
						&				&	Puls.						&	11:01	&			&	12:26	&					\\ \nopagebreak
						&				& RHESSI \mbox{50--100~KeV}		&	11:06	&	11:14	&	11:25	&	 				\\ \nopagebreak			
						&				&	Lace						&	11:07  	&			&	11:09	&					\\ \nopagebreak
						&				&	Fiber						&	11:12	&			&	11:20	&					\\ \nopagebreak
						&				&	Fiber						&	11:43	&			&	11:44	&					\\ \nopagebreak
						&				&	Fiber						&	11:46	&			&	11:47	&					\\ \nopagebreak
						&				&	Zebra						&	11:50	&			&	11:50	&					\\ \nopagebreak
						&				&	Fiber						&	12:00	&			&	12:03	&					\\ \nopagebreak
						&				&	Spikes						&	14:14 	&			&	14:22	&					\\ \nopagebreak
\multicolumn{7}{p{12cm}}{Note: SVTO 4995\,MHz flux exhibits multiple peaks.} \vspace{0.25mm}\\ \nopagebreak
\hline\pagebreak[3]
\rownumber 				& 03.11.2003	&	X3.9						&	09:43	&	09:55	&	11:00	&	N08W77--10488	\\ \nopagebreak
						&				&	CME 						&	09:53	&			&			&	293 (103)		\\ \nopagebreak
						&				&	IV							&	09:57	&			&	10:35 	&					\\ \nopagebreak
						&				&	II							&	09:51	&			&	10:10 	&					\\ \nopagebreak
						&				&	III							&	09:49	&			&			&					\\ \nopagebreak
						&				&	III							&	09:51	&			&			&					\\ \nopagebreak				
						&				& RHESSI \mbox{100--300~KeV}	&	09:48	&	09:49	&	09:55	&	 				\\ \nopagebreak			
						&				&	SVTO 4995\,MHz				&	09:49	&	 		&	10:20	& 					\\ \nopagebreak	
						&				&	Spikes						&	09:48	&			&	09:48	&					\\ \nopagebreak
						&				&	Puls.						&	09:48	&			&	09:49	&					\\ \nopagebreak
						&				&	Fiber						&	09:48	&			&	09:48	&					\\ \nopagebreak
						&				&	Puls.						&	09:53	&			&	09:57	&					\\ \nopagebreak
						&				&	Lace						&	09:57   &			&	09:57	&					\\ \nopagebreak
\multicolumn{7}{p{12cm}}{Note: SVTO flux exhibits multiple peaks.}	\vspace{0.25mm}\\ \nopagebreak			
\hline\pagebreak[3]
\rownumber 				& 04.02.2004	&	C9.9						&	11:12	&	11:18	&	12:15	&	S07W49--10547	\\ \nopagebreak
						&				&	CME 						&	11:19	&			&			&		274 (33)	\\ \nopagebreak
						&				&	IV							&	11:19	&			&	11:26 	&					\\ \nopagebreak
						&				&	II							&	11:16	&			&	11:17 	&					\\ \nopagebreak
						&				&	III							&	11:14	&			&			&					\\ \nopagebreak
						&				&	SVTO 4995\,MHz				&	11:14	&	11:16	&	11:19	&					\\ \nopagebreak
						&				&	Puls.						&	11:16	&			&	11:17	&					\\ \nopagebreak
						&				&	Fiber						&	11:19	&			&	11:24	&					\\ \nopagebreak
						&				&	Puls.						&	11:19	&			&	11:23	&					\\ \nopagebreak
						&				&	Lace						&	11:19  	&			&	11:19	&					\\ \nopagebreak
						&				&	Zebra						&	11:20	&			&	11:20	&					\\ \nopagebreak
						&				&	Spikes						&	11:15 	&			&	11:20	&					\\ \nopagebreak
\multicolumn{7}{p{12cm}}{Note: \NRHGAP.} \vspace{0.25mm}\\ \nopagebreak					
\hline\pagebreak[3]
\rownumber 				& 25.03.2004	&	C3.7						&	12:01	&	12:12	&	12:20	&	N01W19--10577	\\ \nopagebreak
						&				&	IV							&	12:08	&			&	12:14 	&	LASCO Data GAP. \\ \nopagebreak
						&				&	SVTO 4995\,MHz				&	12:05	&	12:06	&	12:07	&					\\ \nopagebreak
						&				&	SVTO 4995\,MHz				&	12:08	&	12:10	&	12:12	&					\\ \nopagebreak
						&				&	Zebra						&	12:07	&			&	12:08	&					\\ \nopagebreak
						&				&	Puls.						&	12:08	&			&	12:13	&					\\ \nopagebreak
						&				&	Zebra						&	12:09	&			&	12:09	&					\\ \nopagebreak
						&				&	Zebra						&	12:11	&			&	12:11	&					\\ \nopagebreak
						&				&	Zebra						&	12:12	&			&	12:12	&					\\ \nopagebreak
						&				&	Spikes						&	12:13 	&			&	12:15 	&					\\ \nopagebreak
\multicolumn{7}{p{12cm}}{Note: \NRHGAP; No EIT Data, MDI is used for spatial localization of the active region. Double SVTO 4995\,MHz peak} \vspace{0.25mm}\\ \nopagebreak					
\hline\pagebreak[3]
\rownumber 				& 30.03.2004	&	C1.7						&	05:37	&	05:41	&	05:46	&	N15E05--10582	\\ \nopagebreak
						&				&	IV							&	05:43	&			&	05:49 	&	LASCO Data GAP. \\ \nopagebreak
						&				&	III							&	05:45	&			&			&					\\ \nopagebreak
						&				& RHESSI \mbox{12--25~KeV}		&	05:36	&	05:40	&	05:47	&	 				\\ \nopagebreak			
						&				& RHESSI \mbox{12--25~KeV}		&	06:00	&	06:05	&	06:10	&	 				\\ \nopagebreak			
						&				&	Puls.						&	05:43	&			&	05:45	&					\\ \nopagebreak
						&				&	Puls.						&	05:47	&			&	05:49	&					\\ \nopagebreak
						&				&	Spikes						&	05:48	&			&	05:50	&					\\ \nopagebreak
\multicolumn{7}{p{12cm}}{Note: \NRHOUT; No EIT Data, MDI is used for spatial localization of the active region. Multiple HXR peaks in two main groups indicating, probably, a double event.} \vspace{0.25mm}\\ \nopagebreak								
\hline\pagebreak[3]
\rownumber 				& 30.03.2004	&	C5.9						&	09:41	&	09:51	&	09:54	&	N15E06--10582	\\ \nopagebreak
						&				&	IV							&	09:45	&			&	09:55 	&	LASCO Data GAP. \\ \nopagebreak
						&				&	SVTO 4995\,MHz				&	09:44	&	09:45	&	09:46	&					\\ \nopagebreak
						&				&	SVTO 4995\,MHz				&	09:51	&	09:52	&	09:55	&					\\ \nopagebreak
						&				&	III							&	09:45	&			&			&					\\ \nopagebreak
						&				&	III							&	09:51	&			&			&					\\ \nopagebreak
						&				&	Puls.						&	09:44	&			&	09:47	&					\\ \nopagebreak
						&				&	Puls.						&	09:51	&			&	09:55	&					\\ \nopagebreak
						&				&	Spikes						&	09:40	&			&	09:41	&					\\ \nopagebreak
						&				&	Spikes						&	09:43	&			&	09:46	&					\\ \nopagebreak
						&				&	Spikes						&	09:48	&			&	09:49	&					\\ \nopagebreak
						&				&	Spikes						&	09:51	&			&	09:55	&					\\ \nopagebreak
						&				&	Zebra						&	09:54	&			&	09:55	&					\\ \nopagebreak
\multicolumn{7}{p{12cm}}{Note: No EIT Data, MDI is used for spatial localization of the active region. Multiple SVTO 4995\,MHz peaks in two main groups indicating, probably, a double event fron the same AR; the second part starts at $\approx$\,09:50 UT.} \vspace{0.25mm}\\ \nopagebreak		
\hline\pagebreak[3]
\rownumber 				& 30.03.2004	&	C4.7						&	12:54	&	13:00	&	13:06	&	N15E05--10582	\\ \nopagebreak
						&				&	CME 						&	GAP		&			&			&	LASCO Data GAP. \\ \nopagebreak
						&				&	IV							&	12:56	&			&	13:03 	&					\\ \nopagebreak
						&				&	III							&	12:56	&			&	13:03 	&					\\ \nopagebreak
						&				&	SVTO 4995\,MHz				&	12:56	&	12:57	&	12:58	&					\\ \nopagebreak
						&				&	SVTO 4995\,MHz				&	13:00	&	13:01	&	13:03	&					\\ \nopagebreak
						&				&	Puls.						&	12:56	&			&	13:03 	&					\\ \nopagebreak
						&				&	Fiber						&	12:57	&			&	12:58	&					\\ \nopagebreak
						&				&	Spikes						&	12:55	&			&	12:58	&					\\ \nopagebreak
						&				&	Spikes						&	13:02	&			&	13:04	&					\\ \nopagebreak
						&				&	Spikes						&	13:08	&			&	13:11	&					\\ \nopagebreak
\multicolumn{7}{p{12cm}}{Note: Following type-III group at 12:56, there is intermittent type-III activity up to 13:03 UT. No EIT Data, MDI is used for spatial localization of the active region. Multiple SVTO 4995\,MHz peaks.} \vspace{0.25mm}\\ \nopagebreak		
\hline\pagebreak[3]
\rownumber 				& 06.04.2004	&	M2.4						&	12:30	&	13:28	&	14:30	&	S18E15--10588	\\ \nopagebreak
						&				&	CME 						&	13:17	&			&			&	167 (halo)		\\ \nopagebreak
						&				&	IV							&	13:16	&			&	13:26 	&					\\ \nopagebreak
						&				&	III							&	13:16	&			&			&					\\ \nopagebreak
						&				&	SVTO 4995\,MHz				&	13:16	&	13:23	&	13:30	&					\\ \nopagebreak
						&				& RHESSI \mbox{50--100~KeV}		&	13:19	&	13:23	&	13:28	&	 				\\ \nopagebreak			
						&				&	Puls.						&	13:16	&			&	13:24	&					\\ \nopagebreak
						&				&	Zebra						&	13:22 	&			&	13:22	&					\\ \nopagebreak
						&				&	Fiber						&	13:25	&			&	13:26	&					\\ \nopagebreak
						&				&	Zebra						&	13:22 	&			&	13:22	&					\\ \nopagebreak
\multicolumn{7}{p{12cm}}{Note: Multiple SVTO 4995\,MHz peaks.} \vspace{0.25mm}\\ \nopagebreak					
\hline\pagebreak[3]                                             		
\rownumber 				& 13.07.2004	&	M5.4						&	08:40	&	08:48	&	10:15	&	N12W52--10646	\\ \nopagebreak
						&				&	CME 						&	08:45	&			&			&	326 (halo)		\\ \nopagebreak
						&				&	IV							&	08:50	&			&	09:48	&					\\ \nopagebreak
						&				&	II							&	08:48	&			&	08:57	&					\\ \nopagebreak
						&				&	III							&	08:46	&			&			&					\\ \nopagebreak
						&				&	SVTO 4995\,MHz				&	08:44	&	08:47	&	08:51	&					\\ \nopagebreak
						&				& RHESSI \mbox{12--25~KeV}		&	08:39	&	08:44	&	08:46	&	 				\\ \nopagebreak			
						&				&	Spikes						&	08:48 	&			&	09:14	&					\\ \nopagebreak
						&				&	Fiber						&	09:04	&			&	09:06	&					\\ \nopagebreak
						&				&	Fiber						&	09:27	&			&	09:36	&					\\ \nopagebreak
						&				&	Fiber						&	09:45	&			&	09:51	&					\\ \nopagebreak
\multicolumn{7}{p{12cm}}{Note: Position Angle from CACTUS. Multiple SVTO 4995\,MHz peaks in a single group.} 	\vspace{0.25mm}\\ \nopagebreak	
\hline\pagebreak[3]
\rownumber 				& 20.07.2004	&	M8.6						&	12:22	&	12:32	&	13:00	&	N11E34--10652	\\ \nopagebreak
						&				&	CME 						&	13:17	&			&			&	334 (halo)		\\ \nopagebreak
						&				&	IV							&	07:00	&			&	15:00 	&					\\ \nopagebreak
						&				&	SVTO 4995\,MHz				&	12:25	&	12:29	&	12:36	&					\\ \nopagebreak
						&				&	II							&	12:33	&			&	12:40 	&					\\ \nopagebreak
						&				&	Spikes						&	12:10 	&			&	14:50 	&					\\ \nopagebreak
						&				&	Puls.						&	12:10	&			&	14:50	&					\\ \nopagebreak
						&				&	Fiber						&	12:56	&			&	13:06	&					\\ \nopagebreak
						&				&	Fiber						&	13:15	&			&	13:16	&					\\ \nopagebreak
\multicolumn{7}{p{12cm}}{Note: The CME onset appears on the 195 \AA ~movies at AR 10652 bound to North West across the disk. Due to this, non radial, propagation the onset time is somewhat uncertain.} 	\vspace{0.25mm}\\ \nopagebreak		

\hline\pagebreak[3]
\rownumber 				& 21.07.2004	&	C8.9						&	05:05	&	05:21	&	06:35	&	N05E24--10652	\\ \nopagebreak
						&				&	CME 						&	05:11	&			&			&	165 (66)		\\ \nopagebreak
						&				&	IV							&	05:41	&			&	10:55 	&					\\ \nopagebreak
						&				&	III							&	07:49	&			&			&					\\ \nopagebreak
						&				&	III							&	07:58	&			&			&					\\ \nopagebreak
						&				&	III							&	08:15	&			&			&					\\ \nopagebreak
						&				&	SVTO 4995\,MHz				&	05:15	&	05:16	&	05:30	&					\\ \nopagebreak
						&				& RHESSI \mbox{12--25~KeV}		&	05:10	&	05:16	&	05:30	&	 				\\ \nopagebreak			
						&				&	Puls.						&	05:42	&			&	09:18	&					\\ \nopagebreak
						&				&	Spikes						&	05:30 	&			&	06:30 	&					\\ \nopagebreak
						&				&	Spikes						&	06:44 	&			&	06:56 	&					\\ \nopagebreak
						&				&	Spikes						&	07:17	&			&	07:28 	&					\\ \nopagebreak
						&				&	Spikes						&	07:34	&			&	07:45	&					\\ \nopagebreak
						&				&	Spikes						&	07:55	&			&	08:22 	&					\vspace{0.25mm}\\ \nopagebreak
\hline\pagebreak[3]                                          
\rownumber 				& 14.01.2005	&	C4.6						&	12:33	&	12:41	&	13:00	&	S07E05--10718	\\ \nopagebreak
% 						&				&	CME 						&	10:05	&			&			&	239 (31)		\\ \nopagebreak
						&				&	IV							&	12:37	&			&	12:48 	&	No CME			\\ \nopagebreak
						&				&	II							&	12:47	&			&	12:50 	&					\\ \nopagebreak
						&				&	III							&	12:37	&			&			&					\\ \nopagebreak
						&				&	SVTO 4995\,MHz				&	12:37	&	12:38	&	12:40	&					\\ \nopagebreak	
						&				&	III							&	12:46	&			&			&					\\ \nopagebreak
						&				& RHESSI \mbox{25--50~KeV}		&	12:35	&	12:39	&	12:42	&	 				\\ \nopagebreak			
						&				&	Puls.						&	12:37	&			&	12:46	&					\\ \nopagebreak
						&				&	Zebra						&	12:37 	&			&	12:37	&					\\ \nopagebreak
						&				& RHESSI \mbox{25--50~KeV}		&	12:42	&	12:43	&	12:55	&	 				\\ \nopagebreak						
						&				&	Zebra						&	12:43	&			&	12:43	&					\\ \nopagebreak
						&				&	Zebra						&	12:44 	&			&	12:44	&					\\ \nopagebreak
						&				&	SVTO 4995\,MHz				&	12:45	&	12:46	&	12:47	&					\\ \nopagebreak													
						&				&	Zebra						&	12:47	&			&	12:48	&					\\ \nopagebreak
						&				&	Spikes						&	12:47	&			&	12:52 	&	 				\\ \nopagebreak	
\multicolumn{7}{p{12cm}}{Note: Multiple HXR peaks.} \vspace{0.25mm}\\ \nopagebreak										
\hline\pagebreak[3]
\rownumber 				& 15.01.2005	&	M8.6						&	05:54	&	06:38	&	8:30	&	N16E04--10720	\\ \nopagebreak
						&				&	CME 						&	06:03	&			&			&	359 (halo)		\\ \nopagebreak
						&				&	IV							&	06:21	&			&	8:40	&					\\ \nopagebreak
						&				&	SVTO 4995\,MHz				&	05:56	&	06:29	&	07:10	&					\\ \nopagebreak
						&				&	Puls.						&	06:07	&			&	06:11	&					\\ \nopagebreak
						&				&	Spikes						&	06:18	&			&	06:24	&					\\ \nopagebreak
						&				&	Puls.						&	06:18	&			&	09:20	&					\\ \nopagebreak
						&				&	Spikes						&	06:38	&			&	06:40	&					\\ \nopagebreak
						&				&	Fiber						&	06:44	&			&	06:45	&					\\ \nopagebreak
						&				&	Spikes						&	06:46 	&			&	06:49	&					\\ \nopagebreak
						&				&	Fiber						&	06:49 	&			&	06:53	&					\\ \nopagebreak
						&				&	Fiber						&	07:05	&			&	08:26	&					\\ \nopagebreak
						&				&	Spikes						&	07:58 	&			&	08:05 	&					\\ \nopagebreak
						&				&	Zebra						&	07:59	&			&	07:59	&					\\ \nopagebreak
						&				&	Zebra						&	08:00	&			&	08:00	&					\\ \nopagebreak
						&				&	Zebra						&	08:03	&			&	08:03	&					\\ \nopagebreak
						&				&	Zebra						&	08:17	&			&	08:17	&					\\ \nopagebreak                             			
\multicolumn{7}{p{12cm}}{Note: Multiple SVTO 4995\,MHz peaks.} \vspace{0.25mm}\\ \nopagebreak					
\hline\pagebreak[3]
\rownumber 				& 17.01.2005	&	X3.8						&	6:59	&	09:52	&	11:00	&	N15W25--10720	\\ \nopagebreak
						&				&	CME 						&	09:06	&			&			&	334 (halo)		\\ \nopagebreak
						&				&	IV							&	08:40	&			&	10:46 	&					\\ \nopagebreak
						&				&	II							&	09:44	&			&	9:48	&					\\ \nopagebreak
						&				&	III							&	09:44	&			&			&					\\ \nopagebreak
						&				& RHESSI \mbox{100--300~KeV}	&	09:36	&	09:50	&	10:30	&	 				\\ \nopagebreak			
						&				&	SVTO 4995\,MHz				&	09:02 	&	09:29	&	10:30	&					\\ \nopagebreak
						&				&	Spikes						&	09:02 	&			&	09:09 	&					\\ \nopagebreak
						&				&	Puls.						&	09:02	&			&	10:27	&					\\ \nopagebreak
						&				&	Spikes						&	09:13 	&			&	09:22 	&					\\ \nopagebreak
						&				&	Fiber						&	09:15	&			&	09:54	&					\\ \nopagebreak
						&				&	Spikes						&	09:32 	&			&	09:34 	&					\\ \nopagebreak
						&				&	CME 						&	09:43	&			&			&	309 (halo)		\\ \nopagebreak				
						&				&	Fiber						&	09:59	&			&	10:49	&					\\ \nopagebreak
						&				&	Spikes						&	10:12	&			&	10:38 	&					\\ \nopagebreak
						&				&	Puls.						&	10:34	&			&	11:20	&					\\ \nopagebreak
						&				&	Fiber						&	11:09	&			&	12:02	&					\\ \nopagebreak
						&				&	Puls.						&	11:33	&			&	12:02	&					\\ \nopagebreak
						&				&	Zebra						&	09:20	&			&	09:20	&					\\ \nopagebreak
						&				&	Zebra						&	10:45 	&			&	10:45	&					\\ \nopagebreak
						&				&	Spikes						&	10:56 	&			&	11:40 	&					\\ \nopagebreak
						&				&	Zebra						&	10:58 	&			&	11:00	&					\\ \nopagebreak
						&				&	Spikes						&	11:51 	&			&	12:01 	&					\\ \nopagebreak
\multicolumn{7}{p{12cm}}{Note: Two CMEs in close succession with SXR flux rising in two stages \citep[see\,][]{Hillaris2011}.} \vspace{0.25mm}\\ \nopagebreak				
\hline\pagebreak[3]
\rownumber 				& 19.01.2005	&	X1.3						&	08:03	&	08:22	&	9:00	&	N15W51--10720	\\ \nopagebreak
						&				&	CME 						&	08:08	&			&			&	320 (halo)		\\\nopagebreak
						&				&	IV							&	08:07	&			&	9:00	&					\\ \nopagebreak
						&				&	II							&	08:11	&			&	8:18 	&					\\ \nopagebreak
						&				&	III							&	08:14	&			&			&					\\ \nopagebreak
						&				& RHESSI \mbox{100--300~KeV}	&	08:12	&	08:26	&	08:38	&	 				\\ \nopagebreak			
						&				&	SVTO 4995\,MHz				&	08:12	&	08:26	&	09:20	&					\\ \nopagebreak
						&				&	Spikes						&	08:05	&			&	08:10 	&					\\ \nopagebreak
						&				&	Puls.						&	08:06	&			&	08:16	&					\\ \nopagebreak
						&				&	Zebra						&	08:09	&			&	08:09	&					\\ \nopagebreak
						&				&	Spikes						&	08:12	&			&	08:15 	&					\\ \nopagebreak
						&				&	Fiber						&	08:20	&			&	08:22	&					\\ \nopagebreak
						&				&	Spikes						&	08:22 	& 			&	09:03	&					\\ \nopagebreak
						&				&	Puls.						&	08:39	&			&	09:38	&					\\ \nopagebreak
						&				&	Spikes						&	09:22	&			&	09:35 	&					\\ \nopagebreak
						&				&	Fiber						&	09:53	&			&	09:54	&					\\ \nopagebreak
						&				&	III							&	10:24	&			&			&					\\ \nopagebreak
						&				&	Puls.						&	10:23	&			&	10:25	&					\\ \nopagebreak
						&				&	Spikes						&	10:23	&			&	10:28 	&					\\ \nopagebreak
\multicolumn{7}{p{12cm}}{Note: Multiple SVTO 4995\,MHz peaks and probably double event with second part starting at $\approx$\,09:15 UT.} \vspace{0.25mm}\\ \nopagebreak
%% See Saldahna et al 2008, I cannot Remember Why !!!
\hline\pagebreak[3]
\rownumber 				& 20.01.2005	&	X7.1						&	06:36	&	07:01	&	07:45	&	N15W51--10720	\\ \nopagebreak
						&				&	CME 						&	06:08	&			&			&	288 (halo)		\\\nopagebreak
						&				&	IV							&	06:36	&			&	08:00	&					\\ \nopagebreak
						&				&	III							&	06:39	&			&			&					\\ \nopagebreak				
						&				&	II							&	06:44	&			&	06:49	&					\\ \nopagebreak
						&				&	III							&	06:44	&			&			&					\\ \nopagebreak
						&				&	III							&	06:57	&			&			&					\\ \nopagebreak	
						&				&	III							&	07:07	&			&			&					\\ \nopagebreak
						&				& RHESSI \mbox{25--50~KeV}		&	06:38	&	06:45	&	07:27	&	 				\\ \nopagebreak			
						&				&	SVTO 4995\,MHz				&	06:38	&	06:49	&	07:30	&					\\ \nopagebreak
						&				&	Spikes						&	06:36 	&			&	06:44 	&					\\ \nopagebreak
						&				&	II							&	06:56	&			&	06:58	&					\\ \nopagebreak				
						&				&	Puls.						&	06:42	&			&	06:47	&					\\ \nopagebreak
						&				&	Spikes						&	06:46	&			&	07:20 	&					\\ \nopagebreak
						&				&	Puls.						&	06:53	&			&	07:02	&					\\ \nopagebreak
						&				&	Fiber						&	06:53	&			&	06:59	&					\\ \nopagebreak
						&				&	Fiber						&	07:05	&			&	07:13	&					\\ \nopagebreak
						&				&	Puls.						&	07:12	&			&	07:30	&					\\ \nopagebreak
						&				&	Fiber						&	07:22	&			&	07:26	&					\\ \nopagebreak
						&				&	Zebra						&	07:28 	&			&	07:28	&					\\ \nopagebreak
						&				&	Fiber						&	07:32	&			&	07:56	&					\\ \nopagebreak
						&				&	Zebra						&	07:34	&			&	07:34	&					\\ \nopagebreak
						&				&	Puls.						&	07:35	&			&	07:50	&					\\ \nopagebreak
						&				&	Spikes						&	07:35	&			&	07:40 	&					\\ \nopagebreak			
\multicolumn{7}{p{12cm}}{Note: Double SVTO 4995\,MHz peak.} \vspace{0.25mm}\\ \nopagebreak
\hline\pagebreak[3]                                             		
\rownumber 				& 13.07.2005	&	M5.0						&	14:01	&	14:49	&	18:15	&	N11W90--10786	\\ \nopagebreak
						&				&	CME 						&	14:12	&			&			&	303 (halo)		\\\nopagebreak
						&				&	IV							&	13:56	&			&	14:23	&					\\ \nopagebreak
						&				& RHESSI \mbox{25--50~KeV}		&	14:12	&			&	    	&	 				\\ \nopagebreak			
						&				&	SVTO 4995\,MHz				&	14:02	&	14:18	&	14:34	&					\\ \nopagebreak
						&				&	Puls.						&	13:52	&			&	14:12	&					\\ \nopagebreak
						&				&	Zebra						&	13:58	&			&	13:59	&					\\ \nopagebreak
						&				&	Fiber						&	14:03 	&			&	14:06	&					\\ \nopagebreak
						&				&	Zebra						&	14:08	&			&	14:08	&					\\ \nopagebreak
						&				&	Spikes						&	14:09	&			&	14:11	&					\\ \nopagebreak
						&				&	Puls.						&	14:16	&			&	14:22	&					\\ \nopagebreak
						&				&	Zebra						&	14:18 	&			&	14:18	&					\\ \nopagebreak
						&				&	Zebra						&	14:19 	&			&	14:19	&					\\ \nopagebreak
						&				&	Zebra						&	14:21 	&			&	14:21	&					\\ \nopagebreak
						&				&	Zebra						&	14:22	&			&	14:22	&					\\ \nopagebreak
\multicolumn{7}{p{12cm}}{Note: No EIT Data, MDI is used for spatial localization of the active region. Activity Extends beyond ARTEMIS-IV Observation Period.} \vspace{0.25mm}\\ \nopagebreak
\hline\pagebreak[3]                                        		
\rownumber 				& 14.07.2005	&	X1.2						&	10:16	&	10:55	&	12:00	&	N11W90--10786	\\ \nopagebreak
						&				&	CME 						&	10:27	&			&			&	296 (halo)		\\\nopagebreak
						&				&	IV							&	10:26	&			&	12:03	&					\\ \nopagebreak
						&				&	III							&	10:32	&			&			&					\\ \nopagebreak
						&				&	III							&	10:38	&			&			&					\\ \nopagebreak
						&				& RHESSI \mbox{25--50~KeV}		&	10:25	&	10:27	&	10:29	&	 				\\ \nopagebreak			
						&				&	SVTO 4995\,MHz				&	10:30	&	10:35	&	11:25	&					\\ \nopagebreak						
						&				& RHESSI \mbox{25--50~KeV}		&	11:00	&	11:04	&	11:25	&	 				\\ \nopagebreak			
						&				&	Puls.						&	10:18	&			&	11:14	&					\\ \nopagebreak
						&				&	Spikes						&	10:32	&			&	10:34	&					\\ \nopagebreak
						&				&	Spikes						&	10:40 	&			&	10:55	&					\\ \nopagebreak
						&				&	Fiber						&	11:10 	&			&	12:12 	&					\\ \nopagebreak
						&				&	Puls.						&	11:26	&			&	11:48	&					\\ \nopagebreak
\multicolumn{7}{p{12cm}}{Note: \NRHGAP. Multiple SVTO 4995\,MHz peaks.}		\vspace{0.25mm}\\ \nopagebreak
\hline\pagebreak[3]
\rownumber 				& 30.07.2005	&	X1.3						&	06:17	&	06:35	&	07:30	&	N12E60--10792	\\ \nopagebreak
						&				&	CME 						&	06:21	&			&			&	050 (halo)		\\\nopagebreak
						&				&	IV							&	06:21	&			&	06:44	&					\\ \nopagebreak
						&				&	II							&	06:26	&			&	06:32	&					\\ \nopagebreak
						&				&	III							&	06:28	&			&			&					\\ \nopagebreak
						&				&	SVTO 4995\,MHz				&	06:20	&	06:32	&	06:45	&					\\ \nopagebreak
						&				& RHESSI \mbox{100--300~KeV}	&	06:27	&	06:32	&	06:45	&	 				\\ \nopagebreak			
						&				&	Spikes						&	06:20	&			&	06:23 	&					\\ \nopagebreak
						&				&	Spikes						&	06:26	&			&	06:39 	&					\\ \nopagebreak
						&				&	Puls.						&	06:29	&			&	06:36	&					\\ \nopagebreak
\multicolumn{7}{p{12cm}}{Note: \NRHOUT. Multiple SVTO 4995\,MHz peaks} 	\vspace{0.25mm}\\ \nopagebreak
\hline\pagebreak[3]
\rownumber 				& 22.8.2005		&	M2.6						&	00:44	&	01:33	&	09:00		& 	S11W65--10798		\\ \nopagebreak
						&				&	CME 						&	05:09	&			&				&	222 (56)			\\\nopagebreak
						&				&	IV							&	05:43	&			&	6:30		&						\\ \nopagebreak
						&				& RHESSI \mbox{6--12~KeV}		&	05:50	&	06:01	&	06:30		&		 				\\ \nopagebreak			
						&				&	Spikes						&	05:40 	&			&	05:46		&						\\ \nopagebreak
						&				&	Puls.						&	05:52 	&			&	05:54		&						\\ \nopagebreak
						&				&	Zebra						&	05:53 	&			&	05:54		&						\\ \nopagebreak
						&				&	Zebra						&	05:59 	&			&	06:03		&						\\ \nopagebreak
						&				&	Spikes						&	06:14	&			&	06:14		&						\\ \nopagebreak
\multicolumn{7}{p{12cm}}{Note: Partial Observation of the end of a type-IV Continuum starting at about 00:44 UT,as reported by the Culgoora Radio--Spectrograph. \NRHOUT.} \vspace{0.25mm}\\ \nopagebreak

\hline\pagebreak[3]
\rownumber 				& 23.8.2005		&	M2.7						&	14:19	&	14:44 	&	16:08		& S11W65--10798			\\ \nopagebreak
						&				&	CME 						&	14:45	&			&				&	230 (halo)			\\\nopagebreak
						&				&	IV							&	14:26	&			&	15:02		&						\\ \nopagebreak
						&				&	SVTO 4995\,MHz				&	14:23	&	14:49	&	15:20		&						\\ \nopagebreak
						&				& RHESSI \mbox{50--100~KeV}		&	14:23	&	14:38	&	15:10 		&						\\ \nopagebreak			
						&				&	Spikes						&	14:26	&			&	15:00 		&						\\ \nopagebreak
\multicolumn{7}{p{12cm}}{Note: Activity Extends beyond ARTEMIS-IV Observation Period. Multiple HXR and SVTO 4995\,MHz peaks} 	\vspace{0.25mm}\\ \nopagebreak
\hline\pagebreak[3]
\rownumber 				& 12.02.2010	&	M8.3						&	11:19	&	11:26	&	11:28	&	N26E11--11046	\\ \nopagebreak
						&				&	CME 						&	11:18	&			&			&	044 (halo)		\\\nopagebreak
						&				&	IV							&	11:26	&			&	11:32	&					\\ \nopagebreak
						&				&	IV							&	12:00	&			&	12:45	&					\\ \nopagebreak
						&				&	II							&	11:26	&			&	11:31	&					\\ \nopagebreak
						&				&	III							&	11:25	&			&			&					\\ \nopagebreak
						&				&	III							&	11:29	&			&			&					\\ \nopagebreak
						&				& RHESSI \mbox{50--100~KeV}		&	11:21	&	11:26	&	11:36	&	 				\\ \nopagebreak			
						&				&	Spikes						&	11:24 	&			&	11:25 	&					\\ \nopagebreak
						&				&	Spikes						&	12:00 	&			&	12:12 	&					\\ \nopagebreak						
						&				&	Puls.						&	11:25	&			&	11:28	&					\vspace{0.25mm}\\ \nopagebreak
\hline\pagebreak[3]
\rownumber 				& 01.08.2010	&	C3.2						&	07:55	&	08:26	&	09:35	&	N20E36--11092	\\ \nopagebreak
						&				&	CME 						&	08:27	&			&			&	084 (halo)		\\\nopagebreak
						&				&	IV							&	08:06	&			&	10:00	&					\\ \nopagebreak
						&				& RHESSI \mbox{100--300~KeV}	&	08:00	&	08:33	&	08:48	&	 				\\ \nopagebreak			
						&				&	 Puls.						&	08:00	&			&	08:03	&					\\ \nopagebreak
						&				&	 Spikes						&	08:08 	&			&	08:09 	&					\\ \nopagebreak
						&				&	 Spikes						&	08:18 	&			&	08:22 	&					\\ \nopagebreak
						&				&	 Puls.						&	08:16	&			&	09:26	&					\\ \nopagebreak
						&				&	 Fiber						&	08:26 	&			&	09:36 	&					\\ \nopagebreak
						&				&	 Spikes						&	09:30 	&			&	09:50 	&					\\ \nopagebreak
						&				&	 Fiber						&	09:47 	&			&	09:48 	&					\\ \nopagebreak
						&				&	 Spikes						&	10:12 	&			&	10:20 	&					\\ \nopagebreak
\multicolumn{7}{p{12cm}}{Note: \NRHGAP.}		\vspace{0.25mm}\\ \nopagebreak
%% type-IIIs, well  Outside the type-IV Duration
\hline
%----------------------------------------------------------------------------------------------------------------------------
\end{longtable}
%\end{landscape}

%% file: TimelineRevision_03.tex
%-------------------------------------------------
\setcounter{table}{0}
\renewcommand{\thetable}{B\arabic{table}}
\setcounter{figure}{0}   
\renewcommand{\thefigure}{B\arabic{figure}} 
\newcommand{\Ht}{0.80\textheight} 
\section{Details on the ARTEMIS-IV Recordings and Accompanying Data} \label{Details}
%-------------------------------------------------
In the following figures we present dynamic spectra and other observational data for the~36 events used in the compilation of table~\ref{Table02} and figures~\ref{AllFiberTimeLineFb},~\ref{AllFiberTimeLinePuls}~and~\ref{AllFiberTimeLineSp}. They are intended to show the temporal and spatial association of type IV continua and their fine structure with flares, microwave and HXR bursts and CMEs, as detailed in section \ref{Obs}. In each Figure we give:

\begin{itemize}

\item{The upper panel shows medium resolution dynamic spectra recorded by the ARTEMIS-IV/ASG in the 650--20 MHz range (cadence of 10~samples/sec); in the 270-450 MHz range the ARTEMIS-IV/SAO high sensitivity--high time resolution spectrum is overlaid (sampling rate 100~samples/sec).}

\item{Soft X--Ray~(SXR)~light curves, obtained from the~Geostationary Operational Environmental Satellites (GOES). The CME Onset times are marked with arrows; they were estimated from the LASCO movies using the linear regression and are included in the on--line LASCO event lists.}

\item{Hard X--Ray~(HXR)~light curves from the Reuven Ramaty High Energy Solar Spectroscopic Imager (RHESSI) archive for events after the beginning of 2003; prior to 2003, HXR data are from the MTI/HXRS and BATSE/GRP experiments.}

\item{Centimetric Radio Flux; 4.995 GHz radio flux profiles from the Radio Solar Telescope Network (RSTN) or, in a few events, from the 2.695 GHz channel of the Trieste solar Radio System (TSTS).}

\item{Timeline plot; relative timing of Pulsations, fiber bursts, zebra stripes and Spike groups associated with~type-IV~continuum.}

\item{Type IV--flare--CME positions; these include, from left to the right: Nan\c cay Radioheliograph (NRH) half power contours at 164, 236, 327 MHz, 410 and 432 MHz (when available), overlaid on Extreme Ultraviolet Imaging Telescope (EIT) images. When EIT data were not available, MDI images were used to identify the associated active region. The flare position from the NOAA/SGD catalogs is schematically shown in the middle panel, for comparison with the direction of the CME launch to the right panel. This is marked graphically in the right panel, which shows the CME measurement position angle (MPA) and, for non-halo CMEs, the angular width, both from the LASCO coronograph event lists and the CACTUS CME catalogue.}

\end{itemize}
%-------------------------------------------------
%-------------------------------------------------
\newcounter{RNUMs}
\regtotcounter{RNUMs} % In order to Use the ToTCount
\newcommand\RNUM{\stepcounter{RNUMs}\arabic{RNUMs}}
%-------------------------------------------------
\begin{table} [t!]

\begin{center}
\caption{List of Event Dates and  Start--End Times.}
\label{Table03}
\begin{tabular}[c]{r c  cc || r c  cc}
\hline
\#  			&	Date		&	Start			&	End		 		& \#  			&	Date		&	Start			&	End		 		\\ 
 				&				& \multicolumn{2}{l}{Universal~Time} 	& 				&				& \multicolumn{2}{l}{Universal~Time} 	\\                 	                                                            	
\hline
%\hline \\ 
\RNUM		& 30.06.1999	&	11:24	&	11:45	& \RNUM		& 13.07.1999	&	05:22	&	07:00	\\ 
\RNUM		& 15.04.2000	&	10:04	&	10:55	& \RNUM		& 15.04.2000	& 	12:13	&	14:43	\\ 
\RNUM		& 30.04.2000	&	07:53	&	09:30	& \RNUM		& 11.07.2000	&	12:12	&	15:20 	\\ 
\RNUM		& 14.07.2000	&	10:03	&	11:30	& \RNUM		& 14.07.2000	&	12:50	&	13:40	\\ 
\RNUM		& 14.07.2000	&	13:44	&	14:30	& \RNUM		& 19.09.2000	&	08:06	&	08:42	\\ 
\RNUM		& 18.11.2000	&	13:02	&	15:00	& \RNUM		& 21.04.2003	&	12:54	&	13:30	\\ 
\RNUM		& 26.10.2003	&	05:57	&	09:10	& \RNUM		& 28.10.2003	&	09:51	&	15:00	\\ 
\RNUM		& 03.11.2003	&	09:43	&	11:00	& \RNUM		& 04.02.2004	&	11:12	&	12:15	\\ 
\RNUM		& 25.03.2004	&	12:01	&	12:20	& \RNUM		& 30.03.2004	&	05:37	&	06:10	\\ 
\RNUM		& 30.03.2004	&	09:41	&	09:55	& \RNUM		& 30.03.2004	&	12:54	&	13:11	\\ 
\RNUM		& 06.04.2004	&	12:30	&	14:30	& \RNUM		& 13.07.2004	&	08:39	&	10:15	\\ 
\RNUM		& 20.07.2004	&	12:10	&	15:00	& \RNUM		& 21.07.2004	&	05:05	&	10:55 	\\
\RNUM		& 14.01.2005	&	12:33	&	13:00	& \RNUM		& 15.01.2005	&	05:54	&	08:40	\\                             			
\RNUM		& 17.01.2005	&	06:59	&	12:00	& \RNUM		& 19.01.2005	&	08:03	&	10:28 	\\
\RNUM		& 20.01.2005	&	06:36	&	07:56	& \RNUM		& 13.07.2005	&	14:01	&	18:15	\\ 
\RNUM		& 14.07.2005	&	10:16	&	12:12	& \RNUM		& 30.07.2005	&	06:17	&	07:30	\\  
\RNUM		& 22.08.2005	&	00:44	&	09:00	& \RNUM		& 23.08.2005	&	14:19	&	16:08	\\ 
\RNUM		& 12.02.2010	&	11:18	&	12:45	& \RNUM		& 01.08.2010	&	07:55	&	10:20 	\\ 
\hline
%----------------------------------------------------------------------------------------------------------------------------
\end{tabular}
\end{center}

\end{table}
%-------------------------------------------------
{\noindent \bl  A concise list of the dates and event start and end times is given in Table \ref{Table03}}

\clearpage

\renewcommand{\event} {30 June 1999}
%--------------------------------------------
\begin{figure}
\begin{center}% trim=0cm 1cm  0cm 1cm,clip,height=\Ht, 
\includegraphics[width=\textwidth]{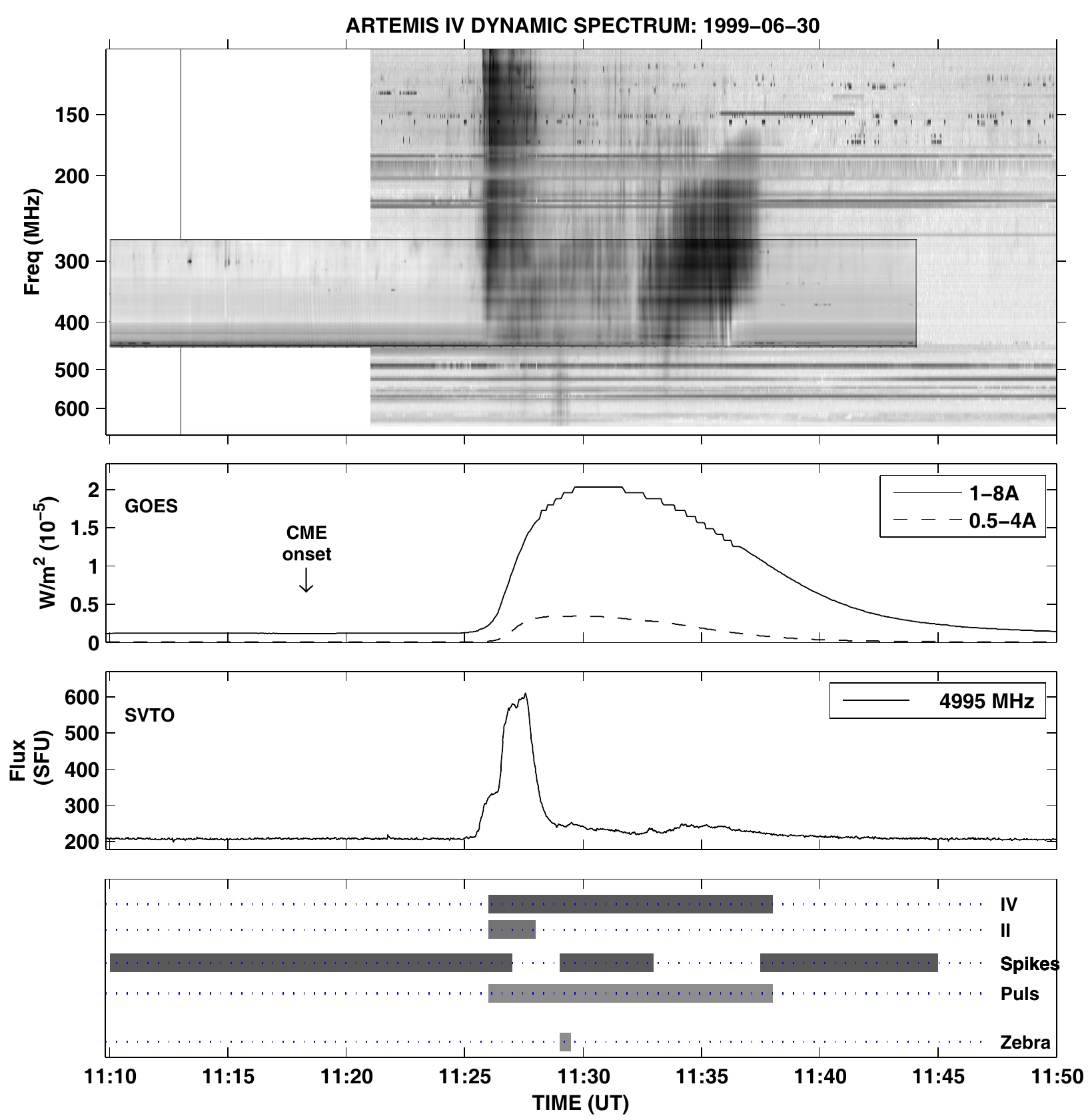}
\includegraphics[trim=0cm 0.0cm  0.0cm 4.0cm,clip,width=0.85\textwidth]{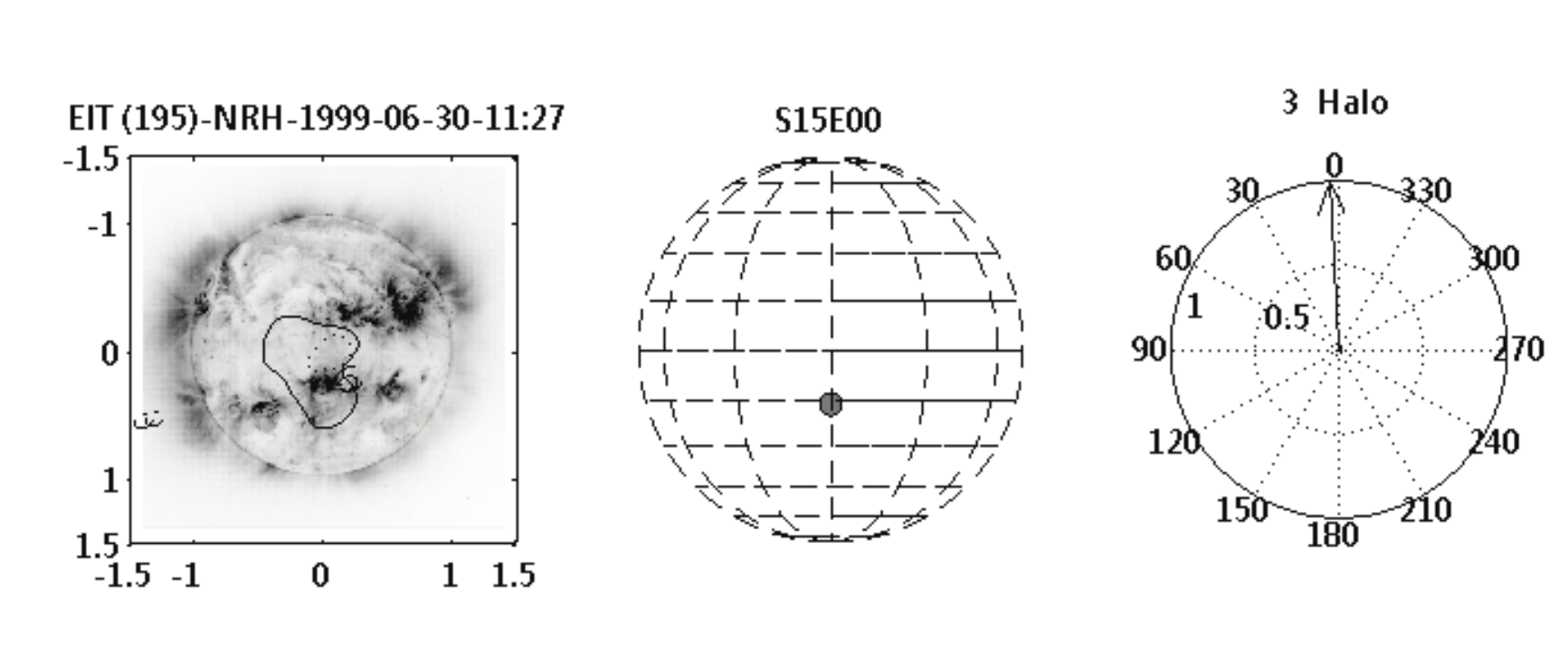}% 
\caption{\event}
\end{center}
\end{figure} % 
%--------------------------------------------
\clearpage
\renewcommand{\event} {13 July 1999}
%--------------------------------------------
\begin{figure}
\begin{center}% trim=0cm 1.5cm  0cm 1.5cm,clip,height=\Ht, 
\includegraphics[width=\textwidth]{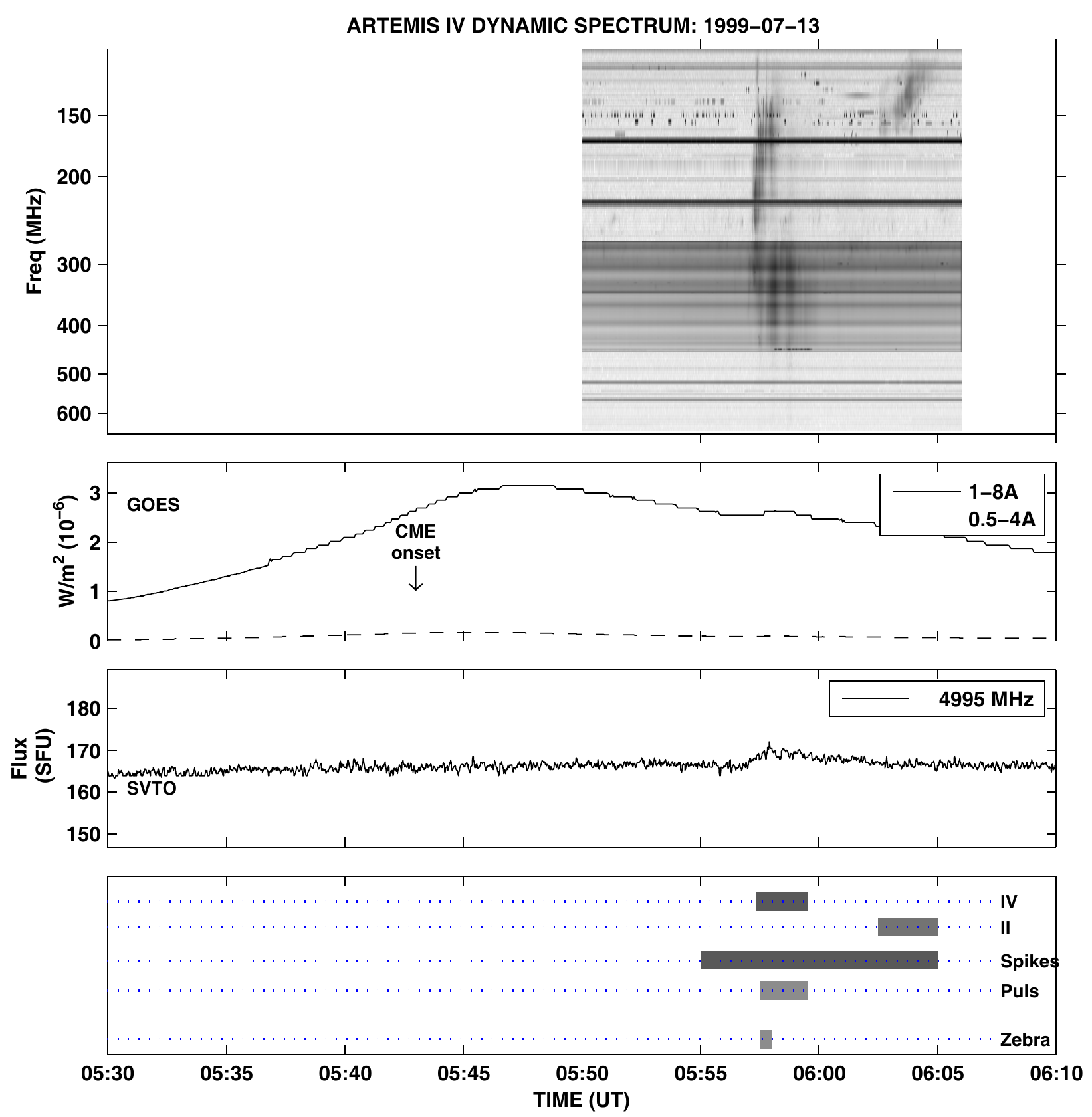}
\includegraphics[trim=0cm 0.0cm  0.0cm 4.0cm,clip,width=0.85\textwidth]{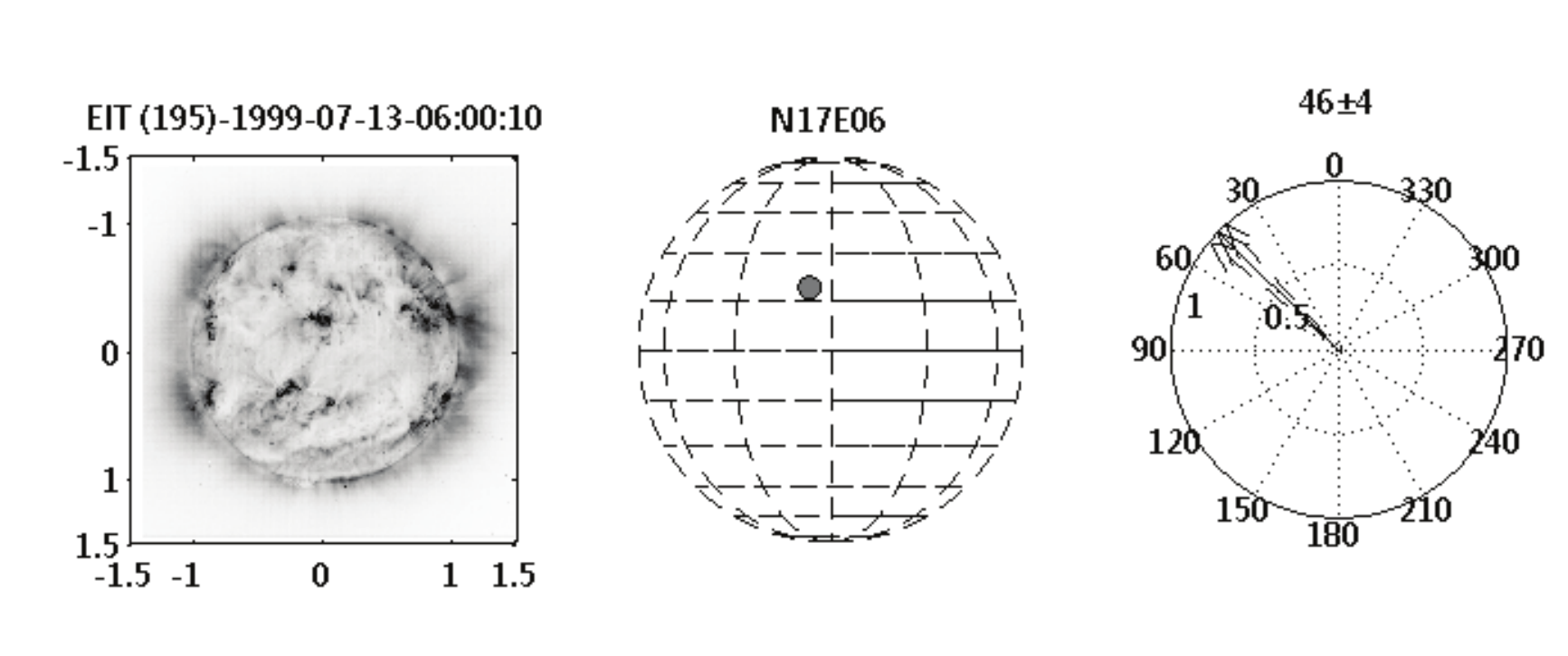}%
\caption{\event}
\end{center}
\end{figure} % 
%--------------------------------------------
\clearpage
\renewcommand{\event} {15 April 2000 Event(A)}
%--------------------------------------------
\begin{figure}
\begin{center}% trim=0cm 1cm  0cm 1cm,clip,height=\Ht, 
\includegraphics[width=\textwidth]{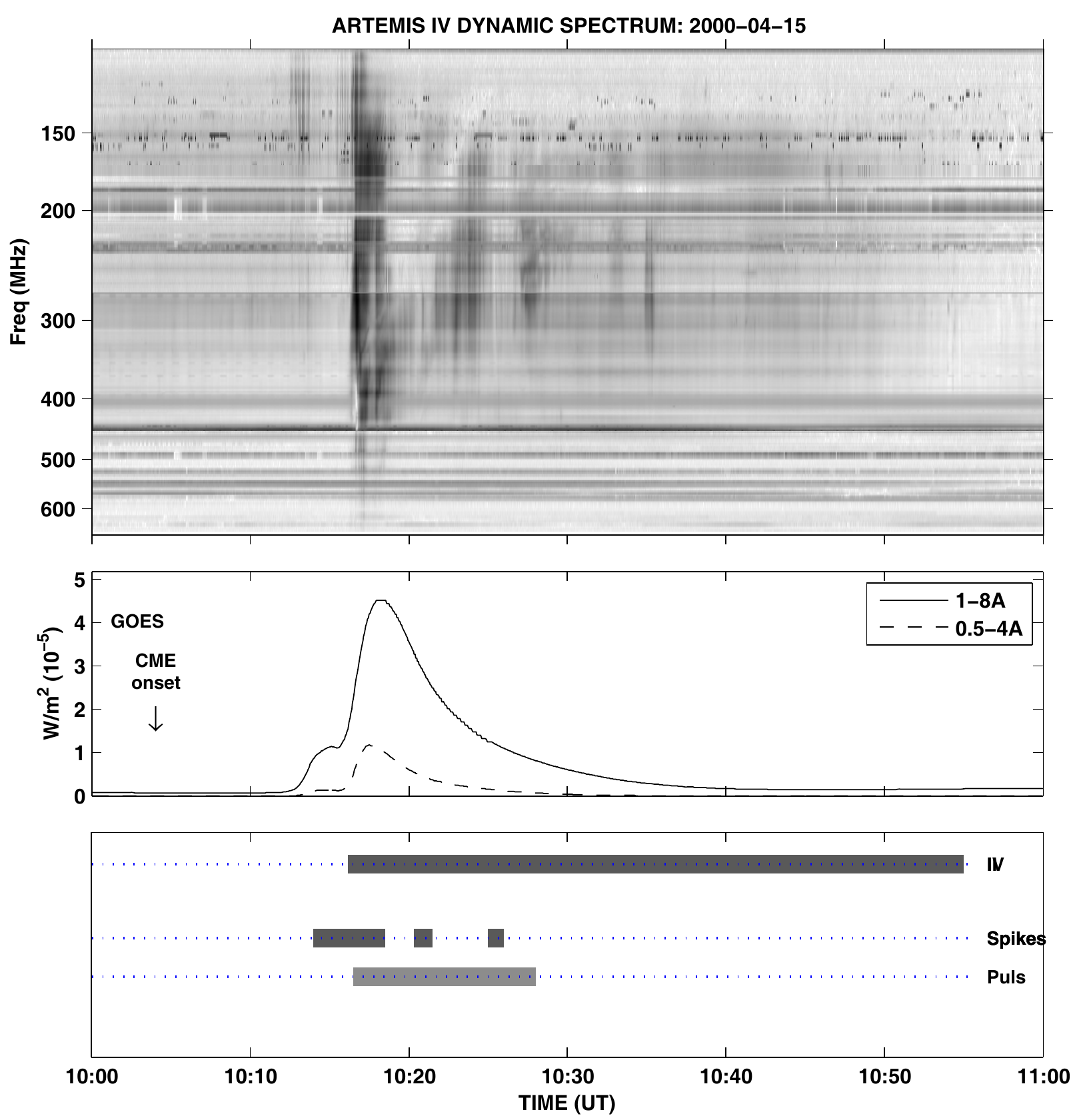}
\includegraphics[trim=0cm 0.0cm  0.0cm 4.0cm,clip,width=0.85\textwidth]{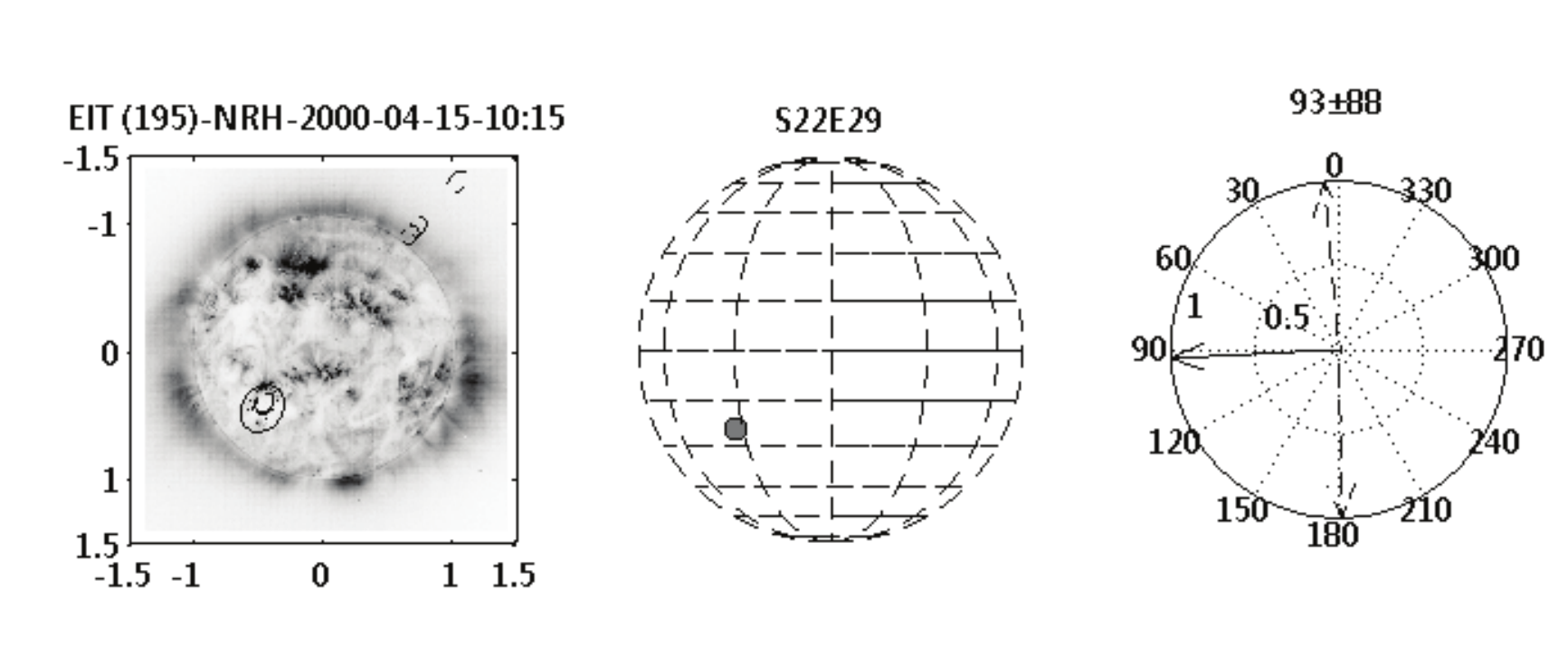}
\caption{\event}
\end{center}
\end{figure} % 
%--------------------------------------------
\clearpage
\renewcommand{\event} {15 April 2000 Event(B)}
%--------------------------------------------
\begin{figure}
\begin{center}% trim=0cm 1cm  0cm 1cm,clip,
\includegraphics[width=\textwidth]{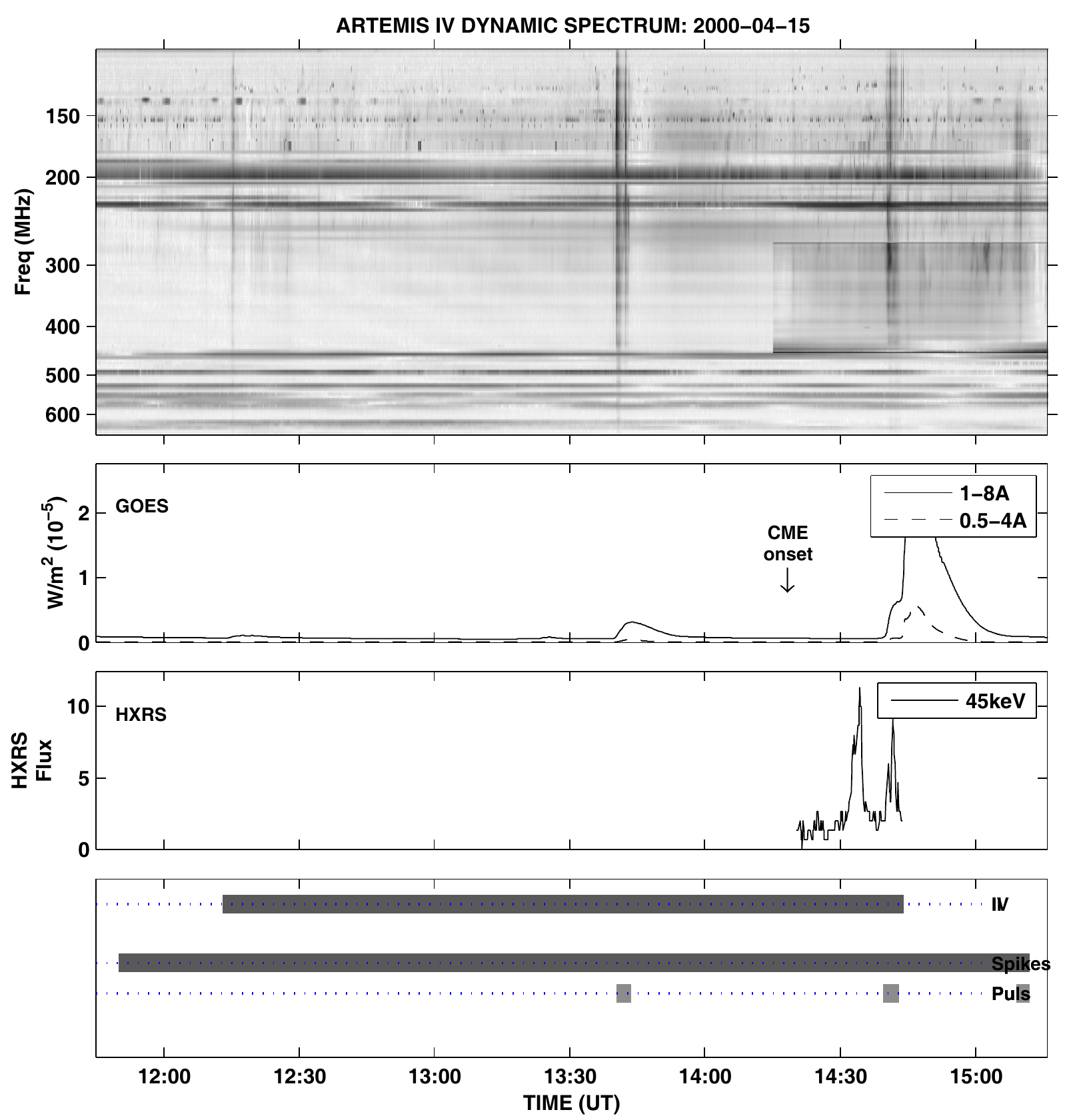}
\includegraphics[trim=0cm 0.0cm  0.0cm 4.0cm,clip,width=0.85\textwidth]{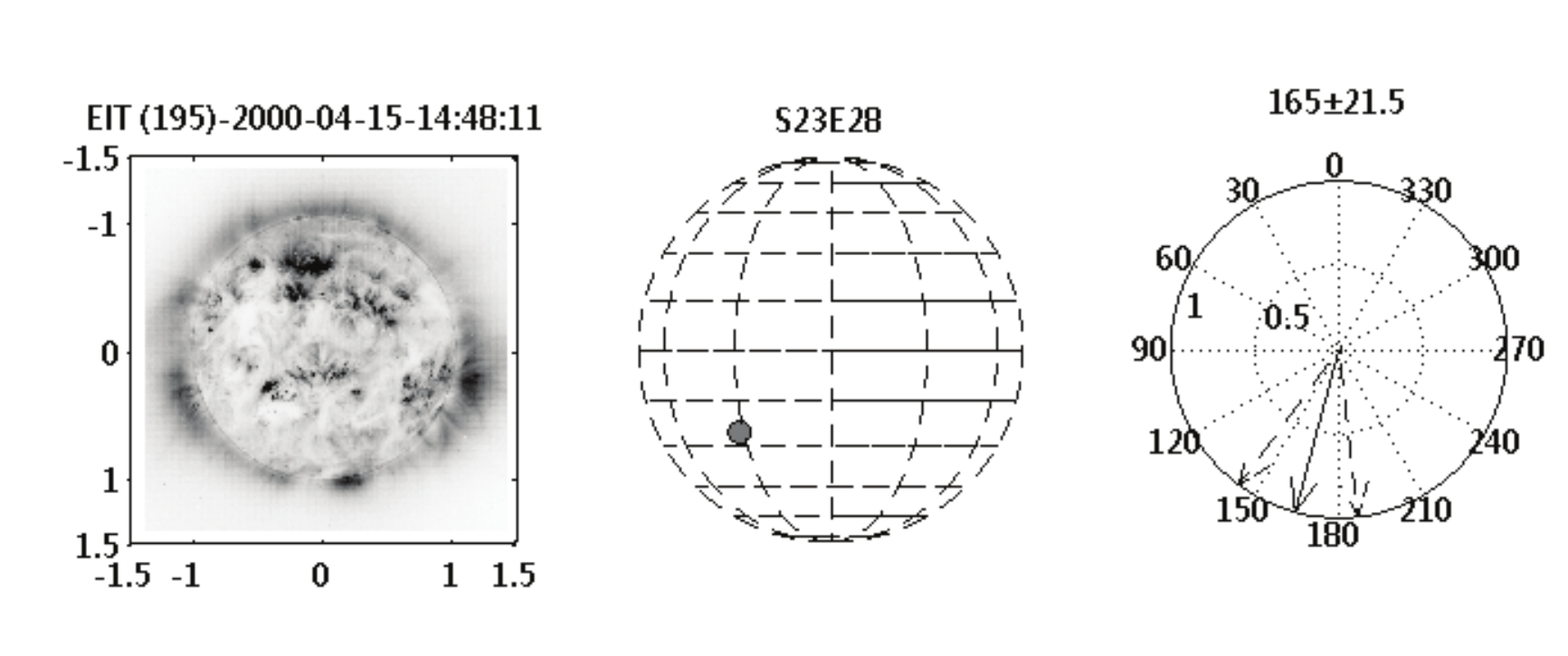}
\caption{\event}
\end{center}
\end{figure} % 
%--------------------------------------------
\clearpage
\renewcommand{\event} {30 April 2000}
%--------------------------------------------
\begin{figure}
\begin{center}% trim=0cm 1cm  0cm 1cm,clip,
\includegraphics[width=\textwidth]{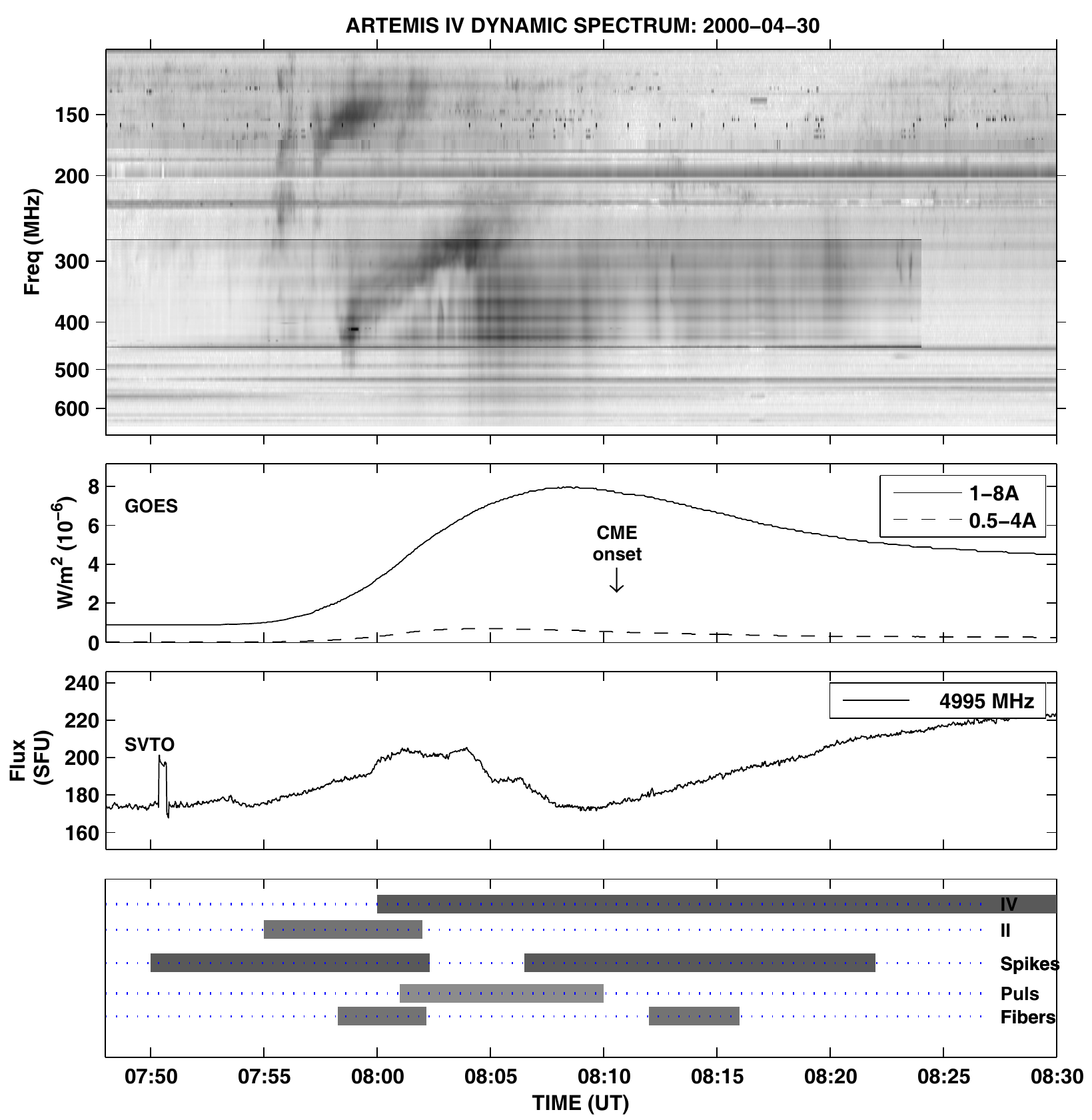}
\includegraphics[trim=0cm 0.0cm  0.0cm 4.0cm,clip,width=0.85\textwidth]{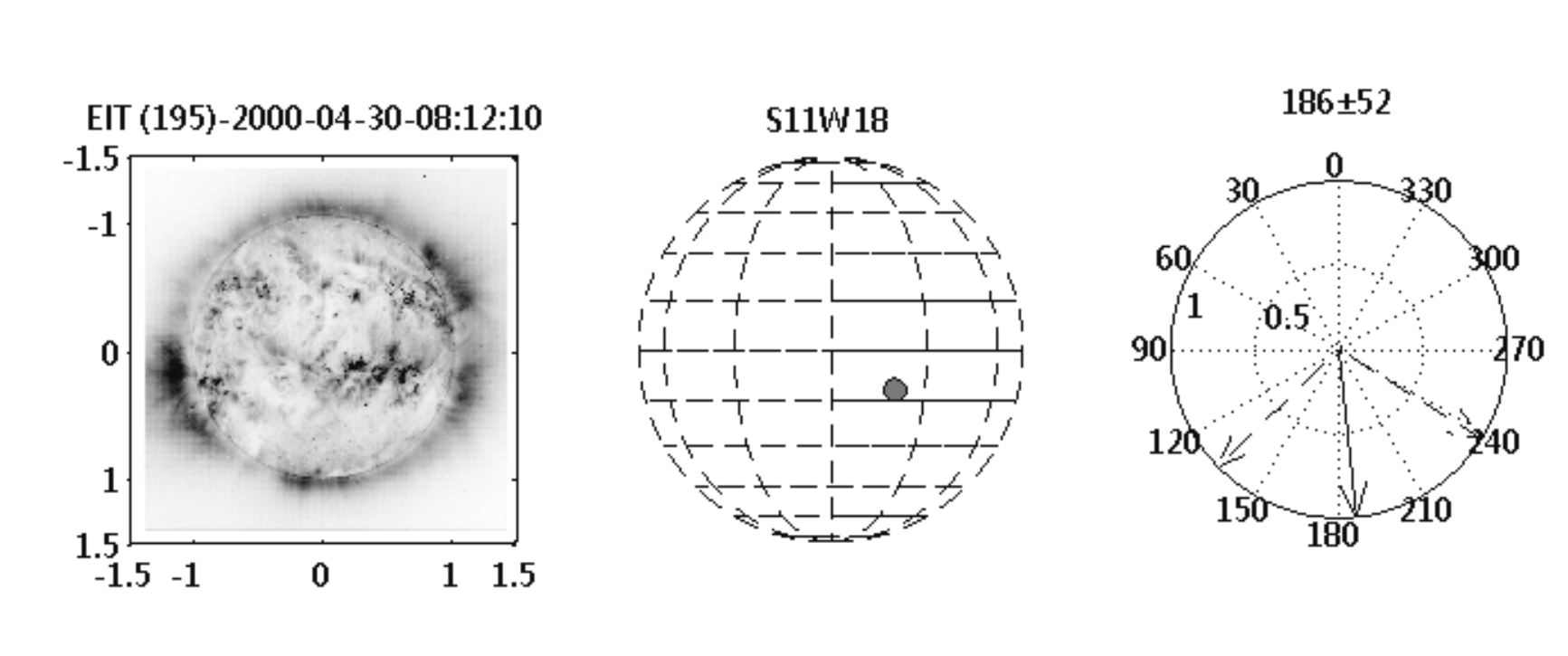}
\caption{\event}
\end{center}
\end{figure} % 
%--------------------------------------------
\clearpage
\renewcommand{\event} {11 July 2000}
%--------------------------------------------
\begin{figure}
\begin{center}% trim=0cm 1cm  0cm 1cm,clip,
\includegraphics[width=\textwidth]{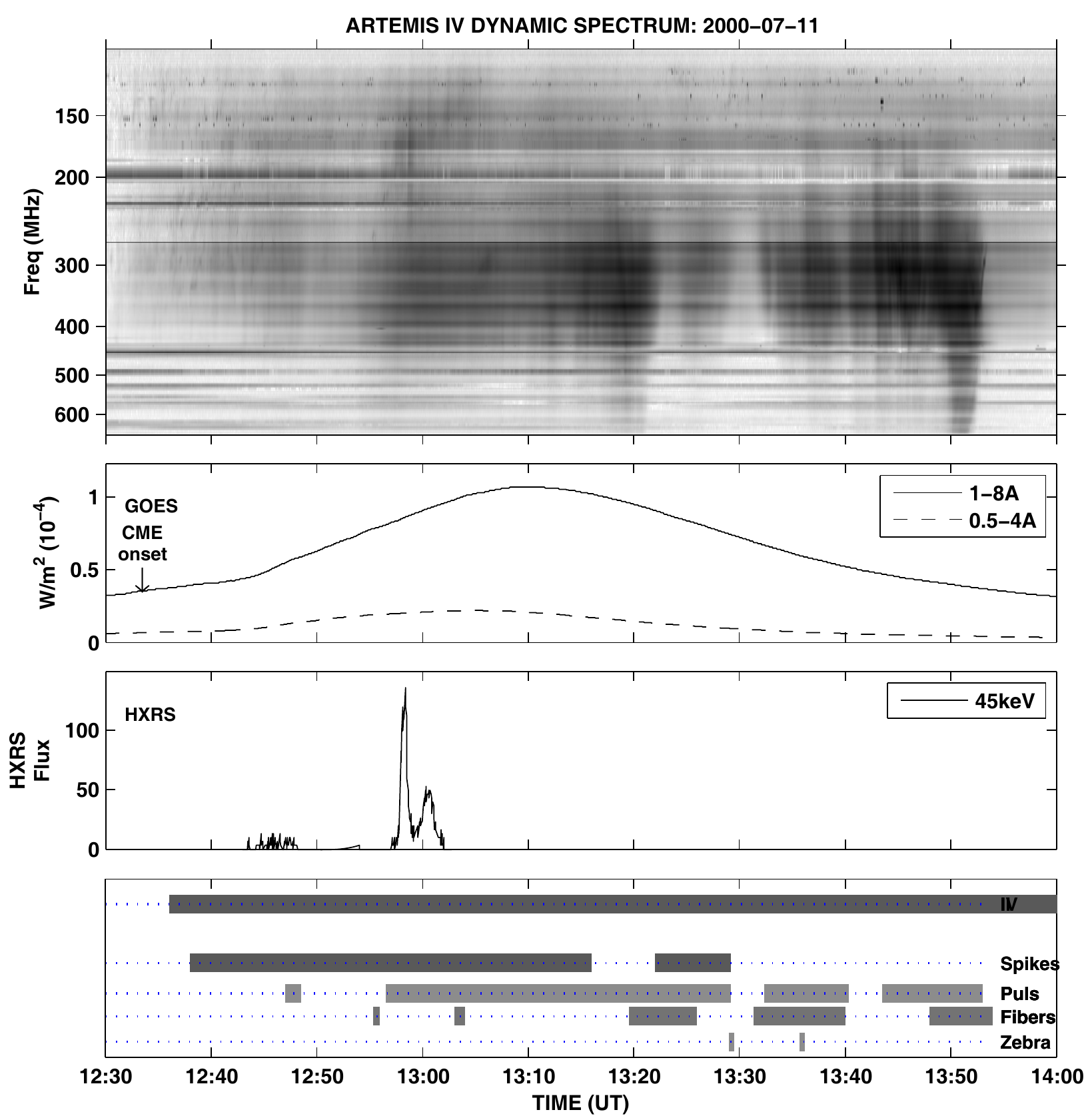}
\includegraphics[trim=0cm 0.0cm  0.0cm 4.0cm,clip,width=0.85\textwidth]{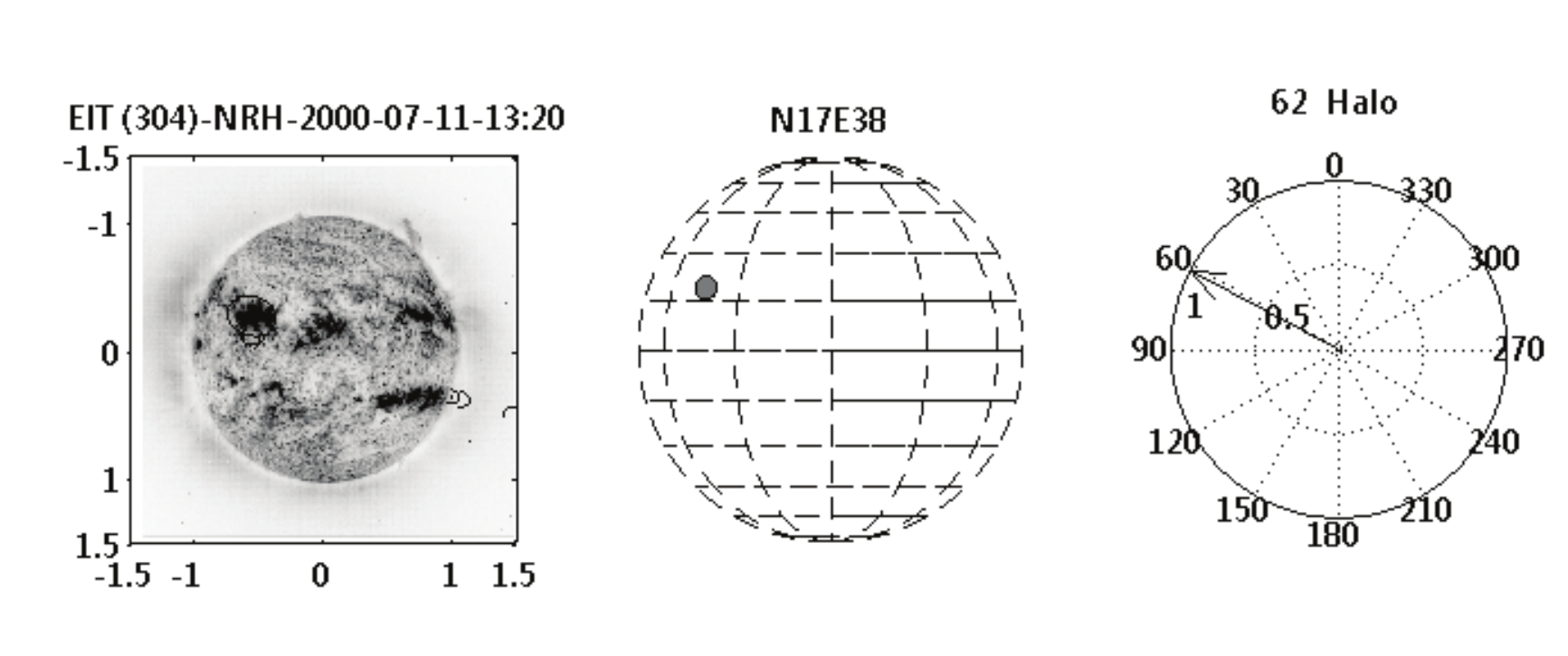}
\caption{\event}
\end{center}
\end{figure} % 
%--------------------------------------------
\clearpage
\renewcommand{\event} {14 July 2000 Event(A)}
%--------------------------------------------
\begin{figure}
\begin{center}% trim=0cm 1cm  0cm 1cm,clip,
\includegraphics[width=\textwidth]{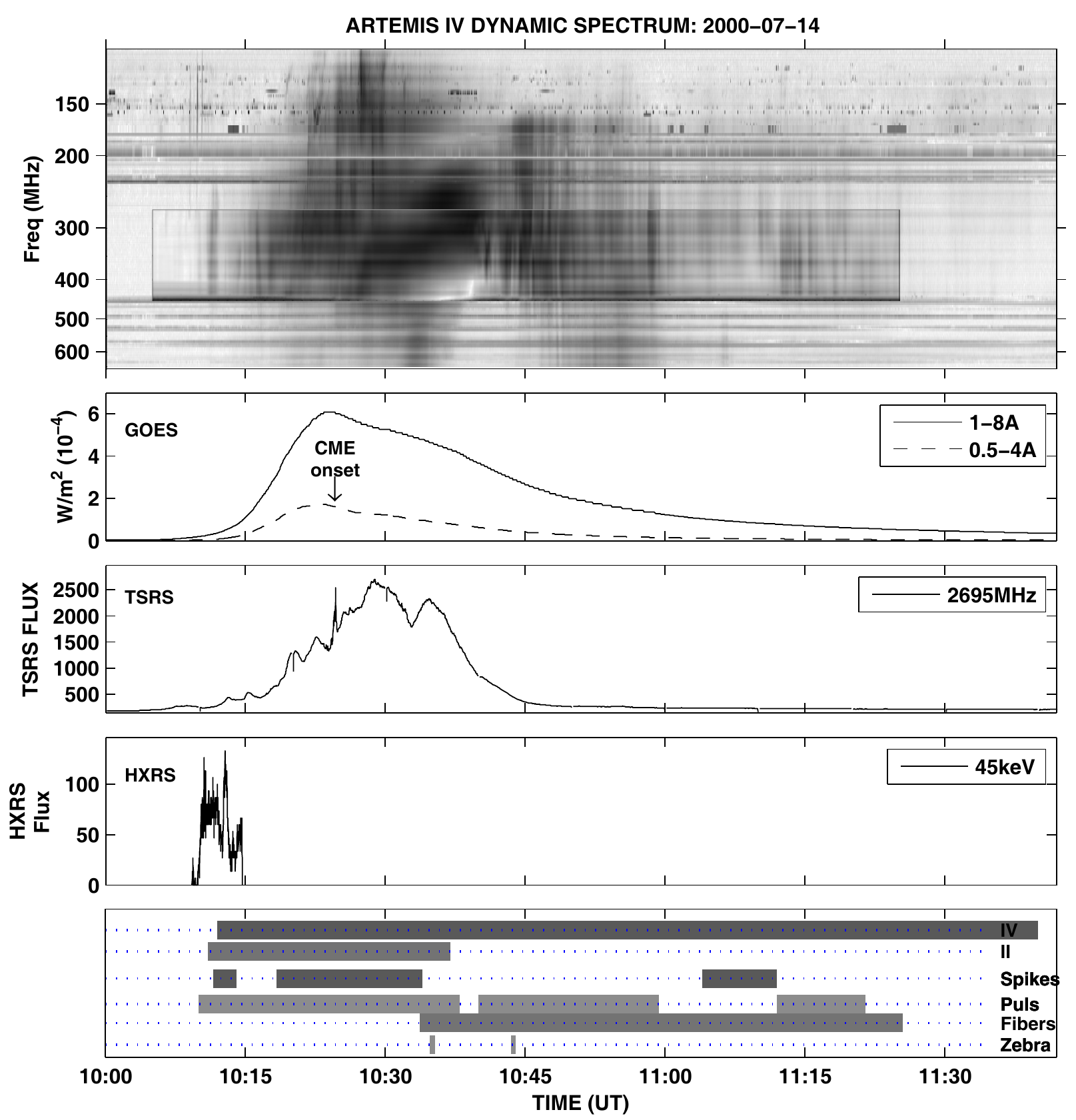}
\includegraphics[trim=0cm 0.0cm  0.0cm 4.0cm,clip,width=0.85\textwidth]{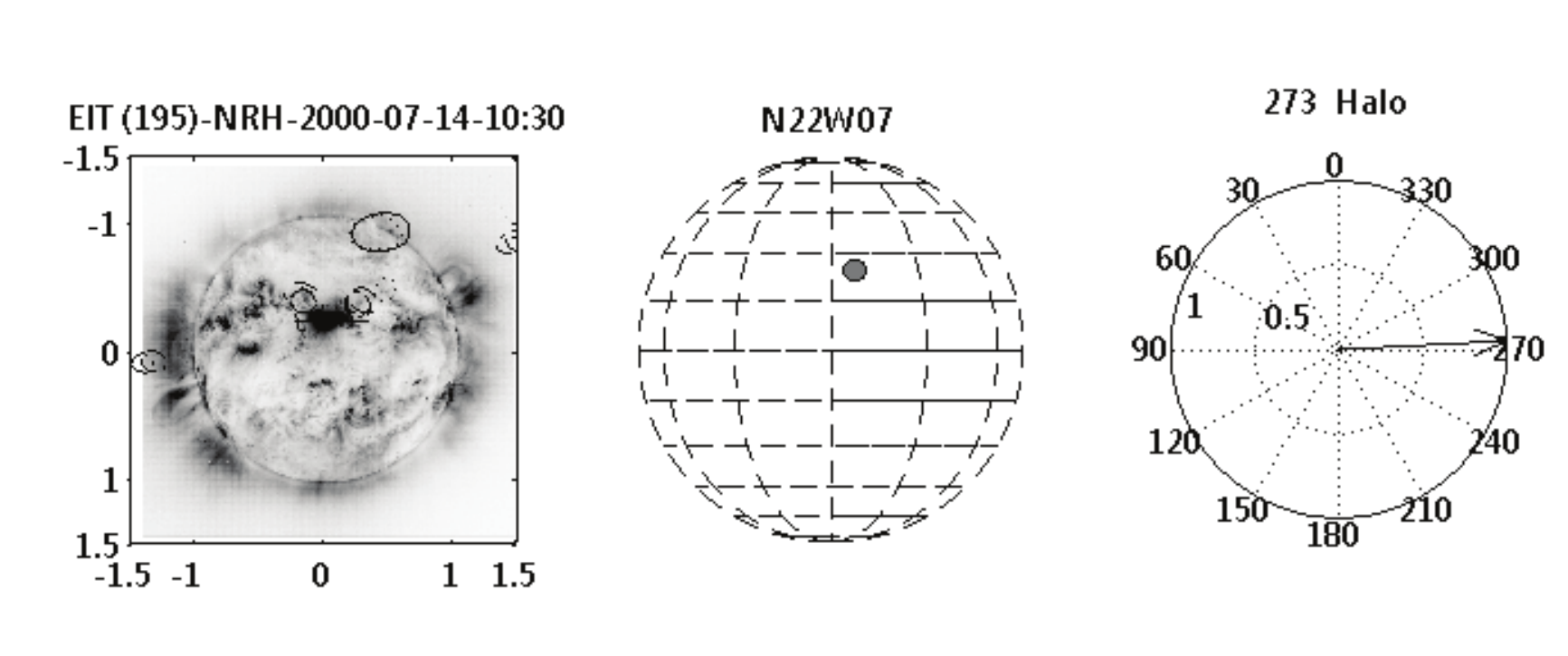}
\caption{\event}
\end{center}
\end{figure} % 
%--------------------------------------------
\clearpage
\renewcommand{\event} {14 July 2000 Event(B)}
%--------------------------------------------
\begin{figure}
\begin{center}% trim=0cm 1cm  0cm 1cm,clip,
\includegraphics[width=\textwidth]{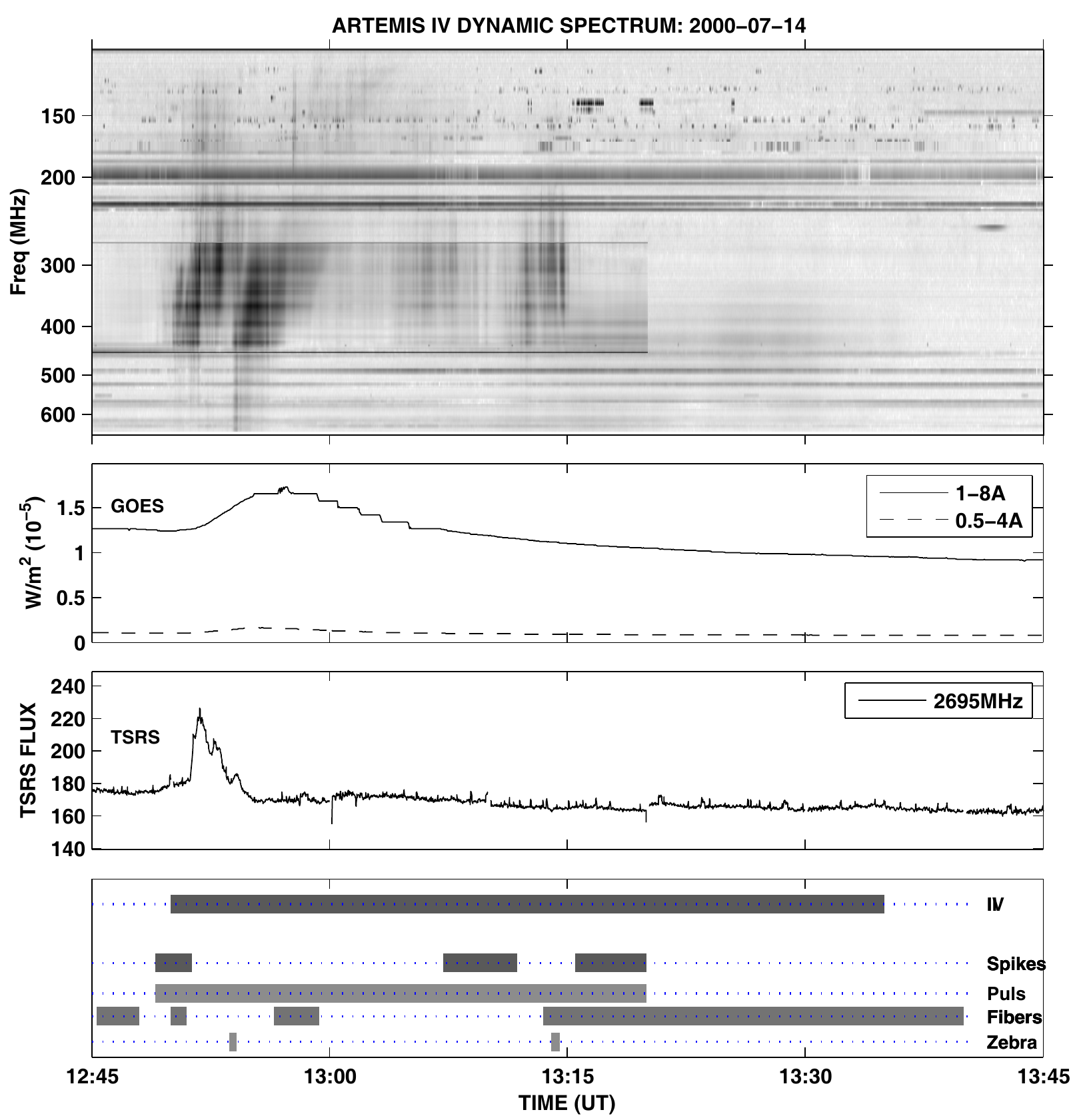}
\includegraphics[trim=0cm 0.0cm  0.0cm 4.0cm,clip,width=0.85\textwidth]{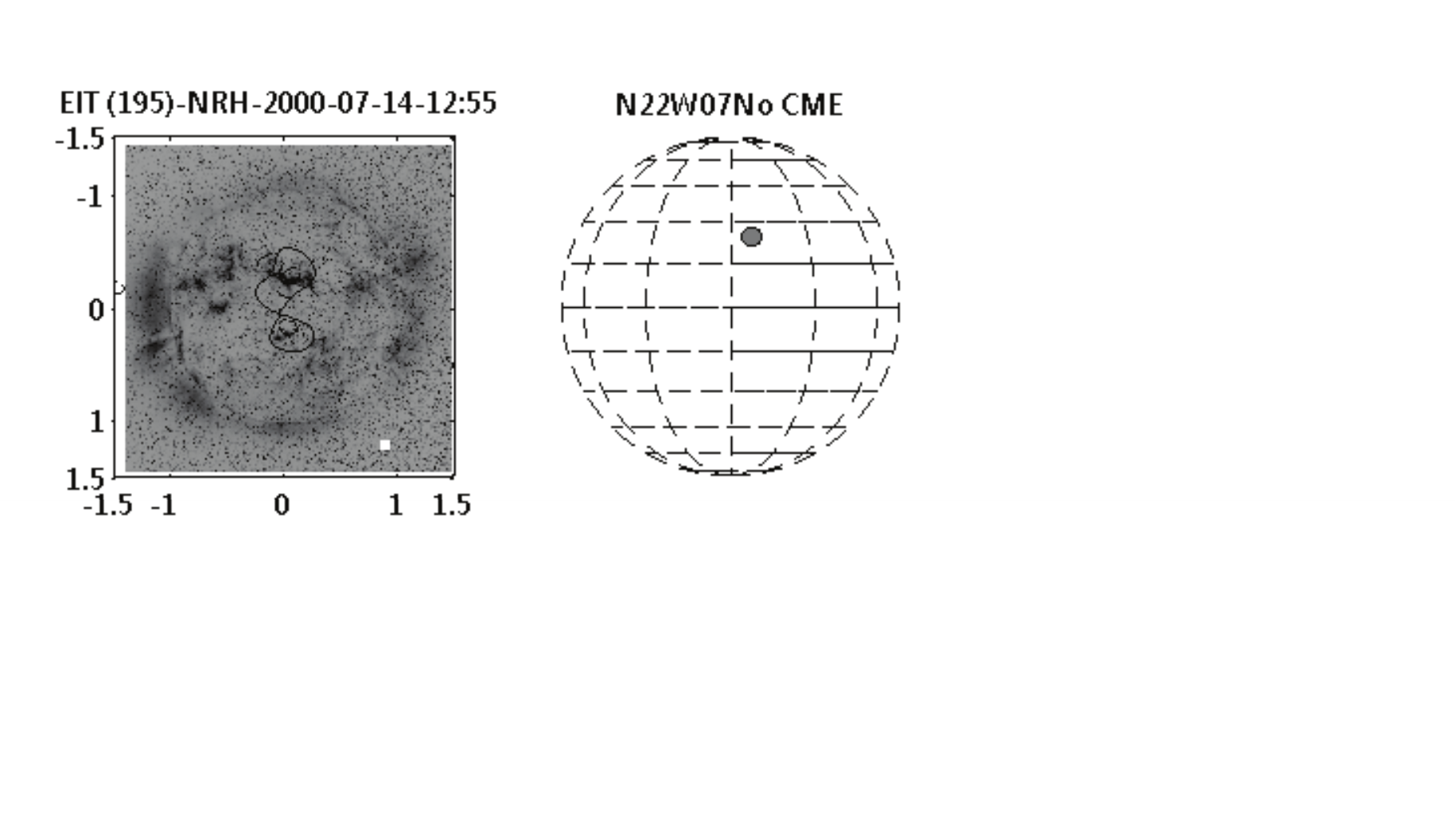}
\caption{\event}
\end{center}
\end{figure} % 
%--------------------------------------------
\clearpage
\renewcommand{\event} {14 July 2000 Event(C)}
%--------------------------------------------
\begin{figure}
\begin{center}% trim=0cm 1cm  0cm 1cm,clip,
\includegraphics[width=\textwidth]{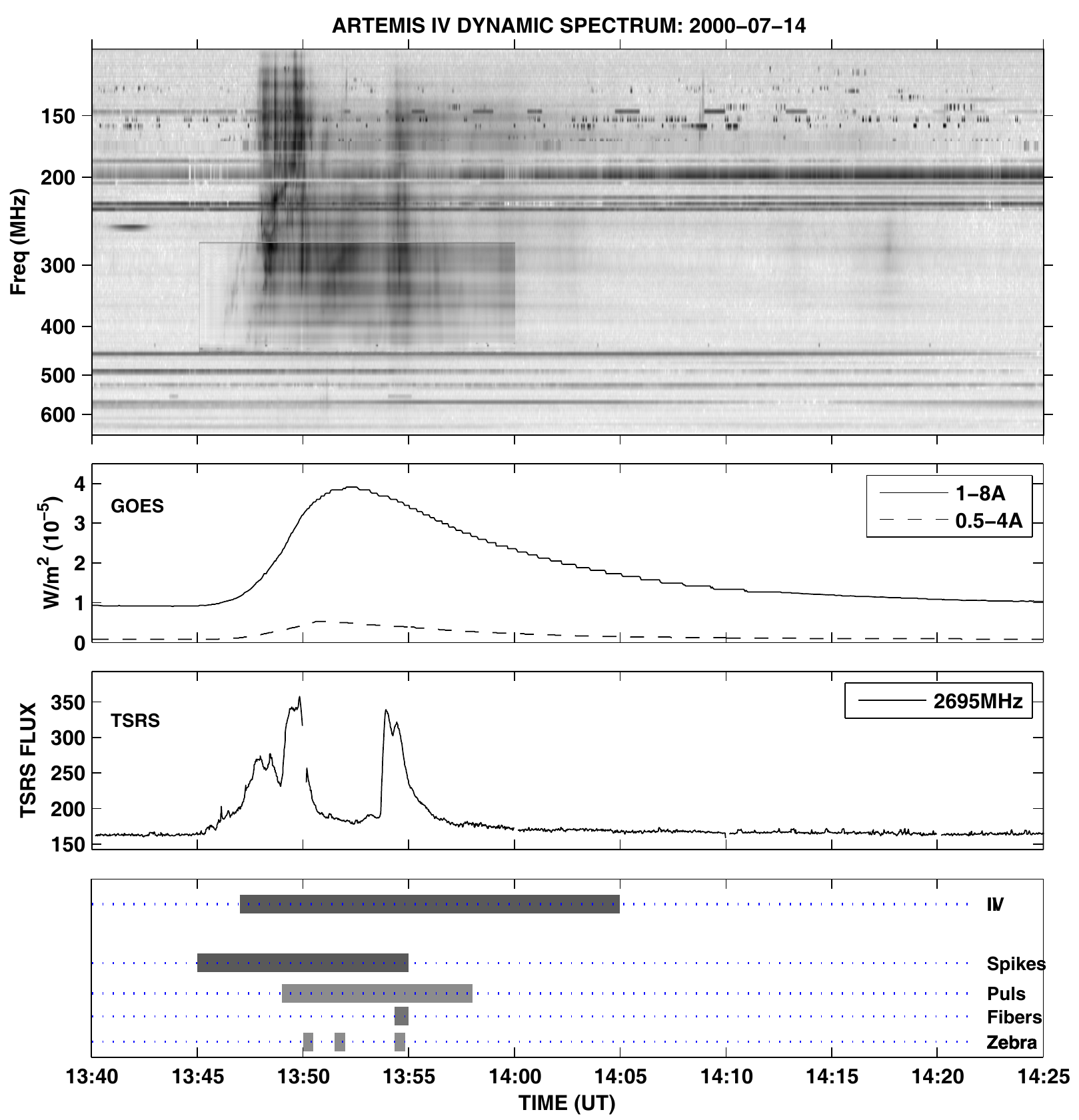}
\includegraphics[trim=0cm 0.0cm  0.0cm 4.0cm,clip,width=0.85\textwidth]{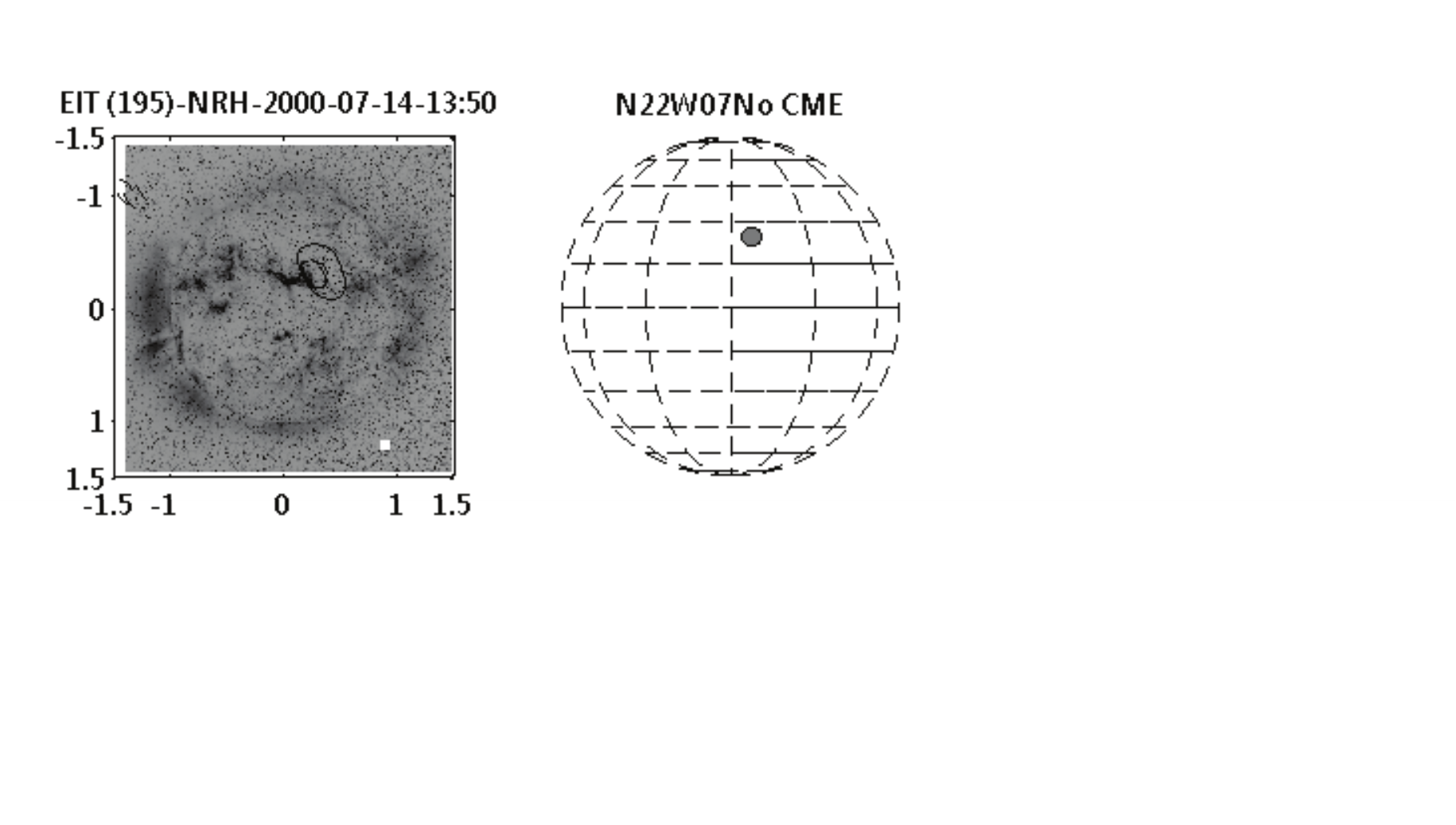}
\caption{\event}
\end{center}
\end{figure} % 
%--------------------------------------------
\clearpage
\renewcommand{\event} {19 September 2000}
%--------------------------------------------
\begin{figure}
\begin{center}% trim=0cm 1cm  0cm 1cm,clip,
\includegraphics[width=\textwidth]{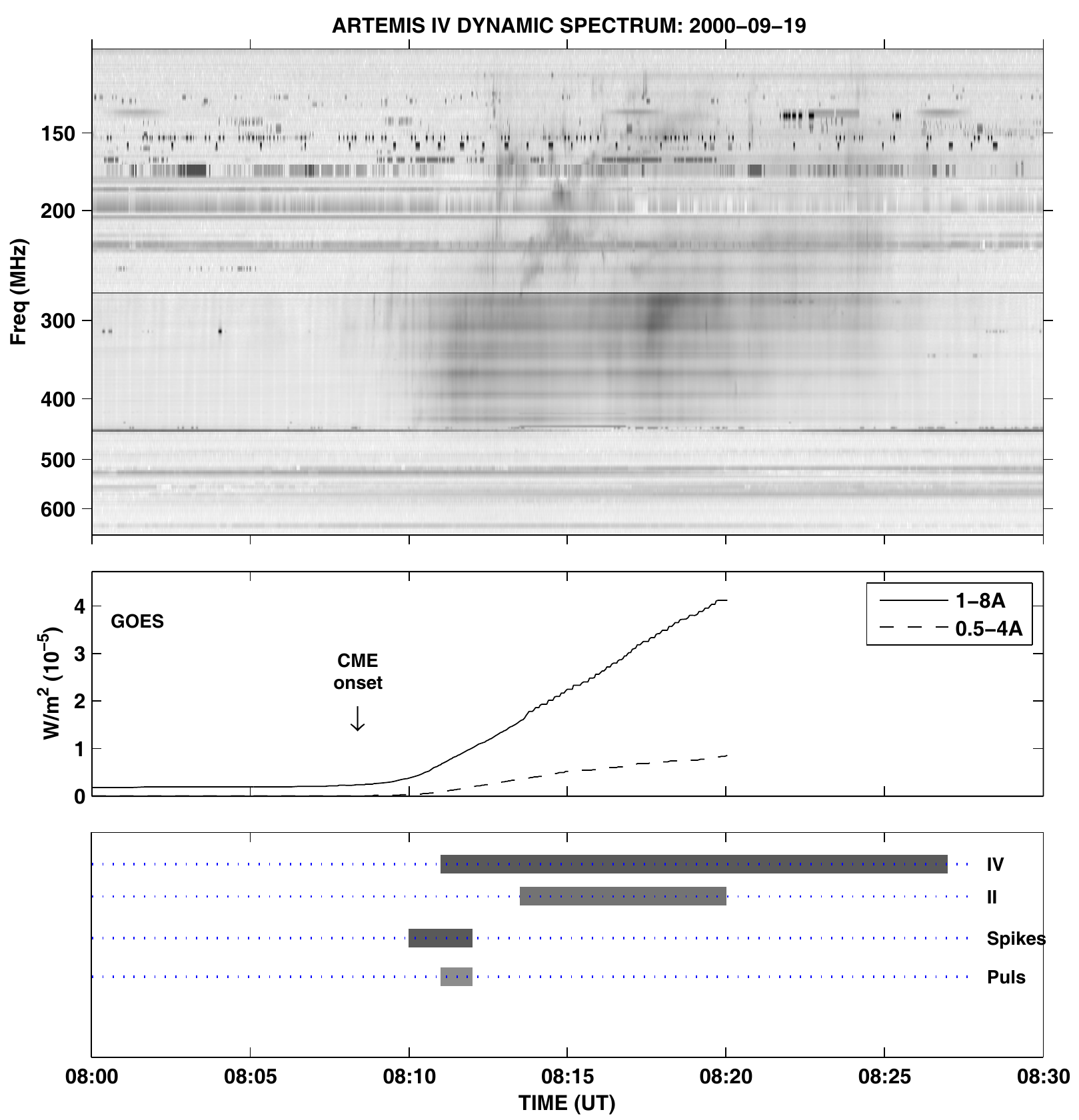}
\includegraphics[trim=0cm 0.0cm  0.0cm 4.0cm,clip,width=0.85\textwidth]{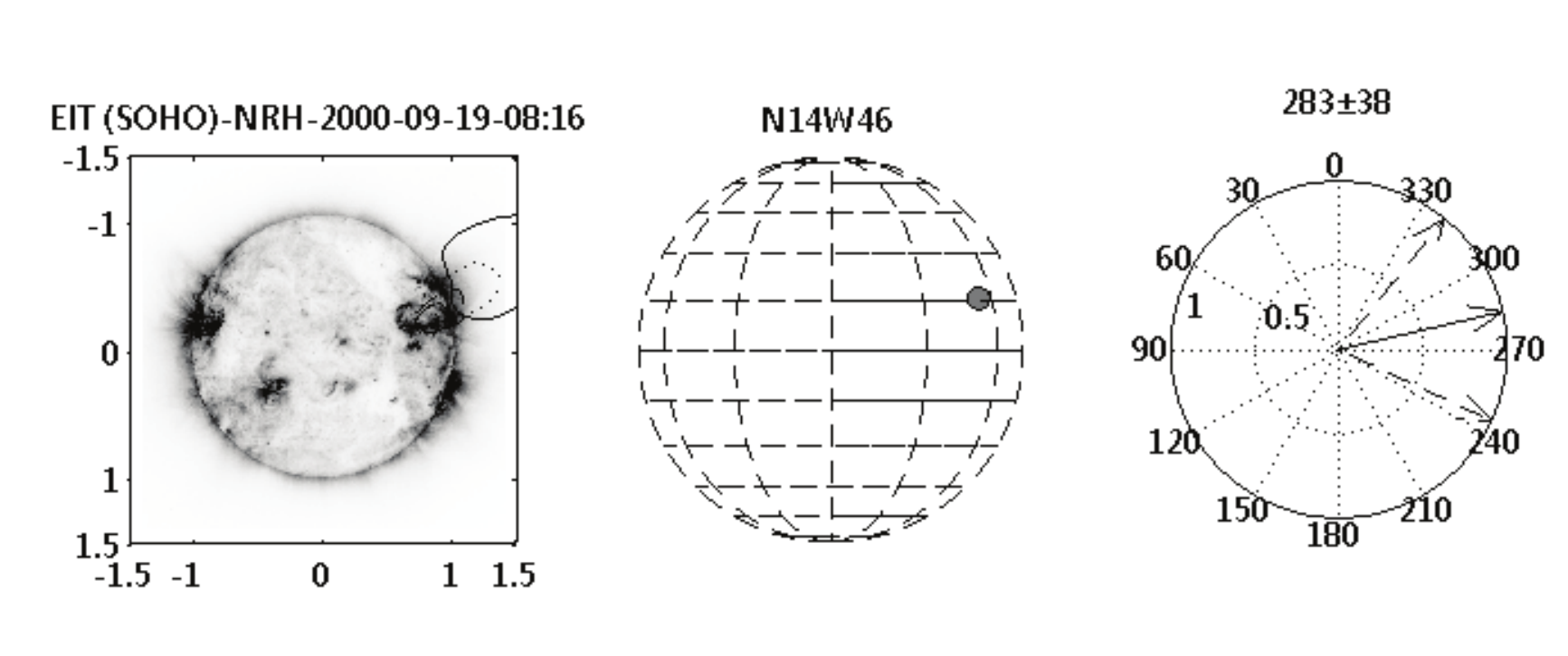}
\caption{\event}
\end{center}
\end{figure} % 
%--------------------------------------------
\clearpage
\renewcommand{\event} {18 November 2000}
%--------------------------------------------
\begin{figure}
\begin{center}% trim=0cm 1cm  0cm 1cm,clip,
\includegraphics[width=\textwidth]{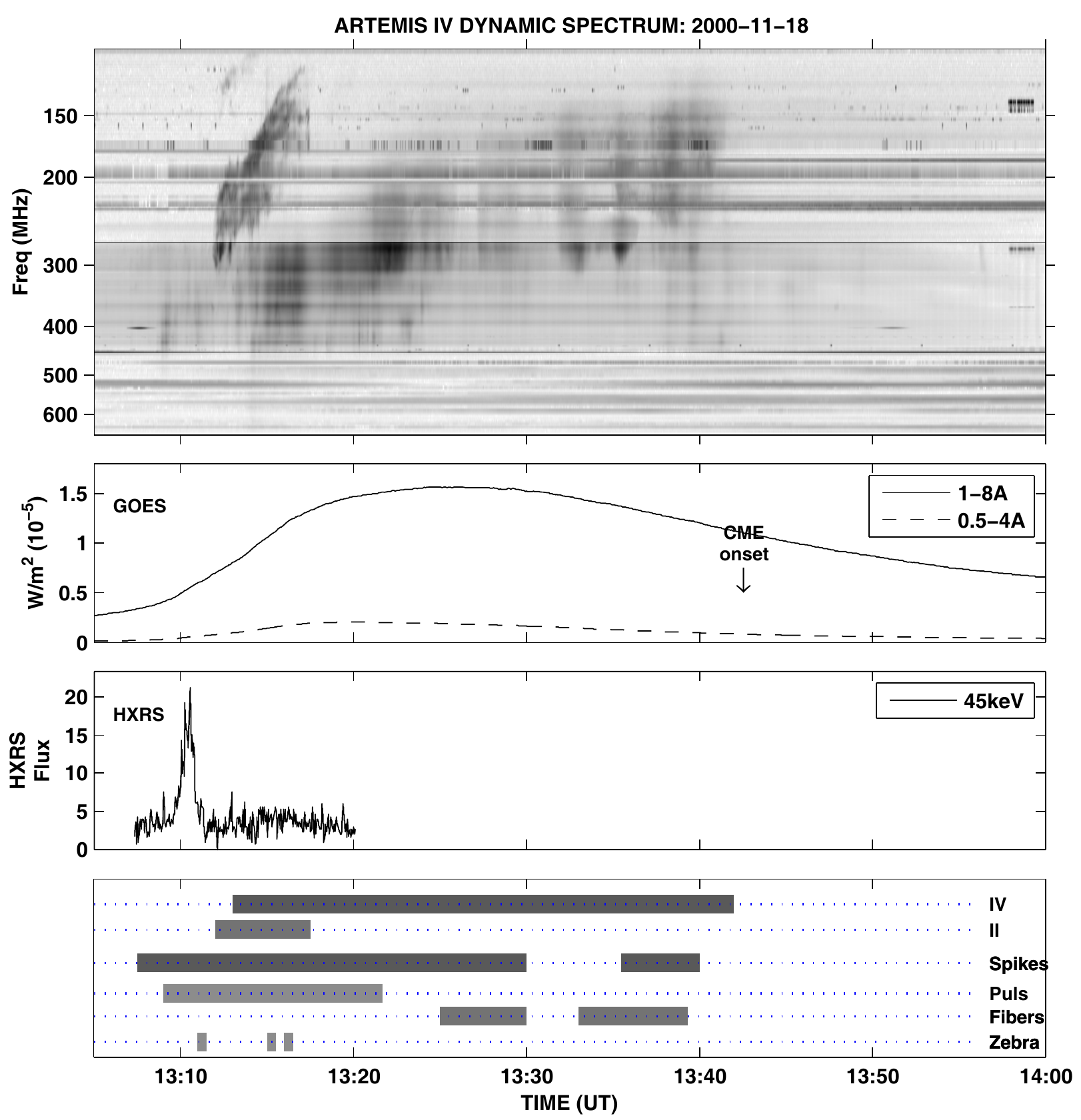}
\includegraphics[trim=0cm 0.0cm  0.0cm 4.0cm,clip,width=0.85\textwidth]{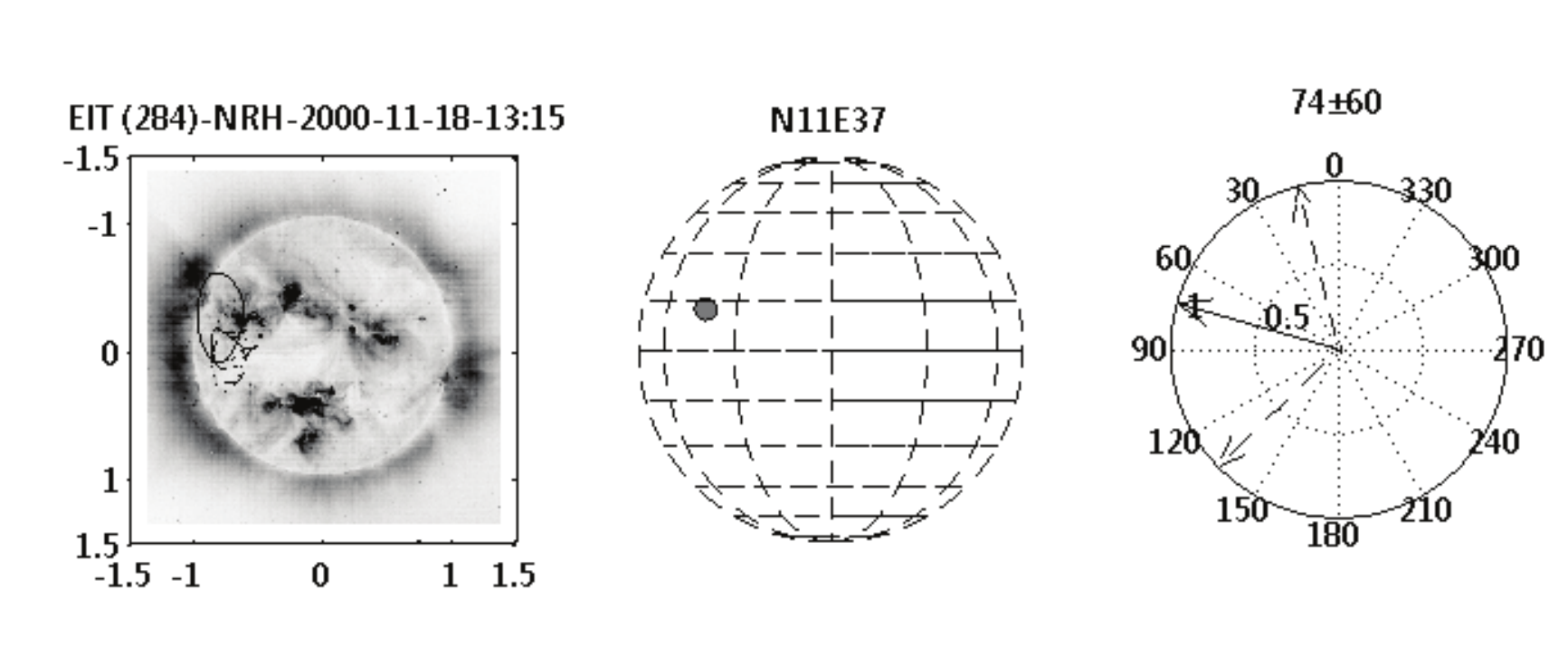}
\caption{\event}
\end{center}
\end{figure} % 
%--------------------------------------------
\clearpage
\renewcommand{\event} {21 April 2003}
%--------------------------------------------
\begin{figure}
\begin{center}% trim=0cm 1cm  0cm 1cm,clip,
\includegraphics[width=\textwidth]{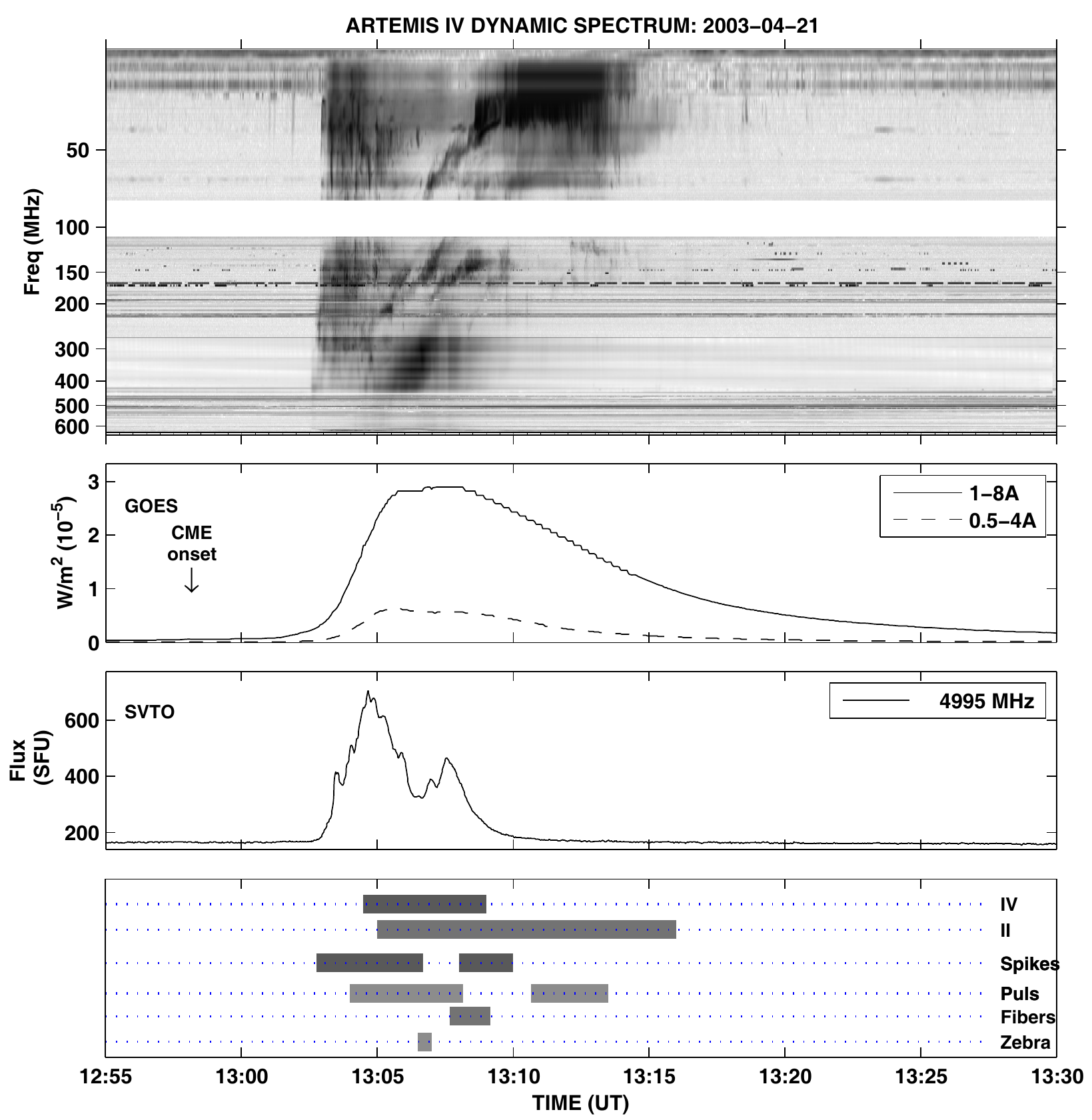}
\includegraphics[trim=0cm 0.0cm  0.0cm 4.0cm,clip,width=0.85\textwidth]{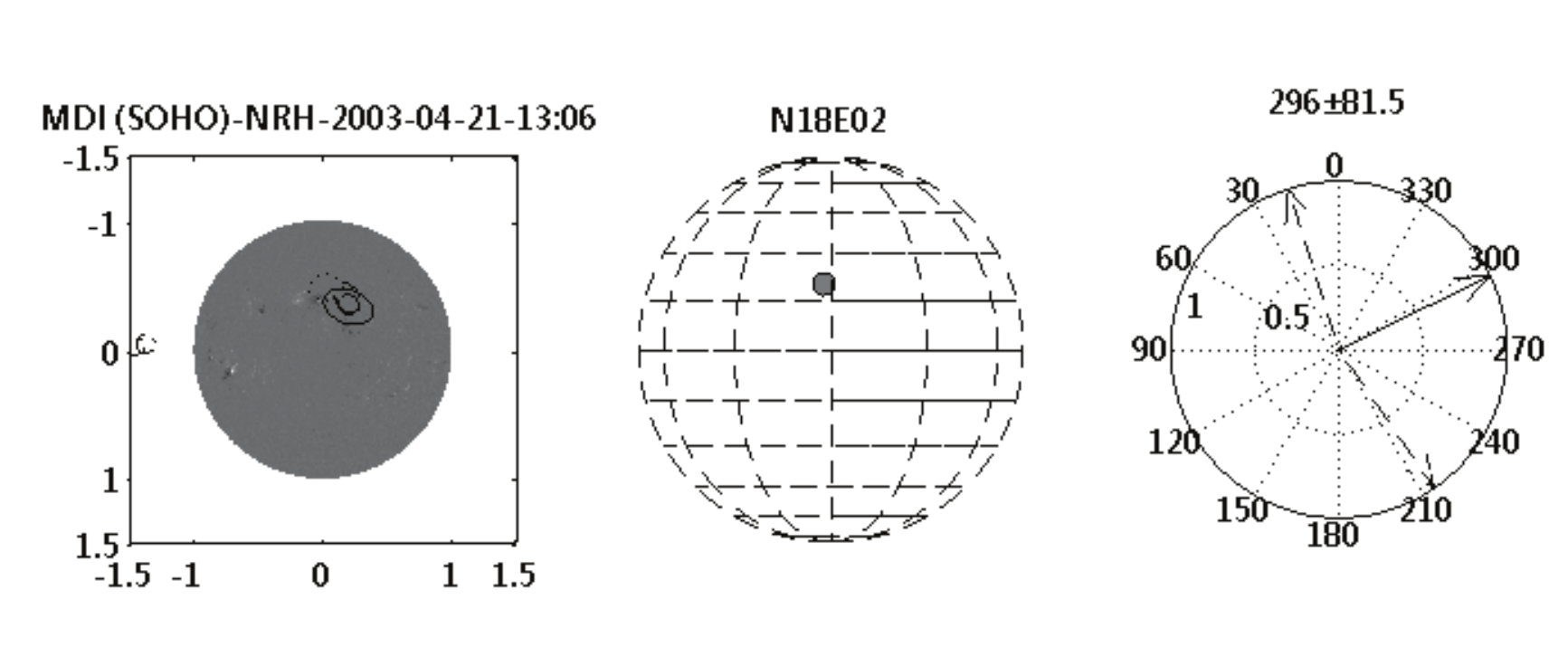}
\caption{\event}
\end{center}
\end{figure} % 
%--------------------------------------------
\clearpage
\renewcommand{\event} {26 October 2003}
%--------------------------------------------
\begin{figure}
\begin{center}% trim=0cm 1cm  0cm 1cm,clip,
\includegraphics[width=\textwidth]{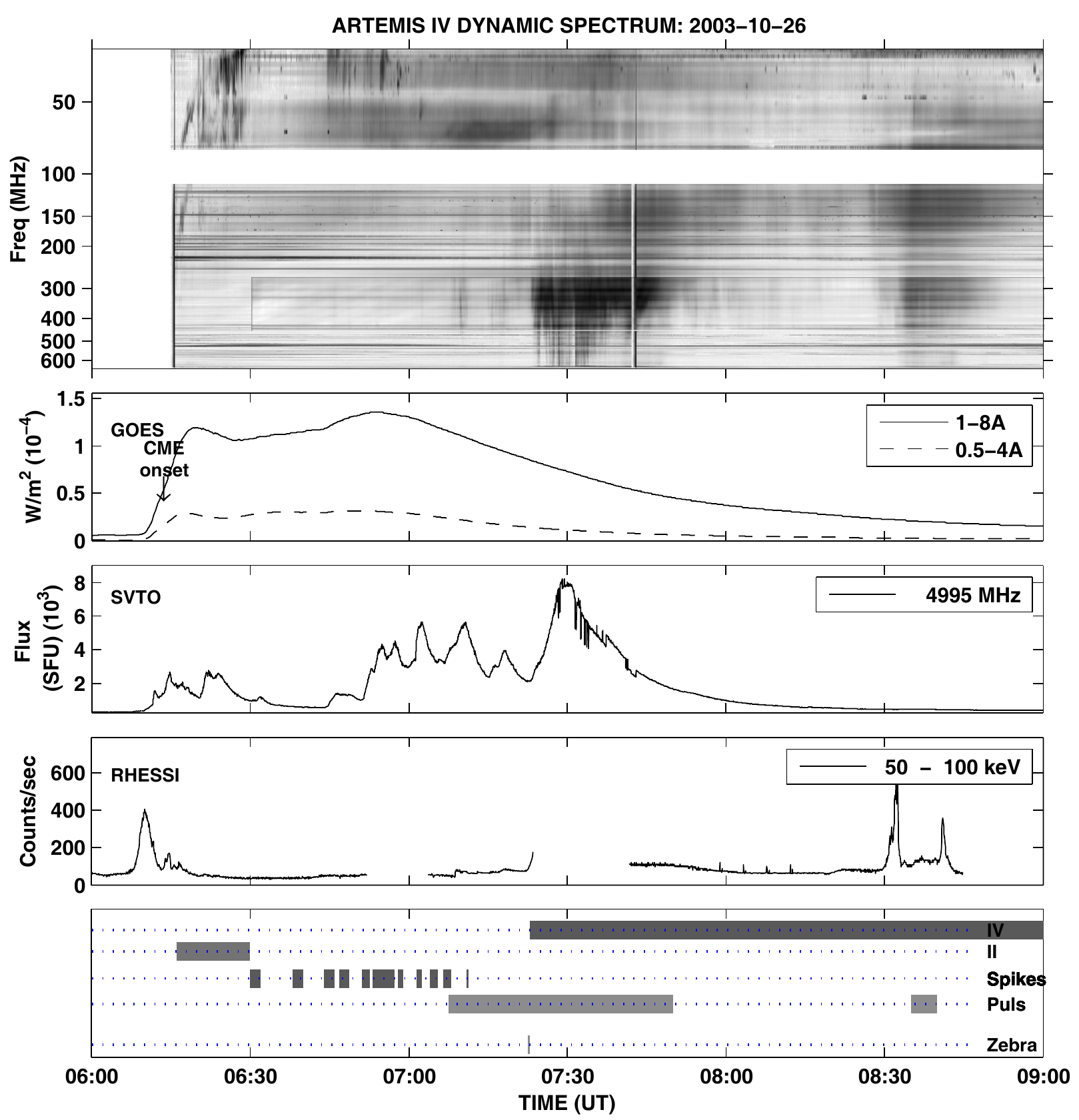}
\includegraphics[trim=0cm 0.0cm  0.0cm 4.0cm,clip,width=0.85\textwidth]{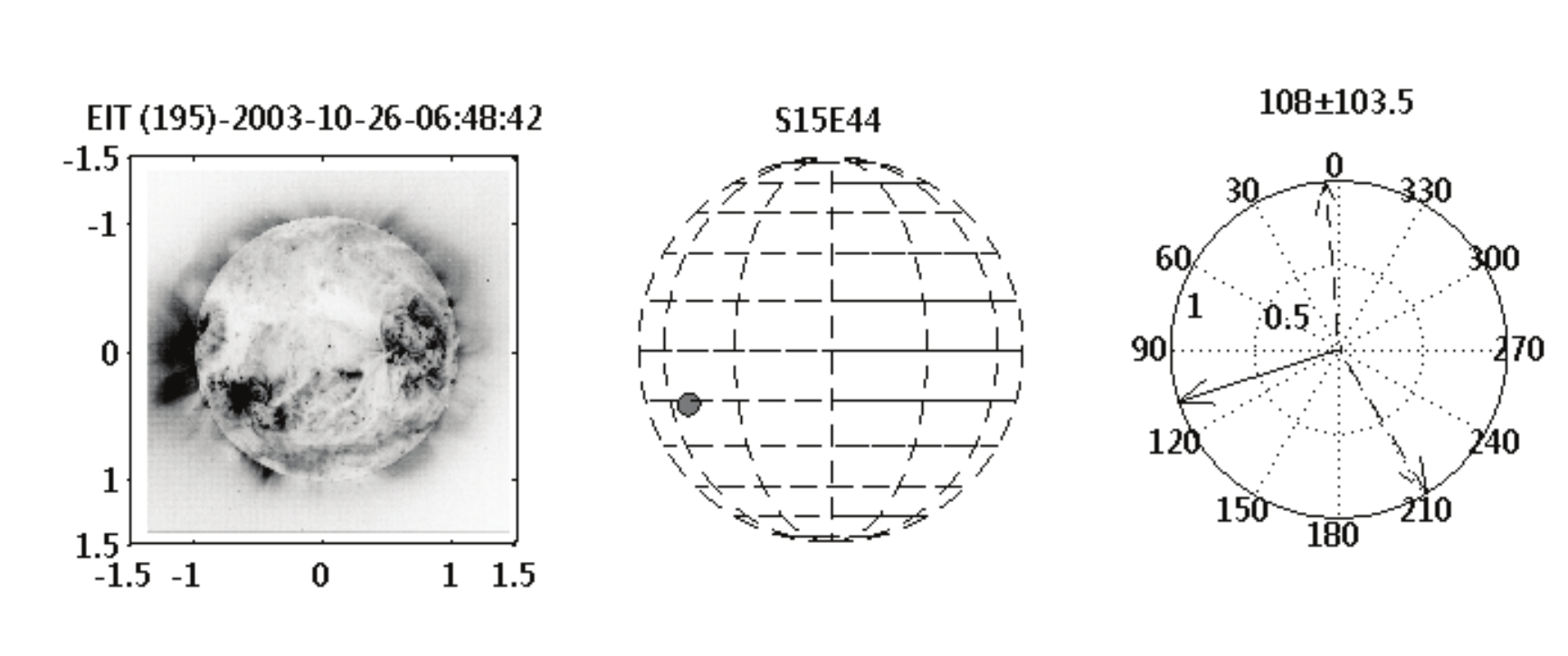}
\caption{\event}
\end{center}
\end{figure} % 
%--------------------------------------------
\clearpage
\renewcommand{\event} {28 October 2003}
%--------------------------------------------
\begin{figure}
\begin{center}% trim=0cm 1cm  0cm 1cm,clip,
\includegraphics[width=\textwidth]{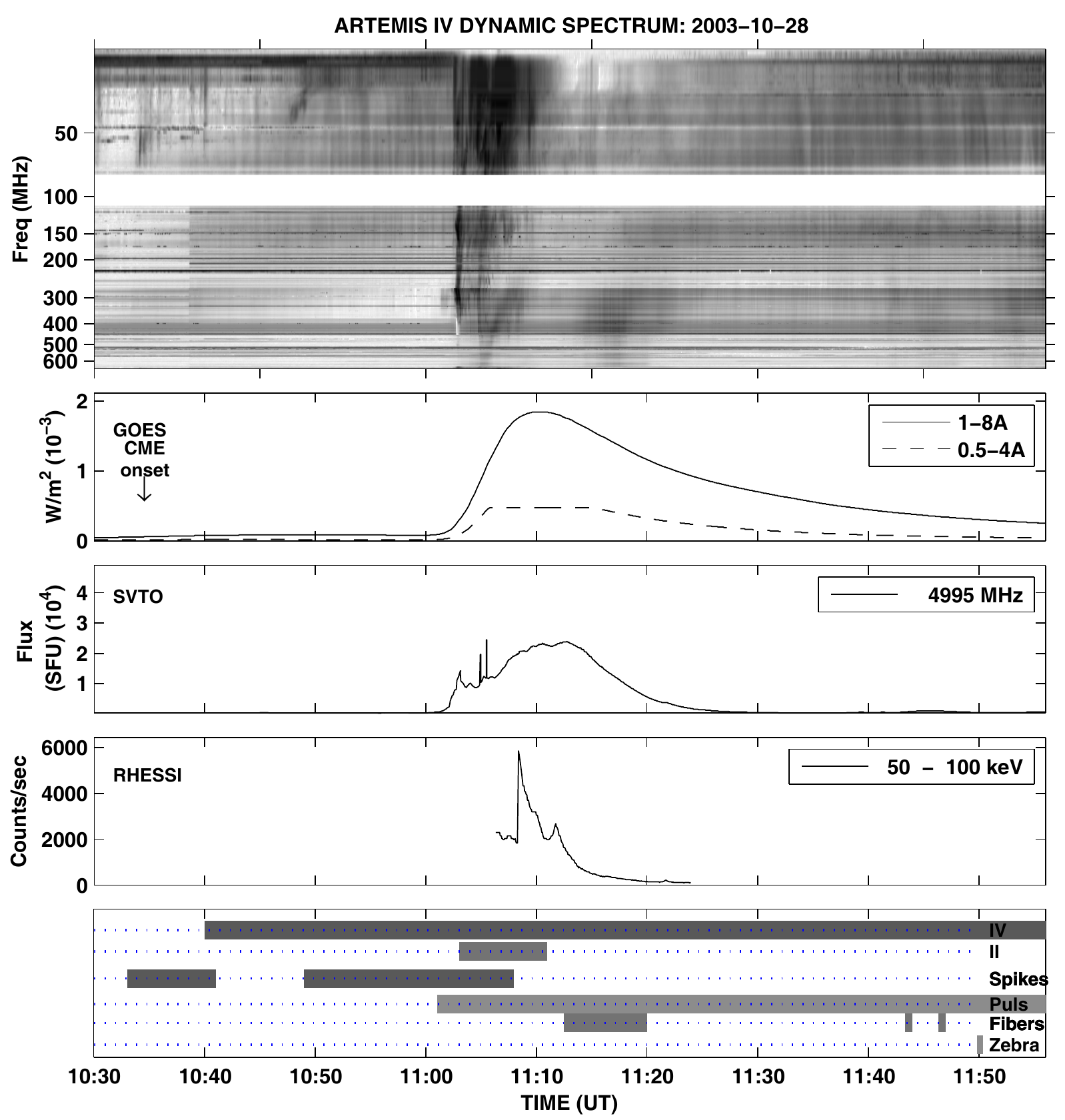}
\includegraphics[trim=0cm 0.0cm  0.0cm 4.0cm,clip,width=0.85\textwidth]{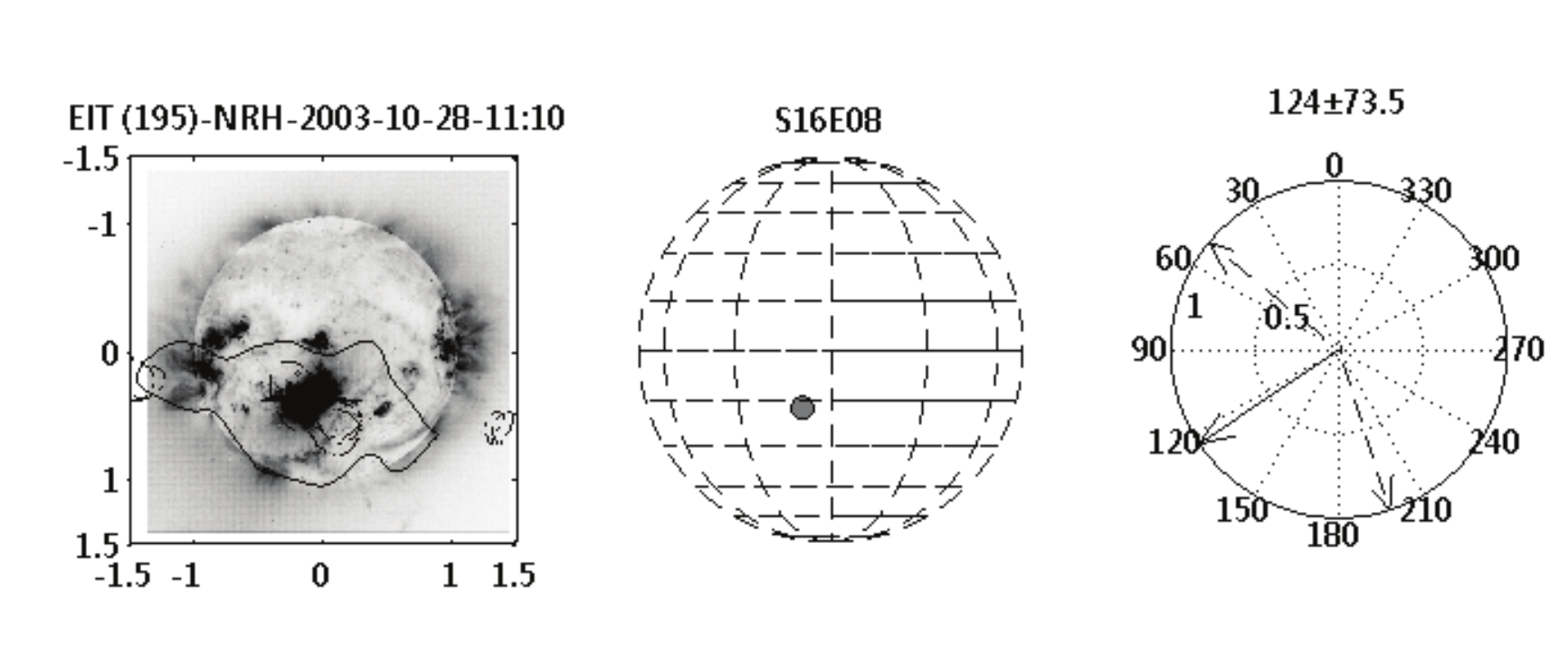}
\caption{\event}
\end{center}
\end{figure} % 
%--------------------------------------------
\clearpage
\renewcommand{\event} {03 November 2003}
%--------------------------------------------
\begin{figure}
\begin{center}% trim=0cm 1cm  0cm 1cm,clip,
\includegraphics[width=\textwidth]{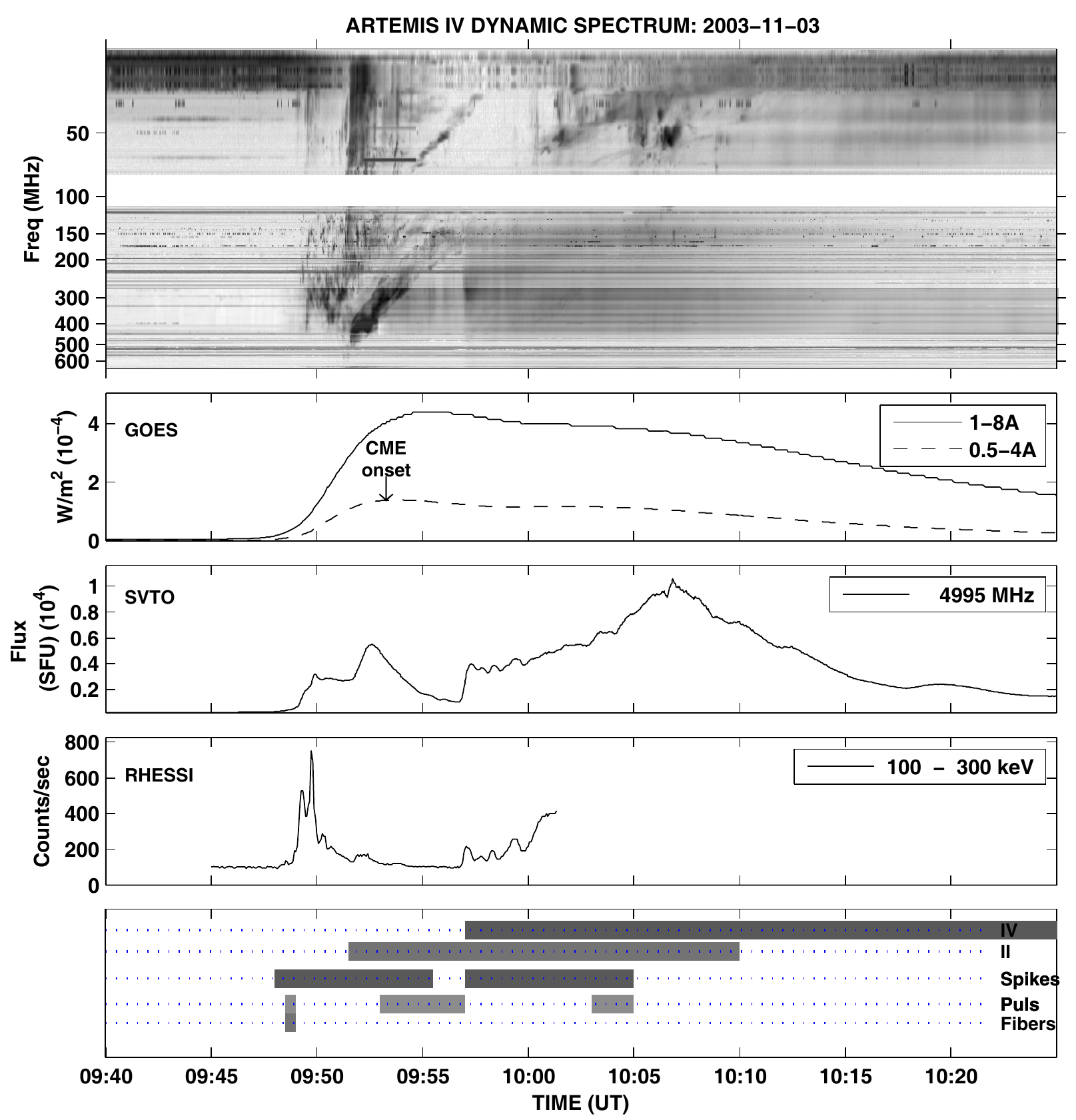}
\includegraphics[trim=0cm 0.0cm  0.0cm 4.0cm,clip,width=0.85\textwidth]{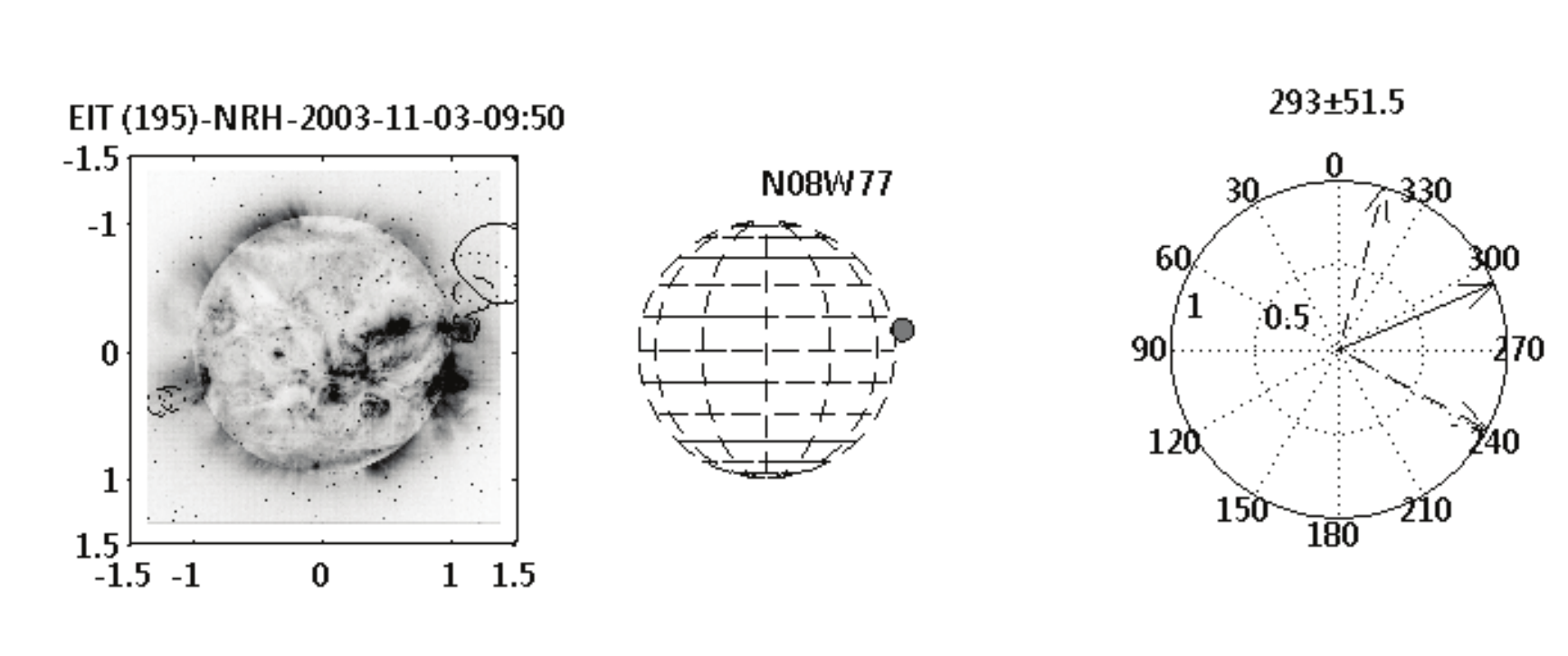}
\caption{\event}
\end{center}
\end{figure} % 
%--------------------------------------------
\clearpage
\renewcommand{\event} {04 February 2004}
%--------------------------------------------
\begin{figure}
\begin{center}% trim=0cm 1cm  0cm 1cm,clip,
\includegraphics[width=\textwidth]{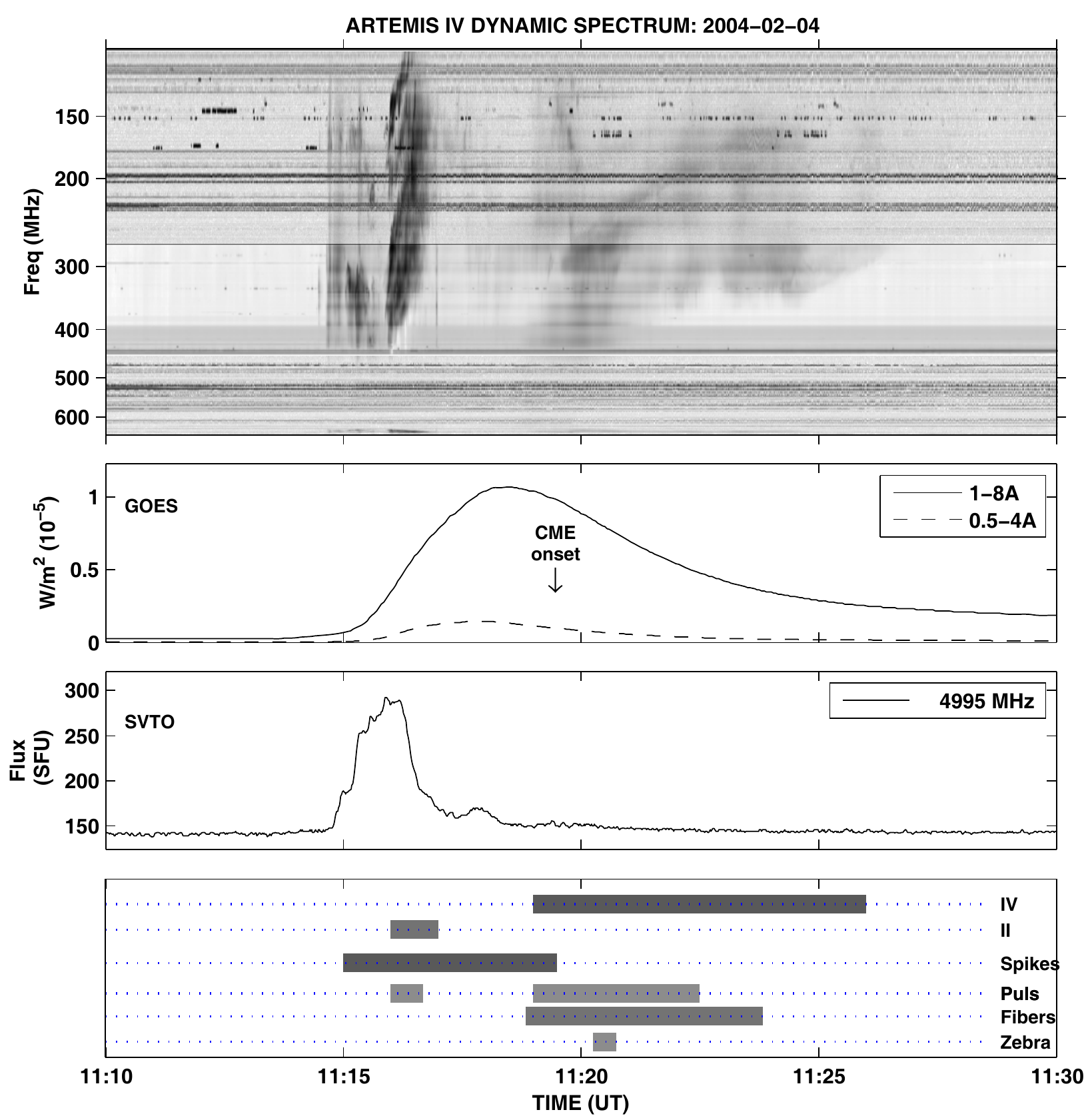}
\includegraphics[trim=0cm 0.0cm  0.0cm 4.0cm,clip,width=0.85\textwidth]{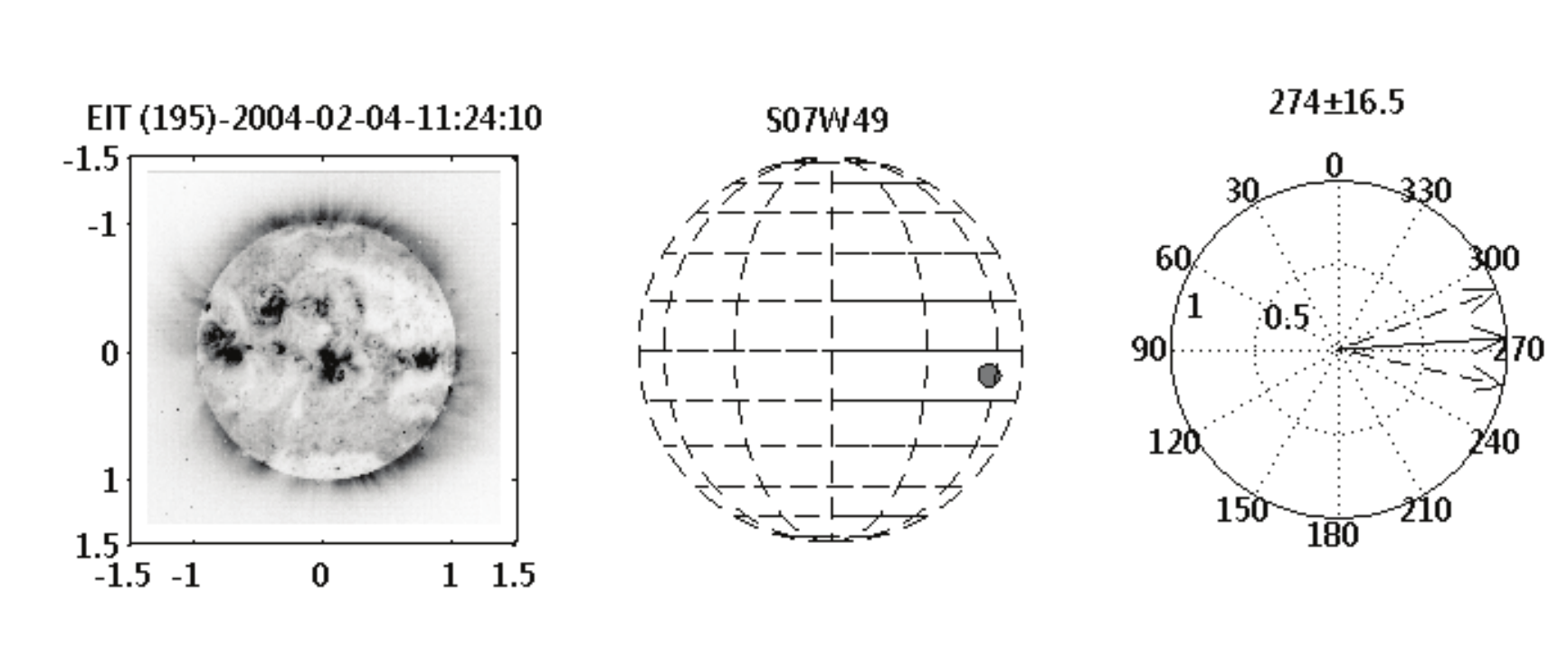}
\caption{\event}
\end{center}
\end{figure} % 
%--------------------------------------------
\clearpage
\renewcommand{\event} {25 March 2004}
%--------------------------------------------
\begin{figure}
\begin{center}% trim=0cm 1cm  0cm 1cm,clip,
\includegraphics[width=\textwidth]{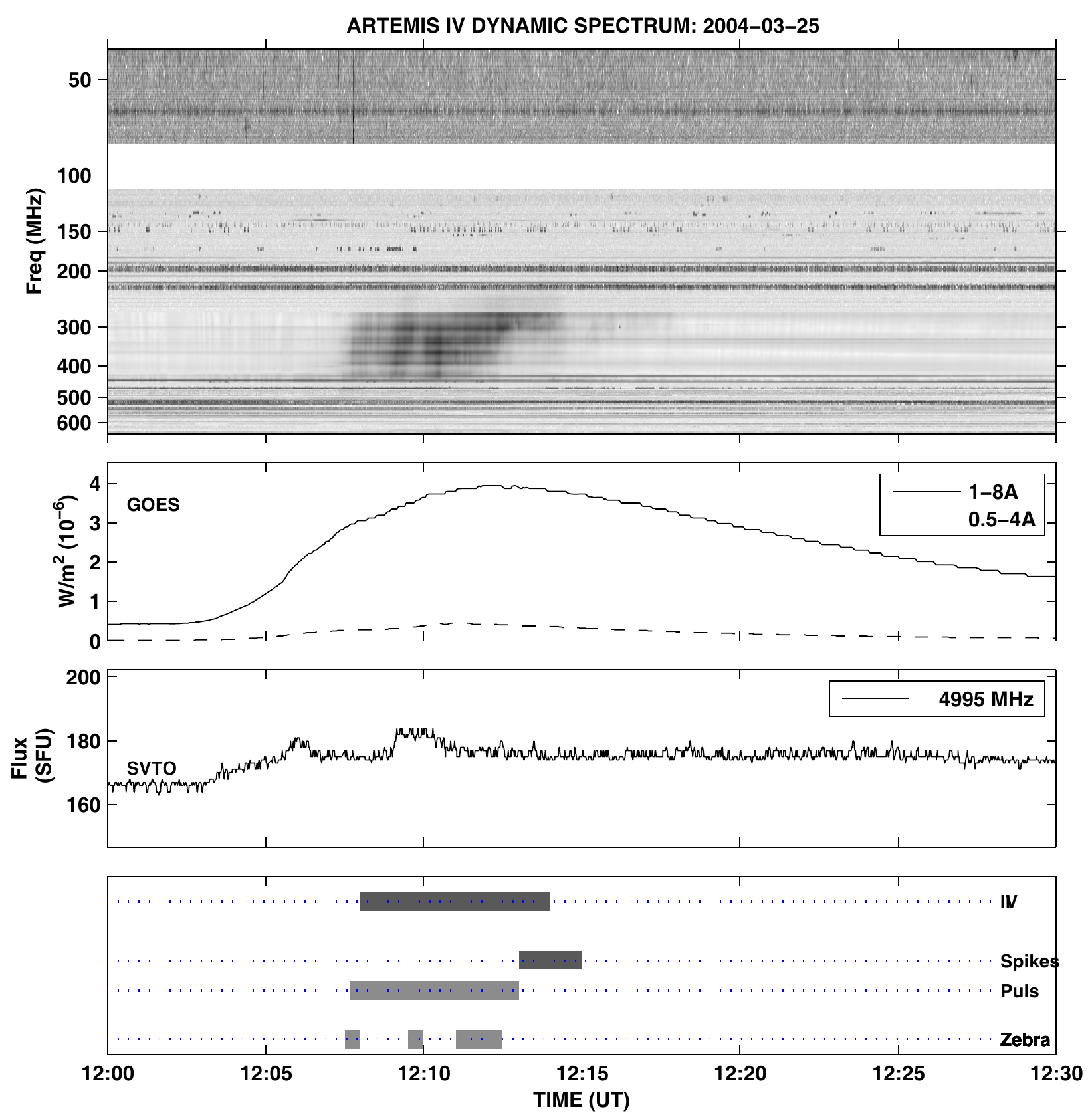}
\includegraphics[trim=0cm 0.0cm  0.0cm 4.0cm,clip,width=0.85\textwidth]{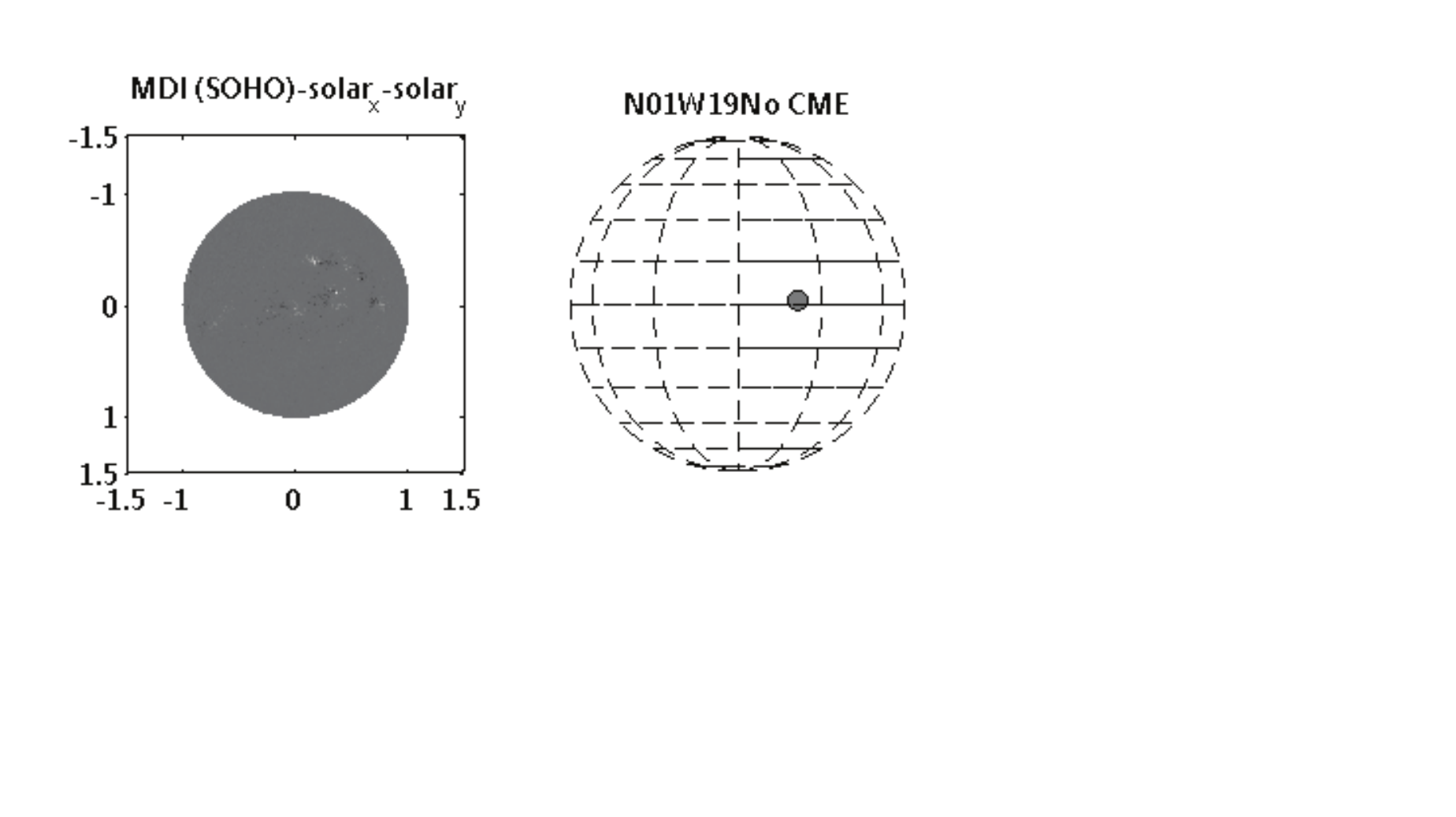}
\caption{\event}
\end{center}
\end{figure} % 
%--------------------------------------------
\clearpage
\renewcommand{\event} {30 March 2004 Event(A)}
%--------------------------------------------
\begin{figure}
\begin{center}% trim=0cm 1cm  0cm 1cm,clip,
\includegraphics[width=\textwidth]{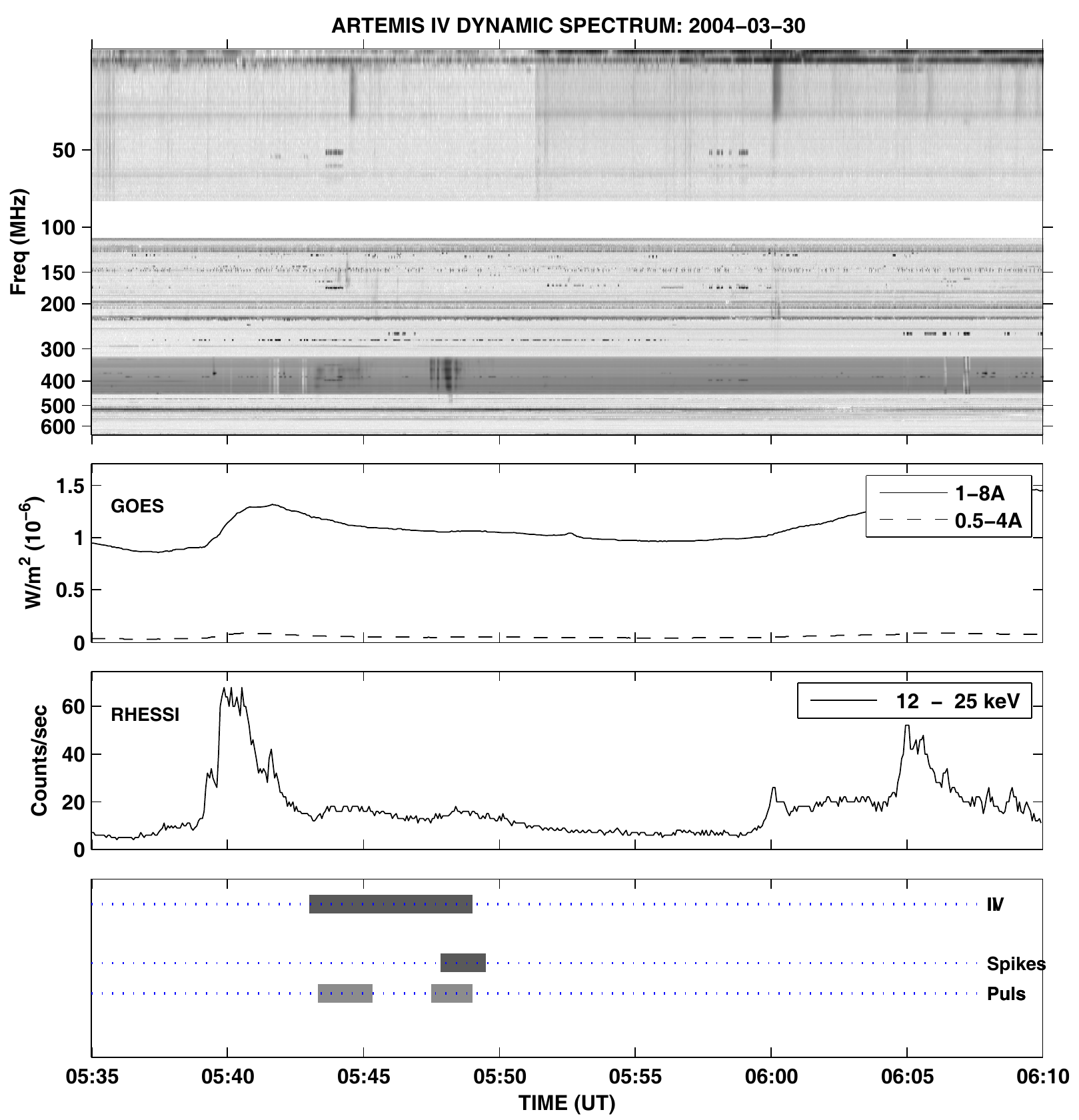}
\includegraphics[trim=0cm 0.0cm  0.0cm 4.0cm,clip,width=0.85\textwidth]{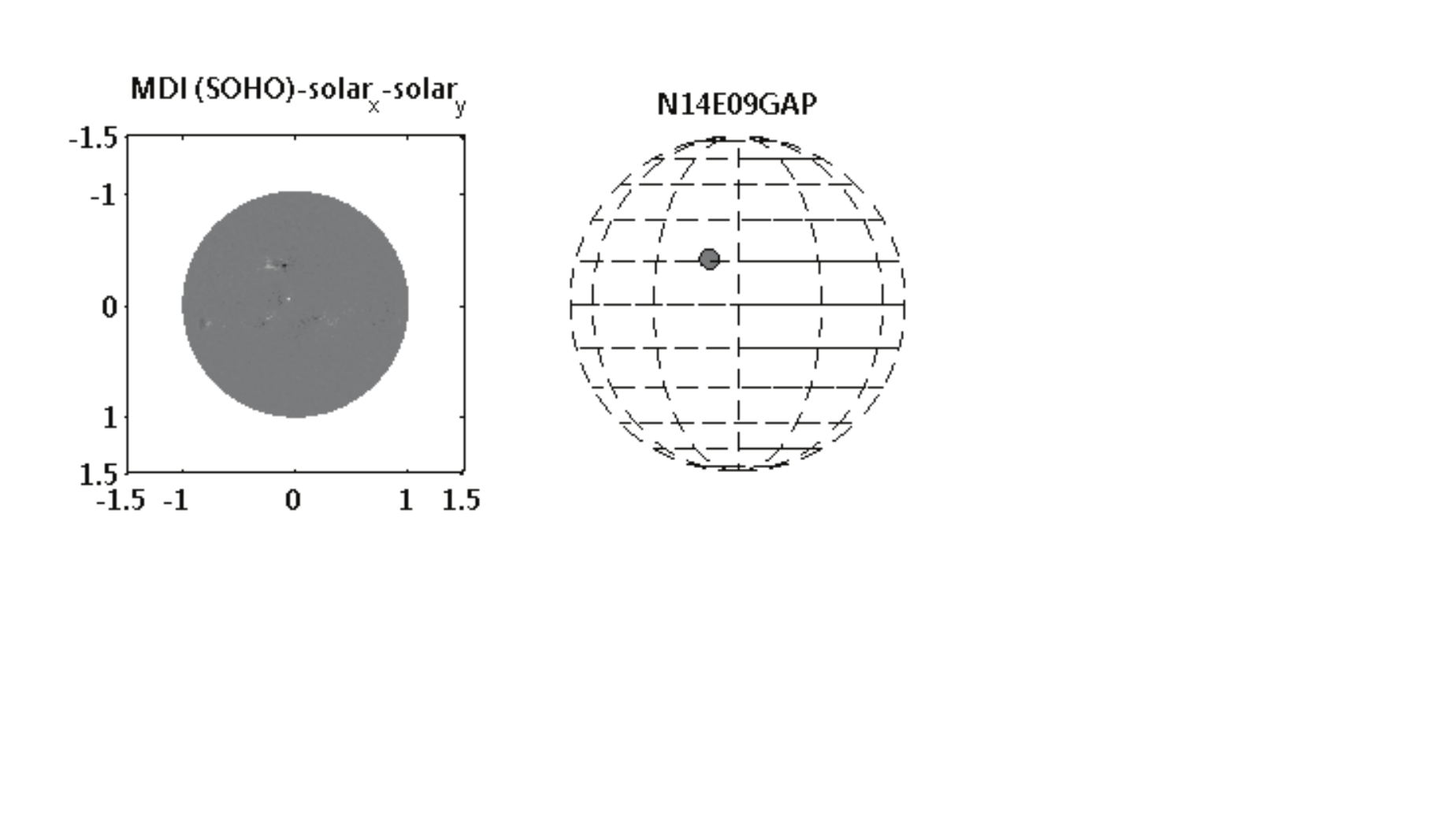}
\caption{\event}
\end{center}
\end{figure} % 
%--------------------------------------------
\clearpage
\renewcommand{\event} {30 March 2004 Event(B)}
%--------------------------------------------
\begin{figure}
\begin{center}% trim=0cm 1cm  0cm 1cm,clip,
\includegraphics[width=\textwidth]{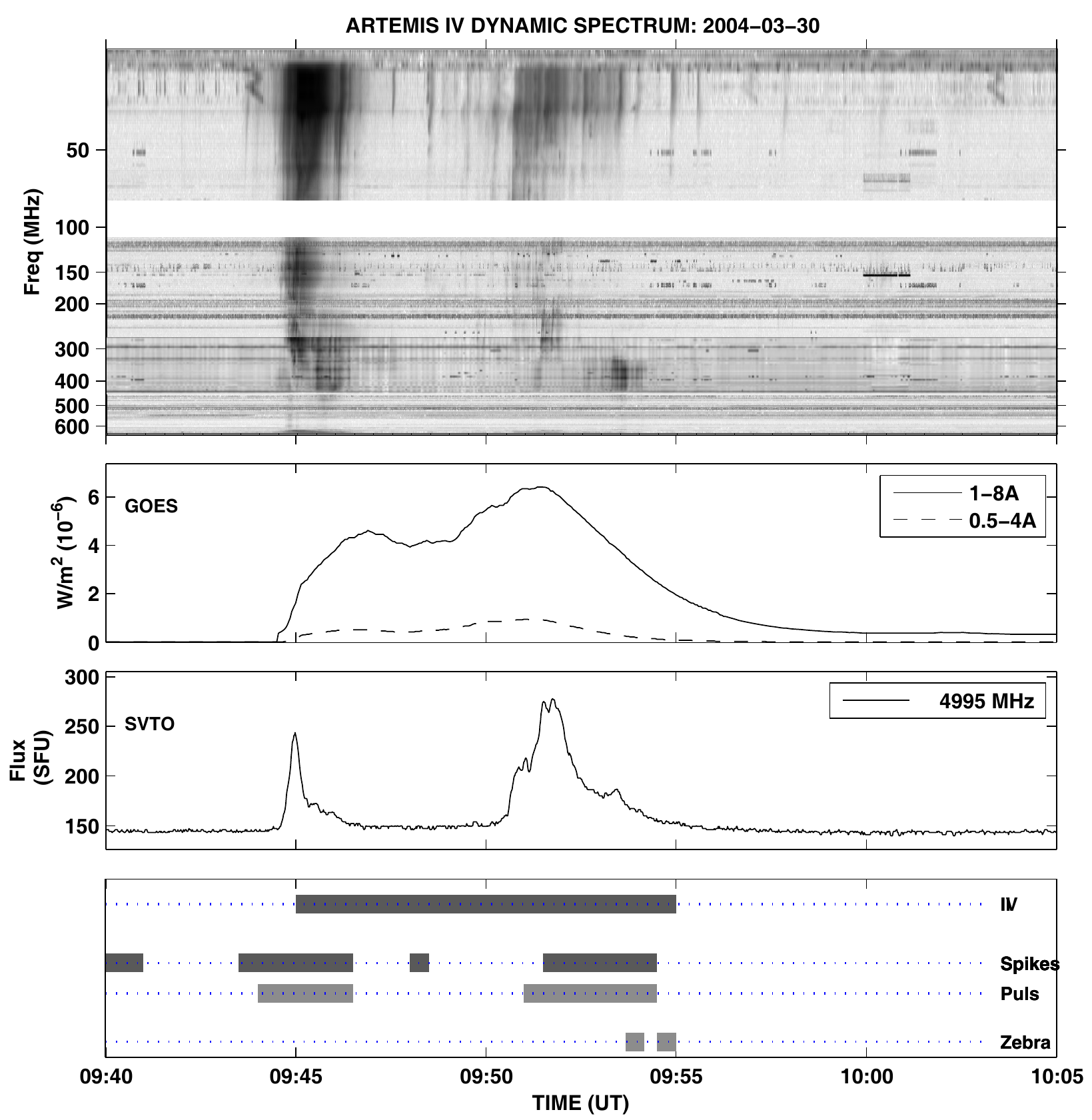}
\includegraphics[trim=0cm 0.0cm  0.0cm 4.0cm,clip,width=0.85\textwidth]{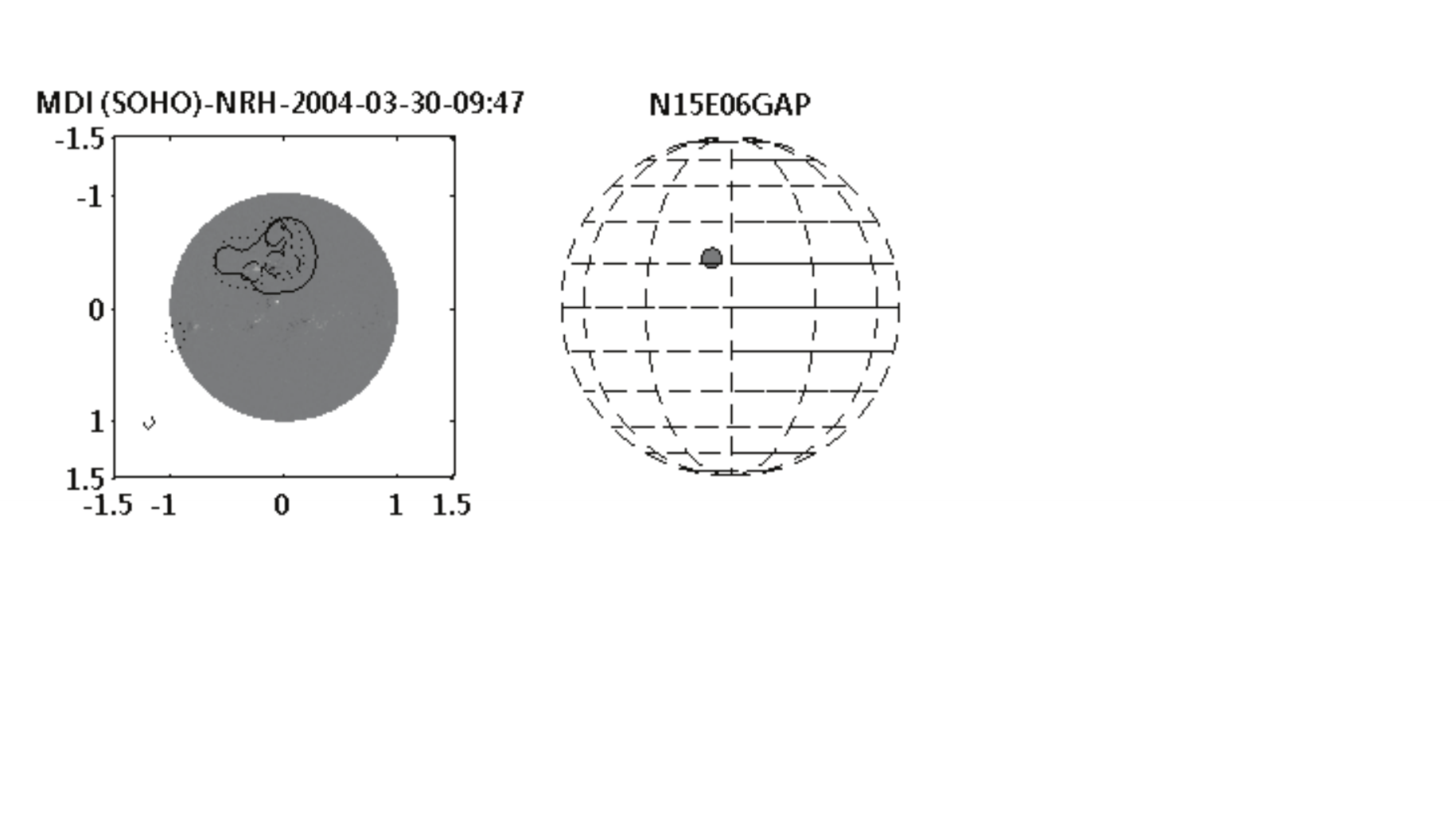}
\caption{\event}
\end{center}
\end{figure} % 
%--------------------------------------------
\clearpage
\renewcommand{\event} {30 March 2004 Event(C)}
%--------------------------------------------
\begin{figure}
\begin{center}% trim=0cm 1cm  0cm 1cm,clip,
\includegraphics[width=\textwidth]{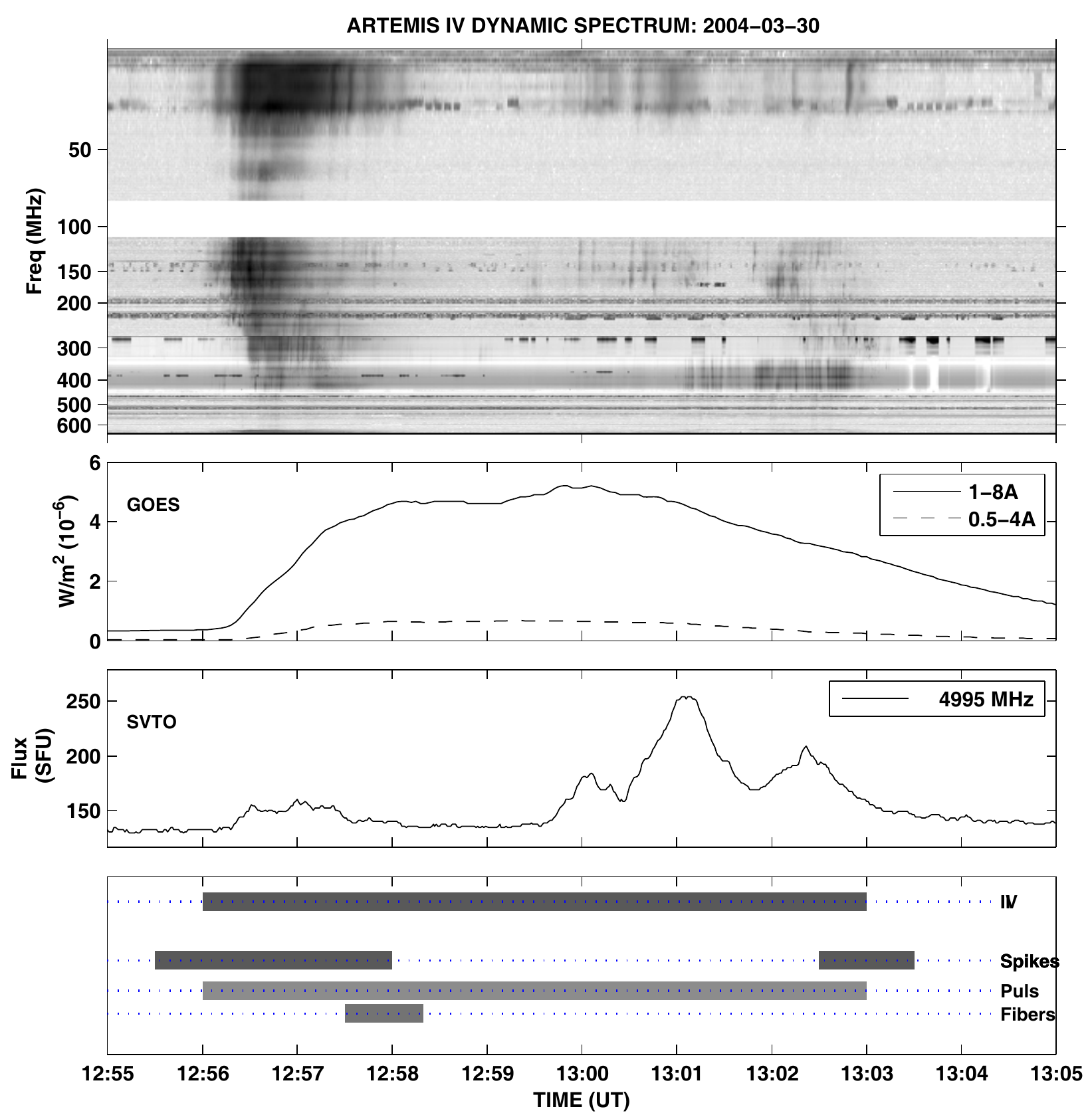}
\includegraphics[trim=0cm 0.0cm  0.0cm 4.0cm,clip,width=0.85\textwidth]{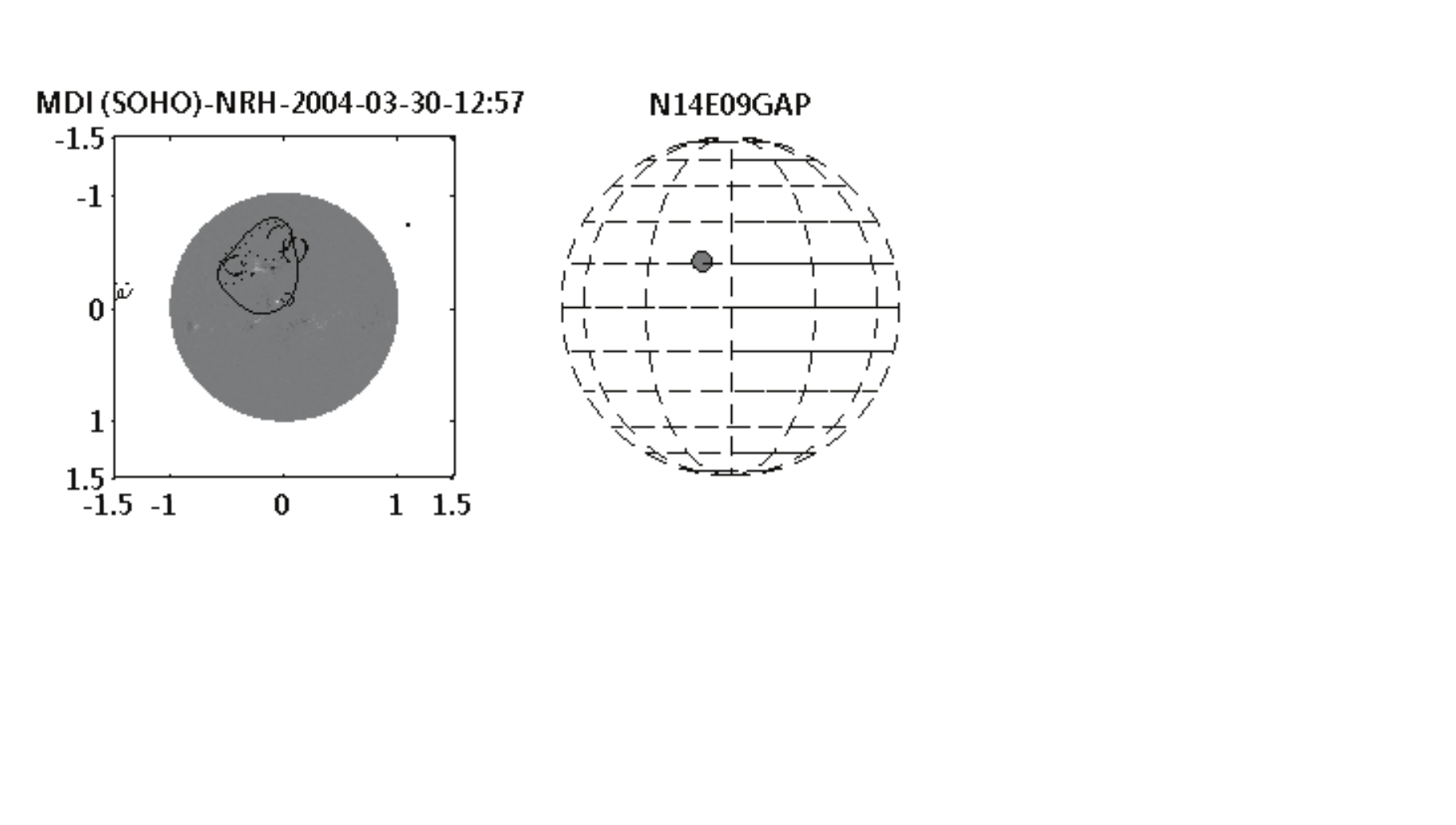}
\caption{\event}
\end{center}
\end{figure} % 
%--------------------------------------------
\clearpage
\renewcommand{\event} {06 April 2004}
%--------------------------------------------
\begin{figure}
\begin{center}% trim=0cm 1cm  0cm 1cm,clip,
\includegraphics[width=\textwidth]{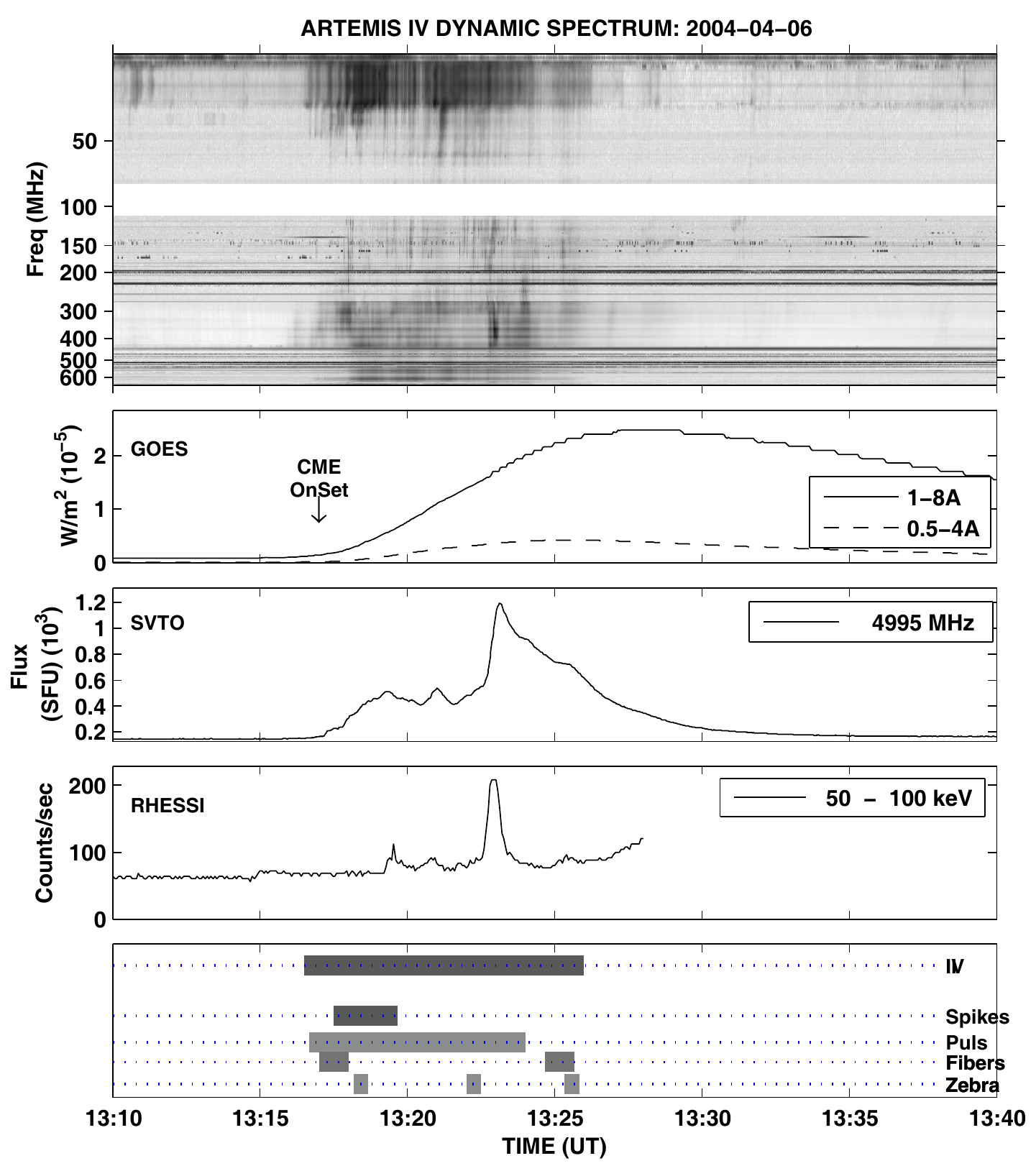}
\includegraphics[trim=0cm 0.0cm  0.0cm 4.0cm,clip,width=0.85\textwidth]{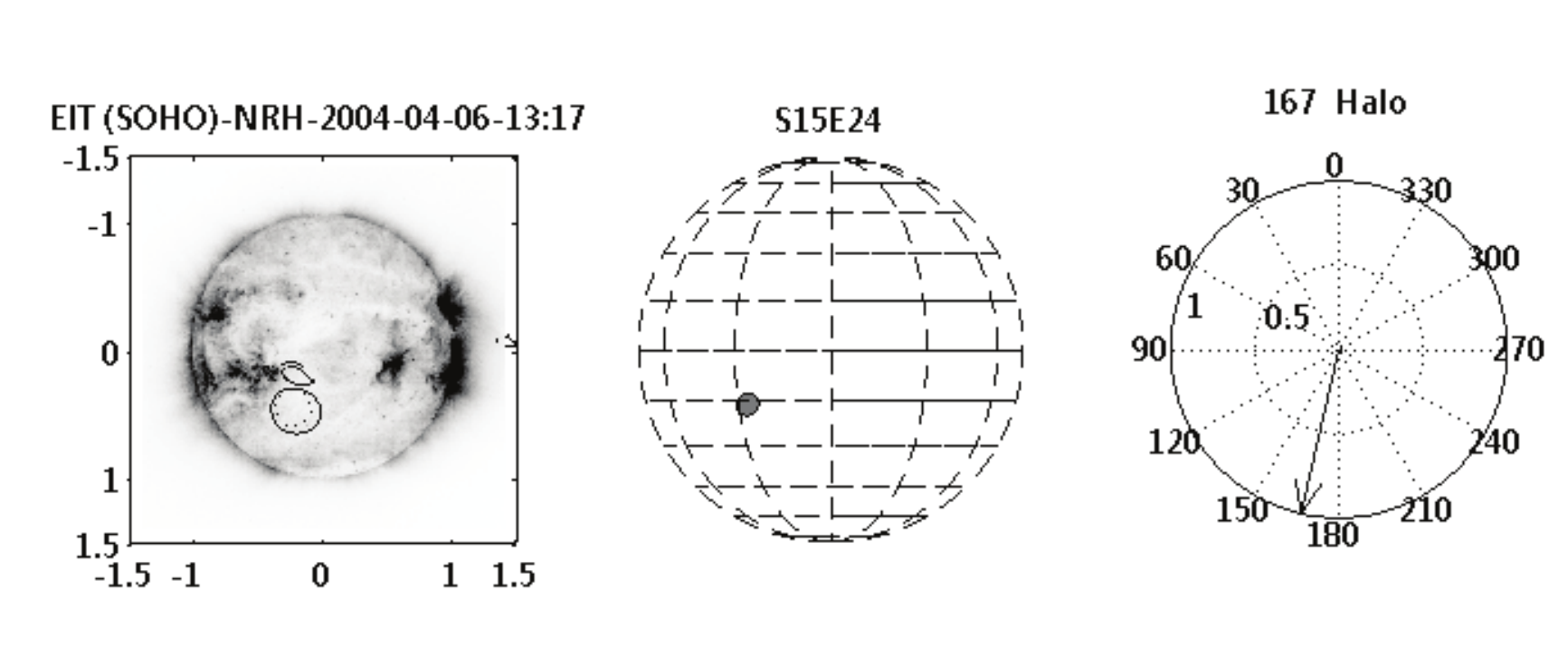}
\caption{\event}
\end{center}
\end{figure} % 
%--------------------------------------------
\clearpage
\renewcommand{\event} {13 July 2004}
%--------------------------------------------
\begin{figure}
\begin{center}% trim=0cm 1cm  0cm 1cm,clip,
\includegraphics[width=\textwidth]{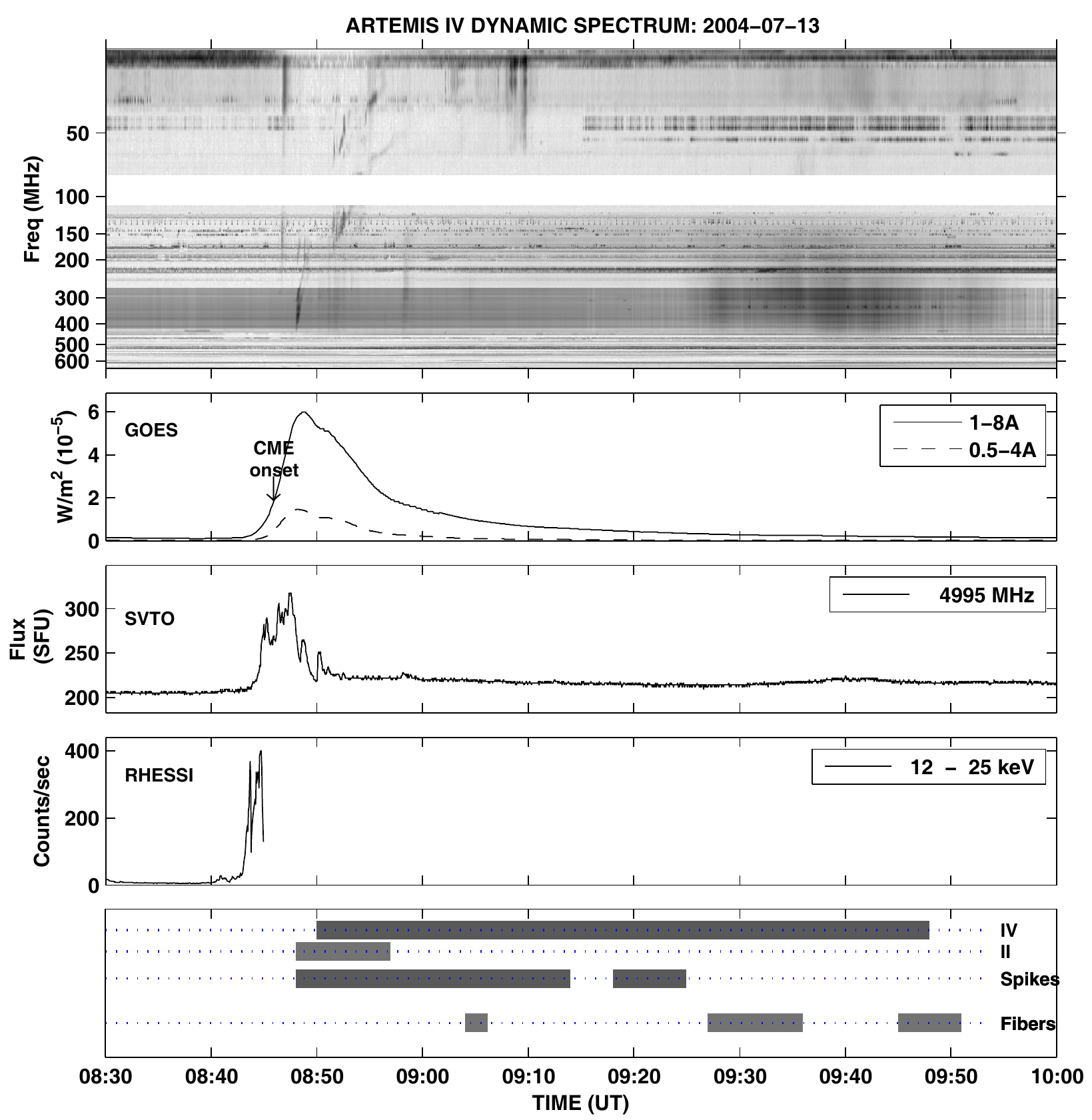}
\includegraphics[trim=0cm 0.0cm  0.0cm 4.0cm,clip,width=0.85\textwidth]{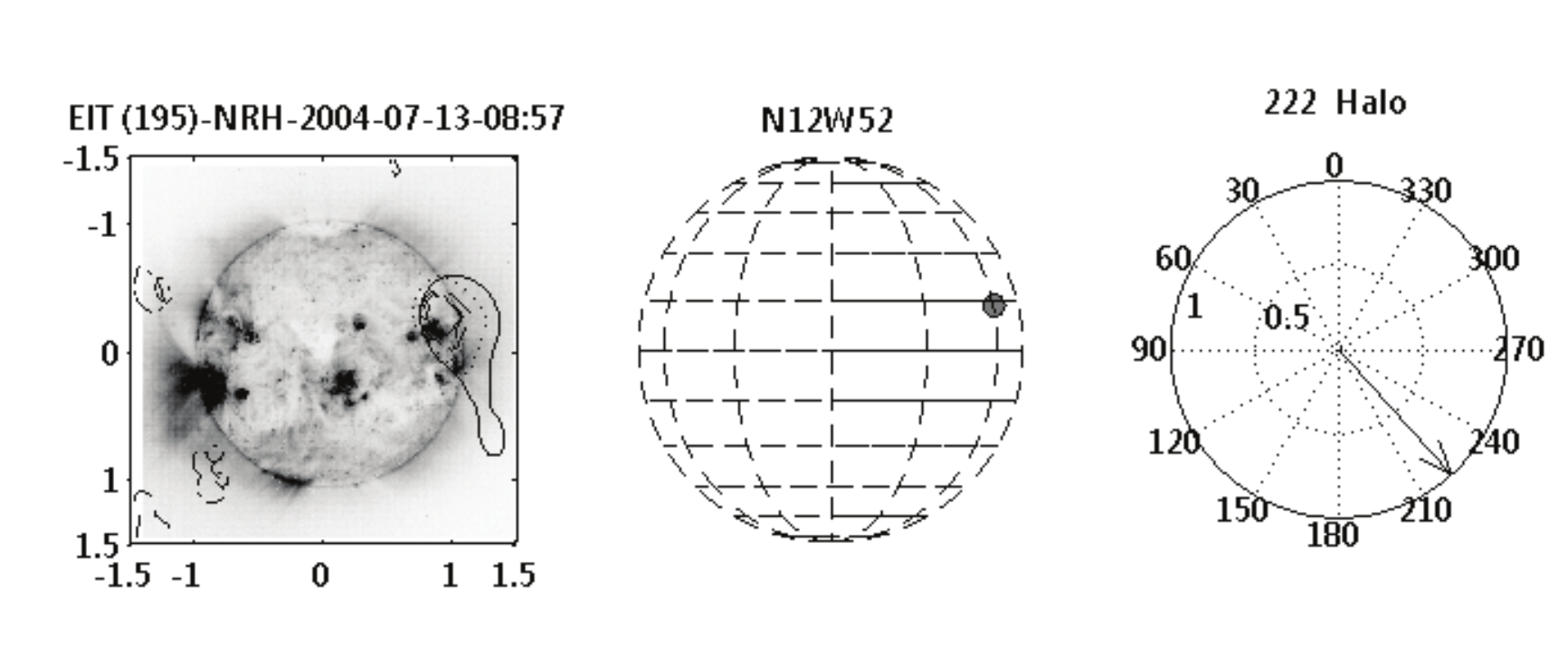}
\caption{\event}
\end{center}
\end{figure} % 
%--------------------------------------------
\clearpage
\renewcommand{\event} {20 July 2004}
%--------------------------------------------
\begin{figure}
\begin{center}% trim=0cm 1cm  0cm 1cm,clip,
\includegraphics[width=\textwidth]{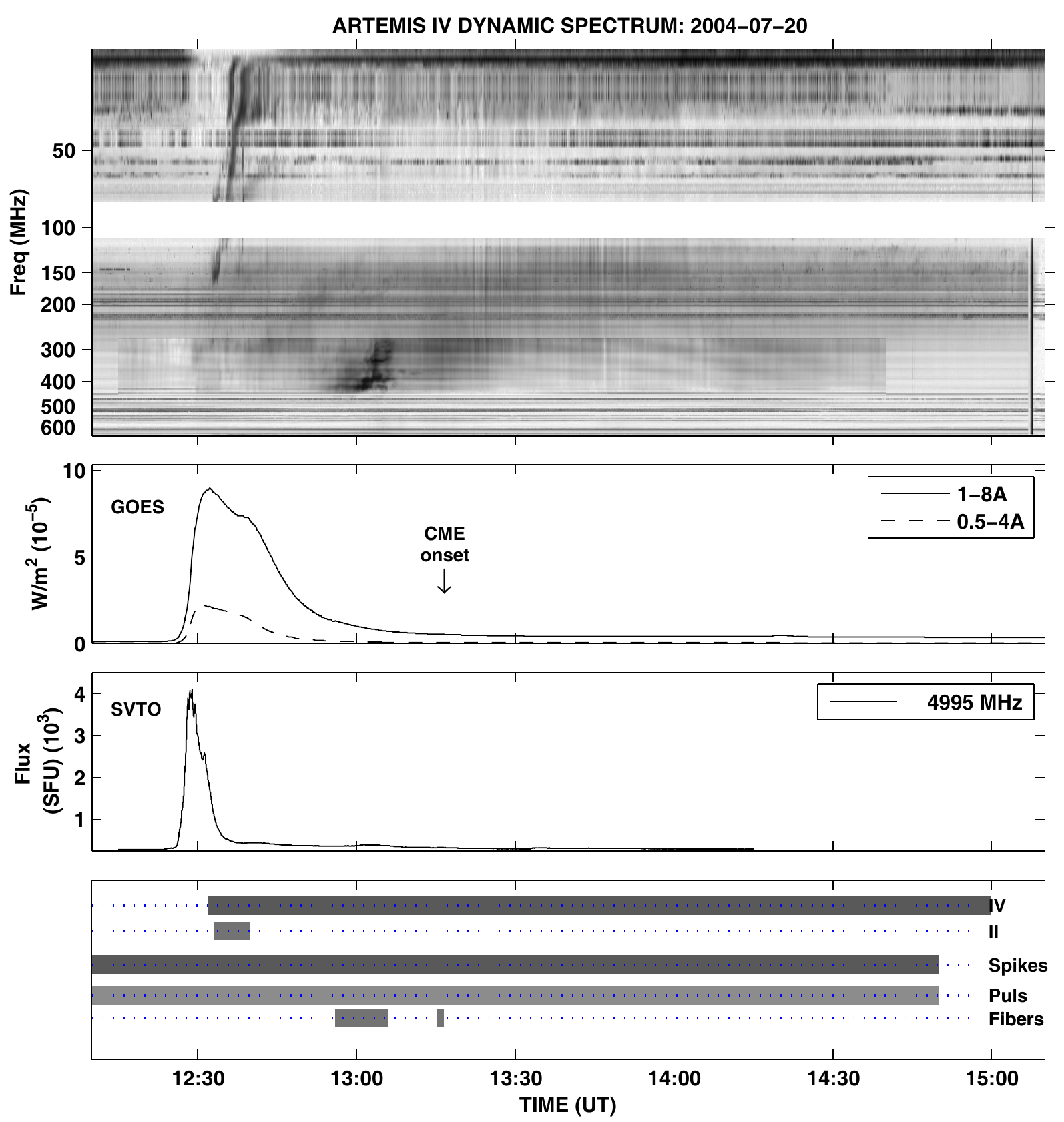}
\includegraphics[trim=0cm 0.0cm  0.0cm 4.0cm,clip,width=0.85\textwidth]{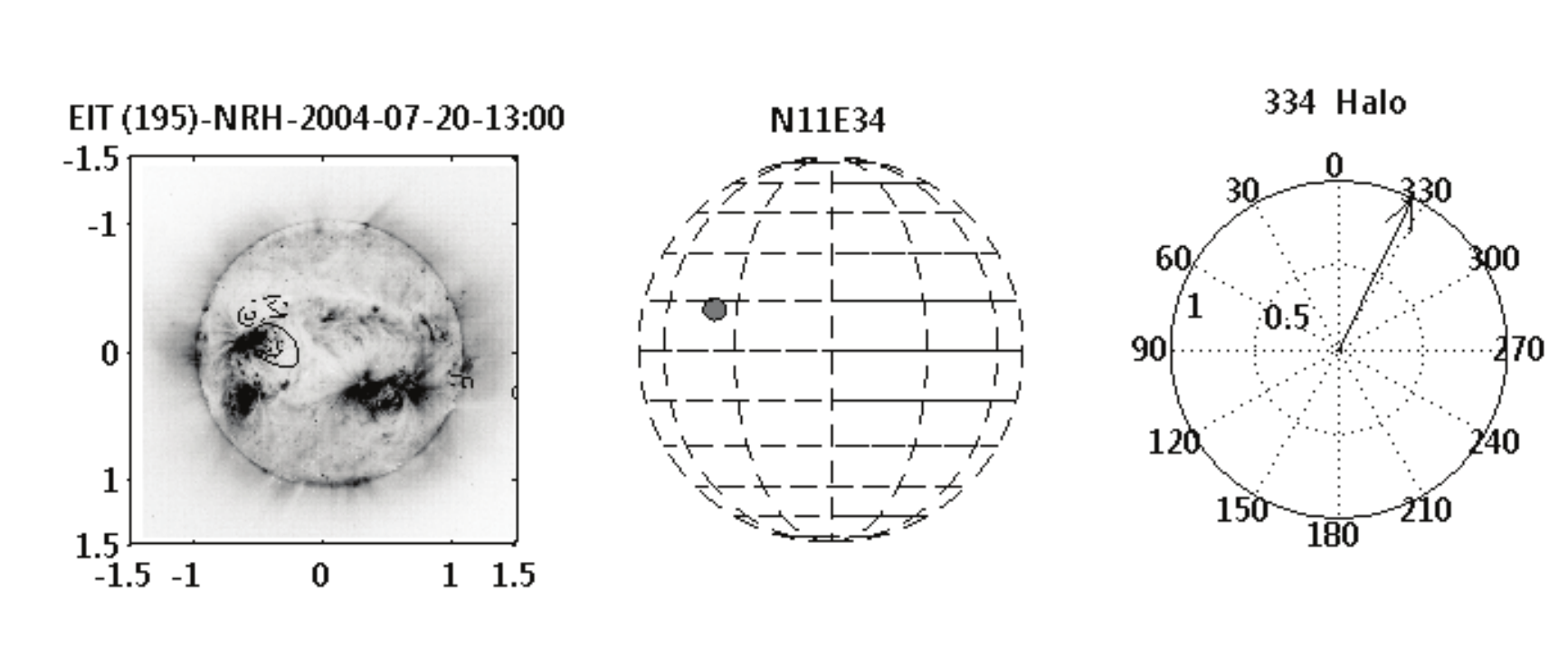}
\caption{\event}
\end{center}
\end{figure} % 
%--------------------------------------------
\clearpage
\renewcommand{\event} {21 July 2004}
%--------------------------------------------
\begin{figure}
\begin{center}% trim=0cm 1cm  0cm 1cm,clip,
\includegraphics[width=\textwidth]{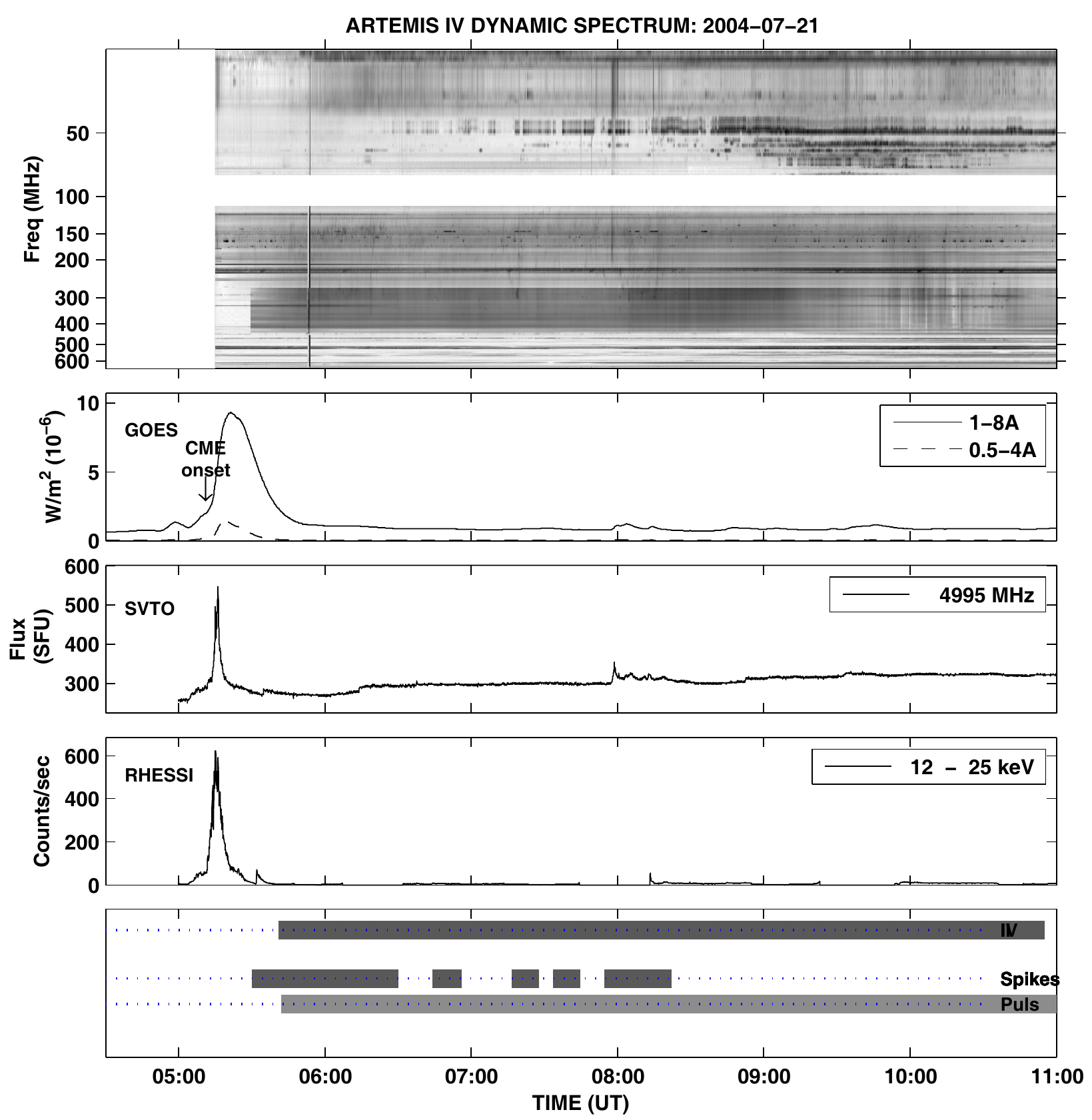}
\includegraphics[trim=0cm 0.0cm  0.0cm 4.0cm,clip,width=0.85\textwidth]{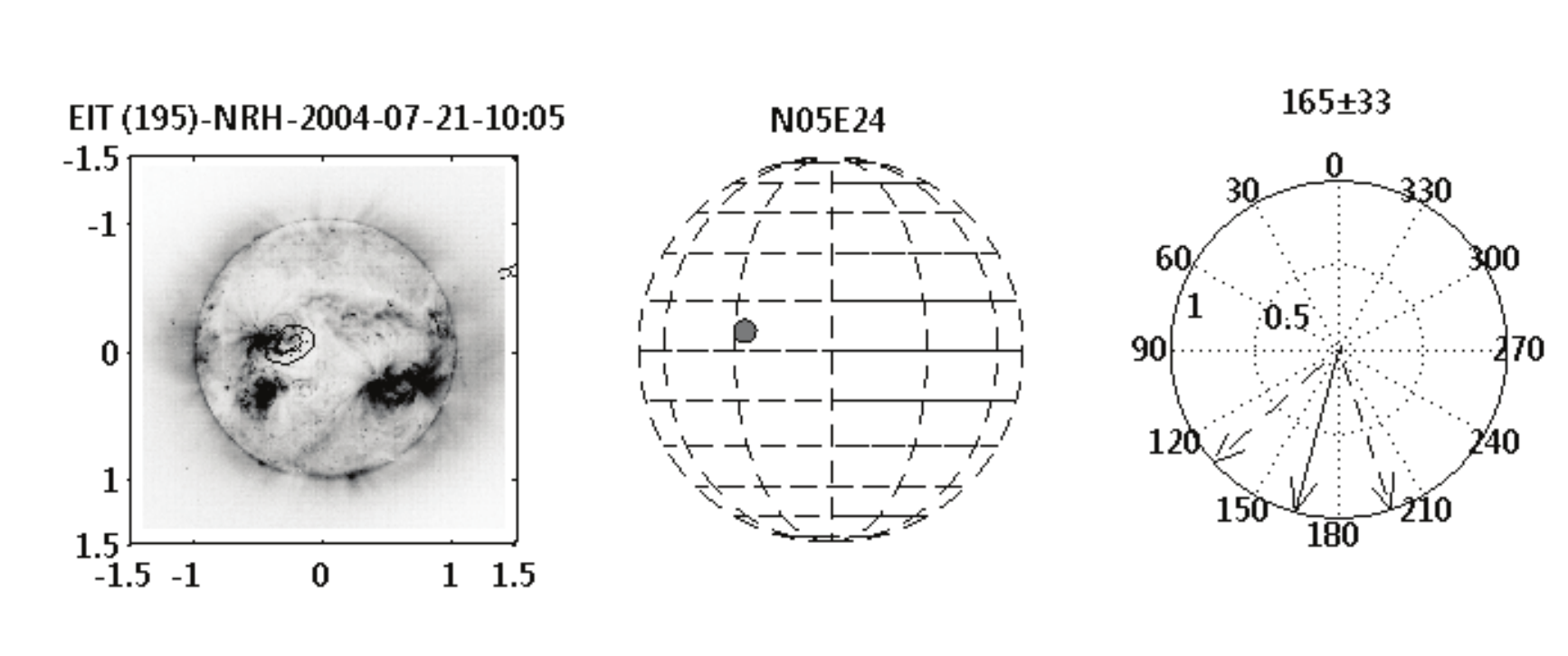}
\caption{\event}
\end{center}
\end{figure} % 
%--------------------------------------------
\clearpage
\renewcommand{\event} {14 January 2005}
%--------------------------------------------
\begin{figure}
\begin{center}% trim=0cm 1cm  0cm 1cm,clip,
\includegraphics[width=\textwidth]{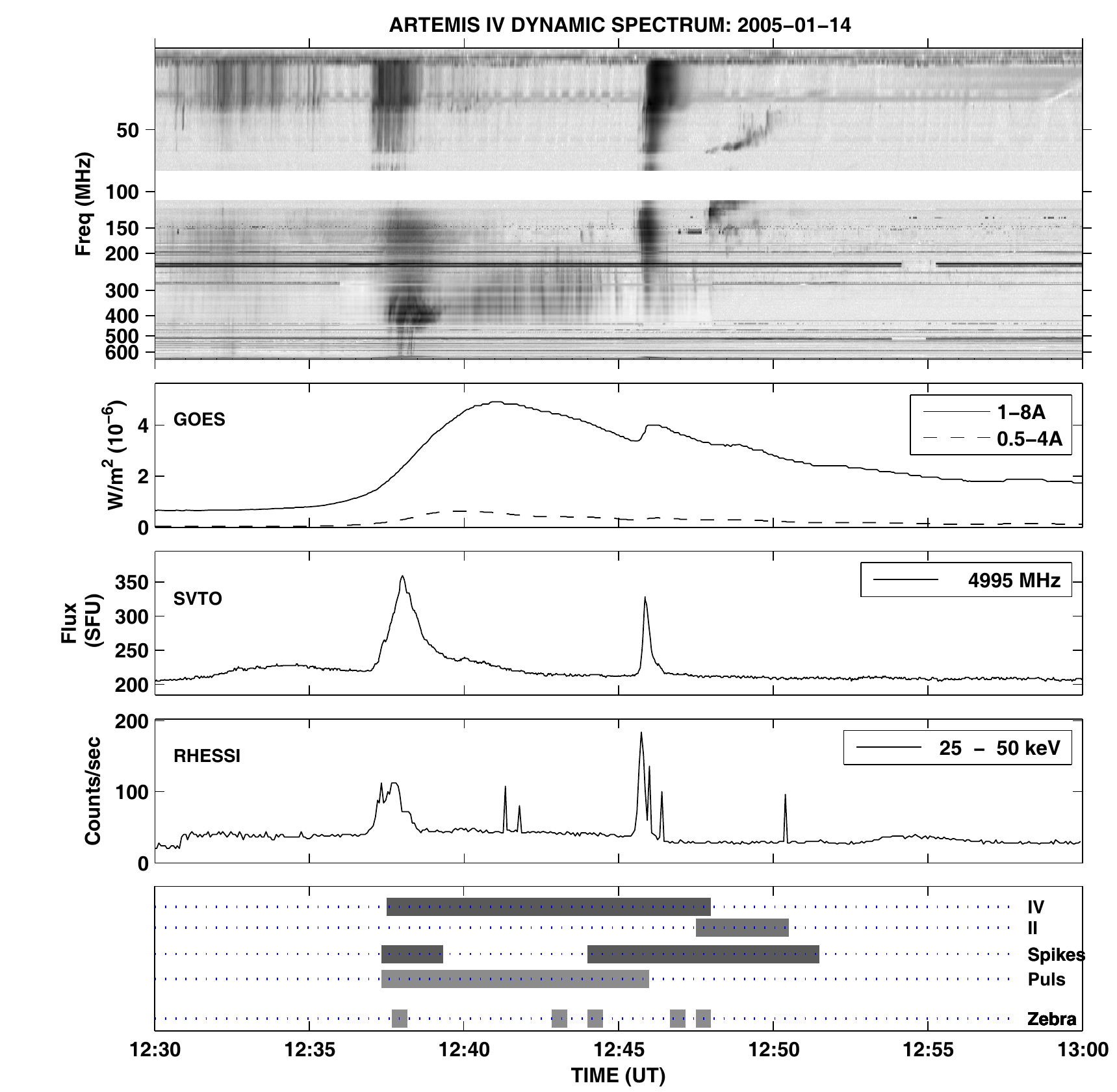}
\includegraphics[trim=0cm 0.0cm  0.0cm 4.0cm,clip,width=0.85\textwidth]{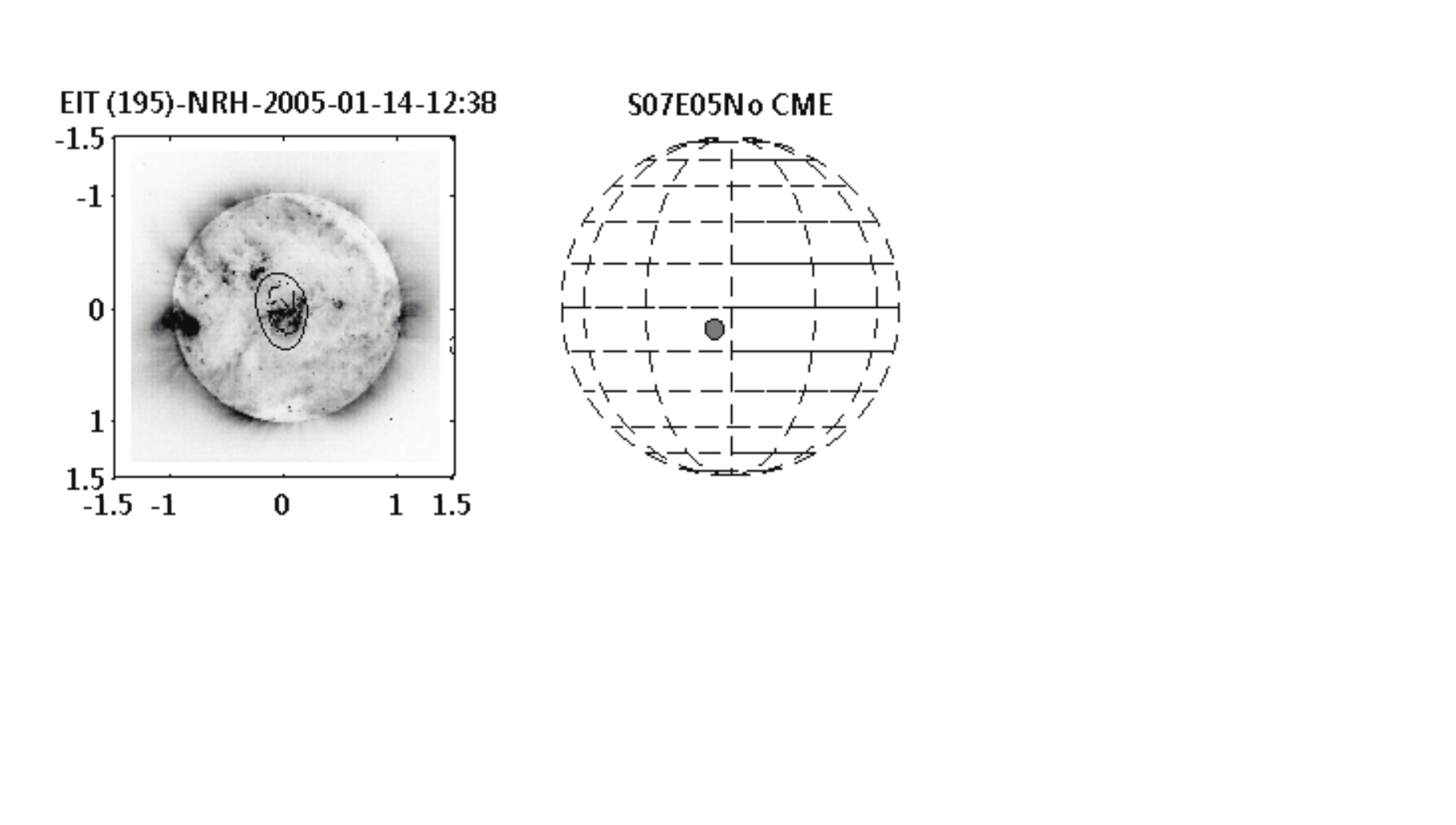}
\caption{\event}
\end{center}
\end{figure} % 
%--------------------------------------------
\clearpage
\renewcommand{\event} {15 January 2005}
%--------------------------------------------
\begin{figure}
\begin{center}% trim=0cm 1cm  0cm 1cm,clip,
\includegraphics[width=\textwidth]{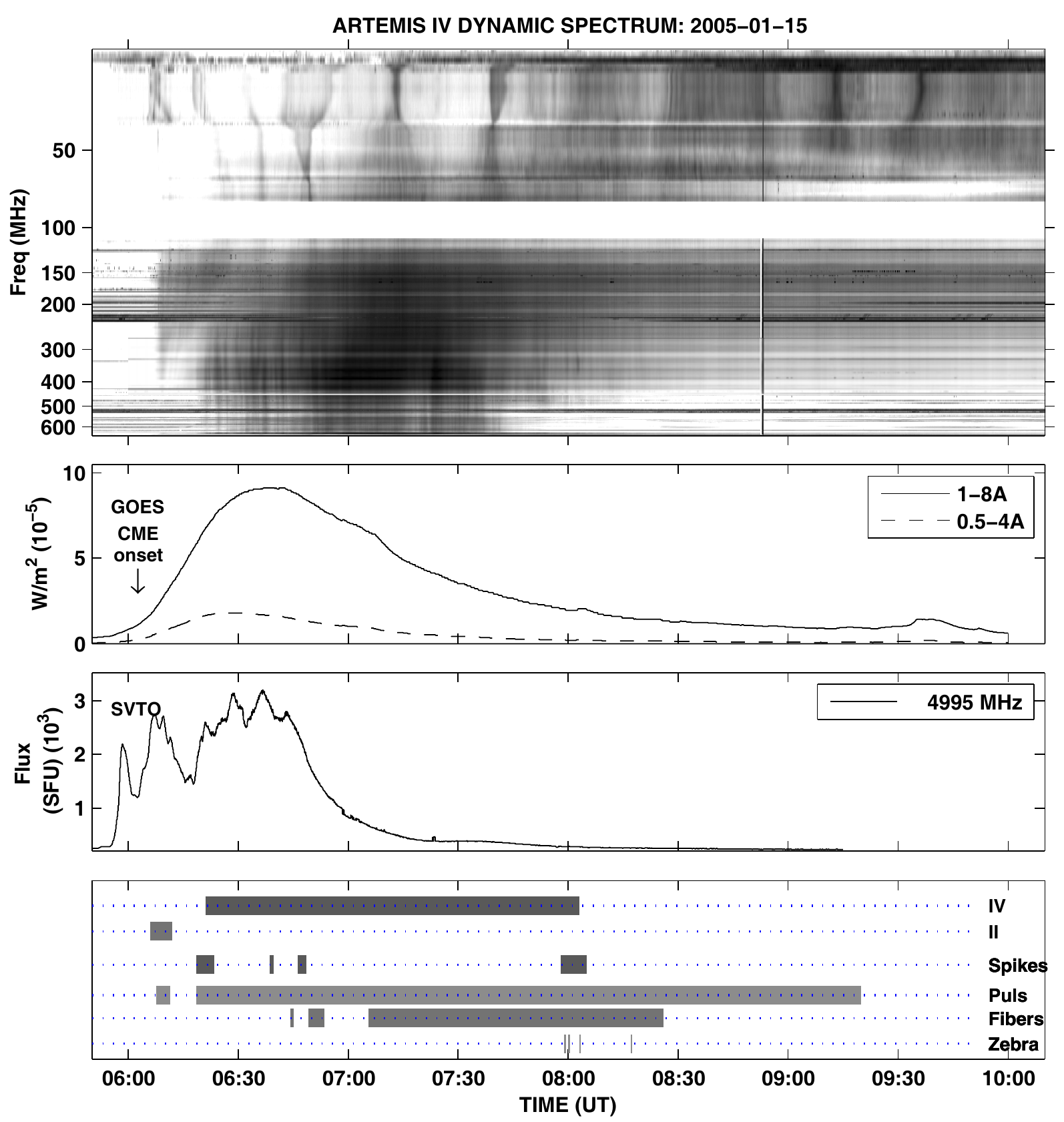}
\includegraphics[trim=0cm 0.0cm  0.0cm 4.0cm,clip,width=0.85\textwidth]{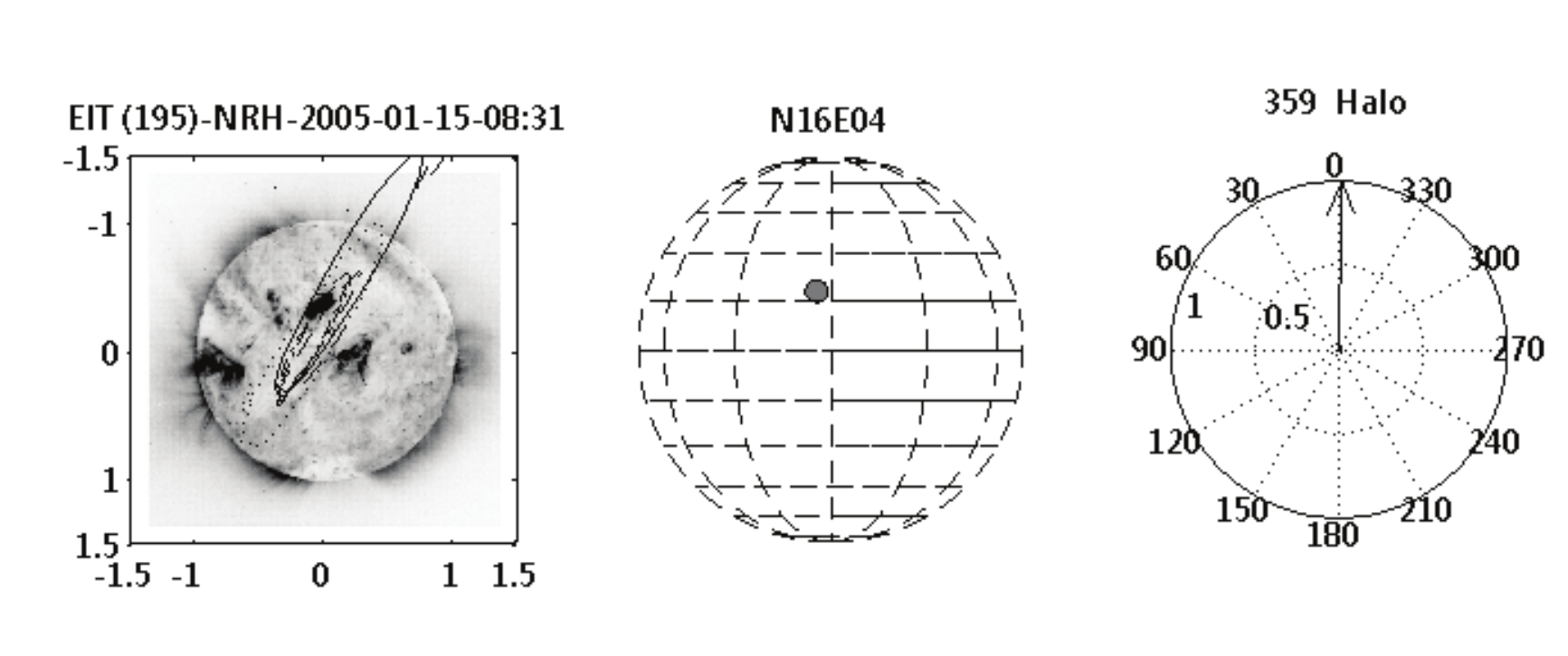}
\caption{\event}
\end{center}
\end{figure} % 
%--------------------------------------------
\clearpage
\renewcommand{\event} {17 January 2005}
%--------------------------------------------
\begin{figure}
\begin{center}% trim=0cm 1cm  0cm 1cm,clip,
\includegraphics[width=\textwidth]{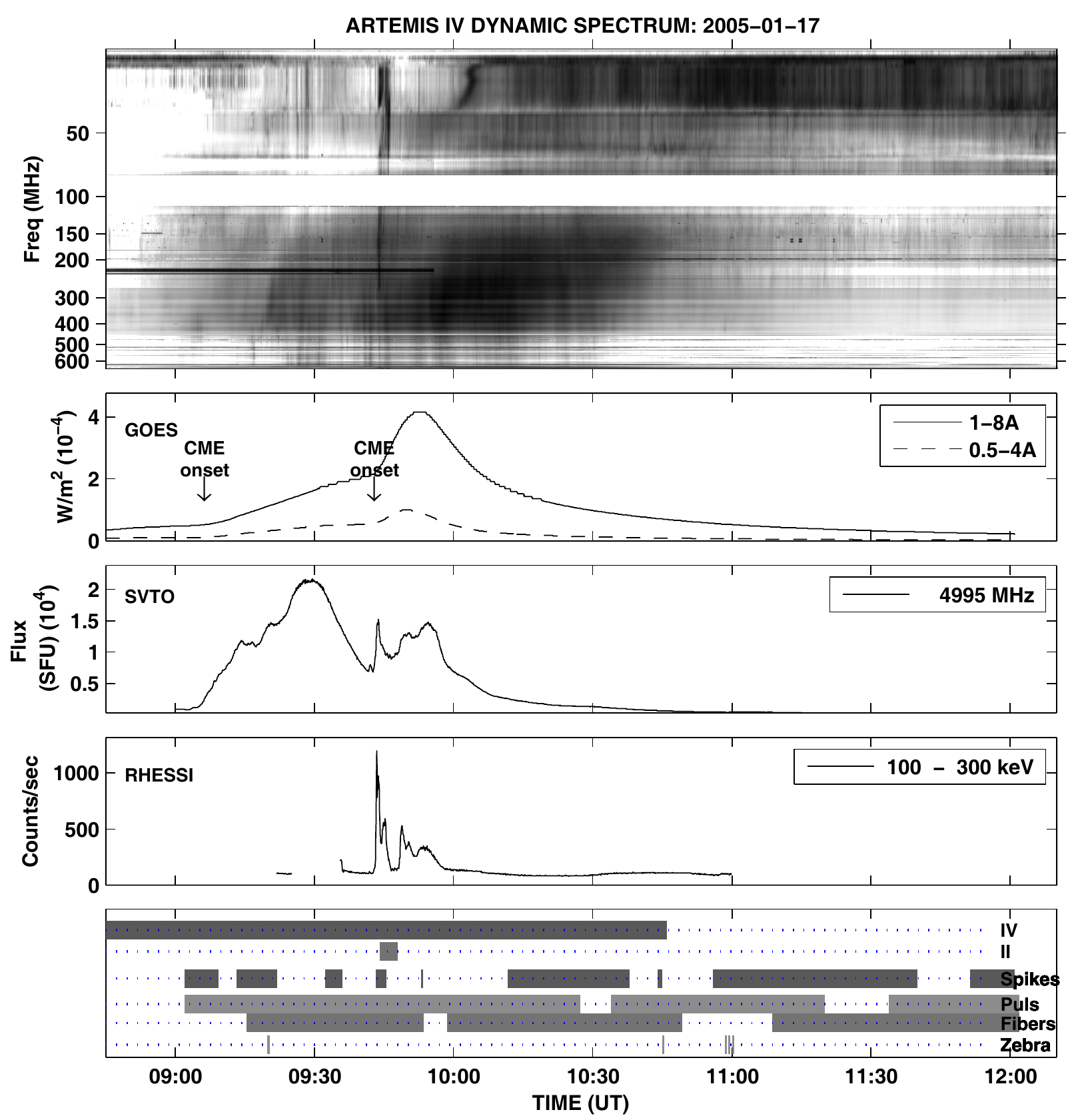}
\includegraphics[trim=0cm 0.0cm  0.0cm 4.0cm,clip,width=0.85\textwidth]{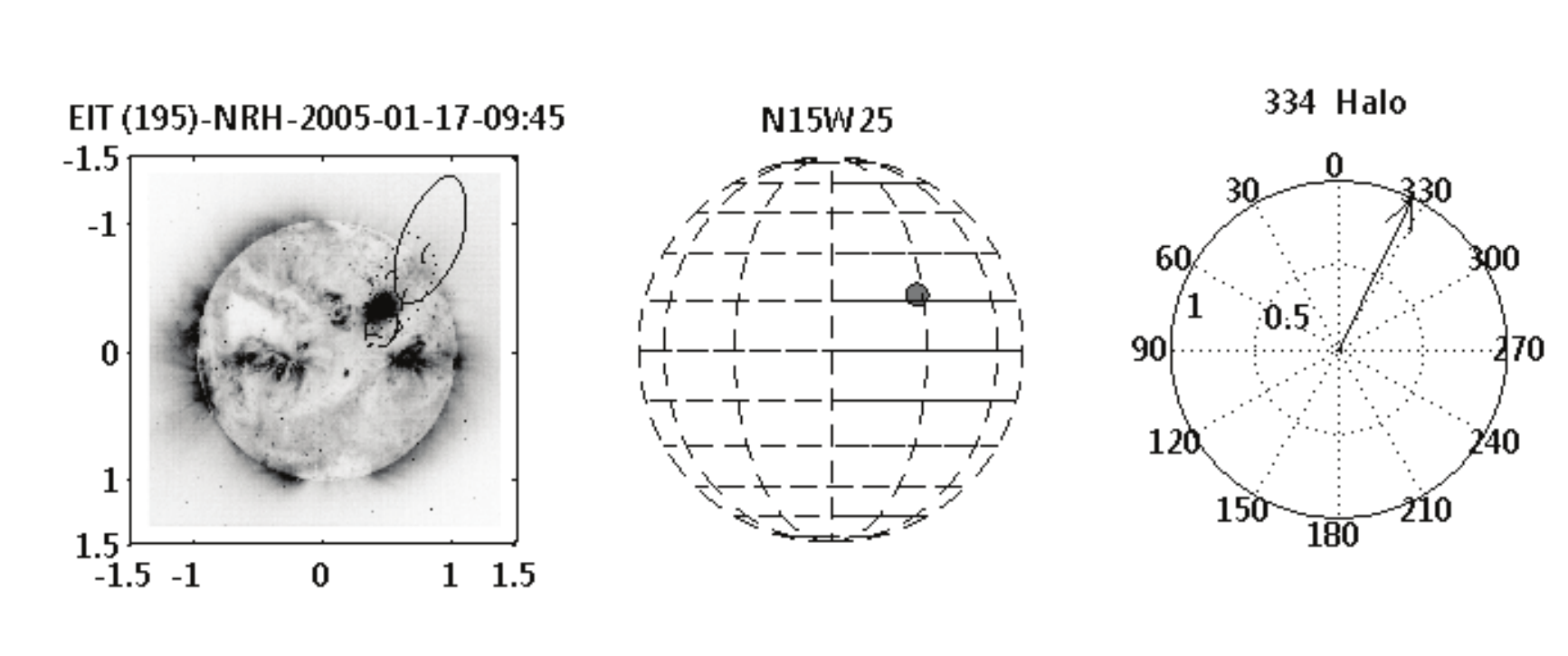}
\caption{\event}
\label{B27}
\end{center}
\end{figure} % 
%--------------------------------------------
\clearpage
\renewcommand{\event} {19 January 2005}
%--------------------------------------------
\begin{figure}
\begin{center}% trim=0cm 1cm  0cm 1cm,clip,
\includegraphics[width=\textwidth]{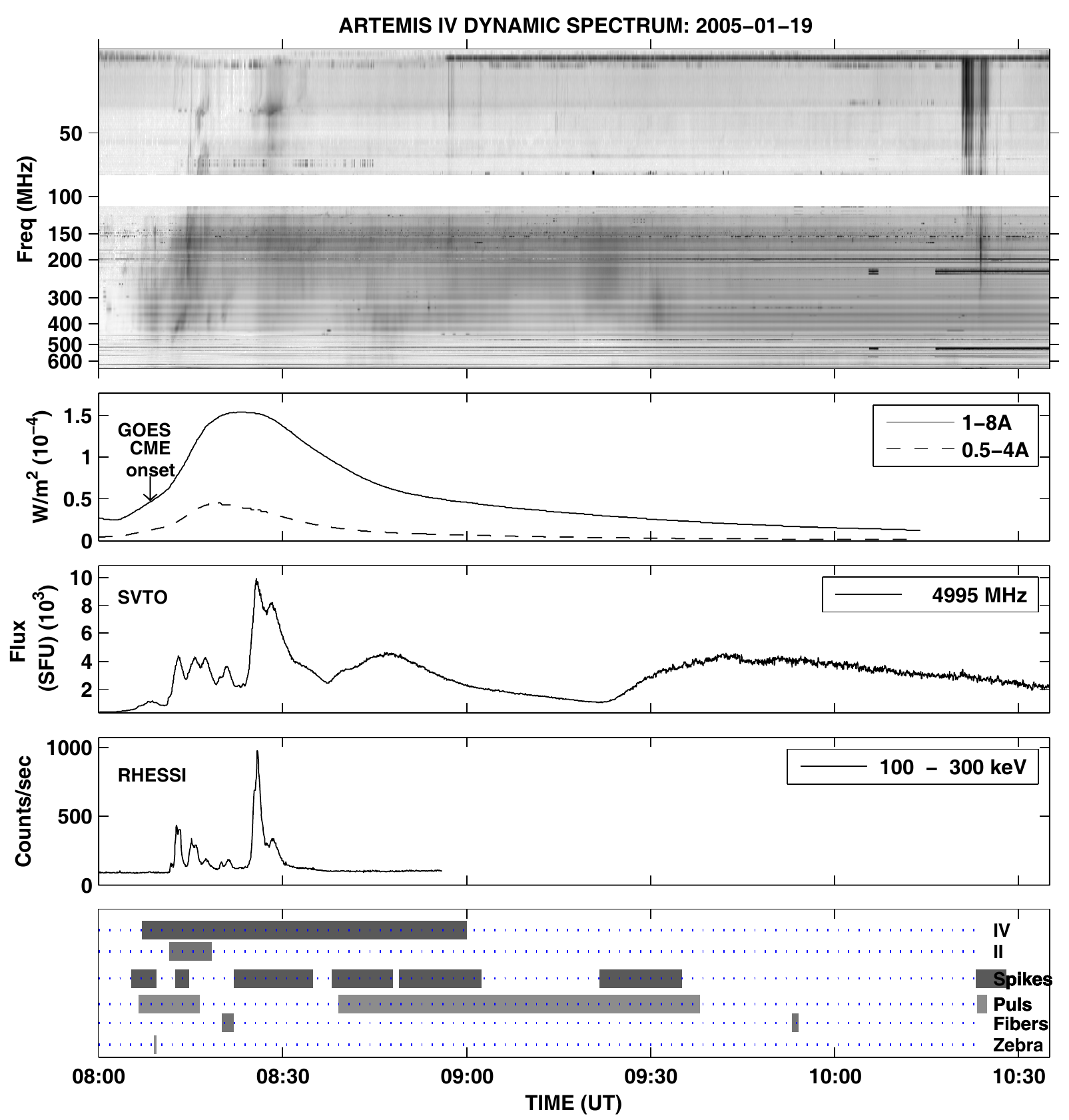}
\includegraphics[trim=0cm 0.0cm  0.0cm 4.0cm,clip,width=0.85\textwidth]{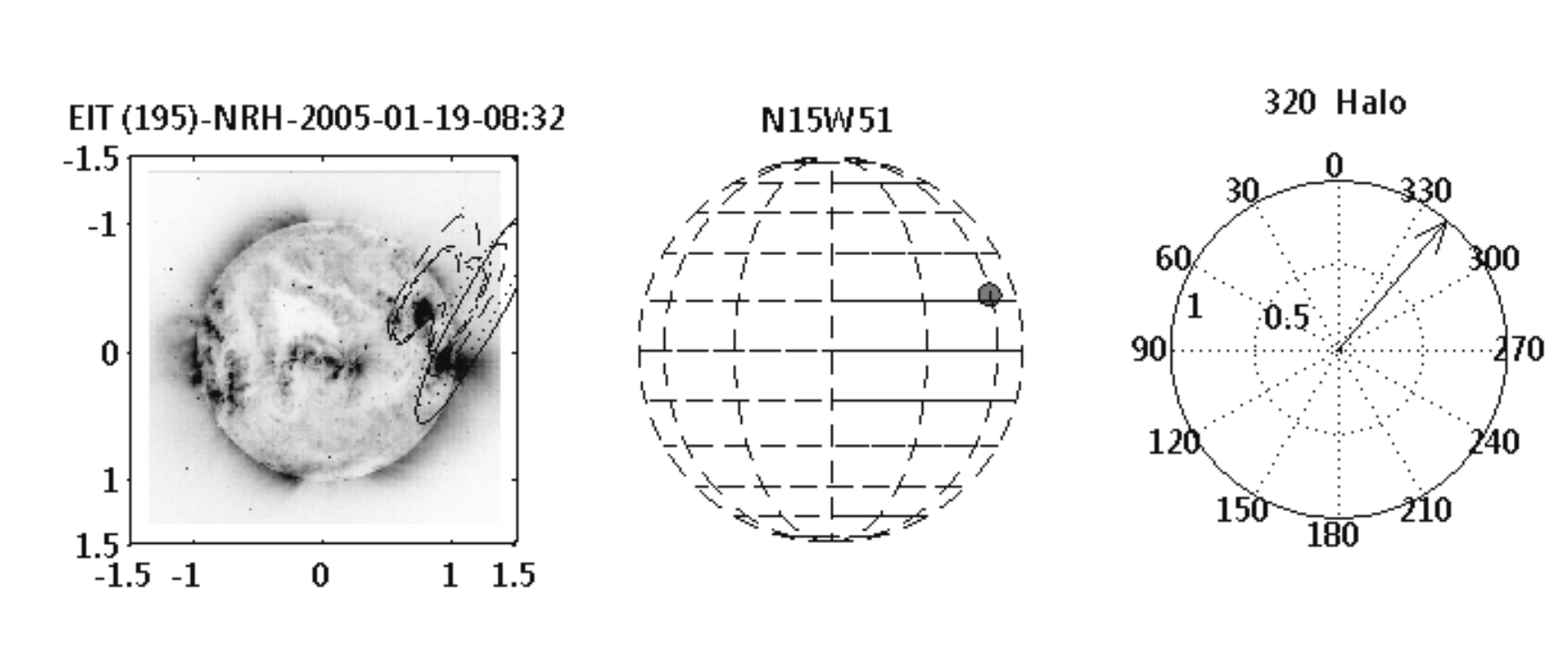}
\caption{\event}
\end{center}
\end{figure} % 
%--------------------------------------------
\clearpage
\renewcommand{\event} {20 January 2005}
%--------------------------------------------
\begin{figure}
\begin{center}% trim=0cm 1cm  0cm 1cm,clip,
\includegraphics[width=\textwidth]{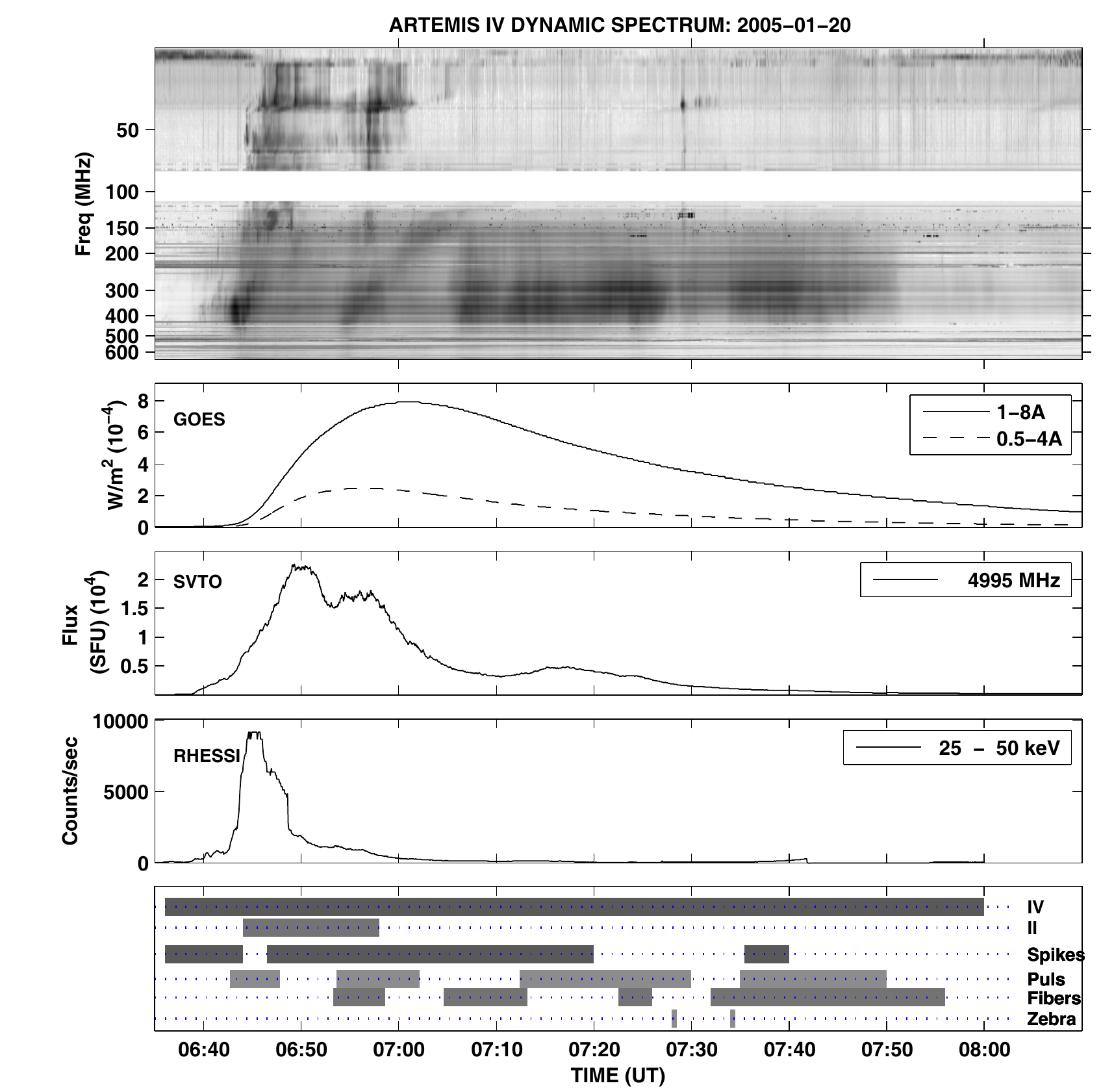}
\includegraphics[trim=0cm 0.0cm  0.0cm 4.0cm,clip,width=0.85\textwidth]{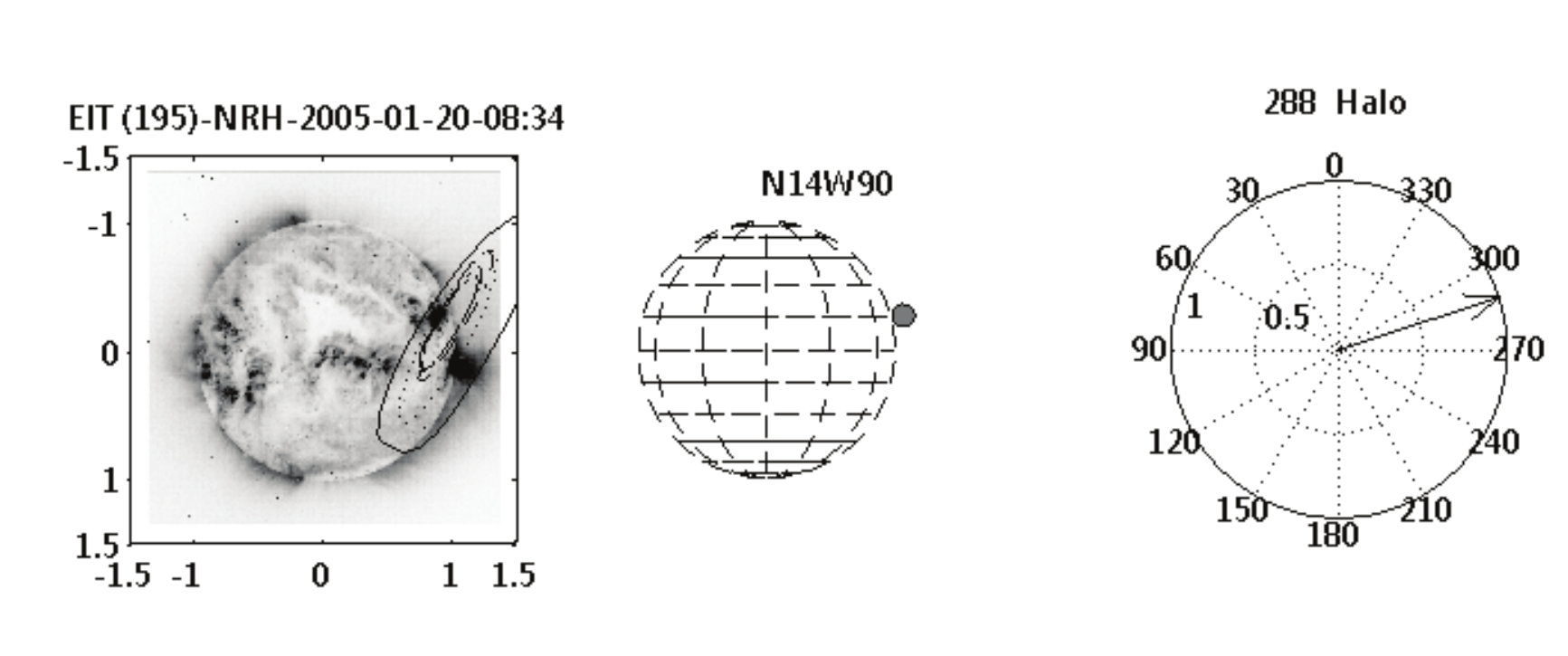}
\caption{\event}
\end{center}
\end{figure} % 
%--------------------------------------------
\clearpage
\renewcommand{\event} {13 July 2005}
%--------------------------------------------
\begin{figure}
\begin{center}% trim=0cm 1cm  0cm 1cm,clip,
\includegraphics[width=\textwidth]{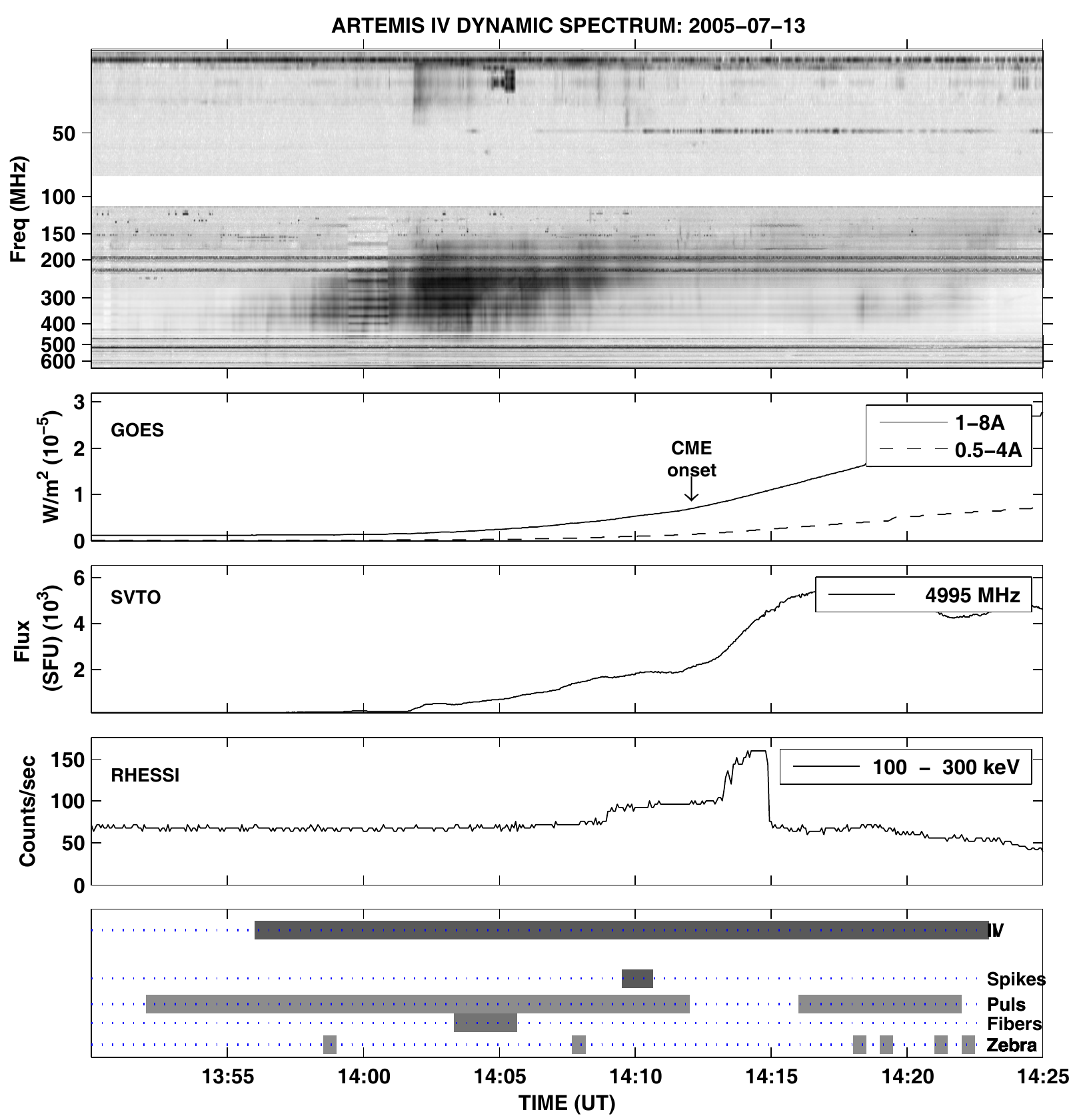}
\includegraphics[trim=0cm 0.0cm  0.0cm 4.0cm,clip,width=0.85\textwidth]{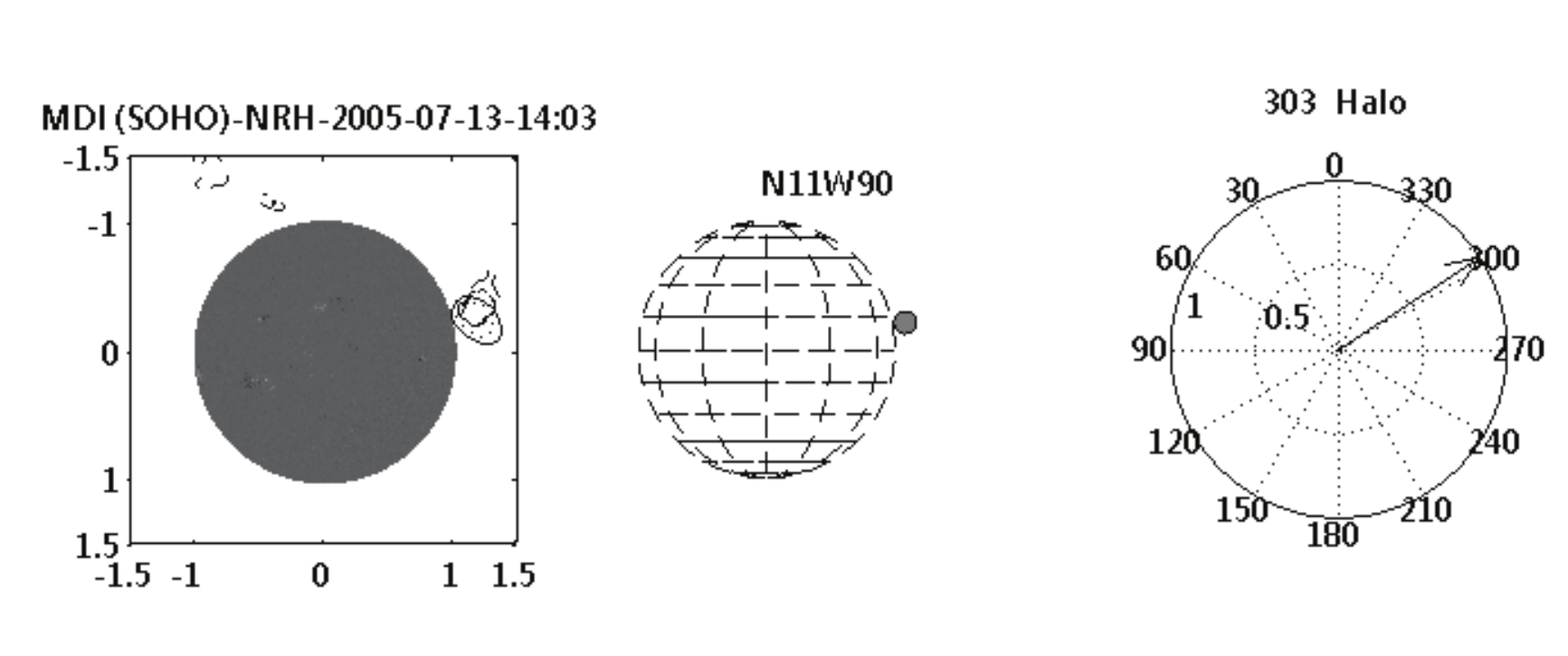}
\caption{\event}
\end{center}
\end{figure} % 
%--------------------------------------------
\clearpage
\renewcommand{\event} {14 July 2005}
%--------------------------------------------
\begin{figure}
\begin{center}% trim=0cm 1cm  0cm 1cm,clip,
\includegraphics[width=\textwidth]{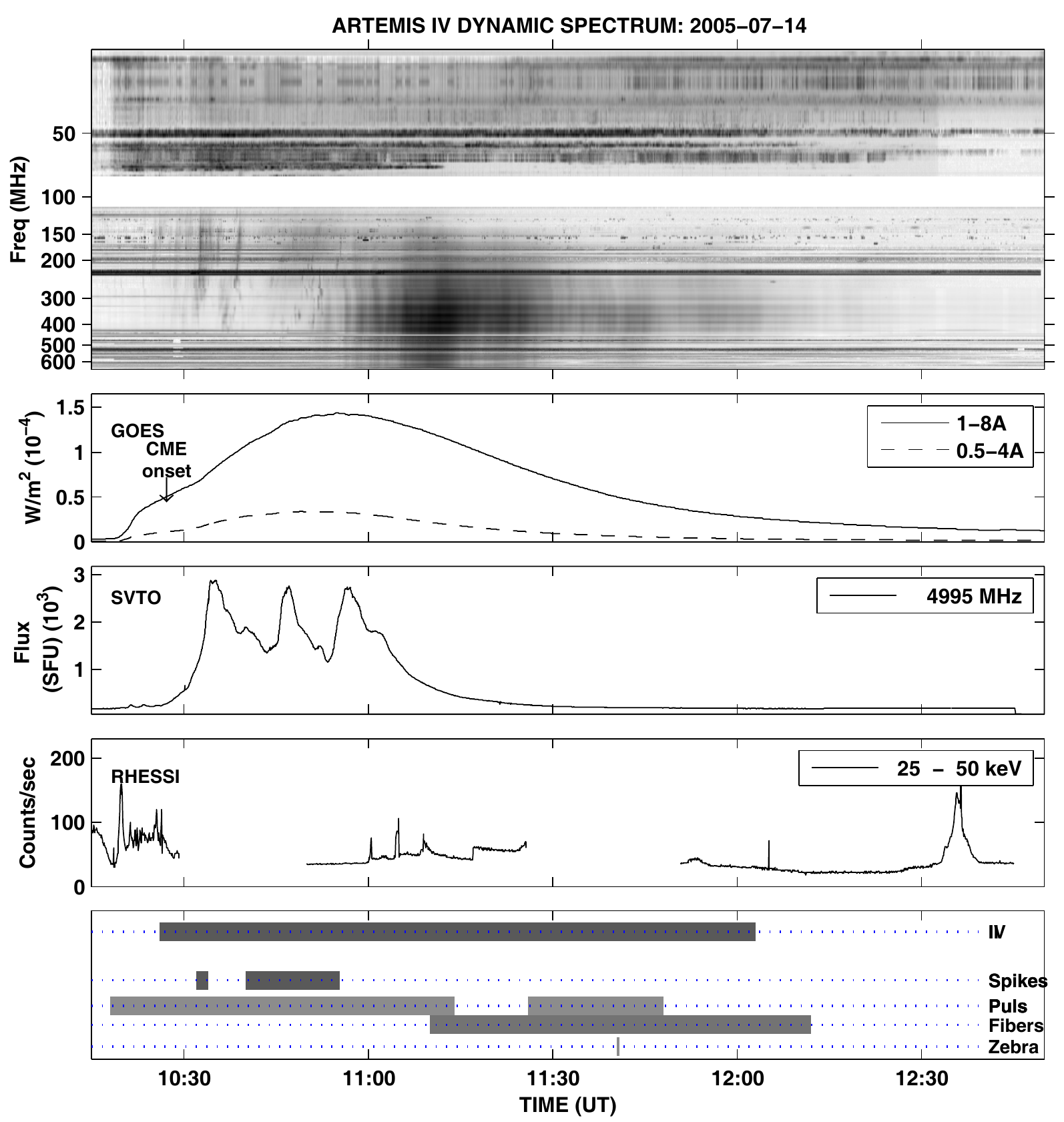}
\includegraphics[trim=0cm 0.0cm  0.0cm 4.0cm,clip,width=0.85\textwidth]{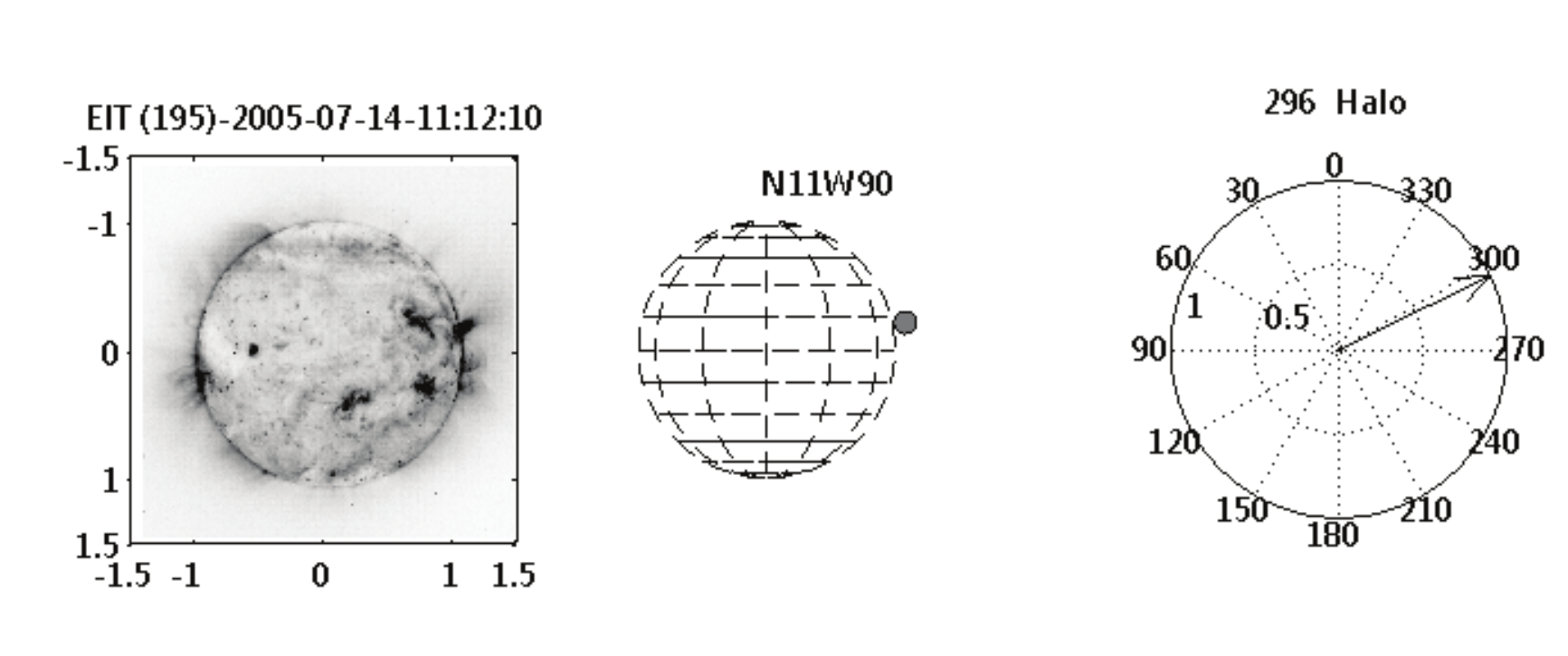}
\caption{\event}
\end{center}
\end{figure} % 
%--------------------------------------------
\clearpage
\renewcommand{\event} {30 July 2005}
%--------------------------------------------
\begin{figure}
\begin{center}% trim=0cm 1cm  0cm 1cm,clip,
\includegraphics[width=\textwidth]{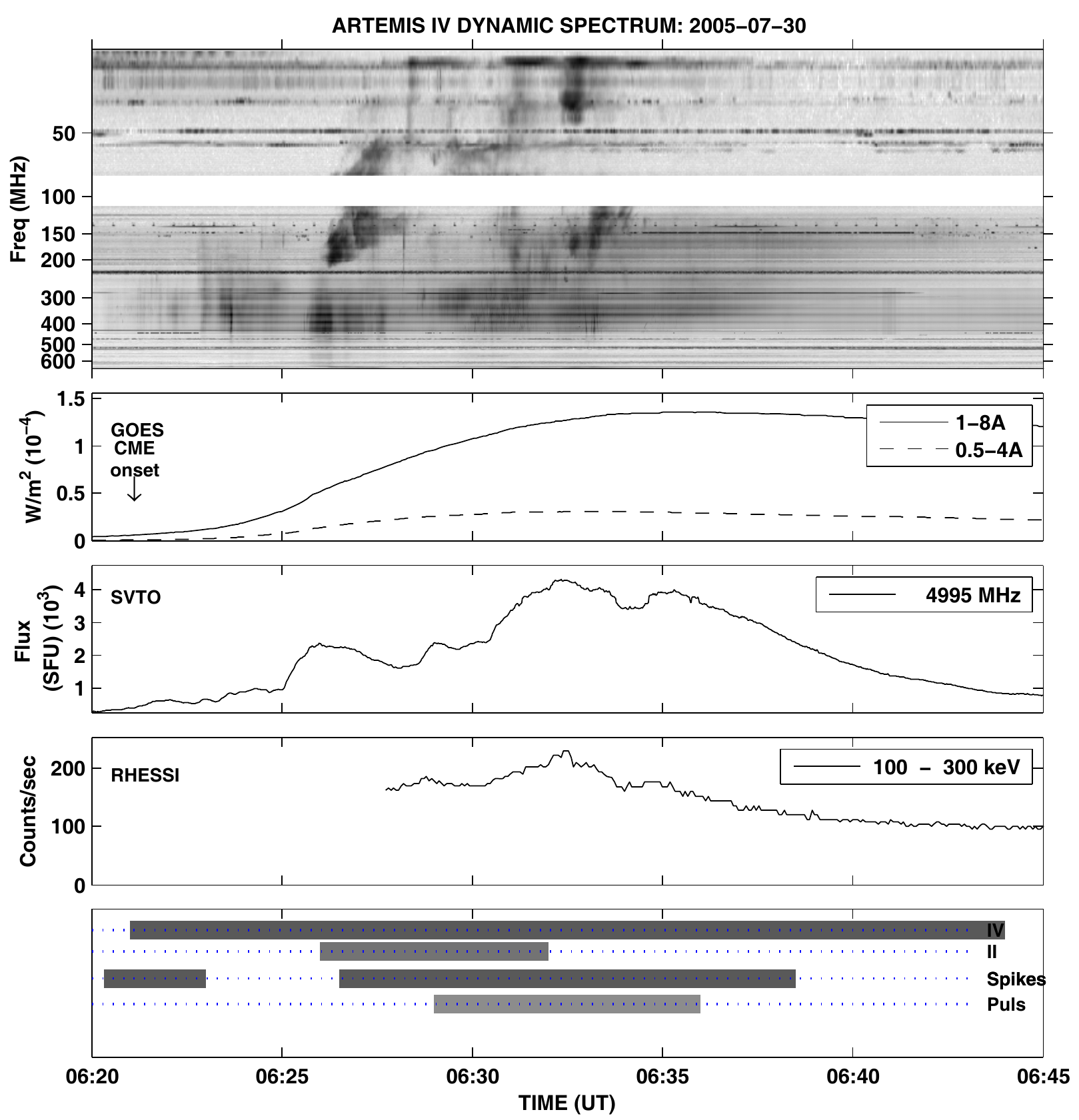}
\includegraphics[trim=0cm 0.0cm  0.0cm 4.0cm,clip,width=0.85\textwidth]{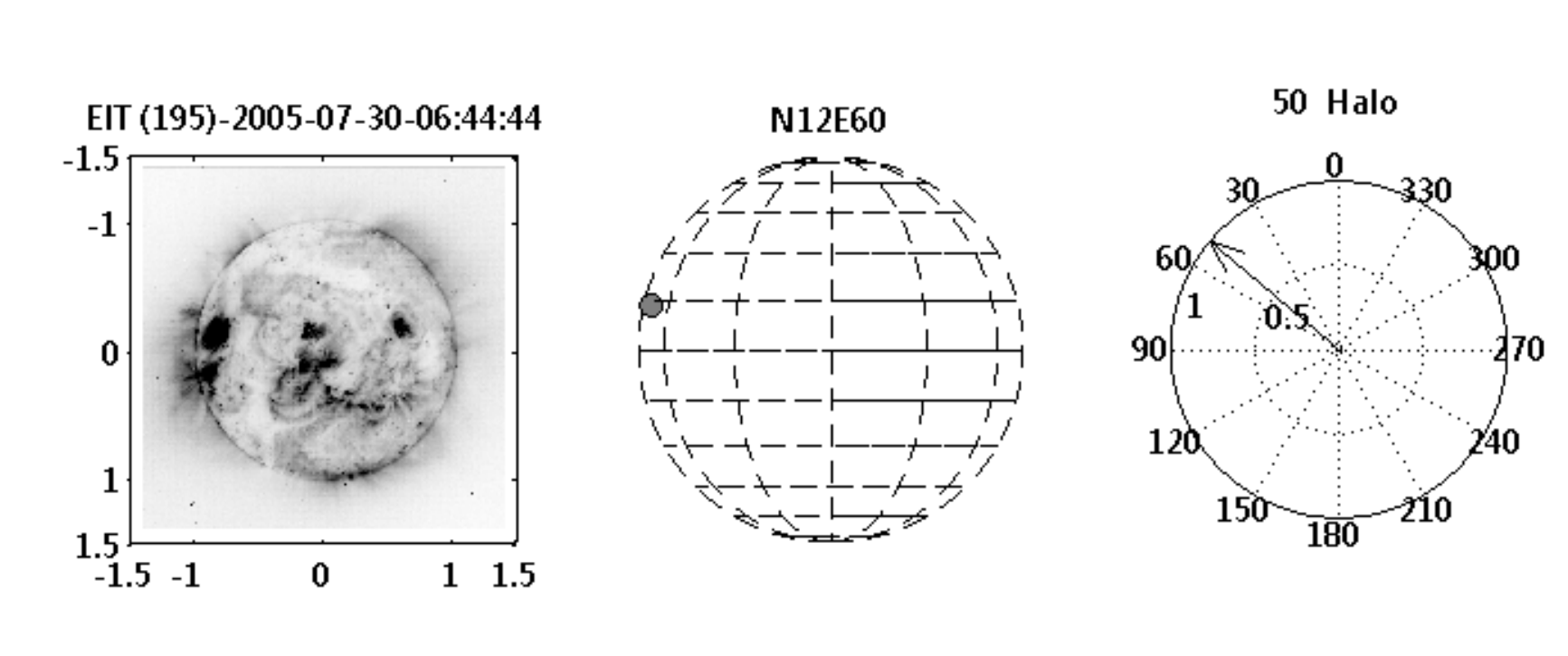}
\caption{\event}
\end{center}
\end{figure} % 
%--------------------------------------------
\clearpage
\renewcommand{\event} {22 August 2005}
%--------------------------------------------
\begin{figure}
\begin{center}% trim=0cm 1cm  0cm 1cm,clip,
\includegraphics[width=\textwidth]{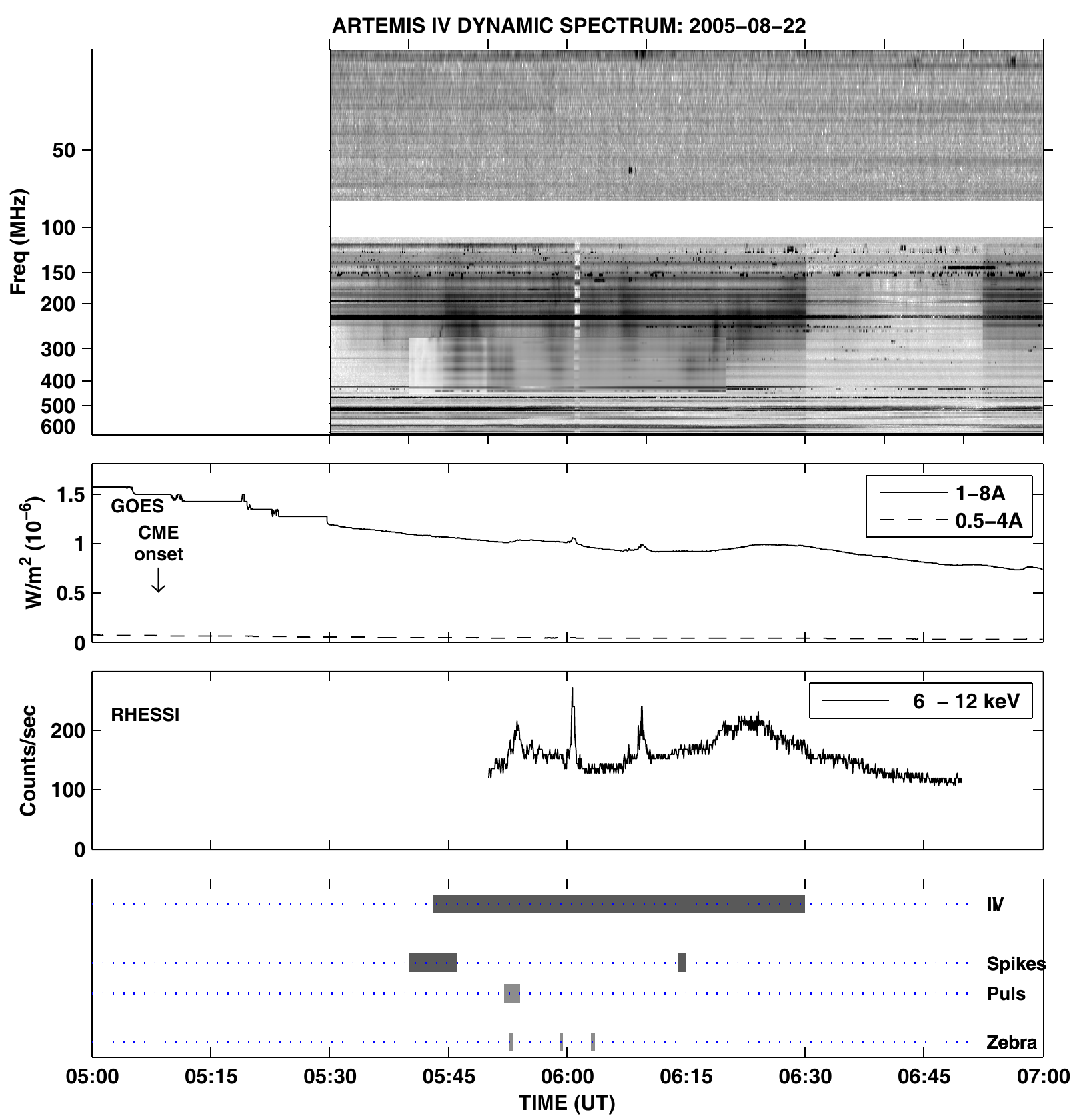}
\includegraphics[trim=0cm 0.0cm  0.0cm 4.0cm,clip,width=0.85\textwidth]{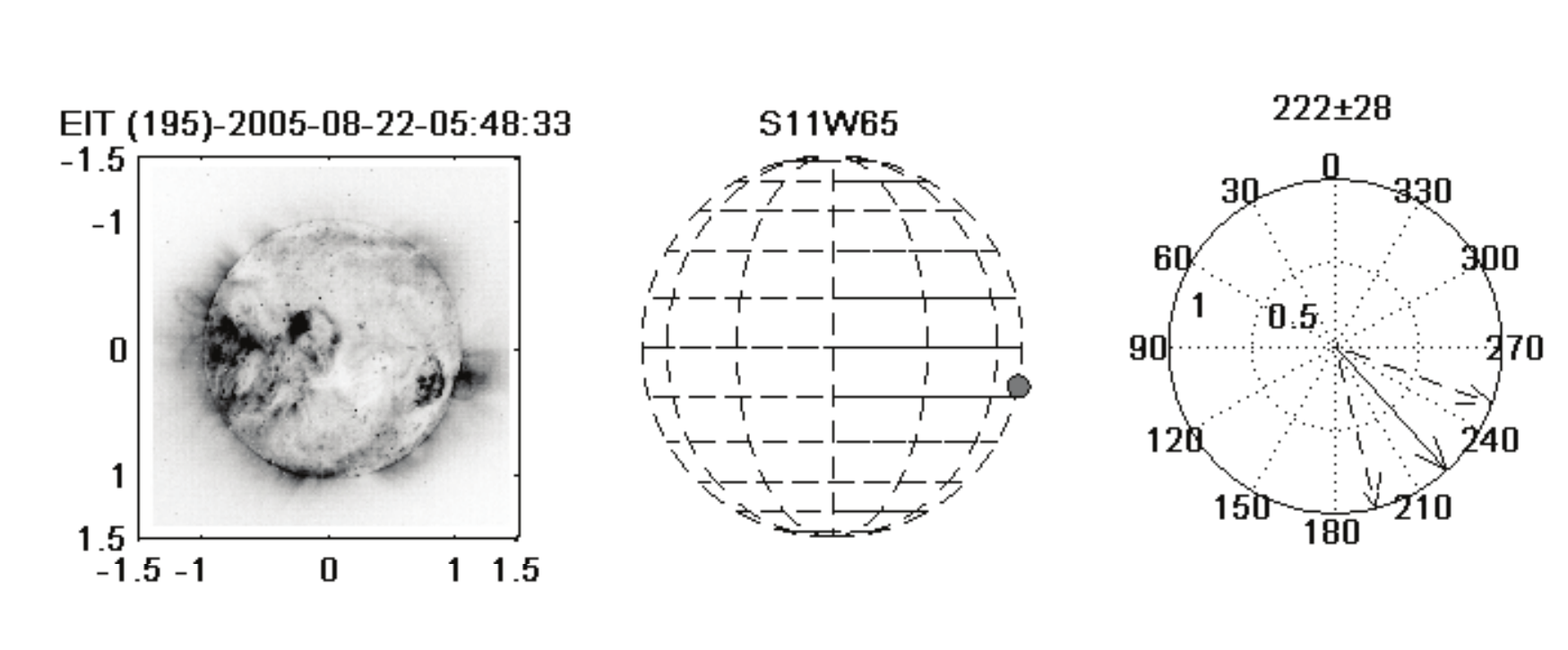}
\caption{\event}
\end{center}
\end{figure} % 
%--------------------------------------------
\clearpage
\renewcommand{\event} {23 August 2005}
%--------------------------------------------
\begin{figure}
\begin{center}% trim=0cm 1cm  0cm 1cm,clip,
\includegraphics[width=\textwidth]{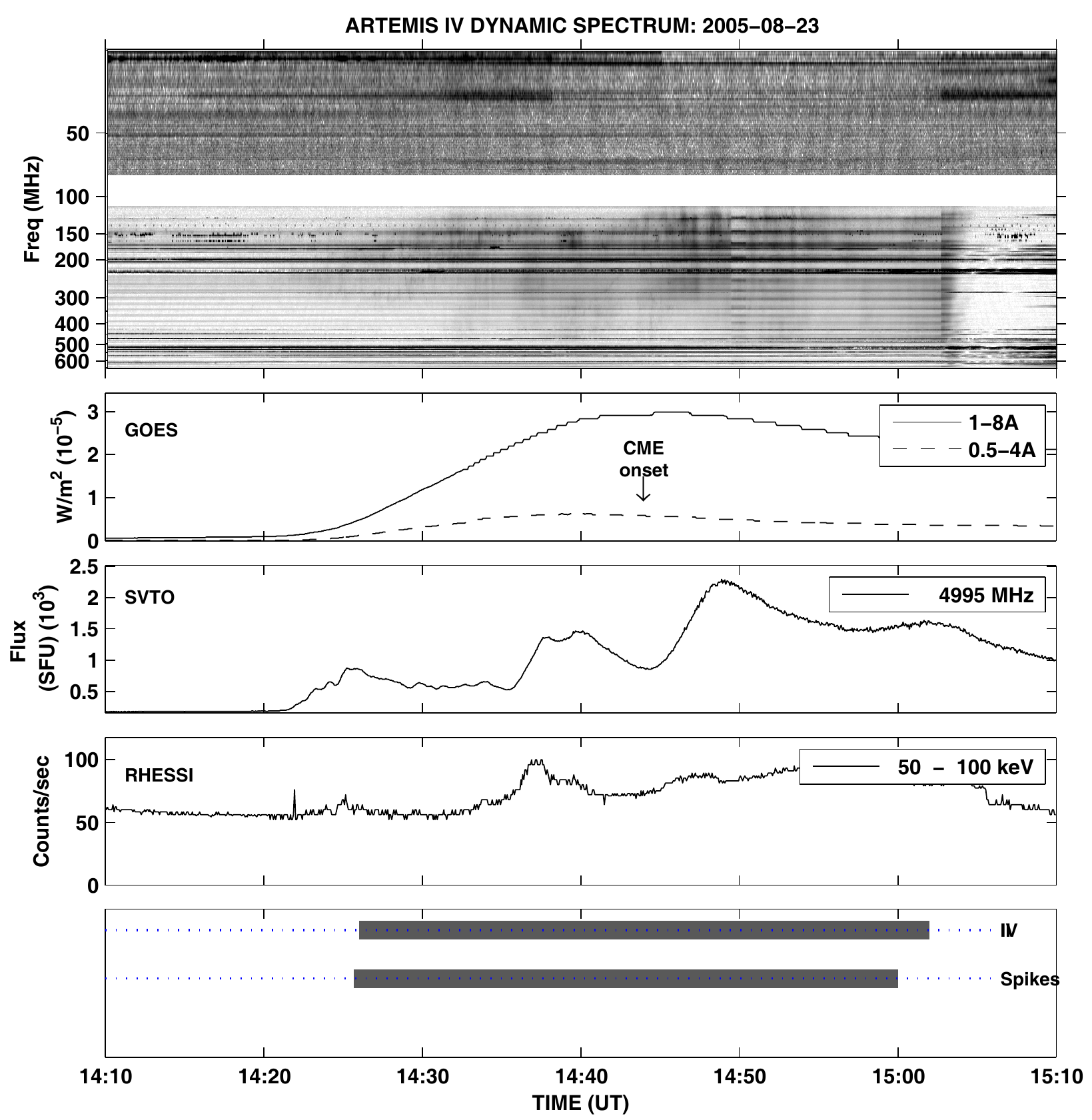}
\includegraphics[trim=0cm 0.0cm  0.0cm 4.0cm,clip,width=0.85\textwidth]{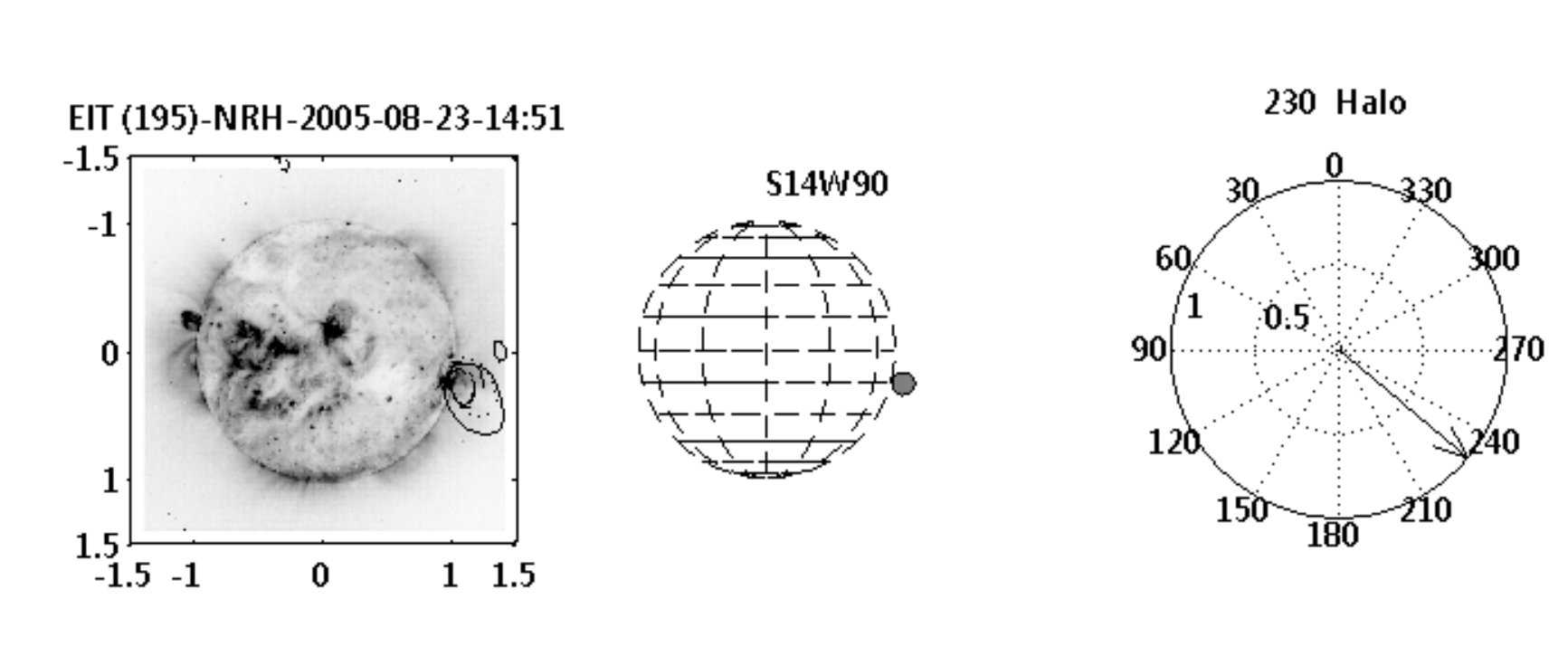}
\caption{\event}
\end{center}
\end{figure} % 
%--------------------------------------------
\clearpage
\renewcommand{\event} {12 February 2010}
%--------------------------------------------
\begin{figure}
\begin{center}% trim=0cm 1cm  0cm 1cm,clip,
\includegraphics[width=\textwidth]{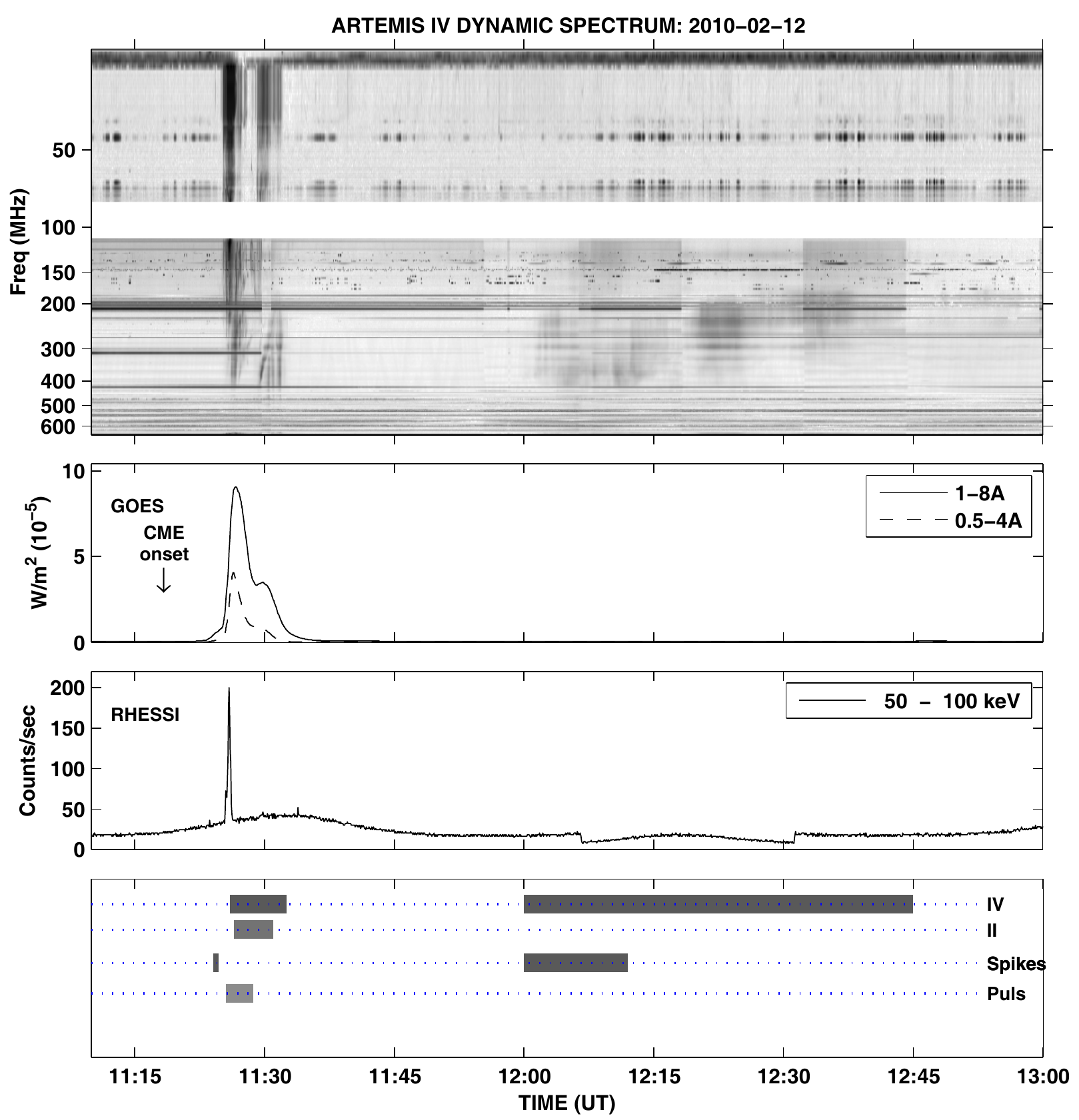}
\includegraphics[trim=0cm 0.0cm  0.0cm 4.0cm,clip,width=0.85\textwidth]{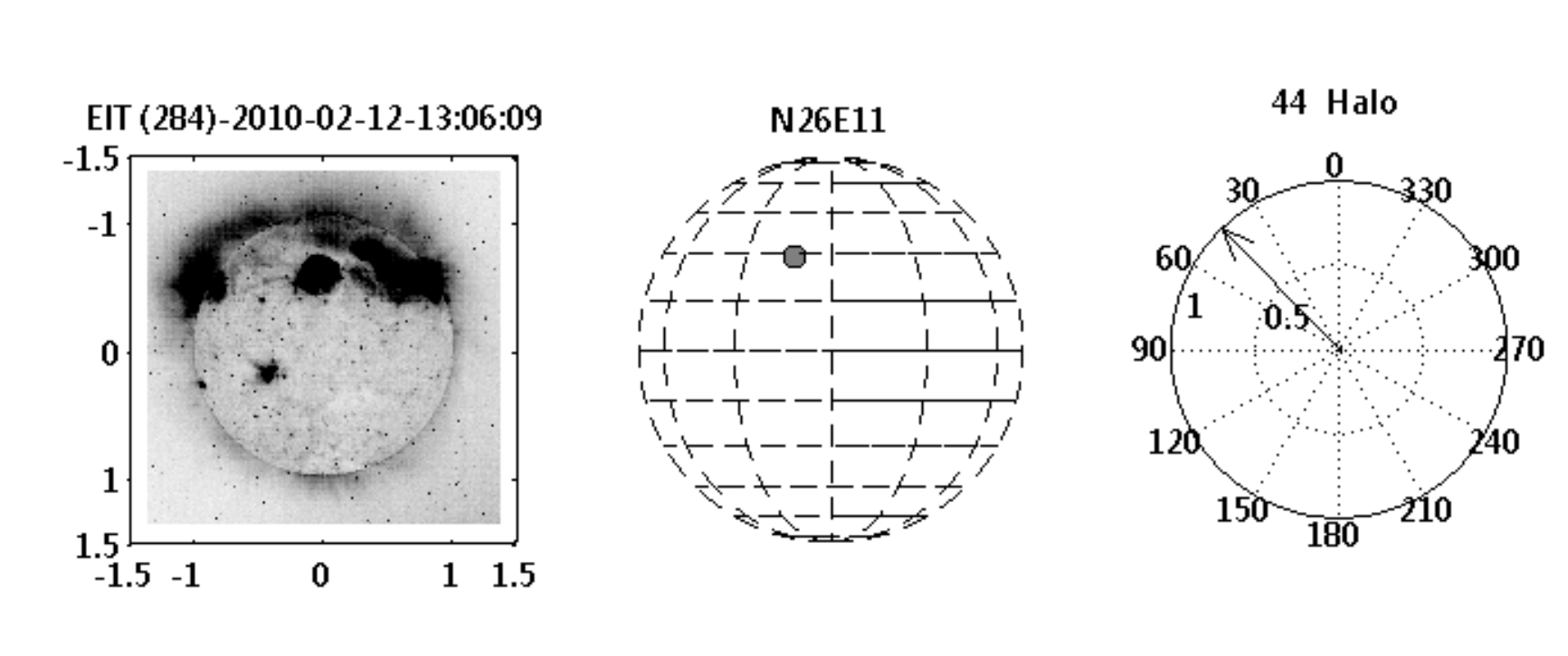}
\caption{\event}
\end{center}
\end{figure} % 
%--------------------------------------------
\clearpage
\renewcommand{\event} {01 August 2010}
%--------------------------------------------
\begin{figure}
\begin{center}% trim=0cm 1cm  0cm 1cm,clip,
\includegraphics[width=\textwidth]{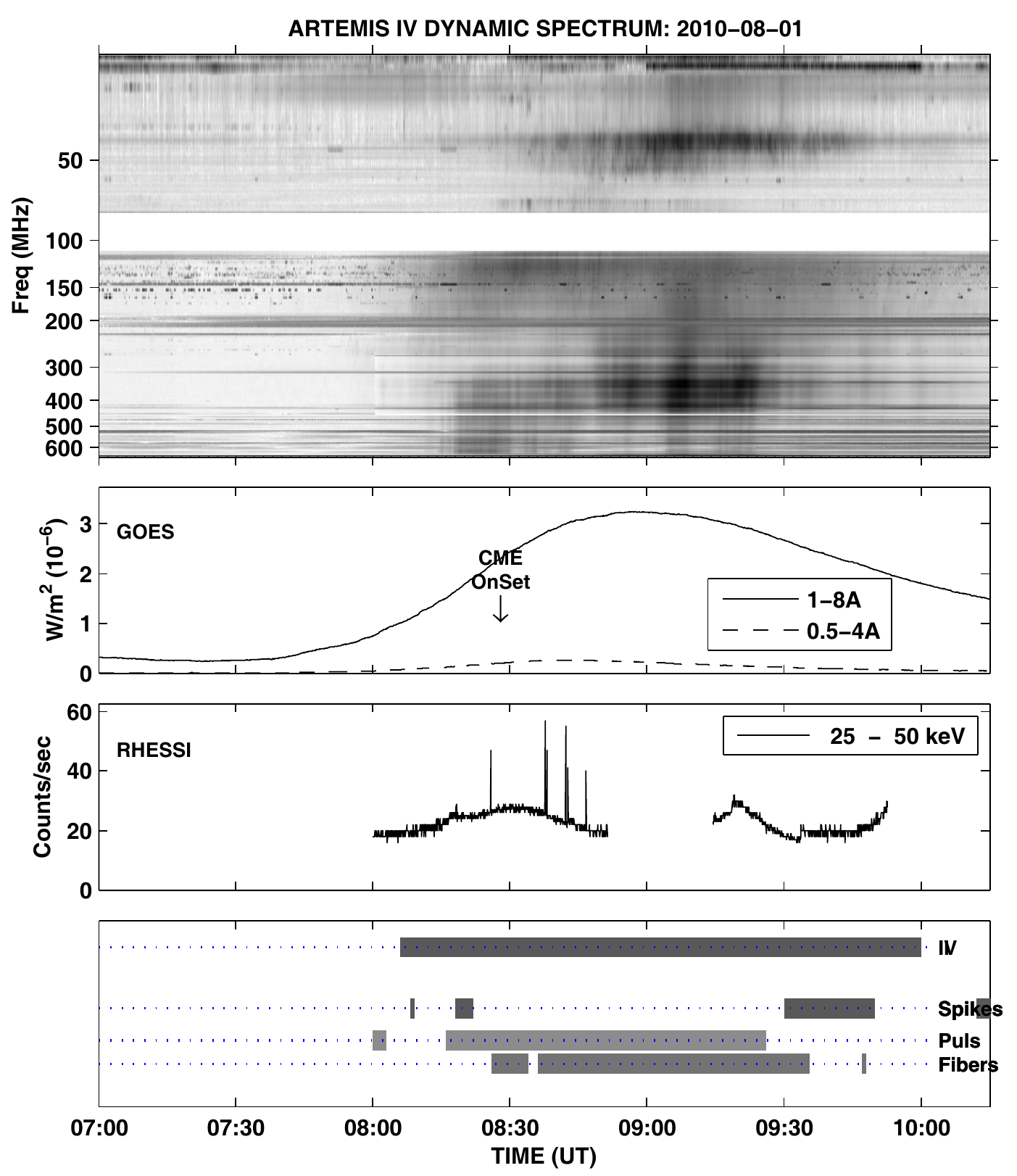}
\includegraphics[trim=0cm 0.0cm  0.0cm 4.0cm,clip,width=0.85\textwidth]{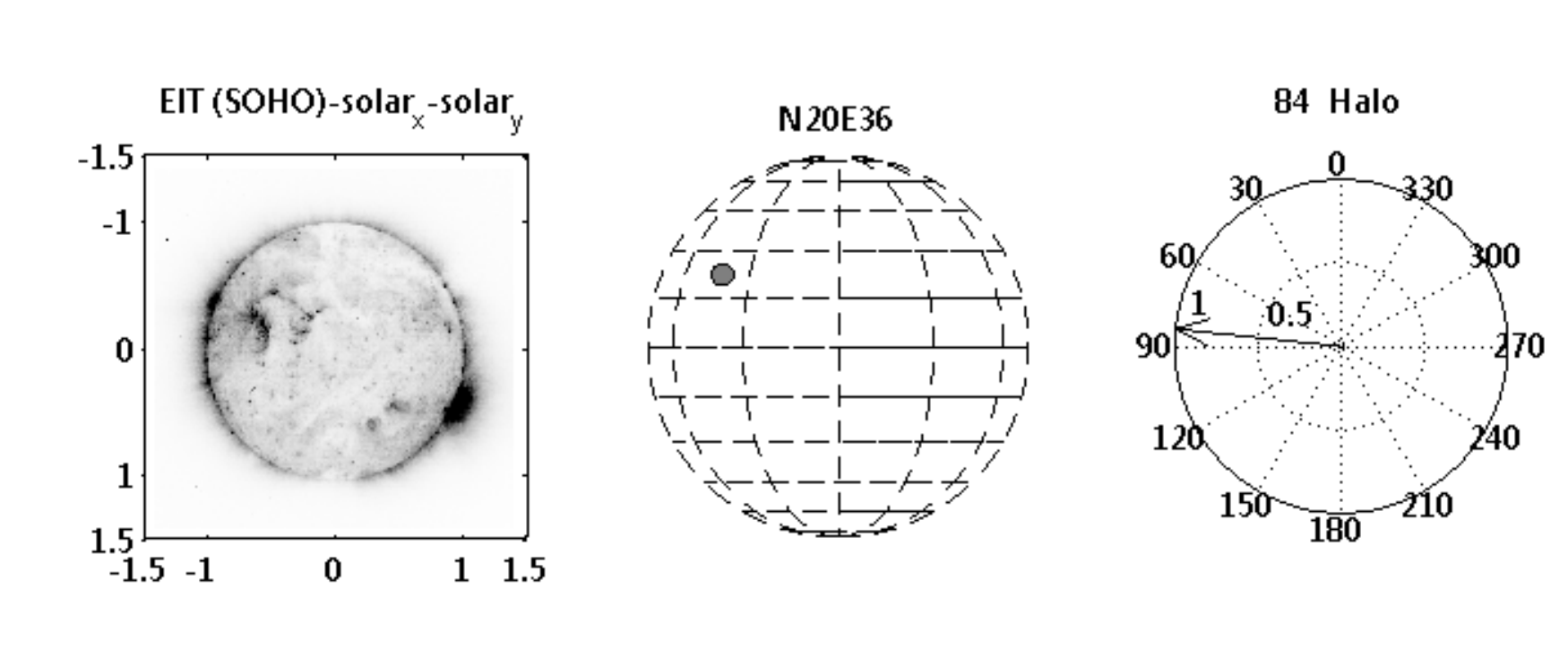}
\caption{\event}
\end{center}
\end{figure} % 
%--------------------------------------------
\clearpage

%% file: 02_SP_Revision_03.bbl
\begin{thebibliography}{65}
% BibTex style file: spr-mp-sola-cnd.bst, 2009-06-03
\ifx \bisbn   \undefined \def \bisbn  #1{ISBN #1}\fi
\ifx \binits  \undefined \def \binits#1{#1} \fi
\ifx \bauthor  \undefined \def \bauthor#1{#1} \fi
\ifx \batitle  \undefined \def \batitle#1{#1} \fi
\ifx \bjtitle  \undefined \def \bjtitle#1{\textit{#1}}\fi
\ifx \bvolume  \undefined \def \bvolume#1{\textbf{#1}}\fi
\ifx \byear  \undefined \def \byear#1{#1} \fi
\ifx \bissue  \undefined \def \bissue#1{#1} \fi
\ifx \bfpage  \undefined \def \bfpage#1{#1} \fi
\ifx \blpage  \undefined \def \blpage #1{#1} \fi
\ifx \burl  \undefined \def \burl#1{\textsf{#1}} \fi
\ifx \href  \undefined \def \href#1#2{#2} \fi
\ifx \doiurl  \undefined \def \doiurl#1{\href{http://dx.doi.org/#1}{#1}} \fi
\ifx \betal  \undefined \def \betal{\textit{et al.}} \fi
\ifx \binstitute  \undefined \def \binstitute#1{#1} \fi
\ifx \bctitle  \undefined \def \bctitle#1{#1} \fi
\ifx \beditor  \undefined \def \beditor#1{#1} \fi
\ifx \bpublisher  \undefined \def \bpublisher#1{#1} \fi
\ifx \bbtitle  \undefined \def \bbtitle#1{\textit{#1}} \fi
\ifx \bedition  \undefined \def \bedition#1{#1} \fi
\ifx \bseriesno  \undefined \def \bseriesno#1{\textbf{#1}} \fi
\ifx \blocation  \undefined \def \blocation#1{#1} \fi
\ifx \bsertitle  \undefined \def \bsertitle#1{\textit{#1}} \fi
\ifx \bsnm \undefined \def \bsnm#1{#1} \fi
\ifx \bsuffix \undefined \def \bsuffix#1{#1} \fi
\ifx \bparticle \undefined \def \bparticle#1{#1} \fi
\ifx \barticle \undefined \def \barticle#1{#1} \fi
\ifx \botherref \undefined \def \botherref #1{#1} \fi
\ifx \url \undefined \def \url#1{\textsf{#1}} \fi
\ifx \bchapter \undefined \def \bchapter#1{#1} \fi
\ifx \bbook \undefined \def \bbook#1{#1} \fi
\ifx \bcomment \undefined \def \bcomment#1{#1} \fi
\ifx \oauthor \undefined \def \oauthor#1{#1} \fi
\ifx \citeauthoryear \undefined \def \citeauthoryear#1{#1} \fi
\def \endbibitem {}

\bibitem[\protect\citeauthoryear{{Allaart} \textit{et~al.}}{1990}]{Allaart90}
\begin{barticle}
\bauthor{\bsnm{{Allaart}}, \binits{M.A.F.}}, \bauthor{\bsnm{{van Nieuwkoop}},
  \binits{J.}}, \bauthor{\bsnm{{Slottje}}, \binits{C.}},
  \bauthor{\bsnm{{Sondaar}}, \binits{L.H.}}:
\byear{1990},
\bjtitle{\solphys}
\bvolume{130},
\bfpage{183}.
\end{barticle}
\endbibitem

\bibitem[\protect\citeauthoryear{{Aurass}}{2007}]{Aurass2007}
\begin{barticle}
\bauthor{\bsnm{{Aurass}}, \binits{H.}}:
\byear{2007},
\bjtitle{Adv. Space Res.}
\bvolume{39},
\bfpage{1407}.
\end{barticle}
\endbibitem

\bibitem[\protect\citeauthoryear{{Aurass} \textit{et~al.}}{2003}]{Aurass03}
\begin{barticle}
\bauthor{\bsnm{{Aurass}}, \binits{H.}}, \bauthor{\bsnm{{Klein}},
  \binits{K.L.}}, \bauthor{\bsnm{{Zlotnik}}, \binits{E.Y.}},
  \bauthor{\bsnm{{Zaitsev}}, \binits{V.V.}}:
\byear{2003},
\bjtitle{\aap}
\bvolume{410},
\bfpage{1001}.
\end{barticle}
\endbibitem

\bibitem[\protect\citeauthoryear{{Aurass} \textit{et~al.}}{2005}]{Aurass2005}
\begin{barticle}
\bauthor{\bsnm{{Aurass}}, \binits{H.}}, \bauthor{\bsnm{{Rausche}},
  \binits{G.}}, \bauthor{\bsnm{{Mann}}, \binits{G.}},
  \bauthor{\bsnm{{Hofmann}}, \binits{A.}}:
\byear{2005},
\bjtitle{\aap}
\bvolume{435},
\bfpage{1137}.
\end{barticle}
\endbibitem

\bibitem[\protect\citeauthoryear{{B{\'a}rta} and
  {Karlick{\'y}}}{2005}]{Barta2005}
\begin{barticle}
\bauthor{\bsnm{{B{\'a}rta}}, \binits{M.}}, \bauthor{\bsnm{{Karlick{\'y}}},
  \binits{M.}}:
\byear{2005},
\bjtitle{Hvar Obs. Bull.}
\bvolume{29},
\bfpage{205}.
\end{barticle}
\endbibitem

\bibitem[\protect\citeauthoryear{{Benz}}{2003}]{Benz03}
\begin{botherref}
\oauthor{\bsnm{{Benz}}, \binits{A.O.}}:
2003,
In: {Klein}, L. (ed.)
\textit{Energy Conversion and Particle Acceleration the Solar Corona, Berlin
  Springer Verlag},
\textit{Lecture Notes Phys.}
\textbf{612},
80.
\end{botherref}
\endbibitem

\bibitem[\protect\citeauthoryear{{Benz} and {Mann}}{1998}]{Benz98}
\begin{barticle}
\bauthor{\bsnm{{Benz}}, \binits{A.O.}}, \bauthor{\bsnm{{Mann}}, \binits{G.}}:
\byear{1998},
\bjtitle{\aap}
\bvolume{333},
\bfpage{1034}.
\end{barticle}
\endbibitem

\bibitem[\protect\citeauthoryear{{Benz}, {Csillaghy}, and
  {Aschwanden}}{1996}]{Benz1996}
\begin{barticle}
\bauthor{\bsnm{{Benz}}, \binits{A.O.}}, \bauthor{\bsnm{{Csillaghy}},
  \binits{A.}}, \bauthor{\bsnm{{Aschwanden}}, \binits{M.J.}}:
\byear{1996},
\bjtitle{\aap}
\bvolume{309},
\bfpage{291}.
\end{barticle}
\endbibitem

\bibitem[\protect\citeauthoryear{{Benz} \textit{et~al.}}{2006}]{Benz2006}
\begin{barticle}
\bauthor{\bsnm{{Benz}}, \binits{A.O.}}, \bauthor{\bsnm{{Perret}}, \binits{H.}},
  \bauthor{\bsnm{{Saint-Hilaire}}, \binits{P.}}, \bauthor{\bsnm{{Zlobec}},
  \binits{P.}}:
\byear{2006},
\bjtitle{Adv. Space Res.}
\bvolume{38},
\bfpage{951}.
\end{barticle}
\endbibitem

\bibitem[\protect\citeauthoryear{{Bernold}}{1980}]{Bernold80}
\begin{barticle}
\bauthor{\bsnm{{Bernold}}, \binits{T.}}:
\byear{1980},
\bjtitle{\aaps}
\bvolume{42},
\bfpage{43}.
\end{barticle}
\endbibitem

\bibitem[\protect\citeauthoryear{Bouratzis \textit{et~al.}}{2010}]{Bouratzis10}
\begin{barticle}
\bauthor{\bsnm{Bouratzis}, \binits{C.}}, \bauthor{\bsnm{Preka-Papadema},
  \binits{P.}}, \bauthor{\bsnm{Hillaris}, \binits{A.}},
  \bauthor{\bsnm{Tsitsipis}, \binits{P.}}, \bauthor{\bsnm{Kontogeorgos},
  \binits{A.}}, \bauthor{\bsnm{Kurt}, \binits{V.}}, \bauthor{\bsnm{Moussas},
  \binits{X.}}:
\byear{2010},
\bjtitle{\solphys}
\bvolume{267},
\bfpage{343}.
\end{barticle}
\endbibitem

\bibitem[\protect\citeauthoryear{{Brueckner}
  \textit{et~al.}}{1995}]{Brueckner95}
\begin{barticle}
\bauthor{\bsnm{{Brueckner}}, \binits{G.E.}}, \bauthor{\bsnm{{Howard}},
  \binits{R.A.}}, \bauthor{\bsnm{{Koomen}}, \binits{M.J.}},
  \bauthor{\bsnm{{Korendyke}}, \binits{C.M.}}, \bauthor{\bsnm{{Michels}},
  \binits{D.J.}}, \bauthor{\bsnm{{Moses}}, \binits{J.D.}},
  \bauthor{\bsnm{\etal}}:
\byear{1995},
\bjtitle{\solphys}
\bvolume{162},
\bfpage{357}.
\end{barticle}
\endbibitem

\bibitem[\protect\citeauthoryear{{Caroubalos}
  \textit{et~al.}}{2001}]{Caroubalos01}
\begin{barticle}
\bauthor{\bsnm{{Caroubalos}}, \binits{C.}}, \bauthor{\bsnm{{Maroulis}},
  \binits{D.}}, \bauthor{\bsnm{{Patavalis}}, \binits{N.}},
  \bauthor{\bsnm{{Bougeret}}, \binits{J.L.}}, \bauthor{\bsnm{{Dumas}},
  \binits{G.}}, \bauthor{\bsnm{{Perche}}, \binits{C.}}, \bauthor{\bsnm{\etal}}:
\byear{2001},
\bjtitle{Exp. Astron.}
\bvolume{11},
\bfpage{23}.
\end{barticle}
\endbibitem

\bibitem[\protect\citeauthoryear{{Caroubalos}
  \textit{et~al.}}{2004}]{Caroubalos04}
\begin{barticle}
\bauthor{\bsnm{{Caroubalos}}, \binits{C.}}, \bauthor{\bsnm{{Hillaris}},
  \binits{A.}}, \bauthor{\bsnm{{Bouratzis}}, \binits{C.}},
  \bauthor{\bsnm{{Alissandrakis}}, \binits{C.E.}},
  \bauthor{\bsnm{{Preka-Papadema}}, \binits{P.}},
  \bauthor{\bsnm{{Polygiannakis}}, \binits{J.}}, \bauthor{\bsnm{\etal}}:
\byear{2004},
\bjtitle{\aap}
\bvolume{413},
\bfpage{1125}.
\end{barticle}
\endbibitem

\bibitem[\protect\citeauthoryear{{Caroubalos}
  \textit{et~al.}}{2006}]{Caroubalos06}
\begin{botherref}
\oauthor{\bsnm{{Caroubalos}}, \binits{C.}}, \oauthor{\bsnm{{Alissandrakis}},
  \binits{C.E.}}, \oauthor{\bsnm{{Hillaris}}, \binits{A.}},
  \oauthor{\bsnm{{Preka-Papadema}}, \binits{P.}},
  \oauthor{\bsnm{{Polygiannakis}}, \binits{J.}}, \oauthor{\bsnm{{Moussas}},
  \binits{X.}}, \oauthor{\bsnm{\etal}}:
2006,
In: {Solomos}, N. (ed.)
\textit{Recent Adv. Astron. Astrophys.},
\textit{Am. Inst. Phys. CS}
\textbf{848},
864.
\end{botherref}
\endbibitem

\bibitem[\protect\citeauthoryear{{Chen} \textit{et~al.}}{2011}]{Chen2011}
\begin{barticle}
\bauthor{\bsnm{{Chen}}, \binits{B.}}, \bauthor{\bsnm{{Bastian}},
  \binits{T.S.}}, \bauthor{\bsnm{{Gary}}, \binits{D.E.}},
  \bauthor{\bsnm{{Jing}}, \binits{J.}}:
\byear{2011},
\bjtitle{\apj}
\bvolume{736},
\bfpage{64}.
\end{barticle}
\endbibitem

\bibitem[\protect\citeauthoryear{{Chernov}}{1990}]{Chernov1990}
\begin{barticle}
\bauthor{\bsnm{{Chernov}}, \binits{G.P.}}:
\byear{1990},
\bjtitle{\solphys}
\bvolume{130},
\bfpage{75}.
\end{barticle}
\endbibitem

\bibitem[\protect\citeauthoryear{{Chernov}}{2005}]{Chernov2005}
\begin{barticle}
\bauthor{\bsnm{{Chernov}}, \binits{G.P.}}:
\byear{2005},
\bjtitle{Plasma Phys. Reports}
\bvolume{31},
\bfpage{314}.
\end{barticle}
\endbibitem

\bibitem[\protect\citeauthoryear{{Chernov}}{2006}]{Chernov2006}
\begin{barticle}
\bauthor{\bsnm{{Chernov}}, \binits{G.P.}}:
\byear{2006},
\bjtitle{Space Sci. Rev.}
\bvolume{127},
\bfpage{195}.
\bcomment{10.1007/s11214-006-9141-7}.
\end{barticle}
\endbibitem

\bibitem[\protect\citeauthoryear{{Chernov}}{2008}]{Chernov2008}
\begin{barticle}
\bauthor{\bsnm{{Chernov}}, \binits{G.P.}}:
\byear{2008},
\bjtitle{Astron. Lett.}
\bvolume{34},
\bfpage{486}.
\end{barticle}
\endbibitem

\bibitem[\protect\citeauthoryear{{Chernov}, {Yan}, and
  {Fu}}{2003}]{Chernov2003}
\begin{barticle}
\bauthor{\bsnm{{Chernov}}, \binits{G.P.}}, \bauthor{\bsnm{{Yan}},
  \binits{Y.H.}}, \bauthor{\bsnm{{Fu}}, \binits{Q.J.}}:
\byear{2003},
\bjtitle{\aap}
\bvolume{406},
\bfpage{1071}.
\end{barticle}
\endbibitem

\bibitem[\protect\citeauthoryear{{Chernov} \textit{et~al.}}{1998}]{Chernov98}
\begin{barticle}
\bauthor{\bsnm{{Chernov}}, \binits{G.P.}}, \bauthor{\bsnm{{Markeev}},
  \binits{A.K.}}, \bauthor{\bsnm{{Poquerusse}}, \binits{M.}},
  \bauthor{\bsnm{{Bougeret}}, \binits{J.L.}}, \bauthor{\bsnm{{Klein}},
  \binits{K.L.}}, \bauthor{\bsnm{{Mann}}, \binits{G.}},
  \bauthor{\bsnm{{Aurass}}, \binits{H.}}, \bauthor{\bsnm{{Aschwanden}},
  \binits{M.J.}}:
\byear{1998},
\bjtitle{\aap}
\bvolume{334},
\bfpage{314}.
\end{barticle}
\endbibitem

\bibitem[\protect\citeauthoryear{{Chernov} \textit{et~al.}}{2007}]{Chernov2007}
\begin{barticle}
\bauthor{\bsnm{{Chernov}}, \binits{G.P.}}, \bauthor{\bsnm{{Kaiser}},
  \binits{M.L.}}, \bauthor{\bsnm{{Bougeret}}, \binits{J.L.}},
  \bauthor{\bsnm{{Fomichev}}, \binits{V.V.}}, \bauthor{\bsnm{{Gorgutsa}},
  \binits{R.V.}}:
\byear{2007},
\bjtitle{\solphys}
\bvolume{241},
\bfpage{145}.
\end{barticle}
\endbibitem

\bibitem[\protect\citeauthoryear{{Delaboudini{\`e}re}
  \textit{et~al.}}{1995}]{Delaboudiniere95}
\begin{barticle}
\bauthor{\bsnm{{Delaboudini{\`e}re}}, \binits{J.P.}},
  \bauthor{\bsnm{{Artzner}}, \binits{G.E.}}, \bauthor{\bsnm{{Brunaud}},
  \binits{J.}}, \bauthor{\bsnm{{Gabriel}}, \binits{A.H.}},
  \bauthor{\bsnm{{Hochedez}}, \binits{J.F.}}, \bauthor{\bsnm{{Millier}},
  \binits{F.}}, \bauthor{\bsnm{\etal}}:
\byear{1995},
\bjtitle{\solphys}
\bvolume{162},
\bfpage{291}.
\end{barticle}
\endbibitem

\bibitem[\protect\citeauthoryear{{Elgaroy}}{1986}]{Elgaroy1986}
\begin{barticle}
\bauthor{\bsnm{{Elgaroy}}, \binits{O.}}:
\byear{1986},
\bjtitle{\solphys}
\bvolume{104},
\bfpage{41}.
\end{barticle}
\endbibitem

\bibitem[\protect\citeauthoryear{{F{\'a}rn{\'{\i}}k}, {Garcia}, and
  {Karlick{\'y}}}{2001}]{Farnik2001}
\begin{barticle}
\bauthor{\bsnm{{F{\'a}rn{\'{\i}}k}}, \binits{F.}}, \bauthor{\bsnm{{Garcia}},
  \binits{H.}}, \bauthor{\bsnm{{Karlick{\'y}}}, \binits{M.}}:
\byear{2001},
\bjtitle{\solphys}
\bvolume{201},
\bfpage{357}.
\end{barticle}
\endbibitem

\bibitem[\protect\citeauthoryear{{Fernandes}
  \textit{et~al.}}{2003}]{Fernandes2003}
\begin{barticle}
\bauthor{\bsnm{{Fernandes}}, \binits{F.C.R.}}, \bauthor{\bsnm{{Krishan}},
  \binits{V.}}, \bauthor{\bsnm{{Andrade}}, \binits{M.C.}},
  \bauthor{\bsnm{{Cecatto}}, \binits{J.R.}}, \bauthor{\bsnm{{Freitas}},
  \binits{D.C.}}, \bauthor{\bsnm{{Sawant}}, \binits{H.S.}}:
\byear{2003},
\bjtitle{\adv}
\bvolume{32},
\bfpage{2545}.
\end{barticle}
\endbibitem

\bibitem[\protect\citeauthoryear{{Fishman} \textit{et~al.}}{1982}]{Fishman1982}
\begin{botherref}
\oauthor{\bsnm{{Fishman}}, \binits{G.J.}}, \oauthor{\bsnm{{Meegan}},
  \binits{C.A.}}, \oauthor{\bsnm{{Parnell}}, \binits{T.A.}},
  \oauthor{\bsnm{{Wilson}}, \binits{R.B.}}:
1982,
In: {Lingenfelter}, R.E., {Hudson}, H.S., {Worrall}, D.M. (eds.)
\textit{Gamma Ray Transients and Related Astrophys. Phenomena},
\textit{Am. Inst. Phys. CS}
\textbf{77},
443.
\end{botherref}
\endbibitem

\bibitem[\protect\citeauthoryear{{Fishman} \textit{et~al.}}{1984}]{Fishman1984}
\begin{botherref}
\oauthor{\bsnm{{Fishman}}, \binits{G.J.}}, \oauthor{\bsnm{{Meegan}},
  \binits{C.A.}}, \oauthor{\bsnm{{Parnell}}, \binits{T.A.}},
  \oauthor{\bsnm{{Wilson}}, \binits{R.B.}}, \oauthor{\bsnm{{Paciesas}},
  \binits{W.}}:
1984,
In: {Woosley}, S.E. (ed.)
\textit{Am. Inst. Phys. CS}
\textbf{115},
651.
\end{botherref}
\endbibitem

\bibitem[\protect\citeauthoryear{{Fu} \textit{et~al.}}{2004}]{Fu04}
\begin{barticle}
\bauthor{\bsnm{{Fu}}, \binits{Q.J.}}, \bauthor{\bsnm{{Yan}}, \binits{Y.H.}},
  \bauthor{\bsnm{{Liu}}, \binits{Y.Y.}}, \bauthor{\bsnm{{Wang}}, \binits{M.}},
  \bauthor{\bsnm{{Wang}}, \binits{S.J.}}:
\byear{2004},
\bjtitle{\cjaa}
\bvolume{4},
\bfpage{176}.
\end{barticle}
\endbibitem

\bibitem[\protect\citeauthoryear{{Gopalswamy}
  \textit{et~al.}}{2009}]{Gopalswamy09}
\begin{barticle}
\bauthor{\bsnm{{Gopalswamy}}, \binits{N.}}, \bauthor{\bsnm{{Yashiro}},
  \binits{S.}}, \bauthor{\bsnm{{Michalek}}, \binits{G.}},
  \bauthor{\bsnm{{Stenborg}}, \binits{G.}}, \bauthor{\bsnm{{Vourlidas}},
  \binits{A.}}, \bauthor{\bsnm{{Freeland}}, \binits{S.}},
  \bauthor{\bsnm{{Howard}}, \binits{R.}}:
\byear{2009},
\bjtitle{Earth Moon and Planets}
\bvolume{104},
\bfpage{295}.
\end{barticle}
\endbibitem

\bibitem[\protect\citeauthoryear{{Guedel} and {Benz}}{1988}]{Guedel88}
\begin{barticle}
\bauthor{\bsnm{{Guedel}}, \binits{M.}}, \bauthor{\bsnm{{Benz}}, \binits{A.O.}}:
\byear{1988},
\bjtitle{\aaps}
\bvolume{75},
\bfpage{243}.
\end{barticle}
\endbibitem

\bibitem[\protect\citeauthoryear{{Guidice} \textit{et~al.}}{1981}]{Guidice81}
\begin{botherref}
\oauthor{\bsnm{{Guidice}}, \binits{D.A.}}, \oauthor{\bsnm{{Cliver}},
  \binits{E.W.}}, \oauthor{\bsnm{{Barron}}, \binits{W.R.}},
  \oauthor{\bsnm{{Kahler}}, \binits{S.}}:
1981,
In: \textit{Bull. Amer. Astron. Soc.}
\textbf{13},
553.
\end{botherref}
\endbibitem

\bibitem[\protect\citeauthoryear{{Hillaris}
  \textit{et~al.}}{2011}]{Hillaris2011}
\begin{barticle}
\bauthor{\bsnm{{Hillaris}}, \binits{A.}}, \bauthor{\bsnm{{Malandraki}},
  \binits{O.}}, \bauthor{\bsnm{{Klein}}, \binits{K.L.}},
  \bauthor{\bsnm{{Preka-Papadema}}, \binits{P.}}, \bauthor{\bsnm{{Moussas}},
  \binits{X.}}, \bauthor{\bsnm{{Bouratzis}}, \binits{C.}},
  \bauthor{\bsnm{{Mitsakou}}, \binits{E.}}, \bauthor{\bsnm{{Tsitsipis}},
  \binits{P.}}, \bauthor{\bsnm{{Kontogeorgos}}, \binits{A.}}:
\byear{2011},
\bjtitle{\solphys}
\bvolume{273},
\bfpage{493}.
\end{barticle}
\endbibitem

\bibitem[\protect\citeauthoryear{{Isliker} and {Benz}}{1994}]{Isliker94}
\begin{barticle}
\bauthor{\bsnm{{Isliker}}, \binits{H.}}, \bauthor{\bsnm{{Benz}},
  \binits{A.O.}}:
\byear{1994},
\bjtitle{\aaps}
\bvolume{104},
\bfpage{145}.
\end{barticle}
\endbibitem

\bibitem[\protect\citeauthoryear{{Ji{\v r}i{\v c}ka}, {Karlick{\'y}}, and
  {M{\'e}sz{\'a}rosov{\'a}}}{2002}]{Jiricka02}
\begin{botherref}
\oauthor{\bsnm{{Ji{\v r}i{\v c}ka}}, \binits{K.}},
  \oauthor{\bsnm{{Karlick{\'y}}}, \binits{M.}},
  \oauthor{\bsnm{{M{\'e}sz{\'a}rosov{\'a}}}, \binits{H.}}:
2002,
In: {Sawaya-Lacoste}, H. (ed.)
\textit{Solspa 2001, Proc. Second Solar Cycle and Space Weather
  Euroconference},
\textit{ESA--SP}
\textbf{477},
351.
\end{botherref}
\endbibitem

\bibitem[\protect\citeauthoryear{{Ji{\v r}i{\v c}ka}
  \textit{et~al.}}{2001}]{Jiricka01}
\begin{barticle}
\bauthor{\bsnm{{Ji{\v r}i{\v c}ka}}, \binits{K.}},
  \bauthor{\bsnm{{Karlick{\'y}}}, \binits{M.}},
  \bauthor{\bsnm{{M{\'e}sz{\'a}rosov{\'a}}}, \binits{H.}},
  \bauthor{\bsnm{{Sn{\'{\i}}{\v z}ek}}, \binits{V.}}:
\byear{2001},
\bjtitle{\aap}
\bvolume{375},
\bfpage{243}.
\end{barticle}
\endbibitem

\bibitem[\protect\citeauthoryear{{Karlick{\'y}}, {M{\'e}sz{\'a}rosov{\'a}}, and
  {Jel{\'{\i}}nek}}{2013}]{Karlicky2013}
\begin{barticle}
\bauthor{\bsnm{{Karlick{\'y}}}, \binits{M.}},
  \bauthor{\bsnm{{M{\'e}sz{\'a}rosov{\'a}}}, \binits{H.}},
  \bauthor{\bsnm{{Jel{\'{\i}}nek}}, \binits{P.}}:
\byear{2013},
\bjtitle{\aap}
\bvolume{550},
\bfpage{A1}.
\end{barticle}
\endbibitem

\bibitem[\protect\citeauthoryear{{Karlick{\'y}}
  \textit{et~al.}}{2001}]{Karlicky01}
\begin{barticle}
\bauthor{\bsnm{{Karlick{\'y}}}, \binits{M.}}, \bauthor{\bsnm{{B{\'a}rta}},
  \binits{M.}}, \bauthor{\bsnm{{Ji{\v r}i{\v c}ka}}, \binits{K.}},
  \bauthor{\bsnm{{M{\'e}sz{\'a}rosov{\'a}}}, \binits{H.}},
  \bauthor{\bsnm{{Sawant}}, \binits{H.S.}}, \bauthor{\bsnm{{Fernandes}},
  \binits{F.C.R.}}, \bauthor{\bsnm{{Cecatto}}, \binits{J.R.}}:
\byear{2001},
\bjtitle{\aap}
\bvolume{375},
\bfpage{638}.
\end{barticle}
\endbibitem

\bibitem[\protect\citeauthoryear{{Kerdraon} and {Delouis}}{1997}]{Kerdraon97}
\begin{botherref}
\oauthor{\bsnm{{Kerdraon}}, \binits{A.}}, \oauthor{\bsnm{{Delouis}},
  \binits{J.M.}}:
1997,
In: {Trottet}, G. (ed.)
\textit{Coronal Phys. from Radio and Space Observations, Berlin Springer
  Verlag},
\textit{Lecture Notes Phys.}
\textbf{483},
192.
\end{botherref}
\endbibitem

\bibitem[\protect\citeauthoryear{{Klassen}, {Aurass}, and
  {Mann}}{2001}]{Klassen01}
\begin{barticle}
\bauthor{\bsnm{{Klassen}}, \binits{A.}}, \bauthor{\bsnm{{Aurass}},
  \binits{H.}}, \bauthor{\bsnm{{Mann}}, \binits{G.}}:
\byear{2001},
\bjtitle{\aap}
\bvolume{370},
\bfpage{L41}.
\end{barticle}
\endbibitem

\bibitem[\protect\citeauthoryear{{Kontogeorgos}
  \textit{et~al.}}{2006}]{Kontogeorgos}
\begin{barticle}
\bauthor{\bsnm{{Kontogeorgos}}, \binits{A.}}, \bauthor{\bsnm{{Tsitsipis}},
  \binits{P.}}, \bauthor{\bsnm{{Moussas}}, \binits{X.}},
  \bauthor{\bsnm{{Preka-Papadema}}, \binits{G.}}, \bauthor{\bsnm{{Hillaris}},
  \binits{A.}}, \bauthor{\bsnm{{Caroubalos}}, \binits{C.}},
  \bauthor{\bsnm{\etal}}:
\byear{2006},
\bjtitle{\ssr}
\bvolume{122},
\bfpage{169}.
\end{barticle}
\endbibitem

\bibitem[\protect\citeauthoryear{{Kuijpers}}{1975}]{Kuijpers1975}
\begin{barticle}
\bauthor{\bsnm{{Kuijpers}}, \binits{J.}}:
\byear{1975},
\bjtitle{\solphys}
\bvolume{44},
\bfpage{173}.
\end{barticle}
\endbibitem

\bibitem[\protect\citeauthoryear{{Lin} and {The Hessi Team}}{2001}]{Lin01B}
\begin{botherref}
\oauthor{\bsnm{{Lin}}, \binits{R.P.}}, \oauthor{\bsnm{{The Hessi Team}}}:
2001,
In: {Gladysheva}, O.G., {Kocharov}, G.E., {Kovaltsov}, G.A., {Usoskin}, I.G.
  (eds.)
\textit{Internat. Cosmic Ray Conf.}
\textbf{27},
209.
\end{botherref}
\endbibitem

\bibitem[\protect\citeauthoryear{{Lin} \textit{et~al.}}{2002}]{Lin02}
\begin{barticle}
\bauthor{\bsnm{{Lin}}, \binits{R.P.}}, \bauthor{\bsnm{{Dennis}},
  \binits{B.R.}}, \bauthor{\bsnm{{Hurford}}, \binits{G.J.}},
  \bauthor{\bsnm{{Smith}}, \binits{D.M.}}, \bauthor{\bsnm{{Zehnder}},
  \binits{A.}}, \bauthor{\bsnm{{Harvey}}, \binits{P.R.}},
  \bauthor{\bsnm{\etal}}:
\byear{2002},
\bjtitle{\solphys}
\bvolume{210},
\bfpage{3}.
\end{barticle}
\endbibitem

\bibitem[\protect\citeauthoryear{{Mann} \textit{et~al.}}{1989}]{Mann1989}
\begin{barticle}
\bauthor{\bsnm{{Mann}}, \binits{G.}}, \bauthor{\bsnm{{Baumgaertel}},
  \binits{K.}}, \bauthor{\bsnm{{Chernov}}, \binits{G.P.}},
  \bauthor{\bsnm{{Karlicky}}, \binits{M.}}:
\byear{1989},
\bjtitle{\solphys}
\bvolume{120},
\bfpage{383}.
\end{barticle}
\endbibitem

\bibitem[\protect\citeauthoryear{{Messerotti}, {Zlobec}, and
  {Padovan}}{2001}]{Messerotti2001}
\begin{barticle}
\bauthor{\bsnm{{Messerotti}}, \binits{M.}}, \bauthor{\bsnm{{Zlobec}},
  \binits{P.}}, \bauthor{\bsnm{{Padovan}}, \binits{S.}}:
\byear{2001},
\bjtitle{\memsai}
\bvolume{72},
\bfpage{633}.
\end{barticle}
\endbibitem

\bibitem[\protect\citeauthoryear{{M{\'e}sz{\'a}rosov{\'a}}, {Karlick{\'y}}, and
  {Ji{\v r}i{\v c}ka}}{2005}]{Meszarosova05}
\begin{barticle}
\bauthor{\bsnm{{M{\'e}sz{\'a}rosov{\'a}}}, \binits{H.}},
  \bauthor{\bsnm{{Karlick{\'y}}}, \binits{M.}}, \bauthor{\bsnm{{Ji{\v r}i{\v
  c}ka}}, \binits{K.}}:
\byear{2005},
\bjtitle{Hvar Obs. Bull.}
\bvolume{29},
\bfpage{309}.
\end{barticle}
\endbibitem

\bibitem[\protect\citeauthoryear{{M{\'e}sz{\'a}rosov{\'a}}
  \textit{et~al.}}{2008}]{Meszarosova2008}
\begin{barticle}
\bauthor{\bsnm{{M{\'e}sz{\'a}rosov{\'a}}}, \binits{H.}},
  \bauthor{\bsnm{{Karlick{\'y}}}, \binits{M.}}, \bauthor{\bsnm{{Sawant}},
  \binits{H.S.}}, \bauthor{\bsnm{{Fernandes}}, \binits{F.C.R.}},
  \bauthor{\bsnm{{Cecatto}}, \binits{J.R.}}, \bauthor{\bsnm{{de Andrade}},
  \binits{M.C.}}:
\byear{2008},
\bjtitle{\aap}
\bvolume{491},
\bfpage{555}.
\end{barticle}
\endbibitem

\bibitem[\protect\citeauthoryear{{Nindos} and {Aurass}}{2007}]{Nindos07}
\begin{botherref}
\oauthor{\bsnm{{Nindos}}, \binits{A.}}, \oauthor{\bsnm{{Aurass}}, \binits{H.}}:
2007,
In: {K.-L.~Klein \& A.~L.~MacKinnon} (ed.)
\textit{The High Energy Solar Corona: Waves, Eruptions, Particles, Berlin
  Springer Verlag},
\textit{Lecture Notes Phys.}
\textbf{725},
251.
\end{botherref}
\endbibitem

\bibitem[\protect\citeauthoryear{{Nindos} \textit{et~al.}}{2008}]{Nindos08}
\begin{barticle}
\bauthor{\bsnm{{Nindos}}, \binits{A.}}, \bauthor{\bsnm{{Aurass}}, \binits{H.}},
  \bauthor{\bsnm{{Klein}}, \binits{K.L.}}, \bauthor{\bsnm{{Trottet}},
  \binits{G.}}:
\byear{2008},
\bjtitle{\solphys}
\bvolume{253},
\bfpage{3}.
\end{barticle}
\endbibitem

\bibitem[\protect\citeauthoryear{{Ning} \textit{et~al.}}{2008}]{Ning2008}
\begin{barticle}
\bauthor{\bsnm{{Ning}}, \binits{Z.}}, \bauthor{\bsnm{{Wu}}, \binits{H.}},
  \bauthor{\bsnm{{Xu}}, \binits{F.}}, \bauthor{\bsnm{{Meng}}, \binits{X.}}:
\byear{2008},
\bjtitle{\solphys}
\bvolume{250},
\bfpage{107}.
\end{barticle}
\endbibitem

\bibitem[\protect\citeauthoryear{{Oberoi}, {Evarts}, and
  {Rogers}}{2009}]{Oberoi2009}
\begin{barticle}
\bauthor{\bsnm{{Oberoi}}, \binits{D.}}, \bauthor{\bsnm{{Evarts}},
  \binits{E.R.}}, \bauthor{\bsnm{{Rogers}}, \binits{A.E.E.}}:
\byear{2009},
\bjtitle{\solphys}
\bvolume{260},
\bfpage{389}.
\end{barticle}
\endbibitem

\bibitem[\protect\citeauthoryear{{Pick} and {Vilmer}}{2008}]{Pick08}
\begin{barticle}
\bauthor{\bsnm{{Pick}}, \binits{M.}}, \bauthor{\bsnm{{Vilmer}}, \binits{N.}}:
\byear{2008},
\bjtitle{\aapr}
\bvolume{16},
\bfpage{1}.
\end{barticle}
\endbibitem

\bibitem[\protect\citeauthoryear{{Rausche} \textit{et~al.}}{2007}]{Rausche07}
\begin{barticle}
\bauthor{\bsnm{{Rausche}}, \binits{G.}}, \bauthor{\bsnm{{Aurass}},
  \binits{H.}}, \bauthor{\bsnm{{Mann}}, \binits{G.}},
  \bauthor{\bsnm{{Karlick{\'y}}}, \binits{M.}}, \bauthor{\bsnm{{Vocks}},
  \binits{C.}}:
\byear{2007},
\bjtitle{\solphys}
\bvolume{245},
\bfpage{327}.
\end{barticle}
\endbibitem

\bibitem[\protect\citeauthoryear{{Robbrecht} and
  {Berghmans}}{2004}]{Robbrecht2004}
\begin{barticle}
\bauthor{\bsnm{{Robbrecht}}, \binits{E.}}, \bauthor{\bsnm{{Berghmans}},
  \binits{D.}}:
\byear{2004},
\bjtitle{\aap}
\bvolume{425},
\bfpage{1097}.
\end{barticle}
\endbibitem

\bibitem[\protect\citeauthoryear{Robbrecht, Berghmans, and der
  Linden}{2009}]{Robbrecht2009}
\begin{barticle}
\bauthor{\bsnm{Robbrecht}, \binits{E.}}, \bauthor{\bsnm{Berghmans},
  \binits{D.}}, \bauthor{\bparticle{der }\bsnm{Linden}, \binits{R.A.M.V.}}:
\byear{2009},
\bjtitle{The Astrophys. J.}
\bvolume{691}(\bissue{2}),
\bfpage{1222}.
\end{barticle}
\endbibitem

\bibitem[\protect\citeauthoryear{Slottje}{1981}]{Slottje1981}
\begin{botherref}
\oauthor{\bsnm{Slottje}, \binits{C.}}:
1981,
\textit{Dissertation, Utrecht Obs.}.
\end{botherref}
\endbibitem

\bibitem[\protect\citeauthoryear{Webb and Howard}{2012}]{Webb2012}
\begin{barticle}
\bauthor{\bsnm{Webb}, \binits{D.F.}}, \bauthor{\bsnm{Howard}, \binits{T.A.}}:
\byear{2012},
\bjtitle{Living Rev. Solar Phys.}
\bvolume{9}(\bissue{3}).
\end{barticle}
\endbibitem

\bibitem[\protect\citeauthoryear{{Yashiro} \textit{et~al.}}{2001}]{Yashiro01}
\begin{botherref}
\oauthor{\bsnm{{Yashiro}}, \binits{S.}}, \oauthor{\bsnm{{Gopalswamy}},
  \binits{N.}}, \oauthor{\bsnm{{St.~Cyr}}, \binits{O.C.}},
  \oauthor{\bsnm{{Lawrence}}, \binits{G.}}, \oauthor{\bsnm{{Michalek}},
  \binits{G.}}, \oauthor{\bsnm{{Young}}, \binits{C.A.}},
  \oauthor{\bsnm{{Plunkett}}, \binits{S.P.}}, \oauthor{\bsnm{{Howard}},
  \binits{R.A.}}:
2001,
\textit{AGU Spring Meeting Abstracts},
31.
\end{botherref}
\endbibitem

\bibitem[\protect\citeauthoryear{{Yashiro} \textit{et~al.}}{2004}]{Yashiro04}
\begin{barticle}
\bauthor{\bsnm{{Yashiro}}, \binits{S.}}, \bauthor{\bsnm{{Gopalswamy}},
  \binits{N.}}, \bauthor{\bsnm{{Michalek}}, \binits{G.}},
  \bauthor{\bsnm{{St.~Cyr}}, \binits{O.C.}}, \bauthor{\bsnm{{Plunkett}},
  \binits{S.P.}}, \bauthor{\bsnm{{Rich}}, \binits{N.B.}},
  \bauthor{\bsnm{{Howard}}, \binits{R.A.}}:
\byear{2004},
\bjtitle{\jgr}
\bvolume{109}(\bissue{18}),
\bfpage{7105}.
\end{barticle}
\endbibitem

\bibitem[\protect\citeauthoryear{{Zhang} \textit{et~al.}}{2001}]{Zhang2001}
\begin{barticle}
\bauthor{\bsnm{{Zhang}}, \binits{J.}}, \bauthor{\bsnm{{Dere}}, \binits{K.P.}},
  \bauthor{\bsnm{{Howard}}, \binits{R.A.}}, \bauthor{\bsnm{{Kundu}},
  \binits{M.R.}}, \bauthor{\bsnm{{White}}, \binits{S.M.}}:
\byear{2001},
\bjtitle{\apj}
\bvolume{559},
\bfpage{452}.
\end{barticle}
\endbibitem

\bibitem[\protect\citeauthoryear{{Zlotnik} \textit{et~al.}}{2003}]{Zlotnik03}
\begin{barticle}
\bauthor{\bsnm{{Zlotnik}}, \binits{E.Y.}}, \bauthor{\bsnm{{Zaitsev}},
  \binits{V.V.}}, \bauthor{\bsnm{{Aurass}}, \binits{H.}},
  \bauthor{\bsnm{{Mann}}, \binits{G.}}, \bauthor{\bsnm{{Hofmann}},
  \binits{A.}}:
\byear{2003},
\bjtitle{\aap}
\bvolume{410},
\bfpage{1011}.
\end{barticle}
\endbibitem

\bibitem[\protect\citeauthoryear{{Zlotnik} \textit{et~al.}}{2005}]{Zlotnik05}
\begin{barticle}
\bauthor{\bsnm{{Zlotnik}}, \binits{E.}}, \bauthor{\bsnm{{Zaitsev}},
  \binits{V.}}, \bauthor{\bsnm{{Aurass}}, \binits{H.}}, \bauthor{\bsnm{{Mann}},
  \binits{G.}}:
\byear{2005},
\bjtitle{\adv}
\bvolume{35},
\bfpage{1774}.
\end{barticle}
\endbibitem

\bibitem[\protect\citeauthoryear{Zlotnik \textit{et~al.}}{2009}]{Zlotnik2009}
\begin{barticle}
\bauthor{\bsnm{Zlotnik}, \binits{E.}}, \bauthor{\bsnm{Zaitsev}, \binits{V.}},
  \bauthor{\bsnm{Aurass}, \binits{H.}}, \bauthor{\bsnm{Mann}, \binits{G.}}:
\byear{2009},
\bjtitle{Solar Phys.}
\bvolume{255},
\bfpage{273}.
\end{barticle}
\endbibitem

\end{thebibliography}
